\documentclass[a4paper,twoside,12pt]{report}

\usepackage{morewrites}

\usepackage{amsthm,amssymb,amsmath}
\usepackage{IEEEtrantools}

\newcommand{\paren}[1]{\left(#1\right)}
\newcommand{\stareq}{\mathop{=}\limits^{\star}}
\newcommand{\ud}{\;\mathrm{d}}

\newcommand{\finsarg}{(t,\mathbf{x})}
\newcommand{\sinsarg}{(t,\mathbf{x'})}
\newcommand{\retarg}{(t',\mathbf{x'})}
\newcommand{\retardedarg}{(t-\frac{|\mathbf{x}-\mathbf{x'}|}{c},\mathbf{x'})}
\newcommand{\uarg}{(t-\frac{ |\mathbf{x}|}{c})}
\newcommand{\varg}{(t+\frac{ |\mathbf{x}|}{c})}
\newcommand{\targ}{(t)}
\newcommand{\tparg}{(t')}
\newcommand{\rettarg}{(t-\frac{|\mathbf{x}-\mathbf{x'}|}{c})}
\newcommand{\xarg}{(\mathbf{x})}
\newcommand{\xparg}{(\mathbf{x'})}

\newcommand{\poisint}[1]{\Delta^{-1}\left[ #1 \right]}
\newcommand{\gpoisint}[1]{\Delta^{-1}\left[\paren{\frac{\xp}{r_0}}^B #1 \right]}
\newcommand{\retint}[1]{\Box^{-1}_{\mathrm{R}}\left[ #1  \right]}
\newcommand{\gretint}[1]{\Box^{-1}_{\mathrm{R}}\left[\paren{\frac{\xp}{r_0}}^B #1  \right]}

\newcommand{\nearint}[1]{-\frac{1}{4\pi}\int_{|\mathbf{x'}|<\mathcal{R}} \frac{#1}{|\mathbf{x}-\mathbf{x'}|}\ud^3\mathbf{x'}}
\newcommand{\gnearint}[1]{-\frac{1}{4\pi}\int_{|\mathbf{x'}|<\mathcal{R}} \paren{\frac{\xp}{r_0}}^B \frac{#1}{|\mathbf{x}-\mathbf{x'}|}\ud^3\mathbf{x'}}
\newcommand{\farint}[1]{-\frac{1}{4\pi}\int_{\mathcal{R}<|\mathbf{x'}|} \frac{#1}{|\mathbf{x}-\mathbf{x'}|}\ud^3\mathbf{x'}}
\newcommand{\gfarint}[1]{-\frac{1}{4\pi}\int_{\mathcal{R}<|\mathbf{x'}|} \paren{\frac{\xp}{r_0}}^B \frac{#1}{|\mathbf{x}-\mathbf{x'}|}\ud^3\mathbf{x'}}
\newcommand{\wholeint}[1]{-\frac{1}{4\pi}\int_{\mathbb{R }^3}\frac{#1}{|\mathbf{x}-\mathbf{x'}|}\ud^3\mathbf{x'}}
\newcommand{\gwholeint}[1]{-\frac{1}{4\pi}\int_{\mathbb{R }^3} \paren{\frac{\xp}{r_0}}^B \frac{#1}{|\mathbf{x}-\mathbf{x'}|}\ud^3\mathbf{x'}}

\newcommand{\x}{|\mathbf{x}|}
\newcommand{\xp}{|\mathbf{x'}|}
\newcommand{\ylm}{Y^{\ell m}(\theta, \varphi)}
\newcommand{\ylmp}{Y^{\ell m}(\theta', \varphi')}

\newcommand{\A}{\mathrm{A}}
\newcommand{\fp}{\mathop{\mathrm{FP}}\limits_{B=0}}
\newcommand{\fpa}{\mathop{\mathrm{FP}}\limits_{B=0}\mathrm{A}}
\newcommand{\res}{\mathop{\mathrm{Residue}}\limits_{B=0}}
\newcommand{\resa}{\mathop{\mathrm{Residue}}\limits_{B=0}\mathrm{A}}
\newcommand{\ellfpa}{\mathop{\mathrm{FP}}\limits_{j=\ell}\mathrm{A}}

\newcommand{\xb}{\left(\frac{\x}{r_0}\right)^B}
\newcommand{\xpb}{\left(\frac{\xp}{r_0}\right)^B}
\newcommand{\re}{\mathrm{Re}}
\newcommand{\im}{\mathrm{Im}}
\newcommand{\lnx}{\paren{\ln{\frac{\x}{r_0}}}}
\newcommand{\lnxp}{\paren{\ln{\frac{\xp}{r_0}}}}

\newcommand{\persiantheorem}[1]{\noindent\textbf{Theorem #1:}}
\newcommand{\persianproof}{\noindent\textit{Proof:}\\}

\usepackage[top=30mm, bottom=22.5mm, left=25mm, right=25mm]{geometry}
\usepackage[nottoc]{tocbibind}
\usepackage{graphicx}
\begin{document}
\pagenumbering{roman}


\newpage
\thispagestyle{empty}
\begin{center}
\includegraphics[height=2cm,width=2cm]{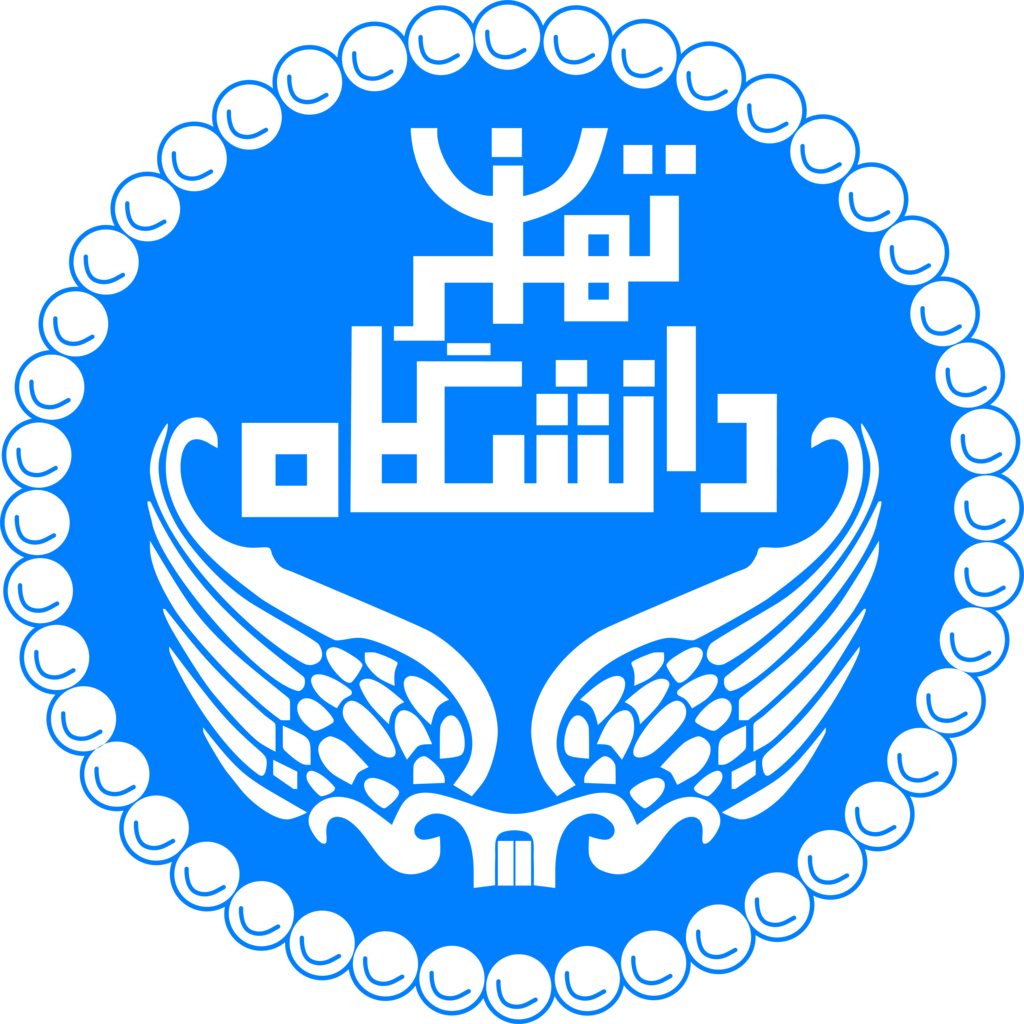}

\vspace*{-2mm}
University of Tehran

\vspace*{-2mm}
College of Science

\vspace*{-2mm}
School of Physics

\vspace*{20mm}
\textbf{{\Huge
Analytical Study of\\ Gravitational Waves}}

\vspace*{15mm}
\textbf{{\large By}}

\textbf{{\Large Abbas Mirahmadi}}

\vspace*{15mm}
\textbf{{\large Under Supervision of}}

\textbf{{\Large  Dr. Nahid Ahmadi}}

\vspace*{20mm}
A Thesis Submitted to the Graduate Studies Office

\vspace*{-2mm}
in Partial Fulfillment of the Requirements for

\vspace*{-2mm}
the Degree of Master of Science in

\vspace*{-2mm}
Physics-Astrophysics

\vspace*{15mm}
February 2019
\end{center}


\newpage
\addcontentsline{toc}{chapter}{Abstract}
\begin{center}
\textbf{\Large{Abstract}}
\end{center}

\vspace{5mm}
{\setlength{\baselineskip}{0.95\baselineskip}
The aim of the present thesis is to review the Blanchet-Damour approach to analytical study of gravitational waves emitted by localized perfect fluid sources. It is assumed these perfect fluids are such that it is possible to define small parameters for asymptotic expansions. Asymptotic expansions in this approach are called post-Minkowskian and post-Newtonian expansions. By plugging these expansions into the Einstein field equation, post-Minkowskian and post-Newtonian equations are obtained. The usual methods for solving these equations are to use the retarded and Poisson integrals. However, they cannot provide solutions up to any arbitrary order in these cases because of their divergence. In fact, these divergences motivated Blanchet and Damour to employ a new approach to solve the post-Minkowskian and post-Newtonian equations. They obtained the general solutions by means of a specific process of analytic continuation. These solutions have some unknown terms that if one determines, the gravitational field is described everywhere in $\mathbb{R }^3$. A matching procedure, which depends deeply on analytic continuation, is used to determine these unknown terms.

\vspace*{5mm}
\par}


\chapter*{Introduction}
\addcontentsline{toc}{chapter}{Introduction}

To detect and analyze gravitational waves, we need prepared templates. These templates can be obtained through either numerical or analytical methods. Although numerical methods are very effective and precise, they are really time-consuming. Hence, we need analytical methods, even if they are approximate. For the first time, it was Einstein himself that employed approximate solution methods
\cite{E1916,E1918}.
In the linear (post-Mainkowskian) approximation, by expanding
$\frac{1}{|\mathbf{x}-\mathbf{x'}|}$
appearing in the retarded integral of the material energy-momentum tensor for far distances from any point within the localized source and then expanding the aforementioned material energy-momentum tensor in powers of
$\frac{1}{c}$
(which is acceptable if the internal speeds inside the source are taken to be very small compared with the speed of light), he found his famous quadrupole formula. But two questions may arise: First, is there any exact wave solution to the Einstein field equation including all its nonlinearities? Second, if the answer to the first question is affirmative, and hence, finding wave solutions are legitimate, what are the approximate wave solutions to the Einstein field equation including all its nonlinearities? The answer to the first question is affirmative
\cite{P1962a,P1962b},
so with no concern we can deal with the second one.

The main method for dealing with the second question is to use post-Minkowskian and post-Newtonian expansions. Although the post-Minkowskian expansion, if one combines it with symmetric-trace-free Cartesian multipole expansion, can provide approximate solutions up to a certain finite order which are valid everywhere in $\mathbb{R }^3$
\cite{B1959,T1980},
there is no guarantee that higher-order approximations are obtainable
\cite{BD1986}.
Furthermore, the post-Newtonian expansion, which is valid in the near zone
\cite{EW1975},
breaks down at orders higher than a specific one because of the divergence of the integrals which itself is due to the behavior of the lower-order approximations in the far zone
\cite{K1980a,K1980b}.
The lack of an algorithm for providing solutions up to any arbitrary order made Blanchet and Damour introduce their own algorithm which is based on the Fock's idea of splitting the problem into two subproblems, one concerning the near zone and the other outside the source, and then using a matching procedure
\cite{F1959}.

Chapter 1 deals with the general framework of the Blanchet-Damour approach. It is mainly based on
\cite{M2007}.
In Chapter 2 we examine the post-Minkowskian approximation. The primary reference for this chapter of the thesis is
\cite{BD1986}.
Chapter 3 is devoted to examining the post-Newtonian approximation. The main source of this chapter is
\cite{PB2002}.
Finally, in Chapter 4 we discuss the matching procedure and obtain the final results of the above-mentioned algorithm. This chapter is primarily based on
\cite{B1998,B2014}.
Moreover, at the end of the thesis two appendices are included. Appendix A contains general requisite formulae, and Appendix B introduces symmetric-trace-free Cartesian tensors.


\tableofcontents


\chapter{Equations Governing the Gravitational Waves}
\pagenumbering{arabic}

In the first section of this chapter the convenient form of the Einstein field equation to study gravitational waves is introduced. In the next section the approximate solutions to this equation, subject to specific conditions, are examined, and finally, in the third section the general framework of the Blancet-Damour approach is presented.

\section{Einstein Field Equation}

Gravitational fields are governed by Einstein field equation
\cite{P2010}
\begin{equation}\label{1.1.1}
G^{\mu\nu}=\frac{8\pi G}{c^4}T^{\mu\nu},
\end{equation}
where
$G^{\mu\nu}$
is the Einstein and
$T^{\mu\nu}$
the material energy-momentum tensor. In these equations the main variables are the componenets of the metric tensor
$g_{\alpha\beta}$. Using Bianchi identity,
$\nabla_\mu G^{\mu\nu}=0$,
from
Eq. (\ref{1.1.1})
 one finds the covariant conservation equation
\begin{equation}\label{1.1.2}
\nabla_\mu T^{\mu\nu}=0.
\end{equation}
Now consider a new set of main variables, the components of
\begin{equation}\label{1.1.3}
\mathfrak{g}^{\mu\nu}=\sqrt{-g} g^{\mu\nu},
\end{equation}
where
$g^{\mu\nu}$
and
$g$
are the inverse metric and the metric determinant respectively. Note that if
$\mathfrak{g}^{\mu\nu}$
(which is called the gothic inverse metric) is known,
$g$
is known too
($\mathrm{det}[\mathfrak{g}^{\mu\nu}]=g$),
and hence,
$g^{\mu\nu}$
can be determined from
Eq. (\ref{1.1.3});
then
$g_{\mu\nu}$
can be obtained by inverting
$[g^{\mu\nu}]$.

Writing the ten tensor components of Einstein field equation in terms of the components of
$\mathfrak{g}^{\mu\nu}$,
we get Landau-Lifshitz formulation of them which are
\cite{PW2014}
\begin{equation}\label{1.1.4}
\partial_{\alpha\beta} H^{\mu\alpha\nu\beta}=\frac{16\pi G}{c^4} \left(-\mathfrak{g} \right) \left( T^{\mu\nu} +  t_{LL}^{\mu\nu} \right),
\end{equation}
where
\begin{equation}\label{1.1.5}
 H^{\mu\alpha\nu\beta}=\mathfrak{g}^{\mu\nu}\mathfrak{g}^{\alpha\beta} - \mathfrak{g}^{\mu\beta}\mathfrak{g}^{\nu\alpha},
\end{equation}
\begin{equation}\label{1.1.6}
\mathfrak{g}=\mathrm{det} [\mathfrak{g}^{\mu\nu}]=g,
\end{equation}
\begin{IEEEeqnarray}{rcl}\label{1.1.7}
 \left(-\mathfrak{g} \right) t_{\mathrm{LL}}^{\mu\nu} & = &  \frac{c^4}{16\pi G}\bigg\{ \partial_\lambda \mathfrak{g}^{\mu\nu} \partial_\alpha \mathfrak{g}^{\lambda\alpha} - \partial_\lambda \mathfrak{g}^{\mu\lambda}\partial_\alpha \mathfrak{g}^{\nu\alpha} + \frac{1}{2}\mathfrak{g}^{\mu\nu}\mathfrak{g}_{\lambda\alpha}\partial_\rho \mathfrak{g}^{\lambda\beta}\partial_\beta \mathfrak{g}^{\alpha\rho}\nonumber \\  
&& \negmedspace {} - \mathfrak{g}^{\mu\lambda}\mathfrak{g}_{\alpha\beta}\partial_\rho \mathfrak{g}^{\nu\beta}\partial_\lambda \mathfrak{g}^{\alpha\rho} - \mathfrak{g}^{\nu\lambda}\mathfrak{g}_{\alpha\beta}\partial_\rho \mathfrak{g}^{\mu\beta}\partial_\lambda \mathfrak{g}^{\alpha\rho} + \mathfrak{g}_{\lambda\alpha}\mathfrak{g}^{\beta\rho}\partial_\beta \mathfrak{g}^{\mu\lambda}\partial_\rho \mathfrak{g}^{\nu\alpha} \nonumber \\
&& \negmedspace {} + \frac{1}{8} \left( 2\mathfrak{g}^{\mu\lambda}\mathfrak{g}^{\nu\alpha} - \mathfrak{g}^{\mu\nu}\mathfrak{g}^{\lambda\alpha}\right) \left( 2\mathfrak{g}_{\beta\rho}\mathfrak{g}_{\sigma\tau} - \mathfrak{g}_{\rho\sigma}\mathfrak{g}_{\beta\tau} \right)\partial_\lambda \mathfrak{g}^{\beta\tau}\partial_\alpha \mathfrak{g}^{\rho\sigma}\bigg\}.
\end{IEEEeqnarray}
$t_{\mathrm{LL}}^{\mu\nu}$
is called the Landau-Lifshitz energy-momentum pseudotensor and by
$\mathfrak{g}_{\mu\nu}$
we mean
\begin{equation}\label{1.1.8}
\mathfrak{g}_{\mu\nu}=\frac{1}{\sqrt{-g}}g_{\mu\nu}.
\end{equation}
Considering the antisymmetry of
$H^{\mu\alpha\nu\beta}$
in the first pair of indices, from
Eq. (\ref{1.1.4})
one easily finds
\begin{equation}\label{1.1.9}
\partial_\mu\left[\left(-\mathfrak{g} \right) \left( T^{\mu\nu} +  t_{\mathrm{LL}}^{\mu\nu} \right)\right]=0,
\end{equation}
which is exactly equivalent to
Eq. ( \ref{1.1.2}).

Eqs. (\ref{1.1.4}) and (\ref{1.1.9})
are valid in any coordinate system. One can simplify these equations by choosing a particular coordinate system. In order to achieve this aim, we choose the one in which
$\mathfrak{g}^{\mu\nu}$
satisfies
\begin{equation}\label{1.1.10}
\partial_\mu \mathfrak{g}^{\mu\nu}=0.
\end{equation}
It can be proven that this choice of the coordinate system is always possible
\cite{PW2014}.
Such a coordinate system is called harmonic gauge and
Eq. (\ref{1.1.10})
itself the harmonic gauge condition. In the harmonic gauge we have
\begin{equation}\label{1.1.12}
\partial_{\alpha\beta} H^{\mu\alpha\nu\beta}=\mathfrak{g}^{\alpha\beta}\partial_{\alpha\beta}\mathfrak{g}^{\mu\nu}-\partial_\alpha \mathfrak{g}^{\mu\beta}\partial_\beta \mathfrak{g}^{\nu\alpha};
\end{equation}
therefore
Eq. (\ref{1.1.4})
becomes
\begin{equation}\label{1.1.13}
\mathfrak{g}^{\alpha\beta}\partial_{\alpha\beta}\mathfrak{g}^{\mu\nu}=\frac{16\pi G}{c^4} \left(-\mathfrak{g} \right) \left( T^{\mu\nu} +   t_{\mathrm{LL\left(H\right)}}^{\mu\nu} \right)+\partial_\alpha\mathfrak{g}^{\mu\beta}\partial_\beta\mathfrak{g}^{\nu\alpha},
\end{equation}
where
$\left(-\mathfrak{g} \right) t_{\mathrm{LL\left(H\right)}}^{\mu\nu}$
($\left(-\mathfrak{g} \right) t_{\mathrm{LL}}^{\mu\nu}$
written in the harmonic gauge) is given by
\begin{IEEEeqnarray}{rcl}\label{1.1.11}
 \left(-\mathfrak{g} \right) t_{\mathrm{LL\left(H\right)}}^{\mu\nu} & = &  \frac{c^4}{16\pi G}\bigg\{\frac{1}{2}\mathfrak{g}^{\mu\nu}\mathfrak{g}_{\lambda\alpha}\partial_\rho \mathfrak{g}^{\lambda\beta}\partial_\beta \mathfrak{g}^{\alpha\rho} - \mathfrak{g}^{\mu\lambda}\mathfrak{g}_{\alpha\beta}\partial_\rho \mathfrak{g}^{\nu\beta}\partial_\lambda \mathfrak{g}^{\alpha\rho} \nonumber \\  
&& \negmedspace {}- \mathfrak{g}^{\nu\lambda}\mathfrak{g}_{\alpha\beta}\partial_\rho \mathfrak{g}^{\mu\beta}\partial_\lambda \mathfrak{g}^{\alpha\rho} + \mathfrak{g}_{\lambda\alpha}\mathfrak{g}^{\beta\rho}\partial_\beta \mathfrak{g}^{\mu\lambda}\partial_\rho \mathfrak{g}^{\nu\alpha} \nonumber \\  
&& \negmedspace {}+ \frac{1}{8} \left( 2\mathfrak{g}^{\mu\lambda}\mathfrak{g}^{\nu\alpha} - \mathfrak{g}^{\mu\nu}\mathfrak{g}^{\lambda\alpha}\right) \left( 2\mathfrak{g}_{\beta\rho}\mathfrak{g}_{\sigma\tau} - \mathfrak{g}_{\rho\sigma}\mathfrak{g}_{\beta\tau} \right)\partial_\lambda \mathfrak{g}^{\beta\tau}\partial_\alpha \mathfrak{g}^{\rho\sigma}\bigg\}.
\end{IEEEeqnarray}
Furthermore,
Eq. (\ref{1.1.9})
can be rewritten as
 \begin{equation}\label{1.1.14}
\partial_\mu\left[\left(-\mathfrak{g} \right) \left( T^{\mu\nu} +  t_{\mathrm{LL\left(H\right)}}^{\mu\nu} \right)\right]=0.
\end{equation}
In order to study gravitational waves, it is more convenient to write
Eqs. (\ref{1.1.10}), (\ref{1.1.13}) and (\ref{1.1.14})
in terms of the components of
\begin{equation}\label{1.1.15}
h^{\mu\nu}=\mathfrak{g}^{\mu\nu} - \eta^{\mu\nu},
\end{equation}
where
$\eta^{\mu\nu}=\mathrm{diag}\left(-1, 1, 1, 1\right)$.
They are
\begin{equation}\label{1.1.18}
\partial_\mu h^{\mu\nu} = 0,
\end{equation}
\begin{equation}\label{1.1.19}
\Box h^{\mu\nu} = \frac{16\pi G}{c^4} \left(-\mathfrak{g} \right) \left( T^{\mu\nu} +   t_{\mathrm{LL\left(H\right)}}^{\mu\nu} \right)+ \partial_\alpha h^{\mu\beta} \partial_\beta h^{\nu\alpha}- h^{\alpha\beta}\partial_{\alpha\beta} h^{\mu\nu},
\end{equation}
\begin{equation}\label{1.1.20}
\partial_\mu\left[\left(-\mathfrak{g} \right) \left( T^{\mu\nu} +  t_{\mathrm{LL\left(H\right)}}^{\mu\nu} \right)\right]=0,
\end{equation}
where in
Eq. (\ref{1.1.19}),
$\Box=\partial_\alpha \partial^\alpha = \eta^{\alpha\beta} \partial_\alpha \partial_\beta$.
Note that although it's not explicitly written,
$\left(-\mathfrak{g} \right) T^{\mu\nu}$
and
$\left(-\mathfrak{g} \right) t_{\mathrm{LL\left(H\right)}}^{\mu\nu}$
are in terms of the components of
$h^{\alpha\beta}$
($\mathfrak{g}$
and
$\mathfrak{g}_{\alpha\beta}$
appearing in
$\left(-\mathfrak{g} \right) T^{\mu\nu}$
and
$\left(-\mathfrak{g} \right) t_{\mathrm{LL\left(H\right)}}^{\mu\nu}$
are the determinant and the inverse of the
$4 \times 4$
matrix
$\left[\eta^{\mu\nu} + h^{\mu\nu}\right]$).
Introducing
\begin{equation}\label{1.1.23}
\tau^{\mu\nu} =  \left(-\mathfrak{g} \right) T^{\mu\nu}+\frac{c^4}{16\pi G}\Lambda^{\mu\nu},
\end{equation}
called the effective energy-momentum pseudotensor, where
\begin{equation}\label{1.1.22}
\Lambda^{\mu\nu} = \frac{16\pi G}{c^4} \left(-\mathfrak{g} \right) t_{\mathrm{LL\left(H\right)}}^{\mu\nu}+ \partial_\alpha h^{\mu\beta} \partial_\beta h^{\nu\alpha}- h^{\alpha\beta}\partial_{\alpha\beta} h^{\mu\nu},
\end{equation}
Eqs. (\ref{1.1.19}) and (\ref{1.1.20})
become
\begin{equation}\label{1.1.27}
\Box h^{\mu\nu} = \frac{16\pi G}{c^4}\tau ^{\mu\nu},
\end{equation}
\begin{equation}\label{1.1.29}
\partial_\mu \tau^{\mu\nu} = 0,
\end{equation}
where to obtain
Eq. (\ref{1.1.29}),
we have used the identity
\begin{equation}\label{1.1.25}
\partial_\mu \left[ \partial_\alpha h^{\mu\beta} \partial_\beta h^{\nu\alpha}- h^{\alpha\beta}\partial_{\alpha\beta} h^{\mu\nu}\right]=0,
\end{equation}
which is itself due to the harmonic gauge condition.
Eqs. (\ref{1.1.18}), (\ref{1.1.27}) and (\ref{1.1.29}) are the ones that we use to study gravitational waves.
Eq. (\ref{1.1.27}),
called the relaxed Einstein field equation,
together with
Eq. (\ref{1.1.18}),
the harmonic gauge condition, are equivalent to the Einstein field equation,
$G^{\mu\nu}=\frac{8\pi G}{c^4}T^{\mu\nu}$.
Also,
Eq. (\ref{1.1.29}),
called the conservation equation in the harmonic gauge, is equivalent to the covariant conservation equation,
$\nabla_\mu T^{\mu\nu}=0$.

The solution to the equation
$\Box f(t,\mathbf{x}) = g(t,\mathbf{x})$,
subject to no-incoming radiation condition (which will be discussed later), is
\begin{IEEEeqnarray}{rcl}\label{1.1.30}
\retint{g\retarg}\finsarg & = &\wholeint{g\retarg} \nonumber \\ 
& = &\wholeint{g\retardedarg},
\end{IEEEeqnarray}
if the integral is convergent. However, the solution to
Eq. (\ref{1.1.27}),
subject to the above-mentioned condition, cannot be given by
$\retint{\frac{16\pi G}{c^4}\tau ^{\mu\nu}}$ since the right-hand side of that equation is actually a functional of the components of
$h^{\alpha\beta}$
and their space-time derivatives, and hence, unknown and impossible to be integrated. That is,
$h^{\mu\nu}=\retint{\frac{16\pi G}{c^4}\tau ^{\mu\nu}}$
is merely the integro-differential form of
Eq. (\ref{1.1.27})
under no-incoming radiation condition. In fact, it is very difficult to find an exact solution to
Eq. (\ref{1.1.27})
that fulfills the harmonic gauge condition too. Thus, we need to apply approximate solution methods.

\section{Approximate Solutions to Einstein Field Equation}

We study the gravitational waves emitted by sources that can be treated as perfect fluids. The material energy-momentum tensor of a perfect fluid reads
\begin{equation}\label{1.2.1}
T^{\mu\nu}=\left(\rho + \frac{p}{c^2}\right)u^\mu u^\nu + pg^{\mu\nu},
\end{equation}
\begin{equation}\label{1.2.2}
u^\mu u_\mu = -c^2.
\end{equation}
We write
$T^{\mu\nu}$
in a coordinate system in which the origin of the spatial coordinates is located within the source. We also assume that
$T^{\mu\nu}\finsarg$
is compactly supported (i.e., there is a positive constant
$r_0$
such that
$T^{\mu\nu} \finsarg = 0$
for
$|\mathbf{x}|>r_0$)
and belongs to
$C^\infty \left(\mathbb{R}^4\right)$.
If we have for
$M$,
the typical mass of the source, and
$d$
the typical radius of the source,
\begin{equation}\label{1.2.3}
\frac{GM}{dc^2} \ll 1,
\end{equation}
or equivalently by using virial theorem,
$\left(\frac{v}{c}\right)^2 \approx \frac{GM}{dc^2}$,
\begin{equation}\label{1.2.4}
\frac{v}{c} \ll 1,
\end{equation}
where
$v$
is a typical internal speed of the source, we can suggest two asymptotic expansions
\begin{equation}\label{1.2.5}
h^{\mu\nu}_{\mathrm{PM}}=\sum_{n=1}^{\infty} G^n h^{\mu\nu}_{\left(n\right)},
\end{equation}
\begin{equation}\label{1.2.6}
h^{\mu\nu}_{\mathrm{PN}}=\sum_{n=2}^{\infty} \frac{1}{c^n} \bar{h}^{\mu\nu}_{\left(n\right)},
\end{equation}
as approximate solutions. (To write more economically, we have absorbed
${\paren{\frac{M}{dc^2}}}^n$
in
$h^{\mu\nu}_{\left(n\right)}$
and
$v^n$
in
$ \bar{h}^{\mu\nu}_{\left(n\right)}$.
Therefore, one would think of
$G$
and
$\frac{1}{c}$
as the small parameters in these two expansions.)
$h^{\mu\nu}_{\mathrm{PM}}$
and
$h^{\mu\nu}_{\mathrm{PN}}$
are called post-Minkowskian and post-Newtonian expansions respectively.
$u^\mu$,
$\rho$
and
$p$
depend on the metric, and the metric itself is a functional of
$h^{\mu\nu}$.
Thus, if we consider
$h^{\mu\nu}_{\mathrm{PM}}$
as an approximate solution for
$h^{\mu\nu}$,
we get the following expansions for
$u^\mu$,
$\rho$
and
$p$\;:
\begin{equation}\label{1.2.7}
u^0_{\mathrm{PM}}=\sum_{n=0}^{\infty} G^n u^0_{\left(n\right)},
\end{equation}
\begin{equation}\label{1.2.8}
u^i_{\mathrm{PM}}=\sum_{n=0}^{\infty} G^n u^i_{\left(n\right)},
\end{equation}
\begin{equation}\label{1.2.9}
\rho_{\mathrm{PM}}=\sum_{n=0}^{\infty} G^n \rho_{\left(n\right)},
\end{equation}
\begin{equation}\label{1.2.10}
p_{\mathrm{PM}}=\sum_{n=0}^{\infty} G^n p_{\left(n\right)},
\end{equation}
and hence,
\begin{equation}\label{1.2.11}
T^{\mu\nu}_{\mathrm{PM}}=\sum_{n=0}^{\infty} G^n T^{\mu\nu}_{\left(n\right)},
\end{equation}
and if we consider
$h^{\mu\nu}_{\mathrm{PN}}$,
we find
\begin{equation}\label{1.2.12}
u^0_{\mathrm{PN}}=\bar{u}^0=\sum_{n=-1}^{\infty} \frac{1}{c^n} \bar{u}^0_{\left(n\right)},
\end{equation}
\begin{equation}\label{1.2.13}
u^i_{\mathrm{PN}}=\bar{u}^i=\sum_{n=0}^{\infty} \frac{1}{c^n} \bar{u}^i_{\left(n\right)},
\end{equation}
\begin{equation}\label{1.2.14}
\rho_{\mathrm{PN}}=\bar{\rho}=\sum_{n=0}^{\infty} \frac{1}{c^n} \bar{\rho}_{\left(n\right)},
\end{equation}
\begin{equation}\label{1.2.15}
p_{\mathrm{PN}}=\bar{p}=\sum_{n=0}^{\infty} \frac{1}{c^n} \bar{p}_{\left(n\right)},
\end{equation}
where the lower limit of the summation in
Eq. (\ref{1.2.12})
is
$n=-1$
due to the fact that
$x^0=ct$.
$T^{\mu\nu}_{\mathrm{PN}}$
also reads
\begin{equation}\label{1.2.16}
T^{\mu\nu}_{\mathrm{PN}}=\bar{T}^{\mu\nu}=\sum_{n=-2}^{\infty} \frac{1}{c^n} \bar{T}^{\mu\nu}_{\left(n\right)},
\end{equation}
where
$\bar{T}^{i\nu}_{\left(-2\right)}$
and
$\bar{T}^{ij}_{\left(-1\right)}$
are equal to zero.

$\mathfrak{g}$
in terms of the components of
$h^{\mu\nu}$
can be straightforwardly obtained as
\begin{equation}\label{1.2.17}
\mathfrak{g}= -1 - h - \frac{1}{2}h^2 + \frac{1}{2}h^{\mu\nu}h_{\mu\nu} + \mathfrak{g}_3 \left(h\right),
\end{equation}
where, introducing
$\eta_{\mu\nu}=\mathrm{diag}\left(-1, 1, 1, 1\right)$,
$h=\eta_{\mu\nu}h^{\mu\nu}$
and
$h_{\mu\nu}=\eta_{\mu\alpha}\eta_{\nu\beta}h^{\alpha\beta}$.
In addition, 
$\mathfrak{g}_3 \left(h\right)$
is the representative of all the terms which are symbolically of the form
$\underbrace{h\;...\;h}_{k\;\text{times}}$
where
$k\ge3$.
Inserting
$h^{\mu\nu}_{\mathrm{PM}}$
for
$h^{\mu\nu}$
in
Eq. (\ref{1.2.17}),
we find
\begin{equation}\label{1.2.18}
\mathfrak{g}_{\mathrm{PM}}=  \sum_{n=0}^{\infty} G^n \mathfrak{g}_{\left(n\right)},
\end{equation}
and inserting
$h^{\mu\nu}_{\mathrm{PN}}$
instead of
$h^{\mu\nu}_{\mathrm{PM}}$
gives
\begin{equation}\label{1.2.19}
\mathfrak{g}_{\mathrm{PN}}=\bar{\mathfrak{g}}=\sum_{n=0}^{\infty} \frac{1}{c^n} \bar{\mathfrak{g}}_{\left(n\right)}.
\end{equation}
It's not difficult to show that
$\Lambda^{\mu\nu}$
in terms of the components of
$h^{\alpha\beta}$
is given by
\begin{equation}\label{1.2.20}
\Lambda^{\mu\nu} = N^{\mu\nu} + O\left(h^3\right),
\end{equation}
where
\begin{IEEEeqnarray}{rcl}\label{1.2.21}
N^{\mu\nu} & = & - h^{\rho\sigma}\partial_{\rho\sigma} h^{\mu\nu} + \frac{1}{2}\partial^\mu h_{\rho\sigma}\partial^\nu h^{\rho\sigma} - \frac{1}{4} \partial^\mu h \partial^\nu h \nonumber \\  
&& \negmedspace {}+ \partial_\sigma h^{\mu\rho}\left(\partial^\sigma h^\nu_\rho + \partial_\rho h^{\nu\sigma}\right) - \partial^\mu h_{\rho\sigma}\partial^\rho h^{\nu\sigma} - \partial^\nu h_{\rho\sigma}\partial^\rho h^{\mu\sigma} \nonumber \\  
&& \negmedspace {}+\eta^{\mu\nu}\left[-\frac{1}{4}\partial_\tau h_{\rho\sigma}\partial^\tau h^{\rho\sigma} + \frac{1}{8}\partial_\rho h \partial^\rho h +\frac{1}{2}\partial_\rho h_{\sigma\tau}\partial^\sigma h^{\rho\tau}\right].
\end{IEEEeqnarray}
In
Eq. (\ref{1.2.21}),
$h^{\mu}_{\nu}=\eta_{\nu\beta}h^{\mu\beta}$
and
$\partial^\mu=\eta^{\mu\nu}\partial_\nu$
The symbolic form of the terms appearing in
$N^{\mu\nu}$
is either
$h\partial^2 h$
or
$\partial h \partial h$.
To obtain
Eq. (\ref{1.2.20}),
we have assumed that
$|h^{\mu\nu}|\ll 1$,
and
$O\left(h^3\right)$
in that equation is the representative of all the terms which are symbolically of the form
$\underbrace{h\;...\;h}_{k\;\text{times}}\partial h \partial h$
where
$k\ge1$.
The two important points about
$\Lambda^{\mu\nu}$
are that it doesn't contain any power of
$G$,
and all its powers of
$c$
come from derivatives with respect to
$\mu=0$
componenet of
$x^\mu$,
$\partial_0=\frac{1}{c}\partial_t$
(to see these, plug the expression for
$\left(-\mathfrak{g} \right) t_{\mathrm{LL\left(H\right)}}^{\mu\nu}$
given by
Eq. (\ref{1.1.11})
into
Eq. (\ref{1.1.22})).
Substituting
$h^{\mu\nu}_{\mathrm{PM}}$
for
$h^{\mu\nu}$
into
Eq. (\ref{1.2.20}),
we get
\begin{equation}\label{1.2.22}
\Lambda^{\mu\nu}_{\mathrm{PM}}=\sum_{n=2}^{\infty} G^n \Lambda^{\mu\nu}_{\left(n\right)}.
\end{equation}
On the other hand, substituting
$h^{\mu\nu}_{\mathrm{PN}}$
for
$h^{\mu\nu}$
results in
\begin{equation}\label{1.2.23}
\Lambda^{\mu\nu}_{\mathrm{PN}}=\bar{\Lambda}^{\mu\nu}=\sum_{n=4}^{\infty} \frac{1}{c^n} \bar{\Lambda}^{\mu\nu}_{\left(n\right)}.
\end{equation}
Now, using the expansions of
$T^{\mu\nu}$,
$\mathfrak{g}$
and
$\Lambda^{\mu\nu}$,
it is obvious that inserting
$h^{\alpha\beta}_{\mathrm{PM}}$
for
$h^{\alpha\beta}$
into
$\tau^{\mu\nu}$
gives
\begin{equation}\label{1.2.24}
\tau^{\mu\nu}_{\mathrm{PM}}=\sum_{n=0}^{\infty} G^n \tau^{\mu\nu}_{\left(n\right)},
\end{equation}
and inserting
$h^{\alpha\beta}_{\mathrm{PN}}$
for
$h^{\alpha\beta}$
gives
\begin{equation}\label{1.2.25}
\tau^{\mu\nu}_{\mathrm{PN}}=\bar{\tau}^{\mu\nu}=\sum_{n=-2}^{\infty} \frac{1}{c^n} \bar{\tau}^{\mu\nu}_{\left(n\right)},
\end{equation}
where
$\bar{T}^{i\nu}_{\left(-2\right)}$
and
$\bar{T}^{ij}_{\left(-1\right)}$
are equal to zero due to
$\bar{\tau}^{i\nu}_{\left(-2\right)}$
and
$\bar{\tau}^{ij}_{\left(-1\right)}$
being equal to zero.

Inserting
$h^{\mu\nu}_{\mathrm{PM}}$
and
$\tau^{\mu\nu}_{\mathrm{PM}}$
for
$h^{\mu\nu}$
and
$\tau^{\mu\nu}$
into
Eqs. (\ref{1.1.18}), (\ref{1.1.27}) and (\ref{1.1.29}),
we get
\begin{equation}\label{1.2.27}
\partial_\mu h^{\mu\nu}_{\paren{n}} = 0 \qquad \text{for}\; n\ge1,
\end{equation}
\begin{equation}\label{1.2.26}
\Box h^{\mu\nu}_{\paren{n}} = \frac{16\pi}{c^4}\tau ^{\mu\nu}_{\paren{n-1}} \qquad \text{for}\; n\ge 1,
\end{equation}
\begin{equation}\label{1.2.28}
\partial_\mu \tau^{\mu\nu}_{\paren{n-1}}= 0 \qquad \text{for}\; n\ge1.
\end{equation}
where to obtain the last equation, we have also made the change
$n\to n-1$.
Since only
$h^{\alpha\beta}_{\paren{m}}$'s
with
$m\le n-1$
appear in
$\tau ^{\mu\nu}_{\paren{n-1}}$,
assuming that we have found the solutions for
$h^{\alpha\beta}_{\paren{m}}$'s
with
$m\le n-1$,
$\tau ^{\mu\nu}_{\paren{n-1}}$
is fully determined. Hence, for a given
$n\ge 1$,
Eq. (\ref{1.2.28})
is already fulfilled, and despite
Eq. (\ref{1.1.27}),
the right-hand side of
Eq. (\ref{1.2.26})
is completely known. The most general solution to the inhomogeneous d'Alembertian equation
$\Box f(t,\mathbf{x}) = g(t,\mathbf{x})$
consists of a surface integral over a surface at infinity and a volume integral over all space. The surface integral vanishes if
\begin{equation}\label{1.2.29}
\lim_{\substack{|\mathbf{x}|\to \infty \\ t+\frac{|\mathbf{x}|}{c} = \mathrm{const}}} \left[ \frac{\partial}{\partial  |\mathbf{x}|}\left(|\mathbf{x}|f(t,\mathbf{x})\right) +\frac{1}{c}\frac{\partial}{\partial t}\left(|\mathbf{x}|f(t,\mathbf{x})\right)\right]=0.
\end{equation}
The above equation is called the no-incoming radiation condition for the function
$f(t,\mathbf{x})$.
The remaining volume integral is nothing but
$\retint{g\retarg}$
which is given by
Eq. (\ref{1.1.30})
and called the retarded integral of
$g(t,\mathbf{x})$.
Of course, the necessary condition for validity of this solution is the convergence of
$\retint{g\retarg}$
which itself rests on the functional dependence of
$g(t,\mathbf{x})$.
Based on the foregoing discussion, the solution to
Eq. (\ref{1.2.26})
subject to no-incoming radiation condition is given by
\begin{equation}\label{1.2.30}
 h^{\mu\nu}_{\paren{n}}=\retint{\frac{16\pi}{c^4}\tau ^{\mu\nu}_{\paren{n-1}}}  \qquad \text{for}\;n\ge 1,
\end{equation}
if the retarded integral converges. This
$ h^{\mu\nu}_{\paren{n}}$
must also satisfy the harmonic gauge condition. As mentioned above,
Eq.(\ref{1.2.28})
is already fulfilled. Therefore, we can use it in checking whether or not
$h^{\mu\nu}_{\paren{n}}$
is divergenceless, where necessary. However, before doing so, note that
$\retint{\frac{16\pi}{c^4}\tau ^{\mu\nu}_{\paren{n-1}}}$
is divergent due to the behavior of
$\Lambda^{\mu\nu}_{\paren{n}}\finsarg$
appearing in
$\tau ^{\mu\nu}_{\paren{n-1}}\finsarg$
at infinity for
$n$'s
greater than some specific value. Thus,
$h^{\mu\nu}_{\paren{n}}$
given in
Eq. (\ref{1.2.30})
isn't a valid solution for any arbitrary order.

On the other hand, substitutions of
$h^{\mu\nu}_{\mathrm{PN}}$
and
$\tau^{\mu\nu}_{\mathrm{PN}}$
for
$h^{\mu\nu}$
and
$\tau^{\mu\nu}$
into
Eqs. (\ref{1.1.18}), (\ref{1.1.27}) and (\ref{1.1.29}),
give
\begin{equation}\label{1.2.32}
\partial_t \bar{h}^{0\nu}_{\paren{n-1}} + \partial_i \bar{h}^{i\nu}_{\paren{n}} = 0 \qquad \text{for}\;n\ge2,
\end{equation}
\begin{equation}\label{1.2.31}
\Delta \bar{ h}^{\mu\nu}_{\paren{n}} = 16\pi G\bar{\tau} ^{\mu\nu}_{\paren{n-4}} + \partial^2_t   \bar{ h}^{\mu\nu}_{\paren{n-2}} \qquad \text{for}\;n\ge 2,
\end{equation}
\begin{equation}\label{1.2.33}
\partial_t \bar{\tau}^{0\nu}_{\paren{n-1}} + \partial_i \bar{\tau}^{i\nu}_{\paren{n}} = 0 \qquad \text{for}\;n\ge -2.
\end{equation}
The term
$\partial^2_t \bar{ h}^{\mu\nu}_{\paren{n-2}}$
in
Eq. (\ref{1.2.31})
is meaningless for
$n=2$
and
$n=3$.
In fact, for these values of
$n$,
this term must not exist. This ambiguity can be resolved by taking
$\bar{ h}^{\mu\nu}_{\paren{0}}$
and
$\bar{ h}^{\mu\nu}_{\paren{1}}$
to be equal to zero. These choices are also compatible with the index of summation in post-Newtonian expansion starting from
$n=2$.
Furthermore, the ambiguity concerning the term
$\partial_t \bar{h}^{0\nu}_{\paren{n-1}}$
in
Eq. (\ref{1.2.32})
for
$n=2$
is now resolved too. By similar reasoning, we must also set
$\bar{\tau}^{\mu\nu}_{\paren{-3}}=0$ (which is not in contrast with the lower limit of the summation in
Eq. (\ref{1.2.25})).

Since only
$\bar{h}^{\alpha\beta}_{\paren{m}}$'s
with
$m\le n-2$
appear in
$\bar{\tau}^{\mu\nu}_{\paren{n-4}}$,
assuming that we have found the solutions for
$\bar{h}^{\alpha\beta}_{\paren{m}}$'s
with
$m\le n-2$,
$\bar{\tau}^{\mu\nu}_{\paren{n-4}}$
is completely known. Because of this, for a given
$n\ge 2$,
the equation
$\partial_t \bar{\tau}^{0\nu}_{\paren{n-5}} + \partial_i \bar{\tau}^{i\nu}_{\paren{n-4}} = 0$ 
(obtained from
Eq. (\ref{1.2.33}))
is already fulfilled,
and the right-hand side of
Eq. (\ref{1.2.31})
is fully determined. The most general solution to the Poisson equation
$\Delta f\left(t,\mathbf{x}\right) = g\left(t,\mathbf{x}\right)$
consists of a surface integral over a surface at infinity and a volume integral over all space. The surface integral is equal to zero if
\begin{equation}\label{1.2.34}
\lim_{\substack{|\mathbf{x}|\to \infty \\ t=\mathrm{const}}} f\finsarg =0.
\end{equation}
The remaining volume integral is nothing but the Poisson integral of
$g\finsarg$
which is given by
\begin{equation}\label{1.2.35}
\poisint{g\sinsarg}\finsarg = \wholeint{g\sinsarg}.
\end{equation}
Of course, the necessary condition for validity of this solution is the convergence of
$\poisint{g\sinsarg}$
which itself rests on the functional dependence of
$g(t,\mathbf{x})$.
Based on the foregoing discussion, the solution to
Eq. (\ref{1.2.31}),
provided that
$\bar{ h}^{\mu\nu}_{\paren{n}}$
tends to zero at infinity, reads
\begin{equation}\label{1.2.36}
\bar{ h}^{\mu\nu}_{\paren{n}}=\poisint{16\pi G\bar{\tau}^{\mu\nu}_{\paren{n-4}}+ \partial^2_t \bar{ h}^{\mu\nu}_{\paren{n-2}}} \qquad \text{for}\;n\ge 2,
\end{equation}
if the Poisson integral converges. This
$\bar{h}^{\mu\nu}_{\paren{n}}$
must fulfill the harmonic gauge condition too. As mentioned earlier,
the equation
$\partial_t \bar{\tau}^{0\nu}_{\paren{n-5}} + \partial_i \bar{\tau}^{i\nu}_{\paren{n-4}} = 0$
is already satisfied, and hence, we are allowed to use
it in checking whether or not
$\partial_t \bar{h}^{0\nu}_{\paren{n-1}} + \partial_i \bar{h}^{i\nu}_{\paren{n}}$
vanishes, where necessary. But, before doing so, note that
$\poisint{16\pi G\bar{\tau}^{\mu\nu}_{\paren{n-4}}+ \partial^2_t \bar{ h}^{\mu\nu}_{\paren{n-2}}}$
is divergent owing to the behavior of
$\bar{\Lambda}^{\mu\nu}_{\paren{n}}\finsarg$
appearing in
$\bar{\tau}^{\mu\nu}_{\paren{n-4}}\finsarg$
at infinity for
$n$'s
greater than some specific value. Therefore,
$h^{\mu\nu}_{\paren{n}}$
given in
Eq. (\ref{1.2.36})
isn't a valid solution for any arbitrary order.

In conclusion, neither post-Minkowskian expansion nor post-Newtonian expansion can be an approximate solution for
$h^{\mu\nu}$
subject to the specified conditions. However, this doesn't mean that it is impossible to find an approximate solution for
$h^{\mu\nu}$.
That approximate solution is provided by Blanchet-Damour approach.

\section{Blanchet-Damour Approach}

Blanchet and Damour considered the following conditions for
$h^{\mu\nu}\finsarg$:
\begin{equation}\label{1.3.1}
t \le -T \longrightarrow \partial _t h^{\mu\nu}\finsarg = 0,
\end{equation}
\begin{equation}\label{1.3.2}
t \le -T \longrightarrow\lim_{\substack{|\mathbf{x}|\to \infty \\ t=\mathrm{const}}} h^{\mu\nu}\finsarg =0,
\end{equation}
\begin{equation}\label{1.3.3}
 h^{\mu\nu}\finsarg \in C^\infty \left(\mathbb{R}^4\right).
\end{equation}
By
$-T$
we mean an instant in the past, and we can assume
$T\to \infty$.
Eq. (\ref{1.3.1})
is called the past-stationarity condition and must be satisfied because real astrophysical systems and hence their
$T^{\mu\nu}\finsarg$'s
are stationary in the remote past.
Eq. (\ref{1.3.2})
demands that the metric  be asymptotically Minkowskian when
$t \le -T$
which is a reasonable condition since we have assumed
$\frac{GM}{dc^2}\ll 1$,
where
$M$
and
$d$
depend on the stationary state of the source at
$t\le -T$.
Furthermore,
Eq. (\ref{1.3.3})
must be fulfilled because
$T^{\mu\nu}\finsarg \in C^\infty \left(\mathbb{R}^4\right)$
by assumption. It is called the smoothness condition for
$h^{\mu\nu}\finsarg$.

Blanchet and Damour employed the post-Minkowskian expansion to describe
$h^{\mu\nu}$
outside the source. Therefore, from now on we denote
$h^{\mu\nu}_{\mathrm{PM}}$
by
$h^{\mu\nu}_{\mathrm{ext}}$,
i.e.,
\begin{equation}\label{1.3.4}
h^{\mu\nu}_{\mathrm{ext}}=\sum_{n=1}^{\infty} G^n h^{\mu\nu}_{\left(n\right)}.
\end{equation}
In the region under discussion,
$T^{\mu\nu}\finsarg=0$,
and hence, the equations governing
$ h^{\mu\nu}$
are
\begin{equation}\label{1.3.4a}
\Box h^{\mu\nu} =\Lambda^{\mu\nu},
\end{equation}
\begin{equation}\label{1.3.4b}
\partial_\mu h^{\mu\nu}= 0 ,
\end{equation}
\begin{equation}\label{1.3.4c}
\partial_\mu \Lambda^{\mu\nu}= 0.
\end{equation}
Substituting
$h^{\mu\nu}_{\mathrm{ext}}$
for
$h^{\mu\nu}$
in these equations, we get
\begin{equation}\label{1.3.5}
\Box h^{\mu\nu}_{\paren{n}} =\Lambda^{\mu\nu}_{\paren{n}} \qquad\text{for}\;n\ge 1,
\end{equation}
\begin{equation}\label{1.3.6}
\partial_\mu h^{\mu\nu}_{\paren{n}} = 0 \qquad \text{for}\;n\ge1,
\end{equation}
\begin{equation}\label{1.3.7}
\partial_\mu \Lambda^{\mu\nu}_{\paren{n}}= 0 \qquad \text{for}\;n\ge1,
\end{equation}
where in
Eqs. (\ref{1.3.5}) and (\ref{1.3.7})
we take
$\Lambda^{\mu\nu}_{\paren{1}}$
to be equal to zero to resolve the ambiguity for
$n=1$.
Since only
$h^{\alpha\beta}_{\paren{m}}$'s
with
$m\le n-1$
appear in
$\Lambda^{\mu\nu}_{\paren{n}}$,
assuming that we have obtained the solutions for
$h^{\alpha\beta}_{\paren{m}}$'s
with
$m\le n-1$,
$\Lambda^{\mu\nu}_{\paren{n}}$
is completely known. Thus, for a given
$n\ge 1$,
Eq. (\ref{1.3.7})
is already fulfilled. The other two equations govern the dynamics of
$h^{\mu\nu}_{\left(n\right)}$,
under the conditions derived from substitution of
$h^{\mu\nu}_{\mathrm{ext}}$
for
$h^{\mu\nu}$
 in
Eqs. (\ref{1.3.1}), (\ref{1.3.2}) and (\ref{1.3.3}),
namely
\begin{equation}\label{1.3.8}
t \le -T \longrightarrow \partial _t h^{\mu\nu}_{\paren{n}}\finsarg = 0 \qquad \text{for}\;n\ge 1,
\end{equation}
\begin{equation}\label{1.3.9}
t \le -T \longrightarrow\lim_{\substack{|\mathbf{x}|\to \infty \\ t=\mathrm{const}}}  h^{\mu\nu}_{\paren{n}}\finsarg = 0 \qquad \text{for}\;n\ge 1,
\end{equation}
\begin{equation}\label{1.3.10}
h^{\mu\nu}_{\paren{n}}\finsarg \; \text{is smooth at}\; |\mathbf{x}|>d \qquad \text{for}\;n\ge 1,
\end{equation}
where to obtain the last condition we have also used the fact that the domain of validity of
$h^{\mu\nu}_{\mathrm{ext}}$
is outside the source.

On the other hand, Blanchet and Damour considered the post-Newtonian expansion (whose coefficients are governed by
Eqs. (\ref{1.2.32}), (\ref{1.2.31}) and (\ref{1.2.33}),
and from now on we denote it by
$\bar{h}^{\mu\nu}$
instead of
$h^{\mu\nu}_{\mathrm{PN}}$)
as the solution for
$h^{\mu\nu}$
at
$|\mathbf{x}|\ll \lambda$
where
$\lambda$
is a typical wavelength of the emitted gravitational wave. The reason behind this is that, to obtain
Eqs. (\ref{1.2.32}), (\ref{1.2.31}) and (\ref{1.2.33}),
we have actually assumed that retardation effects are small, an assumption which is valid only when
$|\mathbf{x}|\ll \lambda$.
Taking this into account, substitution of
$\bar{h}^{\mu\nu}$
for
$h^{\mu\nu}$
in
Eqs. (\ref{1.3.1}) and (\ref{1.3.3})
gives the following conditions for
$\bar{h}^{\mu\nu}_{\paren{n}}$:\footnote{If you wonder why we don't substitute
$\bar{h}^{\mu\nu}$
for
$h^{\mu\nu}$
into
Eq. (\ref{1.3.2}),
note that such a substitution results in
\begin{equation}
t \le -T \longrightarrow\lim_{\substack{|\mathbf{x}|\to \infty \\ t=\mathrm{const}}}\bar{h}^{\mu\nu}_{\paren{n}}\finsarg =0\nonumber,
\end{equation}
which is clearly meaningless due to the domain of validity of post-Newtonian expansion being the region
$|\mathbf{x}|\ll \lambda$.}
\begin{equation}\label{1.3.12}
t \le -T \longrightarrow \partial _t \bar{ h}^{\mu\nu}_{\paren{n}}\finsarg = 0 \qquad \text{for}\;n\ge 2,
\end{equation}
\begin{equation}\label{1.3.13}
\bar{ h}^{\mu\nu}_{\paren{n}}\finsarg \; \text{is smooth at} \; |\mathbf{x}|\ll \lambda \qquad \text{for}\;n\ge 2.
\end{equation}
Last but not least, Blanchet and Damour assumed that there exists a region in which both post-Minkowskian and post-Newtonian expansions are valid (approximate) solutions for
$h^{\mu\nu}$,
or, in other words,
$h^{\mu\nu}_{\text{ext}}\finsarg=\bar{h}^{\mu\nu}\finsarg$.
To explain why such an assumption is sensible, first note that since we have assumed
$\frac{v}{c}\ll1$,
we can write
\begin{equation}\label{1.3.11}
v \ll c \longrightarrow vT \ll cT  \longrightarrow d \ll \lambda,
\end{equation}
where
$T$
is a typical period of the internal motion and hence of the emitted gravitational wave. Moreover, we deduced that the post-Newtonian expansion is valid at
$|\mathbf{x}|\ll \lambda$; hence, it is reasonable to suppose that the domain of validity of post-Newtonian expansion is the region
$|\mathbf{x}|<\mathcal{R}$
where
$\mathcal{R}$
is a positive constant and
$ d < \mathcal{R}\ll \lambda$
(this region is called near zone). This, together with the fact that the post-Minkowskian expansion is valid outside the source, i.e., at
$|\mathbf{x}|>d$,
lead us to the stated assumption.

In Chapters 2 and 3, we will find the general solutions for
$ h^{\mu\nu}_{\paren{n}}$
and
$\bar{ h}^{\mu\nu}_{\paren{n}}$
respectively. Each of these solutions has an unknown part. In order to determine them, in Chapter 4 we will apply a ``matching procedure'', a procedure that we are entitled to use owing to the assumption of existence of  ``matching region'', a region where the domains of validity of post-Minkwskian and post-Newtonian expansions overlap.


\chapter{Post-Minkowskian Approximation}

This chapter is devoted to finding the general solution to post-Minkowskian equations,
Eqs. (\ref{1.3.5}) and (\ref{1.3.6}),
subject to the conditions
(\ref{1.3.8})-(\ref{1.3.10}).
In the first section, the general solution to the first-order problem is obtained. In the next section, construction of the solution to the second-order problem is examined in great detail. This solution consists of two parts fulfilling the harmonic gauge condition separately, the general solution of the source-free d'Alembertian equation and a particular solution to the relaxed Einstein field equation outside the source at second post-Minkowskian order. The most important part of the procedure for finding this particular solution is to use Theorem 2.3 which is based on
\cite{R1949}.
In the third section, we generalized the algorithm discussed in the second section to construct the general solutions up to any arbitrary order, a generalization which is possible due to similarity of the structures of
$h^{\mu\nu}_{\paren{n}}$'s.
Having found all
$h^{\mu\nu}_{\paren{n}}$'s,
in the last section we sum them from
$n=1$
up to infinity in order to obtain
$h^{\mu\nu}_{\mathrm{ext}}$.

\section{First-Order Problem}

\subsection{General Solution of the Equation $\Box f(t,\mathbf{x}) =0$} 

We use the method of separation of variables to find the general solution of the equation
 $\Box f(t,\mathbf{x}) =0$.
Assuming a solution of the form
$f\finsarg=g(t, |\mathbf{x}|) h(\theta, \varphi )$
and taking the constant of separation as
$\ell\paren{\ell+1}$, we get
\begin{equation}\label{2.1.1.1}
\frac{\partial^2}{\partial \theta^2}h + \frac{1}{\tan \theta}\frac{\partial}{\partial \theta}h+\frac{1}{\sin^2 \theta}\frac{\partial^2}{\partial \varphi^2}h=-\ell\left(\ell +1\right)h,
\end{equation}
\begin{equation}\label{2.1.1.2}
\frac{\partial^2}{\partial  |\mathbf{x}|^2}g + \frac{2}{ |\mathbf{x}|}\frac{\partial}{\partial  |\mathbf{x}|}g-\frac{1}{c^2}\frac{\partial^2}{\partial t^2}g=\frac{g}{ |\mathbf{x}|^2}\ell \left(\ell +1\right).
\end{equation}
$\ylm$ is the solution to
Eq. (\ref{2.1.1.1}).
To obtain the solution of
Eq. (\ref{2.1.1.2}),
we rewrite it as
\begin{equation}\label{2.1.1.10}
\left(v-u\right)\frac{\partial^2}{\partial u \partial v}S + \left(\ell +1\right) \left[\frac{\partial}{\partial u}-\frac{\partial}{\partial v}\right]S=0,
\end{equation}
where
$u=t-\frac{ |\mathbf{x}|}{c}$,
$v=t+\frac{ |\mathbf{x}|}{c}$
and
$S=\left(v-u\right)^{-\ell} g$.
The above equation is a particular case of Euler-Poisson-Darboux equation
\begin{equation}\label{2.1.1.11}
\left(v-u\right)\partial_{uv}S + m\partial_u S - n\partial_v S = 0,
\end{equation}
for
$m=n=\ell + 1$. Therefore, the general solution of
Eq. (\ref{2.1.1.10})
is given by
\cite{BD1986}
\begin{equation}\label{2.1.1.12}
S_\ell=\frac{\partial^{2\ell}}{\partial u^\ell \partial v^\ell}\left[\frac{U(u) + V(v)}{v-u}\right],
\end{equation}
where
$U\paren{u}$
and
$V\paren{v}$
are arbitrary smooth functions of
$u$
and
$v$.
Taking this into account and using Leibniz formula (see Appendix A), we find
\begin{IEEEeqnarray}{rcl}\label{2.1.1.14}
g_\ell \left(t, |\mathbf{x}|\right) & = & \frac{\left(-1\right)^\ell c^{\ell +1}\ell !}{2}\sum_{k=0}^{\ell} \frac{\left(2\ell-k\right)!}{2^{\ell-k}\left(\ell-k\right)! k!}\frac{U^{\left(k\right)}( t-\frac{ |\mathbf{x}|}{c}) + \left(-1\right)^k V^{\left(k\right)}( t+\frac{ |\mathbf{x}|}{c})}{c^k |\mathbf{x}|^{\ell - k +1}},
\end{IEEEeqnarray}
where by
$U^{\left(k\right)}( t-\frac{ |\mathbf{x}|}{c})$
and
$V^{\left(k\right)}( t+\frac{ |\mathbf{x}|}{c})$
we mean
$k$th
derivative of
$U\uarg$
and
$V\varg$
with respect to their arguments. Now, using
Eqs. (\ref{B.6}) and (\ref{B.1}),
we can write
\begin{equation}\label{2.1.1.19}
f\finsarg = \sum_{\ell=0}^{\infty}\sum_{m=-\ell}^{\ell} g_\ell(t, |\mathbf{x}|)Y^{\ell m}(\theta, \varphi)= \sum_{\ell=0}^{\infty}\hat{\partial}_L\left(\frac{\hat{U}_L( t-\frac{ |\mathbf{x}|}{c}) + \hat{V}_L( t+\frac{ |\mathbf{x}|}{c})}{ |\mathbf{x}|}\right),
\end{equation}
where
$\hat{U}_L=\frac{c^{\ell +1} \ell !}{2} U \sum_{m=-\ell}^{\ell}\hat{\mathcal{Y}}^{\ell m}_L$,
$\hat{V}_L=\frac{c^{\ell +1} \ell ! }{2} V \sum_{m=-\ell}^{\ell}\hat{\mathcal{Y}}^{\ell m}_L$
and
$L$
denotes the multi-index
$i_{1}...i_{\ell}$
(to become familiar with multi-index notation,
$\hat{\mathcal{Y}}^{\ell m}_L$'s
and symmetric-trace-free [abbreviated as STF] Cartesian tensors, see Appendix B). This solution is mathematically valid everywhere in
$\mathbb{R }^3$
except at
$|\mathbf{x}|=0$,
and since post-Minkowskian expansion is physically valid outside the source, exclusion of
$|\mathbf{x}|=0$
is not a matter of concern.

\subsection{Past-Stationarity and Smoothness Conditions}

First, we impose the past-stationarity condition. Using
Eq. (\ref{B.1}),
the time derivative of the general solution to the equation
$\Box f\left(t,\mathbf{x}\right) =0$
can be written as
\begin{IEEEeqnarray}{rcl}\label{2.1.2.12}
\frac{\partial f\finsarg}{\partial t} & = & \sum_{\ell=0}^{\infty}\sum_{m=0}^{\ell}\frac{\paren{-1}^\ell\paren{\ell+m}!}{2^m m!\paren{\ell-m}!}\frac{\left[\hat{U}^{\left(\ell-m+1\right)}_L( t-\frac{ |\mathbf{x}|}{c})+\paren{-1}^{\ell-m}\hat{V}^{\left(\ell-m+1\right)}_L( t+\frac{ |\mathbf{x}|}{c})\right]}{c^{\ell-m}\x^{m+1}} \hat{n}^L, \nonumber \\  
\end{IEEEeqnarray}
where
$n^i = \frac{x^i}{\x}$.
Since
$1 \le \ell-m+1 \le \ell +1$,
for
$\frac{\partial f\finsarg}{\partial t}$
to be equal to zero everywhere in
$\mathbb{R }^3$
(except at $|\mathbf{x}|=0$)
when
$t \le -T$,
we must have
\begin{equation}\label{2.1.2.13}
\hat{U}^{\left(k\right)}_L( t-\frac{ |\mathbf{x}|}{c})=0,\qquad \forall \ell \ge 0 \; \text{and}\; \forall k \in \left\{1,...,\paren{\ell+1}\right\},
\end{equation}
\begin{equation}\label{2.1.2.14}
\hat{V}^{\left(k\right)}_L( t+\frac{ |\mathbf{x}|}{c})=0,\qquad \forall \ell \ge 0 \; \text{and}\; \forall k \in \left\{1,...,\paren{\ell+1}\right\}.
\end{equation}
These two conditions are fulfilled  if
$\hat{U}_L( t-\frac{ |\mathbf{x}|}{c})$
and
$\hat{V}_L( t+\frac{ |\mathbf{x}|}{c})$
are constant in the specified space-time region. Now note that in this region we have
\begin{equation}\label{2.1.2.16}
 \left[t-\frac{ |\mathbf{x}|}{c}\right] \in \left(-\infty, -T\right);
\end{equation}
therefore,
$\hat{U}_L( t-\frac{ |\mathbf{x}|}{c})$
is constant when
$u < -T$.
$u$
depends on spatial coordinates, but whatever they are,
$\hat{U}_L( t-\frac{ |\mathbf{x}|}{c})$
is constant at
$t \le -T$.
On the other hand,
\begin{equation}\label{2.1.2.15}
 \left[t+\frac{ |\mathbf{x}|}{c}\right] \in \left(-\infty,+ \infty\right);
\end{equation}
thus,
$\hat{V}_L( t+\frac{ |\mathbf{x}|}{c})$
is constant for all values that its argument can take, and hence,
$\hat{U}_L( t-\frac{ |\mathbf{x}|}{c})+\hat{V}_L( t+\frac{ |\mathbf{x}|}{c})$
can be redefined as
$\hat{U}_L( t-\frac{ |\mathbf{x}|}{c})$.
All in all, we get
\begin{equation}\label{2.1.2.17}
f\finsarg=\sum_{\ell=0}^{\infty}\hat{\partial}_L\left(\frac{\hat{U}_L( t-\frac{ |\mathbf{x}|}{c})}{ |\mathbf{x}|}\right),
\end{equation}
where
$\hat{U}_L( t-\frac{ |\mathbf{x}|}{c})$
is constant at
$t \le -T$.
Furthermore, due to smoothness of old
$\hat{U}_L( t-\frac{ |\mathbf{x}|}{c})$,
this new
$\hat{U}_L( t-\frac{ |\mathbf{x}|}{c})$
is also smooth and
$f\finsarg$
given by
Eq. (\ref{2.1.2.17})
thereby satisfies the smoothness condition, as required.

\subsection{Harmonic Gauge Condition}

Considering what we have obtained in previous subsection, the general solution to the equation
 $\Box u^{\mu\nu}(t,\mathbf{x}) =0$
subject to past-stationarity and smoothness conditions is given by
\begin{equation}\label{2.1.3.1}
 u^{\mu\nu}\finsarg=\sum_{\ell=0}^{\infty}\hat{\partial}_L\left(\frac{\hat{U}^{\mu\nu}_L( t-\frac{ |\mathbf{x}|}{c})}{ |\mathbf{x}|}\right).
\end{equation}
The equation governing
$ \partial_\mu u^{\mu\nu}\finsarg$
is
\begin{equation}\label{2.1.3.2}
\Box\left(\partial_\mu u^{\mu\nu}\finsarg\right) =0.
\end{equation}
Since
$u^{\mu\nu}\finsarg$
is a past-stationary function, the general solution to
Eq. (\ref{2.1.3.2})
reads
\begin{equation}\label{2.1.3.3}
 \partial_\mu u^{\mu\nu}\finsarg=\sum_{\ell=0}^{\infty}\hat{\partial}_L\left(\frac{\hat{U}^{\nu}_L( t-\frac{ |\mathbf{x}|}{c})}{ |\mathbf{x}|}\right).
\end{equation}
So there is a possibility that
$\partial_\mu u^{\mu\nu}\finsarg$
is not equal to zero. To examine this, we rewrite
$ u^{0i}\finsarg$
and
$ u^{ij}\finsarg$
in terms of STF Cartesian tensors (at the moment,
$U_{i\left<L\right>} \equiv \hat{U}^{0i}_L$
and
$U_{ij\left<L\right>} \equiv \hat{U}^{ij}_L$
are symmetric and trace-free with respect to
$i_{1}...i_{\ell}$,
not all their indices).
Decomposing
$\hat{U}^{0i}_L$
and
$\hat{U}^{ij}_L$
into terms in the form of the product of
$\delta_{ij}$
or
$\varepsilon_{ijk}$
(which are both invariant under
SO(3))
and a STF tensor (which is the STF part of the product of
$\delta_{ij}$
or
$\varepsilon_{ijk}$
and
$\hat{U}^{0i}_L$
or
$\hat{U}^{ij}_L$
with some indices contracted) with some indices contracted
\cite{DI1991},
and using
\begin{equation}\label{2.1.3.7}
\Delta \left(\frac{F\uarg}{|\mathbf{x}|}\right) = \frac{1}{c^2}\frac{F^{\left(2\right)}\uarg}{|\mathbf{x}|},
\end{equation}
deduced from
$\Box \left(\frac{F\uarg}{|\mathbf{x}|}\right) = 0$,
we get
\begin{IEEEeqnarray}{rcl}\label{2.1.3.5}
 u^{0i}\finsarg & = & \sum_{\ell=0}^{\infty}\partial_{iL} \left( |\mathbf{x}|^{-1}B_L(u)\right) + \sum_{\ell=1}^{\infty}\partial_{L-1} \left( |\mathbf{x}|^{-1}C_{iL-1}(u)\right) \nonumber \\  
&& \negmedspace {}  + \sum_{\ell=1}^{\infty}\varepsilon_{iab}\partial_{aL-1} \left( |\mathbf{x}|^{-1}D_{bL-1}(u)\right),
\end{IEEEeqnarray}
\begin{IEEEeqnarray}{rcl}\label{2.1.3.6}
 u^{ij}\finsarg & = & \sum_{\ell=0}^{\infty}\partial_{ijL} \left( |\mathbf{x}|^{-1}E_L(u)\right) + \sum_{\ell=0}^{\infty}\delta_{ij}\partial_L \left( |\mathbf{x}|^{-1}F_L(u)\right) \nonumber \\  
&& \negmedspace {}+ \sum_{\ell=1}^{\infty}\partial_{L-1(i} \left( |\mathbf{x}|^{-1}G_{j)L-1}(u)\right) + \sum_{\ell=1}^{\infty}\varepsilon_{ab(i}\partial_{j)aL-1} \left( |\mathbf{x}|^{-1}H_{bL-1}(u)\right) \nonumber \\  
&& \negmedspace {} + \sum_{\ell=2}^{\infty}\partial_{L-2} \left( |\mathbf{x}|^{-1}I_{ijL-2}(u)\right) +  \sum_{\ell=2}^{\infty}\partial_{aL-2} \left( |\mathbf{x}|^{-1}\varepsilon_{ab(i}J_{j)bL-2}(u)\right).
\end{IEEEeqnarray}
We also rewrite
$ u^{00}\finsarg$
as
\begin{equation}\label{2.1.3.4}
 u^{00}\finsarg=\sum_{\ell=0}^{\infty}\partial_L \left( |\mathbf{x}|^{-1}A_L(u)\right),
\end{equation}
where
$A_L=\hat{U}^{00}_L$.
The ten tensors
$A_L,...,J_L$,
called moments, are STF and in terms of
$\hat{U}^{00}_L$,
$\hat{U}^{0i}_L$,
$\hat{U}^{ij}_L$,
$\delta_{ij}$,
and
$\varepsilon_{ijk}$.
Since
$\hat{U}^{00}_L$,
$\hat{U}^{0i}_L$
and
$\hat{U}^{ij}_L$
are constant at
$t \le -T$,
so too are
$A_L,...,J_L$.
More importantly, they are algebraically independent (owing to the fact that the set of all STF Cartesian tensors of rank
$\ell$
generates a
($2\ell+1$)-dimensional irreducible representation of
SO(3))
\cite{DI1991}.
Taking the latter into account, noting that
$\varepsilon_{iab}\partial_{ia}f\finsarg=0$
(due to contraction of a symmetric tensor with an antisymmetric one), and using
Eq. (\ref{2.1.3.7}),
$\partial_\mu u^{\mu\nu}= 0$
for
$\nu = 0$
gives
\begin{equation}\label{2.1.3.10}
\frac{A^{\left(1\right)}}{c} + \frac{B^{\left(2\right)}}{c^2}=0,
\end{equation}
\begin{equation}\label{2.1.3.11}
\frac{A^{\left(1\right)}_L}{c}+\frac{B^{\left(2\right)}_L}{c^2}+C_L=0 \qquad \text{for}\;\ell \ge 1,
\end{equation}
while for
$\nu = k$
\begin{equation}\label{2.1.3.12}
\frac{B^{\left(1\right)}}{c}+\frac{E^{\left(2\right)}}{c^2}+F=0,
\end{equation}
\begin{equation}\label{2.1.3.13}
\frac{B^{\left(1\right)}_L}{c}+\frac{E^{\left(2\right)}_L}{c^2}+F_L+\frac{G_L}{2}=0 \qquad \text{for}\;\ell \ge 1,
\end{equation}
\begin{equation}\label{2.1.3.14}
\frac{C^{\left(1\right)}_i}{c}+\frac{G^{\left(2\right)}_i}{2c^2}=0,
\end{equation}
\begin{equation}\label{2.1.3.15}
\frac{C^{\left(1\right)}_L}{c}+\frac{G^{\left(2\right)}_L}{2c^2}+I_L=0 \qquad \text{for}\;\ell \ge 2,
\end{equation}
\begin{equation}\label{2.1.3.16}
\frac{D^{\left(1\right)}_i}{c}+\frac{H^{\left(2\right)}_i}{2c^2}=0,
\end{equation}
\begin{equation}\label{2.1.3.17}
\frac{D^{\left(1\right)}_L}{c}+\frac{H^{\left(2\right)}_L}{2c^2}+\frac{J_L}{2}=0 \qquad \text{for}\;\ell \ge 2.
\end{equation}
Using
Eqs. (\ref{2.1.3.11}), (\ref{2.1.3.13}), (\ref{2.1.3.15}) and (\ref{2.1.3.17}),
we can write
$C_L$,
$G_L$,
$I_L$
and
$J_L$
in terms of the remaining six STF tensors and their derivatives. Hence, imposing the harmonic gauge condition reduces the number of degrees of freedom from ten to six. It is convenient to express these six degrees of freedom by the set
$\{ A_L,B_L,D_L,E_L,F_L,H_L\}$
instead of
$\{ M_L,S_L,W_L,X_L,Y_L,Z_L\}$.
The elements of these two sets are related as follows:
\begin{equation}\label{2.1.3.18}
A_L=M_L-\frac{1}{c}W^{\left(1\right)}_L+\frac{1}{c^2}X^{\left(2\right)}_L+Y_L \qquad\text{for}\;\ell \ge 0,
\end{equation}
\begin{equation}\label{2.1.3.19}
B_L=W_L-\frac{1}{c}X^{\left(1\right)}_L \qquad\text{for}\;\ell \ge 0,
\end{equation}
\begin{equation}\label{2.1.3.20}
D_L=-S_L-\frac{1}{c}Z^{\left(1\right)}_L \qquad \text{for}\;\ell \ge 1,
\end{equation}
\begin{equation}\label{2.1.3.21}
E_L=2X_L \qquad\text{for}\;\ell \ge 0,
\end{equation}
\begin{equation}\label{2.1.3.22}
F_L=-\frac{1}{c}W^{\left(1\right)}_L-\frac{1}{c^2}X^{\left(2\right)}_L-Y_L \qquad \text{for}\;\ell \ge 0,
\end{equation}
\begin{equation}\label{2.1.3.23}
H_L=2Z_L \qquad\text{for}\;\ell \ge 1,
\end{equation}
where
Eqs. (\ref{2.1.3.20}) and (\ref{2.1.3.23})
hold for
$\ell \ge 1$
since
$D_L$
and
$H_L$
in the expressions for
$ u^{0i}$
and
$ u^{ij}$
given by
Eqs. (\ref{2.1.3.5}) and (\ref{2.1.3.6})
have at least one index. Substituting the expressions for
$A_L$,
$B_L$,
$D_L$,
$E_L$,
$F_L$
and
$H_L$,
given by
Eqs. (\ref{2.1.3.18})-(\ref{2.1.3.23})
into
Eqs. (\ref{2.1.3.10})-(\ref{2.1.3.17}),
we get
\begin{equation}\label{2.1.3.26}
Y=0,
\end{equation}
\begin{equation}\label{2.1.3.32}
M=\mathrm{const},
\end{equation}
\begin{equation}\label{2.1.3.28}
M_i=\mathrm{const},
\end{equation}
\begin{equation}\label{2.1.3.30}
S_i=\mathrm{const},
\end{equation}\begin{equation}\label{2.1.3.25}
C_L=-\frac{1}{c}M^{\left(1\right)}_L-\frac{1}{c}Y^{\left(1\right)}_L \qquad \text{for}\;\ell \ge 1,
\end{equation}
\begin{equation}\label{2.1.3.27}
G_L=2Y_L \qquad \text{for}\;\ell \ge 1,
\end{equation}
\begin{equation}\label{2.1.3.29}
I_L=-\frac{1}{c^2}M^{\left(2\right)}_L \qquad \text{for}\;\ell \ge 2,
\end{equation}
\begin{equation}\label{2.1.3.31}
J_L=\frac{1}{c}S^{\left(1\right)}_L \qquad \text{for}\;\ell \ge 2,
\end{equation}
where the most important point about them is that the four moments
$Y$,
$M$,
$M_i$
and
$S_i$
are always constant and not just at
$t \le-T$.

\subsection{General Solution of the First-Order Problem}

Taking into account what we have obtained in the previous subsection, the general solution for the first post-Minkowskian coefficient can be written as
\begin{equation}\label{2.1.4.1}
h^{\mu\nu}_{\paren{1}}\left[M,W\right]=h^{\mu\nu}_{\mathrm{can}\paren{1}}\left[M\right]+\partial^\mu\omega^\nu\left[W\right]+\partial^\nu\omega^\mu\left[W\right]+\eta^{\mu\nu}\partial_\alpha\omega^\alpha\left[W\right],
\end{equation}
where
\begin{equation}\label{2.1.4.2}
h^{00}_{\mathrm{can}\paren{1}}\left[M\right]=\sum_{\ell=0}^{\infty}\partial_L \left( |\mathbf{x}|^{-1}M_L(u)\right),
\end{equation}
\begin{IEEEeqnarray}{rcl}\label{2.1.4.3}
h^{0i}_{\mathrm{can}\paren{1}}\left[M\right] & = & -\frac{1}{c} \sum_{\ell=1}^{\infty}\partial_{L-1} \left( |\mathbf{x}|^{-1}M^{\paren{1}}_{iL-1}(u)\right)\nonumber \\  
&& \negmedspace {} - \sum_{\ell=1}^{\infty}\varepsilon_{iab}\partial_{aL-1} \left( |\mathbf{x}|^{-1}S_{bL-1}(u)\right),
\end{IEEEeqnarray}
\begin{IEEEeqnarray}{rcl}\label{2.1.4.4}
h^{ij}_{\mathrm{can}\paren{1}}\left[M\right] & = & \frac{1}{c^2}  \sum_{\ell=2}^{\infty}\partial_{L-2} \left( |\mathbf{x}|^{-1}M^{\paren{2}}_{ijL-2}(u)\right)\nonumber \\  
&& \negmedspace {} + \frac{2}{c} \sum_{\ell=2}^{\infty}\partial_{aL-2} \left( |\mathbf{x}|^{-1}\varepsilon_{ab(i}S^{\paren{1}}_{j)bL-2}(u)\right),
\end{IEEEeqnarray}
\begin{equation}\label{2.1.4.5}
\omega^0\left[W\right]=\sum_{\ell=0}^{\infty}\partial_L \left( |\mathbf{x}|^{-1}W_L(u)\right),
\end{equation}
\begin{IEEEeqnarray}{rcl}\label{2.1.4.6}
\omega^i\left[W\right] & = &\sum_{\ell=0}^{\infty}\partial_{iL} \left( |\mathbf{x}|^{-1}X_L(u)\right) +  \sum_{\ell=1}^{\infty}\partial_{L-1} \left( |\mathbf{x}|^{-1}Y_{iL-1}(u)\right) \nonumber \\  
&& \negmedspace {}+ \sum_{\ell=1}^{\infty}\varepsilon_{iab}\partial_{aL-1} \left( |\mathbf{x}|^{-1}Z_{bL-1}(u)\right).
\end{IEEEeqnarray}
In these equations,
$M$
denotes the set of
$M_L$
and
$S_L$,
and
$W$
the set of
$W_L$,
$X_L$,
$Y_L$
and
$Z_L$.
Defining new moments from
\begin{equation}\label{2.1.4.7}
M_L=-\frac{4}{c^2}\frac{\left(-1\right)^\ell}{\ell!}M_L^{\mathrm{new}},
\end{equation}
\begin{equation}\label{2.1.4.8}
S_L=-\frac{4}{c^3}\frac{\left(-1\right)^\ell \ell}{\left(\ell+1\right)!}S_L^{\mathrm{new}},
\end{equation}
\begin{equation}\label{2.1.4.9}
W_L=\frac{4}{c^3}\frac{\left(-1\right)^\ell}{\ell!}W_L^{\mathrm{new}},
\end{equation}
\begin{equation}\label{2.1.4.10}
X_L=-\frac{4}{c^4}\frac{\left(-1\right)^\ell}{\ell!}X_L^{\mathrm{new}},
\end{equation}\begin{equation}\label{2.1.4.11}
Y_L=-\frac{4}{c^4}\frac{\left(-1\right)^\ell}{\ell!}Y_L^{\mathrm{new}},
\end{equation}
\begin{equation}\label{2.1.4.12}
Z_L=-\frac{4}{c^3}\frac{\left(-1\right)^\ell \ell}{\left(\ell+1\right)!}Z_L^{\mathrm{new}},
\end{equation}
rewriting
Eqs. (\ref{2.1.4.2})-(\ref{2.1.4.6})
in terms of them, replacing
$M_L^{\mathrm{new}}$
and
$S_L^{\mathrm{new}}$
by
$I_L$
and
$J_L$,
and dropping the superscript
``\textrm{new}''
wherever it appears, we reach
\begin{equation}\label{2.1.4.13}
h^{00}_{\mathrm{can}\paren{1}}\left[I\right]=-\frac{4}{c^2}\sum_{\ell=0}^{\infty}\frac{\left(-1\right)^\ell}{\ell!}\partial_L \left( |\mathbf{x}|^{-1}I_L(u)\right),
\end{equation}
\begin{IEEEeqnarray}{rcl}\label{2.1.4.14}
 h^{0i}_{\mathrm{can}\paren{1}}\left[I\right] & = & \frac{4}{c^3} \sum_{\ell=1}^{\infty}\frac{\left(-1\right)^\ell}{\ell!}\bigg\{\partial_{L-1} \left( |\mathbf{x}|^{-1}I^{\paren{1}}_{iL-1}(u)\right)\nonumber \\  
&& \negmedspace {}+ \frac{\ell}{\ell+1}\varepsilon_{iab}\partial_{aL-1} \left( |\mathbf{x}|^{-1}J_{bL-1}(u)\right)\bigg\},
\end{IEEEeqnarray}
\begin{IEEEeqnarray}{rcl}\label{2.1.4.15}
h^{ij}_{\mathrm{can}\paren{1}}\left[I\right] & = & -\frac{4}{c^4} \sum_{\ell=2}^{\infty}\frac{\left(-1\right)^\ell}{\ell!}\bigg\{\partial_{L-2} \left( |\mathbf{x}|^{-1}I^{\paren{2}}_{ijL-2}(u)\right)\nonumber \\  
&& \negmedspace {} + \frac{2\ell}{\ell+1}\partial_{aL-2} \left( |\mathbf{x}|^{-1}\varepsilon_{ab(i}J^{\paren{1}}_{j)bL-2}(u)\right)\bigg\},
\end{IEEEeqnarray}
\begin{equation}\label{2.1.4.16}
\omega^0\left[W\right]= \frac{4}{c^3}\sum_{\ell=0}^{\infty}\frac{\left(-1\right)^\ell}{\ell!}\partial_L \left( |\mathbf{x}|^{-1}W_L(u)\right),
\end{equation}
\begin{IEEEeqnarray}{rcl}\label{2.1.4.17}
\omega^i\left[W\right] & = & -\frac{4}{c^4}\sum_{\ell=0}^{\infty}\frac{\left(-1\right)^\ell}{\ell!}\partial_{iL} \left( |\mathbf{x}|^{-1}X_L(u)\right) -\frac{4}{c^4} \sum_{\ell=1}^{\infty}\frac{\left(-1\right)^\ell}{\ell!}\bigg\{\partial_{L-1} \left( |\mathbf{x}|^{-1}Y_{iL-1}(u)\right) \nonumber \\  
&& \negmedspace {}+\frac{\ell}{\ell+1}\varepsilon_{iab}\partial_{aL-1} \left( |\mathbf{x}|^{-1}Z_{bL-1}(u)\right)\bigg\}.
\end{IEEEeqnarray}

So far we have found that the general solution for
$ h^{\mu\nu}_{\paren{1}}\finsarg$
can be expressed as
\begin{equation}\label{2.1.4.18}
 h^{\mu\nu}_{\paren{1}}\finsarg=\sum_{\ell=0}^{\infty}\hat{\partial}_L\left(\frac{\hat{U}^{\mu\nu}_L\uarg}{ |\mathbf{x}|}\right),
\end{equation}
where, due to fulfillment of the harmonic gauge condition,
$\hat{U}^{\mu\nu}_L\uarg$'s
are not completely independent, and some of them are subject to more restrictions besides past-stationarity. Using
Eq. (\ref{B.1}),
the structure of
$h^{\mu\nu}_{\paren{1}}\finsarg$
can be written as
\begin{equation}\label{2.1.4.19}
 h^{\mu\nu}_{\paren{1}}\finsarg=\sum_{\ell=0}^{\infty}\sum_{m=0}^{\ell} A_{\ell m }\hat{n}^L \x^{-m-1}\;{}^{\paren{\ell-m}}\hat{U}^{\mu\nu}_L( t-\frac{ |\mathbf{x}|}{c}),
\end{equation}
where
${}^{\paren{\ell-m}}\hat{U}^{\mu\nu}_L\uarg$
means the
($\ell-m$)th
derivative of
$\hat{U}^{\mu\nu}_L\uarg$
with respect to its argument. All derivatives of
$\hat{U}^{\mu\nu}_L\uarg$
vanish at
$t \le -T$
due to its past-stationarity. Thus, it is convenient to rewrite
Eq. (\ref{2.1.4.19})
by considering the decomposition
$\hat{U}^{\mu\nu}_L\uarg = \hat{C}^{\mu\nu}_L + \hat{D}^{\mu\nu}_L\uarg$,
where
$\hat{C}^{\mu\nu}_L $
is always constant and equal to the value of
$\hat{U}^{\mu\nu}_L\uarg$
at
$t \le -T$,
and
$\hat{D}^{\mu\nu}_L\uarg$
is equal to zero when
$t \le -T$.
The result is
\begin{IEEEeqnarray}{rcl}\label{2.1.4.20}
 h^{\mu\nu}_{\paren{1}}\finsarg & = & \sum_{\ell=0}^{\infty}A_{\ell \ell }\hat{n}^L \x^{-\ell-1}\hat{C}^{\mu\nu}_L+\sum_{\ell=0}^{\infty}\sum_{m=0}^{\ell} A_{\ell m }\hat{n}^L \x^{-m-1}\;{}^{\paren{\ell-m}}\hat{D}^{\mu\nu}_L( t-\frac{ |\mathbf{x}|}{c}).\quad
\end{IEEEeqnarray}
After absorbing
$A_{\ell \ell}$
into the redefinition of
$\hat{C}^{\mu\nu}_L $
and
$A_{\ell m}$
into the redefinition of
${}^{\paren{\ell-m}}\hat{D}^{\mu\nu}_L\uarg$
and replacing the new
${}^{\paren{\ell-m}}\hat{D}^{\mu\nu}_L\uarg$
by
$\hat{F}^{\mu\nu}_{L,m}\uarg$,
we get
\begin{equation}\label{2.1.4.21}
 h^{\mu\nu}_{\paren{1}}\finsarg=\sum_{\ell=0}^{\infty}\hat{n}^L \x^{-\ell-1}\hat{C}^{\mu\nu}_L+\sum_{\ell=0}^{\infty}\sum_{m=0}^{\ell}\hat{n}^L\x^{-m-1}\hat{F}^{\mu\nu}_{L,m}\uarg,
\end{equation}
where the first term on the right-hand side is always stationary and the second term past-zero. We assume that
$\hat{C}^{\mu\nu}_L $
and
$\hat{F}^{\mu\nu}_{L,m}$
with
$\ell > \ell_{\mathrm{max}}$
are zero (as we will explain later, this assumption is technically inevitable). The always stationary term cannot become simpler than what it is, but the past-zero term can by using Taylor formula with integral remainder for
$\hat{F}^{\mu\nu}_{L,m}( u) = \hat{F}^{\mu\nu}_{L,m}\uarg$
about
$u_0 = t$,
or equivalently
$\x = 0$,
up to order
$N$,
namely
\begin{IEEEeqnarray}{rcl}\label{2.1.4.22}
\hat{F}^{\mu\nu}_{L,m}\uarg& =& \sum_{j=0}^{N}\frac{\left(-1\right)^j}{c^j j!}\x^j \;{}^{\paren{j}}\!\hat{F}^{\mu\nu}_{L,m}( t)  \nonumber \\  
&& \negmedspace {}+ \frac{\paren{-1}^{N+1}}{c^{N+1}}\x^{N+1}\int_{0}^{1}\frac{\paren{1-\alpha}^{N}}{N!}\;{}^{\paren{N+1}}\!\hat{F}^{\mu\nu}_{L,m}(t-\alpha\frac{\x}{c})\ud\alpha,
\end{IEEEeqnarray}
which, since
$\hat{F}^{\mu\nu}_{L,m}\uarg \in C^\infty \left(\mathbb{R}\right)$,
is valid for any
$u$
and hence any space-time coordinates, and moreover, for any arbitrary
$N$. We then get
\begin{IEEEeqnarray}{rcl}\label{2.1.4.23}
 h^{\mu\nu}_{\paren{1}}\finsarg & =& \sum_{\ell=0}^{\ell_{\mathrm{max}}}\hat{n}^L \x^{-\ell-1}\hat{C}^{\mu\nu}_L +\sum_{\ell=0}^{\ell_{\mathrm{max}}}\sum_{m=0}^{\ell}\sum_{j=0}^{N}\frac{\left(-1\right)^j}{c^j j!}\hat{n}^L\x^{j-m-1} \;{}^{\paren{j}}\!\hat{F}^{\mu\nu}_{L,m}( t) \nonumber \\  
&& \negmedspace {}+\sum_{\ell=0}^{\ell_{\mathrm{max}}}\sum_{m=0}^{\ell} \frac{\paren{-1}^{N+1}}{c^{N+1}}\hat{n}^L\x^{N-m}\int_{0}^{1}\frac{\paren{1-\alpha}^{N}}{N!}\;{}^{\paren{N+1}}\!\hat{F}^{\mu\nu}_{L,m}(t-\alpha\frac{\x}{c})\ud\alpha.\quad
\end{IEEEeqnarray}
Since
\begin{eqnarray}\label{2.1.4.24l}
\lefteqn{\lim_{\substack{|\mathbf{x}|\to 0 \\ t=\mathrm{const}}}\frac{ \frac{\paren{-1}^{N+1}}{c^{N+1}}\hat{n}^L\x^{N-m}\int_{0}^{1}\frac{\paren{1-\alpha}^{N}}{N!}\;{}^{\paren{N+1}}\!\hat{F}^{\mu\nu}_{L,m}(t-\alpha\frac{\x}{c})\ud\alpha}{\x^{N-m}}} \nonumber \\ 
&\!\!\!\!\!\!\!\!=\!\!\!\!\!\!\!\!& -\frac{\paren{-1}^{N+1}}{c^{N+1}}\hat{n}^L\lim_{\substack{|\mathbf{x}|\to 0 \\ t=\mathrm{const}}}\int_{0}^{1}\frac{\paren{1-\alpha}^{N}}{N!}\;{}^{\paren{N+1}}\!\hat{F}^{\mu\nu}_{L,m}(t-\alpha\frac{\x}{c})\ud\alpha \nonumber\\
&\!\!\!\!\!\!\!\!=\!\!\!\!\!\!\!\!& \frac{\paren{-1}^{N+1}}{c^{N+1}\paren{N+1}!}\hat{n}^L\; {}^{\paren{N+1}}\!\hat{F}^{\mu\nu}_{L,m}(t),
\end{eqnarray}
it is obvious the third term on the right-hand side of
Eq. (\ref{2.1.4.23})
is
$\sum_{\ell=0}^{\ell_{\mathrm{max}}}\sum_{m=0}^{\ell}O\!\paren{\x^{N-m}}$
as
$\x \to 0$
(``$O$'' is one of two Landau symbols, called big Oh).
Choosing
$N=N_0 + \ell_{\mathrm{max}} + 1$
where
$N_0 \in \mathbb{N}$,
it becomes
$\sum_{M=N_0+1}^{N_0 +\ell_{\mathrm{max}}+1}O\!\paren{\x^M}$
as
$\x \to 0$,
and hence, the structure of
$h^{\mu\nu}_{\paren{1}}\finsarg$
reads
\begin{equation}\label{2.1.4.26}
 h^{\mu\nu}_{\paren{1}}\finsarg =\sum_{\ell=0}^{\ell_{\mathrm{max}}}\hat{n}^L \x^{-\ell-1}\hat{C}^{\mu\nu}_{\paren{1}L} +\sum_{\ell=0}^{\ell_{\mathrm{max}}}\sum_{a=-\paren{\ell_{\mathrm{max}}+1}}^{N_0 +\ell_{\mathrm{max}}}\hat{n}^L\x^a \hat{G}^{\mu\nu}_{\paren{1}L,a}( t) + R^{\mu\nu}_{\paren{1}}\finsarg,
\end{equation}
where
$N_0$
is an arbitrary nonnegative integer,
$\hat{C}^{\mu\nu}_{\paren{1}L}$
a constant,
$\hat{G}^{\mu\nu}_{\paren{1}L,a}\left( t\right)$
and
$R^{\mu\nu}_{\paren{1}}\finsarg$
are past-zero functions, and
$R^{\mu\nu}_{\paren{1}}\finsarg=\sum_{N=N_0+1}^{N_0 +\ell_{\mathrm{max}}+1}O\!\paren{\x^N}$
as
$\x \to 0$.

\section{Second-Order Problem}

\subsection{Divergence of the Retarded Integral of $\Lambda^{\mu\nu}_{\paren{2}}\finsarg$}

$\Lambda^{\mu\nu}_{\paren{2}}$
is equal to
$N^{\mu\nu}\paren{h_{\paren{1}},h_{\paren{1}}}$
(we have omitted the indices of
$h^{\alpha\beta}_{\paren{1}}$
for notational simplicity). Since
$h^{\alpha\beta}_{\paren{1}}$
consists of an always stationary and a past-zero part, so too does
$\Lambda^{\mu\nu}_{\paren{2}}$.
We have
\begin{IEEEeqnarray}{rcl}\label{2.2.1.1}
 \Lambda^{\mu\nu}_{\paren{2}\mathrm{AS}} & = & N^{\mu\nu}(h_{\paren{1}\mathrm{AS}},h_{\paren{1}\mathrm{AS}}) \nonumber \\  
& = & N^{\mu\nu}(\;\sum_{\ell=0}^{\ell_{\mathrm{max}}}\hat{n}^L \x^{-\ell-1}\hat{C}_{\paren{1}L},\sum_{\ell=0}^{\ell_{\mathrm{max}}}\hat{n}^L \x^{-\ell-1}\hat{C}_{\paren{1}L}\;),
\end{IEEEeqnarray}
\begin{IEEEeqnarray}{rcl}\label{2.2.1.2}
\Lambda^{\mu\nu}_{\paren{2}\mathrm{PZ}}& =&N^{\mu\nu}(h_{\paren{1}\mathrm{AS}},h_{\paren{1}\mathrm{PZ}}) + N^{\mu\nu}(h_{\paren{1}\mathrm{PZ}},h_{\paren{1}\mathrm{AS}}) +N^{\mu\nu}(h_{\paren{1}\mathrm{PZ}},h_{\paren{1}\mathrm{PZ}})\nonumber \\
& =& N^{\mu\nu}(\;\sum_{\ell=0}^{\ell_{\mathrm{max}}}\hat{n}^L \x^{-\ell-1}\hat{C}_{\paren{1}L},\sum_{\ell=0}^{\ell_{\mathrm{max}}}\sum_{a=-\paren{\ell_{\mathrm{max}}+1}}^{N_0 +\ell_{\mathrm{max}}}\hat{n}^L\x^a \hat{G}_{\paren{1}L,a}\left( t\right)\;)\nonumber \\  
&& \negmedspace {} +  N^{\mu\nu}(\;\sum_{\ell=0}^{\ell_{\mathrm{max}}}\hat{n}^L \x^{-\ell-1}\hat{C}_{\paren{1}L}, R_{\paren{1}}\finsarg\;)\nonumber \\  
&& \negmedspace {}+N^{\mu\nu}(\;\sum_{\ell=0}^{\ell_{\mathrm{max}}}\sum_{a=-\paren{\ell_{\mathrm{max}}+1}}^{N_0 +\ell_{\mathrm{max}}}\hat{n}^L\x^a \hat{G}_{\paren{1}L,a}\left( t\right),\sum_{\ell=0}^{\ell_{\mathrm{max}}}\hat{n}^L \x^{-\ell-1}\hat{C}_{\paren{1}L}\;)\nonumber \\  
&& \negmedspace {} + N^{\mu\nu}(\;R_{\paren{1}}\finsarg,\sum_{\ell=0}^{\ell_{\mathrm{max}}}\hat{n}^L \x^{-\ell-1}\hat{C}_{\paren{1}L}\;)\nonumber \\  
&& \negmedspace {} +N^{\mu\nu}(\;\sum_{\ell=0}^{\ell_{\mathrm{max}}}\sum_{a=-\paren{\ell_{\mathrm{max}}+1}}^{N_0 +\ell_{\mathrm{max}}}\hat{n}^L\x^a \hat{G}_{\paren{1}L,a}\left( t\right),\sum_{\ell=0}^{\ell_{\mathrm{max}}}\sum_{a=-\paren{\ell_{\mathrm{max}}+1}}^{N_0 +\ell_{\mathrm{max}}}\hat{n}^L\x^a \hat{G}_{\paren{1}L,a}\left( t\right)\;)\nonumber \\  
&& \negmedspace {}+ N^{\mu\nu}(\;\sum_{\ell=0}^{\ell_{\mathrm{max}}}\sum_{a=-\paren{\ell_{\mathrm{max}}+1}}^{N_0 +\ell_{\mathrm{max}}}\hat{n}^L\x^a \hat{G}_{\paren{1}L,a}\left( t\right), R_{\paren{1}}\finsarg\;)\nonumber \\  
&& \negmedspace {}+ N^{\mu\nu}(\;R_{\paren{1}}\finsarg, \sum_{\ell=0}^{\ell_{\mathrm{max}}}\sum_{a=-\paren{\ell_{\mathrm{max}}+1}}^{N_0 +\ell_{\mathrm{max}}}\hat{n}^L\x^a \hat{G}_{\paren{1}L,a}\left( t\right)\;)\nonumber \\  
&& \negmedspace {}+ N^{\mu\nu}(\;R_{\paren{1}}\finsarg, R_{\paren{1}}\finsarg\;).
\end{IEEEeqnarray}
Using
Eq. (\ref{1.2.21}),
the structure of
$\Lambda^{\mu\nu}_{\paren{2}\mathrm{AS}}\xarg$
can be obtained as
$\sum_{q=0}^{q_{\mathrm{max}}}\sum_{a=a_{\mathrm{min}}}^{-4} n^Q \x^a C^{\mu\nu}_{Q,a}$
where
$q_{\mathrm{max}}$
and
$a_{\mathrm{min}}$
depend on
$\ell_{\mathrm{max}}$.
$C^{\mu\nu}_{Q,a}$,
which is a constant, is not STF with respect to the indices
$Q$,
and hence, in this structure
$n^Q$
cannot be replaced by
$\hat{n}^Q$
with the use of
Eq. (\ref{B.2}).
However, since
$n^Q$
is a well-behaved function of
$\theta$
and
$\varphi$,
as explained in Appendix B, it can be expanded in STF tensors
$\hat{n}^P$
(namely by using
Eq. (\ref{B.11})).
Therefore, the structure of
$\Lambda^{\mu\nu}_{\paren{2}\mathrm{AS}}\xarg$
can be rewritten as
$\sum_{p=0}^{p_{\mathrm{max}}}\sum_{a=a_{\mathrm{min}}}^{-4}\hat{n}^P\x^a  \hat{C}'^{\;\mu\nu}_{P,a}$.

The structure of
$\Lambda^{\mu\nu}_{\paren{2}\mathrm{PZ}}\finsarg$
consists of two parts. The first, third and fifth terms appearing on the right-hand side of the second equality in
Eq. (\ref{2.2.1.2})
are of the form
$\sum_{q=0}^{q_{\mathrm{max}}}\sum_{a=a'_{\mathrm{min}}}^{a'_{\mathrm{max}}}n^Q\x^a E^{\mu\nu}_{Q,a}\targ$
which can be converted into
$\sum_{p=0}^{p_{\mathrm{max}}}\sum_{a=a'_{\mathrm{min}}}^{a'_{\mathrm{max}}}\hat{n}^P\x^a\hat{F}^{\mu\nu}_{P,a}\targ$
by reasoning analogous to that used to rewrite the structure of
$\Lambda^{\mu\nu}_{\paren{2}\mathrm{AS}}\xarg$.
To determine the structure of the remaining terms of
$\Lambda^{\mu\nu}_{\paren{2}\mathrm{PZ}}\finsarg$,
first we must compute the time and the space derivatives of
$R_{\paren{1}}\finsarg$.
Differentiating the explicit form of
$R_{\paren{1}}\finsarg$
given in the third term on the right-hand side of
Eq. (\ref{2.1.4.23}),
one finds that, as
$\x \to 0$,
the time derivative of
$R_{\paren{1}}\finsarg$
is still
$\sum_{N=N_0+1}^{N_0 +\ell_{\mathrm{max}}+1}O\!\paren{\x^N}$,
but the spatial derivatives are
$\sum_{N=N_0}^{N_0 +\ell_{\mathrm{max}}+1}O\!\paren{\x^N}$.
Therefore, the second, fourth, sixth and seventh terms are
$\sum_{N=M_{\mathrm{min}}}^{M_{\mathrm{max}}}O\!\paren{\x^N}$
as
$\x \to 0$,
where due to multiplication of the terms containing negative power of
$\x$
and
$R_{\paren{1}}\finsarg$,
the lower limit of the summation is negative unless we choose
$N_0$
to be a large enough number; we do so (we will discuss this later). Furthermore, it is obvious that the eighth and last term is
$\sum_{N=M'_{\mathrm{min}}}^{M'_{\mathrm{max}}}O\!\paren{\x^N}$
when
$\x \to 0$
(where, evidently, the lower limit of the summation is nonnegative). All in all, the structure of
$\Lambda^{\mu\nu}_{\paren{2}}\finsarg$
reads
\begin{equation}\label{2.2.1.3}
\Lambda^{\mu\nu}_{\paren{2}}\finsarg =\sum_{\ell=0}^{\ell_{\mathrm{max}}}\sum_{k=k_{\mathrm{min}}}^{-4}\hat{n}^L \x^k\hat{C}^{\mu\nu}_{\paren{2}L,k} +\sum_{q=0}^{q_{\mathrm{max}}}\sum_{a=a_{\mathrm{min}}}^{a_{\mathrm{max}}}\hat{n}^Q\x^a\hat{F}^{\mu\nu}_{\paren{2}Q,a}\targ + R^{\mu\nu}_{\paren{2}}\finsarg,
\end{equation}
where
$a_{\mathrm{max}} > 0$,
$\hat{C}^{\mu\nu}_{\paren{2}L,k}$
is a constant,
$\hat{F}^{\mu\nu}_{\paren{2}Q,a}\targ$
and
$R^{\mu\nu}_{\paren{2}}\finsarg$
are past-zero, and
$R^{\mu\nu}_{\paren{2}}\finsarg$
is
$\sum_{N=N_{\mathrm{min}}}^{N_{\mathrm{max}}}O\!\paren{\x^N}$
with
$N_{\mathrm{min}}\ge 0$
as
$\x \to 0$.\footnote{In addition, a crucial, albeit obvious, point is that
$\hat{C}^{\mu\nu}_{\paren{2}L,k}$
and
$\hat{F}^{\mu\nu}_{\paren{2}Q,a}\targ$
appearing in
Eq. (\ref{2.2.1.3})
are zero unless there exist a particular relation between
$k$
and
$\ell$
as well as
$a$
and
$q$.
These relations will play key roles in the future.}

Using
Eq. (\ref{2.2.1.3}),
the retarded integral of
$\Lambda^{\mu\nu}_{\paren{2}}\finsarg$
is given by
\begin{IEEEeqnarray}{rcl}\label{2.2.1.4}
\retint{\Lambda^{\mu\nu}_{\paren{2}}\retarg} & = &  \poisint{\sum_{\ell=0}^{\ell_{\mathrm{max}}}\sum_{k=k_{\mathrm{min}}}^{-4}\hat{n}'^L \xp^k\hat{C}^{\mu\nu}_{\paren{2}L,k}} 
 \nonumber \\  
&& \negmedspace {}+\retint{\sum_{q=0}^{q_{\mathrm{max}}}\sum_{a=a_{\mathrm{min}}}^{a_{\mathrm{max}}}\hat{n}'^Q\xp^a\hat{F}^{\mu\nu}_{\paren{2}Q,a}\paren{t'} }  \nonumber \\  
&& \negmedspace {} + \retint{R^{\mu\nu}_{\paren{2}}\retarg}.
\end{IEEEeqnarray}
If
$\retint{\Lambda^{\mu\nu}_{\paren{2}}\retarg}$
converges, it is a particular solution of the equation
$\Box h^{\mu\nu}_{\paren{2}}\finsarg = \Lambda^{\mu\nu}_{\paren{2}}\finsarg$.
To decide whether
$\retint{\Lambda^{\mu\nu}_{\paren{2}}\retarg}$
is convergent or not, first it is necessary to consider two following theorems:

\persiantheorem{2.1}
Suppose that
$q \in \mathbb{N}$,
$a \in \mathbb{R}$,
$p  \in \mathbb{R}^{\ge 0}$,
$g\finsarg$
is a bounded function in
$\mathbb{R}^4$,
and
$a > -q-3$.
Then
\begin{equation}\label{2.2.1.5}
I_< = \nearint{\hat{n}'^Q \xp^a \paren{\ln{\xp}}^p g\sinsarg},
\end{equation}
where
$0 < \mathcal{R} <\infty$
converges when
$\x > 0$.

\persianproof
First, we assume
$p=0$
and
$g\finsarg=1$.
Then
$I_<$
becomes
\begin{equation}\label{2.2.1.6}
I_< = \nearint{\hat{n}'^Q \xp^a }.
\end{equation}
Since for
$\xp < \x$
we have
\begin{equation}\label{2.2.1.7}
\frac{1}{|\mathbf{x}-\mathbf{x'}|}= \frac{1}{\x}\sum_{\ell=0}^{\infty}\frac{4\pi}{2\ell+1}\paren{\frac{\xp}{\x}}^\ell \sum_{m=-\ell}^{\ell}\left[\ylmp\right]^*\ylm,
\end{equation}
and for
$\x < \xp$
\begin{equation}\label{2.2.1.8}
\frac{1}{|\mathbf{x}-\mathbf{x'}|}= \frac{1}{\xp}\sum_{\ell=0}^{\infty}\frac{4\pi}{2\ell+1}\paren{\frac{\x}{\xp}}^\ell \sum_{m=-\ell}^{\ell}\left[\ylm\right]^*\ylmp,
\end{equation}
at
$\x < \mathcal{R}$,
$I_<$
given in
Eq. (\ref{2.2.1.6})
can be rewritten as
\begin{IEEEeqnarray}{rcl}\label{2.2.1.9}
I_< & = & -\frac{1}{4\pi}\int_{|\mathbf{x'}|<\x} \frac{\hat{n}'^Q \xp^a }{|\mathbf{x}-\mathbf{x'}|}\ud^3\mathbf{x'} -\frac{1}{4\pi}\int_{\x<|\mathbf{x'}|<\mathcal{R}} \frac{\hat{n}'^Q \xp^a}{|\mathbf{x}-\mathbf{x'}|}\ud^3\mathbf{x'} \nonumber \\  
& = & \sum_{\ell=0}^{\infty}\frac{-1}{2\ell+1}\frac{1}{\x^{\ell+1}}\int_{0}^{\x}\xp^{a+\ell+2}\ud\xp\sum_{m=-\ell}^{\ell}\ylm \int \hat{n}'^Q\left[\ylmp\right]^*\ud\Omega' \nonumber \\  
&& \negmedspace {} + \sum_{\ell=0}^{\infty}\frac{-1}{2\ell+1}\x^{\ell}\int_{\x}^{\mathcal{R}}\xp^{a-\ell+1}\ud\xp\sum_{m=-\ell}^{\ell}\left[\ylm\right]^* \int \hat{n}'^Q\ylmp\ud\Omega'. \nonumber \\ 
\end{IEEEeqnarray}
Using
Eq. (\ref{B.3}),
this becomes
\begin{equation}\label{2.2.1.10}
I_< = -\frac{1}{2q+1}\frac{\hat{n}^Q}{\x^{q+1}}\int_{0}^{\x}\xp^{a+q+2}\ud\xp - \frac{1}{2q+1}\hat{n}^Q \x^q \int_{\x}^{\mathcal{R}}\xp^{a-q+1}\ud\xp.
\end{equation}
The integrals appearing in the above equation are evaluated as follows:
\begin{equation}\label{2.2.1.11}
\int_{0}^{\x}\xp^{a+q+2}\ud\xp =
\begin{cases}
\frac{\xp^{a+q+3}}{a+q+3}\Big|^{\x}_{0}, &\qquad a + q\neq -3,\\
\\
\ln{\xp}\Big|^{\x}_{0}, &\qquad a + q = -3,
\end{cases}
\end{equation}
\begin{equation}\label{2.2.1.12}
\int_{\x}^{\mathcal{R}}\xp^{a-q+1}\ud\xp=
\begin{cases}
\frac{\xp^{a-q+2}}{a-q+2}\Big|_{\x}^{\mathcal{R}}, &\qquad a - q\neq -2,\\
\\
\ln{\xp}\Big|_{\x}^{\mathcal{R}}, &\qquad a - q = -2.
\end{cases}
\end{equation}
Since we have assumed
$\x < \mathcal{R}$,
and the theorem is stated for
$\x>0$,
it is obvious that, despite the fact that the integral in
Eq. (\ref{2.2.1.12})
is convergent regardless of the values of
$a$
and
$q$,
the integral in
Eq. (\ref{2.2.1.11})
is convergent only if
$a+q+3>0$.
Thus, for
$a>-q-3$
and
$0<\x<\mathcal{R}$,
we have
\begin{equation}\label{2.2.1.13}
I_<=
\begin{cases}
- \frac{1}{2q+1}\hat{n}^Q \left[ \frac{\x^{a+2}}{a+q+3}-\frac{\x^{a+2}}{a-q+2}+\x^q\frac{\mathcal{R}^{a-q+2}}{a-q+2}\right], &\qquad a-q\neq-2,\\
\\
- \frac{1}{2q+1}\hat{n}^Q \left[ \frac{\x^{a+2}}{a+q+3} + \x^q \ln{\frac{\mathcal{R}}{\x}}\right], &\qquad a - q = -2.
\end{cases}
\end{equation}
On the other hand, when
$\x>\mathcal{R}$
(implying
$\xp<\x$),
$I_<$
given in
Eq. (\ref{2.2.1.6})
becomes
\begin{IEEEeqnarray}{rcl}\label{2.2.1.14}
I_< =  \sum_{\ell=0}^{\infty}\frac{-1}{2\ell+1}\frac{1}{\x^{\ell+1}}\int_{0}^{\mathcal{R}}\xp^{a+\ell+2}\ud\xp\sum_{m=-\ell}^{\ell}\ylm \int \hat{n}'^Q\left[\ylmp\right]^*\ud\Omega'.\nonumber \\
\end{IEEEeqnarray}
Using
Eq. (\ref{B.3}),
this can be written as
\begin{equation}\label{2.2.1.15}
I_< =
\begin{cases}
 -\frac{1}{2q+1}\frac{\hat{n}^Q}{\x^{q+1}}\frac{\xp^{a+q+3}}{a+q+3}\Big|^{\mathcal{R}}_{0}, &\qquad a + q\neq -3,\\
\\
-\frac{1}{2q+1}\frac{\hat{n}^Q}{\x^{q+1}}\ln{\xp}\Big|^{\mathcal{R}}_{0}, &\qquad a + q = -3.
\end{cases}
\end{equation}
As it can be seen,
$I_<$
is convergent for
$a+q+3>0$
and equal to
\begin{equation}\label{2.2.1.16}
I_< =   -\frac{1}{2q+1}\frac{\hat{n}^Q}{\x^{q+1}}\frac{\mathcal{R}^{a+q+3}}{a+q+3}.
\end{equation}
We are now in a position to examine the convergence of the general case.
$I_<$
can be expressed as
\begin{IEEEeqnarray}{rcl}\label{2.2.1.17}
I_< & = &\nearint{\hat{n}'^Q\xp^a\paren{\ln{\xp}}^p g\sinsarg} \nonumber \\  
& = &  \nearint{\hat{n}'^Q\xp^{a-\varepsilon}\xp^\varepsilon\paren{\ln{\xp}}^p g\sinsarg},
\end{IEEEeqnarray}
where
$\varepsilon \ll 1$.
Based on the above results, when
$\x > 0$,
$\nearint{\hat{n}'^Q\xp^{a-\varepsilon}}$
is convergent for
$a-\varepsilon>-q-3$.
Further,
$\xp^\varepsilon \paren{\ln{\xp}}^p g\sinsarg$
is bounded at
$\xp<\mathcal{R}$,
even as
$\xp \to 0$
because
\begin{equation}\label{2.2.1.18}
\lim_{\xp \to 0}\xp^\varepsilon \paren{\ln{\xp}}^p =0,\qquad \forall \varepsilon>0 \; \text{and}\;\forall p \ge 0;
\end{equation}
hence, when
$\x>0$,
$I_<$
is convergent for
$a-\varepsilon>-q-3$,
and since
$\varepsilon \ll 1$,
it can be concluded that the convergence takes place when
$a>-q-3$.
\hfill
$\blacksquare$

\persiantheorem{2.2}
Suppose that
$q \in \mathbb{N}$,
$a \in \mathbb{R}$,
$p  \in \mathbb{R}^{\ge 0}$,
$g\finsarg$
is a bounded function in
$\mathbb{R}^4$,
and
$a < q-2$.
Then
\begin{equation}\label{2.2.1.19}
I_> = \farint{\hat{n}'^Q \xp^a \paren{\ln{\xp}}^p g\sinsarg},
\end{equation}
where
$0 < \mathcal{R} <\infty$
converges when
$\x < \infty$.

\persianproof
First, we assume
$p=0$
and
$g\finsarg=1$.
Then
$I_>$
becomes
\begin{equation}\label{2.2.1.20}
I_> = \farint{\hat{n}'^Q \xp^a }.
\end{equation}
Using
Eqs. (\ref{2.2.1.7}) and (\ref{2.2.1.8}),
at
$\x > \mathcal{R}$,
$I_>$ given in
Eq. (\ref{2.2.1.20})
can be rewritten as
\begin{IEEEeqnarray}{rcl}\label{2.2.1.21}
I_>& = &  -\frac{1}{4\pi}\int_{\mathcal{R}<|\mathbf{x'}|<\x} \frac{\hat{n}'^Q \xp^a}{|\mathbf{x}-\mathbf{x'}|}\ud^3\mathbf{x'} -\frac{1}{4\pi}\int_{\x<|\mathbf{x'}|} \frac{\hat{n}'^Q \xp^a }{|\mathbf{x}-\mathbf{x'}|}\ud^3\mathbf{x'} \nonumber \\  
& = & \sum_{\ell=0}^{\infty}\frac{-1}{2\ell+1}\frac{1}{\x^{\ell+1}}\int_{\mathcal{R}}^{\x}\xp^{a+\ell+2}\ud\xp\sum_{m=-\ell}^{\ell}\ylm \int \hat{n}'^Q\left[\ylmp\right]^*\ud\Omega' \nonumber \\  
&& \negmedspace {} + \sum_{\ell=0}^{\infty}\frac{-1}{2\ell+1}\x^{\ell}\int_{\x}^{\infty}\xp^{a-\ell+1}\ud\xp\sum_{m=-\ell}^{\ell}\left[\ylm\right]^* \int \hat{n}'^Q\ylmp\ud\Omega'.\nonumber \\ 
\end{IEEEeqnarray}
Now, by means of
Eq. (\ref{B.3}),
we can write
\begin{equation}\label{2.2.1.22}
I_> = -\frac{1}{2q+1}\frac{\hat{n}^Q}{\x^{q+1}}\int_{\mathcal{R}}^{\x}\xp^{a+q+2}\ud\xp - \frac{1}{2q+1}\hat{n}^Q \x^q \int_{\x}^{\infty}\xp^{a-q+1}\ud\xp.
\end{equation}
The integrals appearing in the above equation are evaluated as follows:
\begin{equation}\label{2.2.1.23}
\int_{\mathcal{R}}^{\x}\xp^{a+q+2}\ud\xp =
\begin{cases}
\frac{\xp^{a+q+3}}{a+q+3}\Big|^{\x}_{\mathcal{R}} &\qquad a + q\neq -3,\\
\\
\ln{\xp}\Big|^{\x}_{\mathcal{R}}, &\qquad a + q = -3,
\end{cases}
\end{equation}
\begin{equation}\label{2.2.1.24}
\int_{\x}^{\infty}\xp^{a-q+1}\ud\xp=
\begin{cases}
\frac{\xp^{a-q+2}}{a-q+2}\Big|_{\x}^{\infty}, &\qquad a - q\neq -2,\\
\\
\ln{\xp}\Big|_{\x}^{\infty}, &\qquad a - q = -2.
\end{cases}
\end{equation}
Since we have assumed
$\x > \mathcal{R}$,
and the theorem is stated for
$\x<\infty$,
it is evident that, in spite of the fact that the integral in
Eq. (\ref{2.2.1.23})
is convergent irrespective of the values of
$a$
and
$q$,
the integral in
Eq. (\ref{2.2.1.24})
converges on if
$a-q+2<0$.
Therefore, for
$a<q-2$
and
$\mathcal{R}<\x<\infty$,
we have
\begin{equation}\label{2.2.1.25}
I_>=
\begin{cases}
- \frac{1}{2q+1}\hat{n}^Q \left[-\frac{\x^{a+2}}{a-q+2}+ \frac{\x^{a+2}}{a+q+3}-\frac{1}{\x^{q+1}}\frac{\mathcal{R}^{a+q+3}}{a+q+3}\right], &\qquad a+q\neq-3,\\
\\
- \frac{1}{2q+1}\hat{n}^Q \left[- \frac{\x^{a+2}}{a-q+2} + \frac{1}{\x^{q+1}} \ln{\frac{\x}{\mathcal{R}}}\right], &\qquad a + q = -3.
\end{cases}
\end{equation}
On the other hand, when
$\x<\mathcal{R}$
(implying
$\x<\xp$),
$I_>$
given in
Eq. (\ref{2.2.1.20})
becomes
\begin{IEEEeqnarray}{rcl}\label{2.2.1.26}
I_> =  \sum_{\ell=0}^{\infty}\frac{-1}{2\ell+1}\x^\ell \int_{\mathcal{R}}^{\infty}\xp^{a-\ell+1}\ud\xp\sum_{m=-\ell}^{\ell}\left[\ylm\right]^* \int \hat{n}'^Q\ylmp\ud\Omega',\nonumber \\  
\end{IEEEeqnarray}
which with the use of
Eq. (\ref{B.3})
yields
\begin{equation}\label{2.2.1.27}
I_> =
\begin{cases}
 -\frac{1}{2q+1}\hat{n}^Q\x^q\frac{\xp^{a-q+2}}{a-q+2}\Big|^{\infty}_{\mathcal{R}}, &\qquad a - q\neq -2,\\
\\
-\frac{1}{2q+1}\hat{n}^Q\x^q\ln{\xp}\Big|^{\infty}_{\mathcal{R}}, &\qquad a - q = -2.
\end{cases}
\end{equation}
As it can be seen,
$I_>$
is convergent for
$a-q+2<0$
and equals
\begin{equation}\label{2.2.1.28}
I_> =  \frac{1}{2q+1}\hat{n}^Q\x^q\frac{\mathcal{R}^{a-q+2}}{a-q+2}
\end{equation}
Now we are in a position to examine the convergence of the general case.
$I_>$
can be expressed as
\begin{IEEEeqnarray}{rcl}\label{2.2.1.29}
I_> & = &\farint{\hat{n}'^Q\xp^a\paren{\ln{\xp}}^p g\sinsarg} \nonumber \\  
& = &  \farint{\hat{n}'^Q\xp^{a+\varepsilon}\xp^{-\varepsilon}\paren{\ln{\xp}}^p g\sinsarg},
\end{IEEEeqnarray}
where
$\varepsilon \ll 1$.
Based on the above results, when
$\x<\infty$,
$\farint{\hat{n}'^Q\xp^{a-\varepsilon}}$
is convergent for
$a+\varepsilon<q-2$.
Furthermore,
$\xp^{-\varepsilon} \paren{\ln{\xp}}^p g\sinsarg$
is bounded at
$\mathcal{R}<\xp$,
even as
$\xp \to \infty$
because
\begin{equation}\label{2.2.1.30}
\lim_{\xp \to \infty}\xp^{-\varepsilon} \paren{\ln{\xp}}^p =0,\qquad \forall \varepsilon>0 \;\text{and} \;\forall p \ge 0;
\end{equation}
thus, when
$\x<\infty$,
$I_>$
is convergent for
$a+\varepsilon<q-2$,
and since
$\varepsilon \ll 1$,
we can say that the convergence takes place when
$a<q-2$.
\hfill
$\blacksquare$

Taking Theorems 2.1 and 2.2 into account, it is evident that, when
$0<\x<\infty$,
\begin{equation}
\poisint{\hat{n}'^Q\xp^a\paren{\ln{\xp}}^p g\sinsarg}=\wholeint{\hat{n}'^Q\xp^a\paren{\ln{\xp}}^p g\sinsarg}\nonumber
\end{equation}
is convergent for
$-q-3<a<q-2$.
In particular, by summing
Eqs. (\ref{2.2.1.13}) and (\ref{2.2.1.28})
in the region
$0<\x<\mathcal{R}$,
and
Eqs. (\ref{2.2.1.16}) and (\ref{2.2.1.25})
in the region
$\mathcal{R}<\x<\infty$,
we get
\begin{equation}\label{2.2.1.32}
\poisint{\hat{n}'^Q \xp^a}=\frac{\hat{n}^Q \x^{a+2}}{\paren{a-q+2}\paren{a+q+3}},
\end{equation}
and since
$\Delta\poisint{g\sinsarg}=g\finsarg$,
we reach
\begin{equation}\label{2.2.1.33}
\Delta\left[\frac{\hat{n}^Q \x^{a+2}}{\paren{a-q+2}\paren{a+q+3}}\right]=\hat{n}^Q \x^a,
\end{equation}
which is valid not only for
$-q-3<a<q-2$,
but  also for any
$a \in \mathbb{R}$
and
$q \in \mathbb{N}$
subject to
$a-q+2\neq0$
and
$a+q+3\neq0$.
Therefore, we have:\\
1. If
$a-q+2\neq0$
and
$a+q+3\neq0$,
$\frac{\hat{n}^Q \x^{a+2}}{\paren{a-q+2}\paren{a+q+3}}$
is a particular solution of the equation
$\Delta f\finsarg=\hat{n}^Q \x^a$.\\
2. If
$-q-3<a<q-2$,
$\poisint{\hat{n}'^Q \xp^a}$
converges. Further,
$\lim_{\x \to \infty}\poisint{\hat{n}'^Q \xp^a}\finsarg=0$
 if
$a<-2$.
Thus, if
$-q-3<a<q-2$,
the solution of the equation
$\Delta f\finsarg=\hat{n}^Q \x^a$
subject to
$\lim_{\x \to \infty} f\finsarg=0$
is
$\poisint{\hat{n}'^Q \xp^a}$.
Otherwise,
$\poisint{\hat{n}'^Q \xp^a}$,
when converges, is just a particular solution for the equation
$\Delta f\finsarg=\hat{n}^Q \x^a$.

We now examine the convergence of
$\retint{\Lambda^{\mu\nu}_{\paren{2}}\retarg}$
as follows:\\
1.
$R^{\mu\nu}_{\paren{2}}\finsarg$
is a past-zero function. Therefore, by taking
$q=0$,
$a=0$,
$p=0$
and
$g\sinsarg=R^{\mu\nu}_{\paren{2}}\retardedarg$
in Theorem 2.1, we can conclude the convergence of
$\retint{R^{\mu\nu}_{\paren{2}}\retarg}$.\\
2.
$\hat{F}^{\mu\nu}_{\paren{2}Q,a}\targ$
is a past-zero function. We take
$g\sinsarg=\hat{F}^{\mu\nu}_{\paren{2}Q,a}\paren{t-\frac{|\mathbf{x}-\mathbf{x'}|}{c}}$.
Since in general
$\retint{\sum_{q=0}^{q_{\mathrm{max}}}\sum_{a=a_{\mathrm{min}}}^{a_{\mathrm{max}}}\hat{n}'^Q\xp^a\hat{F}^{\mu\nu}_{\paren{2}Q,a}\paren{t'} }  \nonumber$
has
$a$'s
and
$q$'s
that don't fulfill
$a>-q-3$,
the divergence of this term can be deduced from Theorem 2.1.\\
3.
$\hat{C}^{\mu\nu}_{\paren{2}L,k}$
is a constant. In general,
$k$'s
and
$\ell$'s
appearing in the Poisson integral
$\poisint{\sum_{\ell=0}^{\ell_{\mathrm{max}}}\sum_{k=k_{\mathrm{min}}}^{-4}\hat{n}'^L \xp^k\hat{C}^{\mu\nu}_{\paren{2}L,k}} $
don't satisfy
$-\ell-3<k<\ell-2$.
Hence, based on Theorems 2.1 and 2.2, this term is divergent.\\
Taking 1, 2 and 3 into account,
$\retint{\Lambda^{\mu\nu}_{\paren{2}}\retarg}$
is divergent and thus cannot be a particualr solution to the equation
$\Box h^{\mu\nu}_{\paren{2}}\finsarg=\Lambda^{\mu\nu}_{\paren{2}}\finsarg$.
In order to find a particular solution, another method must be employed.

\subsection{Particular Solution of the Relaxed Einstein Field Equation in the Second-Order Problem and Its Structure}

To provide a particular solution of the equation
$\Box h^{\mu\nu}_{\paren{2}}\finsarg=\Lambda^{\mu\nu}_{\paren{2}}\finsarg$,
we first need to establish the following theorem:

\persiantheorem{2.3}
Consider the following equation:
\begin{equation}\label{2.2.2.1}
\mathbf{L}_i f\finsarg=g\finsarg, \qquad\text{where}\; \mathbf{L}_i=
\begin{cases}
\Delta, &\qquad i=1,\\
\Box, &\qquad i=2.
\end{cases}
\end{equation}
Suppose that
\begin{equation}\label{2.2.2.2}
\mathbf{L}_i^{-1}\left[\paren{\frac{\xp}{r_0}}^B g\sinsarg\right]=
\begin{cases}
\poisint{\paren{\frac{\xp}{r_0}}^B g\sinsarg}, &\qquad i=1,\\ \\
\retint{\paren{\frac{\xp}{r_0}}^B g\retarg}, &\qquad i=2,
\end{cases}\nonumber
\end{equation}
where
$B$
is a complex number and
$r_0$
an arbitrary constant, is analytic in some original domain and can be analytically continued down to a punctured neighborhood of
$B=0$.
Then a particular solution to
Eq. (\ref{2.2.2.1})
is given by
\begin{equation}\label{2.2.2.3}
f_{\mathrm{part}}\finsarg=\mathop{\mathrm{FP}}\limits_{B=0}\mathrm{A}\mathbf{L}_i^{-1}\left[\paren{\frac{\xp}{r_0}}^B g\sinsarg\right],
\end{equation}
where
$\mathrm{A}\mathbf{L}_i^{-1}\left[\paren{\frac{\xp}{r_0}}^B g\sinsarg\right]$
denotes the analytic continuation of
$\mathbf{L}_i^{-1}\left[\paren{\frac{\xp}{r_0}}^B g\sinsarg\right]$
and
$\mathop{\mathrm{FP}}\limits_{B=0}\mathrm{A}\mathbf{L}_i^{-1}\left[\paren{\frac{\xp}{r_0}}^B g\sinsarg\right]$
the coefficient of zeroth power of
$B$
in Laurent expansion of
$\mathrm{A}\mathbf{L}_i^{-1}\left[\paren{\frac{\xp}{r_0}}^B g\sinsarg\right]$
about
$B=0$.

\persianproof
By assumption, the analytic continuation of
$\mathbf{L}_i^{-1}\left[\paren{\frac{\xp}{r_0}}^B g\sinsarg\right]$
is defined in a punctured neighborhood of
$B=0$,
and hence, by using Laurent expansion,
in this region we can write
\begin{equation}\label{2.2.2.4}
\mathrm{A}\mathbf{L}_i^{-1}\left[\paren{\frac{\xp}{r_0}}^B g\sinsarg\right]=\sum_{j=-|j_0|}^{\infty}C_j\finsarg B^j,
\end{equation}
where
$j_0$
is an integer and can be infinite. Applying the operator
$\mathbf{L}_i$
to both sides of the above equation, we get
\begin{equation}\label{2.2.2.5}
\mathbf{L}_i\mathrm{A}\mathbf{L}_i^{-1}\left[\paren{\frac{\xp}{r_0}}^B g\sinsarg\right]=\sum_{j=-|j_0|}^{\infty}\mathbf{L}_i C_j\finsarg B^j.
\end{equation}
In the original domain of analyticity of
$\mathbf{L}_i^{-1}\left[\paren{\frac{\xp}{r_0}}^B g\sinsarg\right]$,
the left-hand side of
Eq. (\ref{2.2.2.5})
reads
\begin{equation}\label{2.2.2.6}
\mathbf{L}_i\mathbf{L}_i^{-1}\left[\paren{\frac{\xp}{r_0}}^B g\sinsarg\right]=\paren{\frac{\x}{r_0}}^B g\finsarg.
\end{equation}
Now by virtue of identity theorem for analytic functions, in the region where the analytic continuations of
$\mathbf{L}_i^{-1}\left[\paren{\frac{\xp}{r_0}}^B g\sinsarg\right]$
and
$\paren{\frac{\x}{r_0}}^B g\finsarg$
are both defined, we can write
\begin{equation}\label{2.2.2.7}
\mathbf{L}_i\mathrm{A}\mathbf{L}_i^{-1}\left[\paren{\frac{\xp}{r_0}}^B g\sinsarg\right]=\mathrm{A}\left[\paren{\frac{\x}{r_0}}^B g\finsarg\right].
\end{equation}
$\paren{\frac{\x}{r_0}}^B$
is an entire function; therefore,
$\mathrm{A}\left[\paren{\frac{\x}{r_0}}^B g\finsarg\right]=\paren{\frac{\x}{r_0}}^B g\finsarg$.
In a neighborhood of
$B=0$,
$\paren{\frac{\x}{r_0}}^B g\finsarg$
has the following Taylor expansion:
\begin{equation}\label{2.2.2.8}
\paren{\frac{\x}{r_0}}^B g\finsarg=e^{B\ln{\frac{\x}{r_0}}}g\finsarg=g\finsarg\sum_{j=0}^{\infty}\frac{1}{j!}\paren{\ln{\frac{\x}{r_0}}}^j B^j;
\end{equation}
thus, combining
Eqs. (\ref{2.2.2.5}) and (\ref{2.2.2.8}),
we get
\begin{equation}\label{2.2.2.9}
\mathbf{L}_i C_j\finsarg=
\begin{cases}
0, &\qquad j<0,\\
\frac{1}{j!}\paren{\ln{\frac{\x}{r_0}}}^j g\finsarg,&\qquad j \ge 0.
\end{cases}
\end{equation}
In particular, for
$j=0$
we find
\begin{equation}\label{2.2.2.10}
\mathbf{L}_i C_0\finsarg=g\finsarg,
\end{equation}
where
$C_0\finsarg$
is nothing but
$\mathop{\mathrm{FP}}\limits_{B=0}\mathrm{A}\mathbf{L}_i^{-1}\left[\paren{\frac{\xp}{r_0}}^B g\sinsarg\right]$.
\hfill
$\blacksquare$

We now make two important remarks in regard to Theorem 2.3:\\
1. We know that the equation
$\mathbf{L}_i f\finsarg=g\finsarg$
has an infinite number of particular solutions. Then two natural questions about
$\mathop{\mathrm{FP}}\limits_{B=0}\mathrm{A}\mathbf{L}_i^{-1}\left[\paren{\frac{\xp}{r_0}}^B g\sinsarg\right]$
arise:\\
a) Suppose that by employing some technique,
$\mathbf{L}_i^{-1}\left[\paren{\frac{\xp}{r_0}}^B g\sinsarg\right]$
can be analytically continued down to a punctured neighborhood of
$B=0$,
and by another down to a neighborhood of
$B=0$, including
$B=0$
(we call this latter analytical continuation ``maximal''). Are
$\mathop{\mathrm{FP}}\limits_{B=0}\mathrm{A}\mathbf{L}_i^{-1}\left[\paren{\frac{\xp}{r_0}}^B g\sinsarg\right]$'s
corresponding to these two analytic continuations different? The answer is negative because, as a consequence of identity theorem, these two analytic continuations must be the same wherever they are both defined, including punctured neighborhood of
$B=0$.
This takes place only when, for each
$n$,
the coefficient of
$B^n$
in Laurent expansion of the first analytic continuation equates to the corresponding coefficient in
Laurent expansion of the second analytic continuation. Hence, the coefficients of
$B^0$,
or
$\mathop{\mathrm{FP}}\limits_{B=0}\mathrm{A}\mathbf{L}_i^{-1}\left[\paren{\frac{\xp}{r_0}}^B g\sinsarg\right]$'s,
must be the same.\\
b) If
$\mathbf{L}_i^{-1}\left[\!\ g\sinsarg\right]$
converges, thereby being a particular solution to the equation
$\mathbf{L}_i f\finsarg\!=g\finsarg$,
is it identical to
$\mathop{\mathrm{FP}}\limits_{B=0}\mathrm{A}\mathbf{L}_i^{-1}\left[\paren{\frac{\xp}{r_0}}^B g\sinsarg\right]$?
The answer is affirmative since
\begin{IEEEeqnarray}{rcl}\label{2.2.2.18}
\mathop{\mathrm{FP}}\limits_{B=0}\mathrm{A}\mathbf{L}_i^{-1}\left[\paren{\frac{\xp}{r_0}}^B g\sinsarg\right] & = & \mathop{\mathrm{FP}}\limits_{B=0}\mathbf{L}_i^{-1}\left[\paren{\frac{\xp}{r_0}}^B g\sinsarg\right] \nonumber \\  
& = & \mathop{\mathrm{FP}}\limits_{B=0}\mathbf{L}_i^{-1}\left[\paren{\sum_{j=0}^{\infty}\frac{1}{j!}\paren{\ln{\xp}}^j B^j} g\sinsarg\right]\nonumber \\ 
& = & \mathbf{L}_i^{-1}\left[\ g\sinsarg\right].
\end{IEEEeqnarray}
2. To apply this theorem, we need to examine first if its hypotheses, i.e., analyticity of
$\mathbf{L}_i^{-1}\left[\paren{\frac{\xp}{r_0}}^B g\sinsarg\right]$
in some original region and the possibility of analytic continuation of
$\mathbf{L}_i^{-1}\left[\paren{\frac{\xp}{r_0}}^B g\sinsarg\right]$ down to a punctured neighborhood of
$B=0$,
are fulfilled. In order to examine the possibility of analytic continuation, we try to analytically continue
$\mathbf{L}_i^{-1}\left[\paren{\frac{\xp}{r_0}}^B g\sinsarg\right]$
by means of some technique. Although the analytic continuation of a complex function is independent of the technique used and thereby unique, choosing the most convenient technique is important and the most convenient technique relies on the functional dependence of
$g\finsarg$.
On the other hand, the examination of analyticity of
$\mathbf{L}_i^{-1}\left[\paren{\frac{\xp}{r_0}}^B g\sinsarg\right]$
in some original region can be done irrespective of what the functional dependence of
$g\finsarg$
is. To see this, first note that
$\mathbf{L}_i^{-1}\left[\paren{\frac{\xp}{r_0}}^B g\sinsarg\right]$
can be rewritten as
\begin{IEEEeqnarray}{rcl}\label{2.2.2.11}
\mathbf{L}_i^{-1}\left[\paren{\frac{\xp}{r_0}}^B g\sinsarg\right] & = &  \mathbf{L}_i^{-1}\left[\paren{\frac{\xp}{r_0}}^x g\sinsarg \cos{(y\ln{\frac{\xp}{r_0}})}\right]\nonumber \\  
&& \negmedspace {} + i\:  \mathbf{L}_i^{-1}\left[\paren{\frac{\xp}{r_0}}^x g\sinsarg \sin{(y\ln{\frac{\xp}{r_0}})}\right],
\end{IEEEeqnarray}
where
$x=\mathrm{Re}B$
and
$y=\mathrm{Im}B$.
We respectively denote the real and imaginary parts of
$\mathbf{L}_i^{-1}\left[\paren{\frac{\xp}{r_0}}^B g\sinsarg\right]$
by
$u\paren{x,y}$
and
$v\paren{x,y}$.
$u\paren{x,y}$
and
$v\paren{x,y}$
must be both defined in some common region, that is to say, both
$\mathbf{L}_i^{-1}\left[\paren{\frac{\xp}{r_0}}^x g\sinsarg \cos{(y\ln{\frac{\xp}{r_0}})}\right]$
and
$ \mathbf{L}_i^{-1}\left[\paren{\frac{\xp}{r_0}}^x g\sinsarg \sin{(y\ln{\frac{\xp}{r_0}})}\right]$
must converge in some common region. A subregion of that common region in which Cauchy-Riemann conditions
\begin{equation}\label{2.2.2.12-2.2.2.13}
\frac{\partial u\paren{x,y}}{\partial x}=\frac{\partial v\paren{x,y}}{\partial y} \qquad \text{and}\qquad \frac{\partial u\paren{x,y}}{\partial y}=-\frac{\partial v\paren{x,y}}{\partial x}\nonumber
\end{equation}
are satisfied is the domain of analyticity of
$\mathbf{L}_i^{-1}\left[\paren{\frac{\xp}{r_0}}^B g\sinsarg\right]$.
Since
\begin{equation}\label{2.2.2.14}
\frac{\partial u\paren{x,y}}{\partial x}=\mathbf{L}_i^{-1}\left[\paren{\frac{\xp}{r_0}}^x\paren{\ln{\frac{\xp}{r_0}}} g\sinsarg \cos{(y\ln{\frac{\xp}{r_0}})}\right],
\end{equation}
\begin{equation}\label{2.2.2.15}
\frac{\partial u\paren{x,y}}{\partial y}=- \mathbf{L}_i^{-1}\left[\paren{\frac{\xp}{r_0}}^x\paren{\ln{\frac{\xp}{r_0}}} g\sinsarg \sin{(y\ln{\frac{\xp}{r_0}})}\right],
\end{equation}
\begin{equation}\label{2.2.2.16}
\frac{\partial v\paren{x,y}}{\partial x}= \mathbf{L}_i^{-1}\left[\paren{\frac{\xp}{r_0}}^x\paren{\ln{\frac{\xp}{r_0}}} g\sinsarg \sin{(y\ln{\frac{\xp}{r_0}})}\right],
\end{equation}
\begin{equation}\label{2.2.2.17}
\frac{\partial v\paren{x,y}}{\partial y}=\mathbf{L}_i^{-1}\left[\paren{\frac{\xp}{r_0}}^x\paren{\ln{\frac{\xp}{r_0}}} g\sinsarg \cos{(y\ln{\frac{\xp}{r_0}})}\right],
\end{equation}
Cauchy-Riemann conditions are always fulfilled in the common region of convergence of 
$\mathbf{L}_i^{-1}\left[\paren{\frac{\xp}{r_0}}^x\paren{\ln{\frac{\xp}{r_0}}} g\sinsarg \cos{(y\ln{\frac{\xp}{r_0}})}\right]$
and
$\mathbf{L}_i^{-1}\left[\paren{\frac{\xp}{r_0}}^x\paren{\ln{\frac{\xp}{r_0}}} g\sinsarg \sin{(y\ln{\frac{\xp}{r_0}})}\right]$.
Thus, the region in which the integrals
$\mathbf{L}_i^{-1}\left[\paren{\frac{\xp}{r_0}}^x\paren{\ln{\frac{\xp}{r_0}}}^p g\sinsarg \cos{(y\ln{\frac{\xp}{r_0}})}\right]$
and
$\mathbf{L}_i^{-1}\left[\paren{\frac{\xp}{r_0}}^x\paren{\ln{\frac{\xp}{r_0}}}^p g\sinsarg \sin{(y\ln{\frac{\xp}{r_0}})}\right]$
with
$p=0,1$
all converge is the original domain of analyticity of
$\mathbf{L}_i^{-1}\left[\paren{\frac{\xp}{r_0}}^B g\sinsarg\right]$.
Since
$\frac{1}{r_0^x}$
and
$\ln{r_0}$
don't affect the convergence, we can say that the original domain of analyticity is a region in which the integrals
$\mathbf{L}_i^{-1}\left[\xp^x \paren{\ln{\xp}}^p g\sinsarg \cos{(y\ln{\frac{\xp}{r_0}})}\right]$
and
$\mathbf{L}_i^{-1}\left[\xp^x \paren{\ln{\xp}}^p g\sinsarg \sin{(y\ln{\frac{\xp}{r_0}})}\right]$
with
$p=0,1$
are all convergent.

Due to the variety of
$k$'s
and
$\ell$'s
in the structure of
$\Lambda^{\mu\nu}_{\paren{2}\mathrm{AS}}\xarg$,
$\gretint{\Lambda^{\mu\nu}_{\paren{2}}\retarg}$ 
not only is not analytic in any original domain, but also is not defined anywhere in the entire complex plane (we will discuss this later). Thus, the preliminary hypothesis of Theorem 2.3 is not met and
$\fpa \gretint{\Lambda^{\mu\nu}_{\paren{2}}\retarg}$
is not thereby a particular solution to the equation
$\Box  h^{\mu\nu}_{\paren{2}}\finsarg=\Lambda^{\mu\nu}_{\paren{2}}\finsarg$.
However, fortunately
$\Box$
is a linear operator. Hence, if
$f_{\mathrm{part,1}}\finsarg$
is a particular solution to the equation
$\Box f\finsarg=g_1\finsarg$
and
$f_{\mathrm{part,2}}\finsarg$
a particular solution to the equation
$\Box f\finsarg=g_2\finsarg$,
then
$f_{\mathrm{part,1}}\finsarg+f_{\mathrm{part,2}}\finsarg$
is a particular solution to the equation
$\Box f\finsarg=g_1\finsarg+g_2\finsarg$.
Therefore, since the constituents of
$\Lambda^{\mu\nu}_{\paren{2}}\finsarg$
are of the general forms
$\hat{n}^Q \x^a C$
(where
$C$
is a constant),
$\hat{n}^Q \x^a F\targ$
(where
$F\targ$
is past-zero) and
$\x^{N+1}g\finsarg$
(where
$g\finsarg$
is smooth and past-zero), to obtain a particular solution for the equation
$\Box  h^{\mu\nu}_{\paren{2}}\finsarg=\Lambda^{\mu\nu}_{\paren{2}}\finsarg$,
it is enough to find a particular solution to each of the three equations
$\Box f\finsarg=\hat{n}^Q \x^a C$,
$\Box f\finsarg=\hat{n}^Q \x^a \paren{\ln{\frac{\x}{r_0}}}^p F\targ$
(where
$r_0$
is the same constant as in Theorem 2.3) and
$\Box f\finsarg=\x^{N+1}g\finsarg$
for arbitrary values of
$a$,
$q$,
$p$
and
$N$
where
$a \in \mathbb{Z}$
and
$p,q,N \in \mathbb{N}$
(values of
$a$,
$q$,
$p$
and
$N$
in the structure of
$\Lambda^{\mu\nu}_{\paren{2}}\finsarg$
are not arbitrary [in particular,
$p$
equals zero], but other values of them appear at higher orders).

Among an infinite number of particular solutions to the equation
$\Box f\finsarg=\hat{n}^Q \x^a C$,
there exist time-independent particular solutions too. Such solutions fulfill the equation
$\Delta f(\mathbf{x})=\hat{n}^Q \x^a C$.
Considering remark 2 following Theorem 2.3 and using Theorems 2.1 and 2.2, it is apparent that the original domain of analyticity of
$\poisint{\xpb \hat{n}'^Q\xp^a C}$
is the vertical strip
$-a-q-3 < \re B < -a+q-2$.
Using
Eq. (\ref{2.2.1.32}),
$\poisint{\xpb \hat{n}'^Q\xp^a C}$
in this region is given by
\begin{equation}\label{2.2.2.19}
\poisint{\xpb \hat{n}'^Q\xp^a C}=\xb \frac{\hat{n}^Q\x^{a+2}C}{\left(B+a-q+2\right)\left(B+a+q+3\right)}.
\end{equation}
By virtue of identity theorem, the equality between analytic continuations of each side of
Eq. (\ref{2.2.2.19})
holds in the region where both are defined. Then we have
\begin{equation}\label{2.2.2.20}
\mathrm{A}\poisint{\xpb \hat{n}'^Q\xp^a C}=\xb \frac{\hat{n}^Q\x^{a+2}C}{\left(B+a-q+2\right)\left(B+a+q+3\right)}.
\end{equation}
From
Eq. (\ref{2.2.2.20})
it is obvious that the singularities of
$\mathrm{A}\poisint{\xpb \hat{n}'^Q\xp^a C}$
are at the points with
$B\in \mathbb{Z}$.
Therefore,
$\mathrm{A}\poisint{\xpb \hat{n}'^Q\xp^a C}$
is analytic in some punctured neighborhood of
$B=0$,
or, in other words,
$\poisint{\xpb \hat{n}'^Q\xp^a C}$
can be analytically continued down to a punctured neighborhood of
$B=0$.
Thus, a particular solution to the equation
$\Delta f(\mathbf{x})=\hat{n}^Q \x^a C$
can be given by
$\fpa\poisint{\xpb \hat{n}'^Q\xp^a C}$.

In order to determine the structure of
$\fpa\poisint{\xpb \hat{n}'^Q\xp^a C}$,
we must first assess whether
$B=0$
is a singularity of
$\mathrm{A}\poisint{\xpb \hat{n}'^Q\xp^a C}$
or not.
Eq. (\ref{2.2.2.20})
implies that
$B=0$
is a singular point when at least one of the following equalities holds:
\begin{equation}\label{2.2.2.21}
a-q+2=0 \longrightarrow a=q-2,
\end{equation}
\begin{equation}\label{2.2.2.22}
a+q+3=0 \longrightarrow a=-q-3.
\end{equation}
These two equalities cannot hold simultaneously since if they can, for given values of
$a$
and
$q$,
we conclude the false proposition that
$q$
(or
$a$)
is not an integer.

If
$\mathrm{A}\poisint{\xpb \hat{n}'^Q\xp^a C}$
is analytic at
$B=0$,
we have
\begin{equation}\label{2.2.2.23}
\fpa\poisint{\xpb \hat{n}'^Q\xp^a C}=\frac{\hat{n}^Q\x^{a+2}C}{\left(a-q+2\right)\left(a+q+3\right)},
\end{equation}
and if
$B=0$
is a singularity of
$\mathrm{A}\poisint{\xpb \hat{n}'^Q\xp^a C}$,\footnote{In this case, whether
Eq. (\ref{2.2.2.21})
or
Eq. (\ref{2.2.2.22})
holds, it can be easily shown that this singularity is a pole (not a removable one).}
taking into account the remark about this singularity following
Eqs. (\ref{2.2.2.21}) and (\ref{2.2.2.22}),
we reach
\begin{equation}\label{2.2.2.24}
\fpa\poisint{\xpb \hat{n}'^Q\xp^a C}=
\begin{cases}
\frac{\hat{n}^Q\x^{a+2}\paren{\ln{\frac{\x}{r_0}}}C}{a-q+2}, \qquad a=-q-3,\\
\frac{\hat{n}^Q\x^{a+2}\paren{\ln{\frac{\x}{r_0}}}C}{a+q+3}, \qquad a=q-2.
\end{cases}
\end{equation}

Based on the foregoing discussion, a particualr solution to the equation
$\Box  h^{\mu\nu}_{\paren{2}\mathrm{AS}}\finsarg=\Lambda^{\mu\nu}_{\paren{2}\mathrm{AS}}(\mathbf{x})$
is given by
\begin{IEEEeqnarray}{rcl}\label{2.2.2.25a}
u^{\mu\nu}_{\paren{2}\mathrm{AS}}\xarg & = &  \sum_{\ell=0}^{\ell_{\mathrm{max}}}\sum_{k=k_{\mathrm{min}}}^{-4}\fpa\poisint{\xpb \hat{n}'^L\xp^k \hat{C}^{\mu\nu}_{\paren{2}L,k}} \nonumber \\  
& = & \sum_{\ell=0}^{\ell_{\mathrm{max}}}\sum_{k=k_{\mathrm{min}}}^{-4}\fpa\left[\gnearint{\hat{n}'^L\xp^k \hat{C}^{\mu\nu}_{\paren{2}L,k}}\right] \nonumber \\  
&& \negmedspace {} + \sum_{\ell=0}^{\ell_{\mathrm{max}}}\sum_{k=k_{\mathrm{min}}}^{-4}\fpa\left[\gfarint{\hat{n}'^L\xp^k \hat{C}^{\mu\nu}_{\paren{2}L,k}}\right].
\end{IEEEeqnarray}
Since all integrals
$\gnearint{\hat{n}'^L\xp^k \hat{C}^{\mu\nu}_{\paren{2}L,k}}$
with different values of
$k$
and
$\ell$
are analytic in some right half-plane and all integrals
$\gfarint{\hat{n}'^L\xp^k \hat{C}^{\mu\nu}_{\paren{2}L,k}}$
with different values of
$k$
and
$\ell$
are analytic in some left half-plane, we can write
\begin{IEEEeqnarray}{rcl}\label{2.2.2.25b}
u^{\mu\nu}_{\paren{2}\mathrm{AS}}\xarg & = & \fpa\left[\gnearint{\sum_{\ell=0}^{\ell_{\mathrm{max}}}\sum_{k=k_{\mathrm{min}}}^{-4}\hat{n}'^L\xp^k \hat{C}^{\mu\nu}_{\paren{2}L,k}}\right] \nonumber \\  
&& \negmedspace {}+ \fpa\left[\gfarint{\sum_{\ell=0}^{\ell_{\mathrm{max}}}\sum_{k=k_{\mathrm{min}}}^{-4}\hat{n}'^L\xp^k \hat{C}^{\mu\nu}_{\paren{2}L,k}}\right] \nonumber \\  
&=&\fpa\left[\gnearint{\Lambda^{\mu\nu}_{\paren{2}\mathrm{AS}}(\mathbf{x'})}\right] \nonumber \\  
&& \negmedspace {}+ \fpa\left[\gfarint{\Lambda^{\mu\nu}_{\paren{2}\mathrm{AS}}(\mathbf{x'})}\right].
\end{IEEEeqnarray}
The right-hand side of the last equality in the above equation cannot be equated with
$\fpa\poisint{\xpb \Lambda^{\mu\nu}_{\paren{2}\mathrm{AS}}(\mathbf{x'})}$
since
$\poisint{\xpb \Lambda^{\mu\nu}_{\paren{2}\mathrm{AS}}(\mathbf{x'})}$
is not analytic in any original domain due to original domains of analyticity of
$\poisint{\xpb \hat{n}'^L\xp^k \hat{C}^{\mu\nu}_{\paren{2}L,k}}$'s
with different values of 
$k$
and
$\ell$
appearing in the structure of
$\Lambda^{\mu\nu}_{\paren{2}\mathrm{AS}}\xarg$
(namely regions which are vertical strips, each with a width of
$2\ell+1$
units and centered on the line
$\re B=-k-\frac{5}{2}$)
not overlapping. In fact, because the above-mentioned vertical strips are also the domains of definition of
$\poisint{\xpb \hat{n}'^L\xp^k \hat{C}^{\mu\nu}_{\paren{2}L,k}}$'s,
$\poisint{\xpb \Lambda^{\mu\nu}_{\paren{2}\mathrm{AS}}(\mathbf{x'})}$
is not even defined anywhere in the complex plane (consequently, as stated before,
$\retint{\xpb \Lambda^{\mu\nu}_{\paren{2}}(t',\mathbf{x'})}$
is not defined anywhere in the complex plane too). Therefore, instead of
$u^{\mu\nu}_{\paren{2}\mathrm{AS}}\xarg = \fpa\poisint{\xpb \Lambda^{\mu\nu}_{\paren{2}\mathrm{AS}}(\mathbf{x'})}$
we write
\begin{equation}\label{2.2.2.25c}
u^{\mu\nu}_{\paren{2}\mathrm{AS}}\xarg \stareq \fpa\poisint{\xpb \Lambda^{\mu\nu}_{\paren{2}\mathrm{AS}}(\mathbf{x'})},
\end{equation}
where by the sign
$\stareq$
we mean that the left-hand side of this sign equals the right-hand side of it provided that
$\int_{\mathbb{R }^3}$
appearing on the right-hand side is written as the sum of
$\int_{|\mathbf{x'}|<\mathcal{R}}$
and
$\int_{\mathcal{R}<|\mathbf{x'}|}$.

It can be shown that
$a$'s
and
$q$'s
which appear in the structure of
$\Lambda^{\mu\nu}_{\paren{2}\mathrm{AS}}$,
or, in other words,
$a$'s
and
$q$'s
for which
$\hat{C}^{\mu\nu}_{\paren{2}Q,a}$
is nonzero, fulfill neither
Eq. (\ref{2.2.2.21})
nor
Eq. (\ref{2.2.2.22}).
Accordingly, the structure of
$ u^{\mu\nu}_{\paren{2}\mathrm{AS}}$
reads
\begin{equation}\label{2.2.2.26}
 u^{\mu\nu}_{\paren{2}\mathrm{AS}}=\sum_{\ell=0}^{\ell_{\mathrm{max}}}\sum_{k=k'_{\mathrm{min}}}^{-2}\hat{n}^L\x^k \hat{C}'^{\;\mu\nu}_{\paren{2}L,k}.
\end{equation}

We next examine the equation
$\Box f\finsarg=\hat{n}^Q \x^a \paren{\ln{\frac{\x}{r_0}}}^p F\targ$.
Considering remark 2 following Theorem 2.3 and using Theorem 2.1, one finds that the original domain of analyticity of
$\retint{\xpb \hat{n}'^Q \xp^a \lnxp^p F\tparg}$
is the half-plane
$\re B > -a-q-3$
(we have used Theorem 2.1 due to
$F\targ$
being past-zero). Now note that everywhere in
$\mathbb{R }^3$
(except at
$\x=0$), we have
\begin{IEEEeqnarray}{rcl}\label{2.2.2.29}
\Box\left[\xb\hat{n}^Q\x^{a+2}F\targ\right] & = & \frac{1}{r_0^B}\left(\Delta - \frac{1}{c^2}\partial_t^2\right)\left[\hat{n}^Q \x^{B+a+2} F\targ\right] \nonumber \\  
& = & \left(B+a-q+2\right)\left(B+a+q+3\right)\xb\hat{n}^Q\x^a F\targ \nonumber \\  
&& \negmedspace {} -\frac{1}{c^2}\xb\hat{n}^Q\x^{a+2}F^{\paren{2}}\targ.
\end{IEEEeqnarray}
With the assumptions that
$B+a-q+2\neq 0$
and
$B+a+q+3\neq 0$,
we get
\begin{IEEEeqnarray}{rcl}\label{2.2.2.30}
\xb\hat{n}^Q\x^a F\targ & = & \Box\left[\xb\frac{\hat{n}^Q\x^{a+2}F\targ}{\left(B+a-q+2\right)\left(B+a+q+3\right)}\right] \nonumber \\  
&& \negmedspace {} +\frac{1}{c^2}\left[\xb\frac{\hat{n}^Q\x^{a+2}F^{\paren{2}}\targ}{\left(B+a-q+2\right)\left(B+a+q+3\right)}\right].
\end{IEEEeqnarray}
As it can be seen, the structure of the last term on the right-hand side of
Eq. (\ref{2.2.2.30})
is similar to the structure of the left-hand side of it. Therefore, assuming
$B+a-q+4\neq 0$
and
$B+a+q+5\neq 0$,
we can write
\begin{IEEEeqnarray}{rcl}\label{2.2.2.31}
\xb\hat{n}^Q\x^a F\targ & = & \Box\!\left[\!\xb\!\frac{\hat{n}^Q\x^{a+2}F\targ}{\left(B\!+\!a\!-\!q\!+\!2\right)\left(B\!+\!a\!+\!q\!+3\right)}\!\right] \nonumber \\  
&& \negmedspace {} \!+\!\Box\!\left[\!\frac{1}{c^2}\xb\!\frac{\hat{n}^Q\x^{a+4}F^{\paren{2}}\targ}{\mathop{\prod}\limits_{k=0}^{1}\left(B\!+\!a\!-\!q\!+\!2\!+\!2k\right)\left(B\!+\!a\!+\!q\!+\!3\!+\!2k\right)}\!\right]\nonumber \\  
&& \negmedspace {}\!+\!\frac{1}{c^4}\!\left[\!\xb\!\frac{\hat{n}^Q\x^{a+4}F^{\paren{4}}\targ}{\mathop{\prod}\limits_{k=0}^{1}\left(B\!+\!a\!-\!q\!+\!2\!+\!2k\right)\left(B\!+\!a\!+\!q\!+\!3\!+\!2k\right)}\!\right].\nonumber \\ 
\end{IEEEeqnarray}
By repeating this process of substitution into the last term on the right-hand side
$s-1$
more times, we reach
\begin{IEEEeqnarray}{rcl}\label{2.2.2.32}
\xb\hat{n}^Q\x^a F\targ & = & \Box\!\left[\!\sum_{m=0}^{s}\frac{1}{c^{2m}}\!\xb\!\frac{\hat{n}^Q\x^{a+2+2m}F^{\paren{2m}}\targ}{\mathop{\prod}\limits_{k=0}^{m}\left(B\!+\!a\!-\!q\!+\!2\!+\!2k\right)\left(B\!+\!a\!+\!q\!+\!3\!+\!2k\right)}\!\right] \nonumber \\  
&& \negmedspace {}\!+\!\frac{1}{c^{2s+2}}\!\left[\!\xb\!\frac{\hat{n}^Q\x^{a+2+2s}F^{\paren{2s+2}}\targ}{\mathop{\prod}\limits_{k=0}^{s}\left(B\!+\!a\!-\!q\!+\!2\!+\!2k\right)\left(B\!+\!a\!+\!q\!+\!3\!+\!2k\right)}\!\right].\nonumber \\ 
\end{IEEEeqnarray}
Equality in the above equation holds if we have
$B+a-q+2+2k\neq0$
and
$B+a+q+3+2k\neq0$
for each
$0 \le k \le s$.
Applying
$\Box^{-1}_{\mathrm{R}}$
to both sides of
Eq. (\ref{2.2.2.32}),
the left-hand side of this equation becomes
$\retint{\xpb \hat{n}'^Q \xp^a F\tparg}$
which is defined in the half-plane
$\re B>-q-a-3$.
In the first term on the right-hand side of it, the retarded integral
$\retint{\Box'\left[\xpb \hat{n}'^Q\xp^{a+2+2m}F^{\paren{2m}}\tparg\right]}$
appears. We have
\begin{eqnarray}\label{2.2.2.33}
\lefteqn{\Box\left[\xb \hat{n}^Q\x^{a+2+2m}F^{\paren{2m}}\targ\right]} \nonumber \\ 
&\!\!\!\!\!\!\!\!=\!\!\!\!\!\!\!\!&\left(B\!+\!a\!-\!q\!+\!2\!+\!2m\right)\left(B\!+\!a\!+\!q\!+\!3\!+\!2m\right) \xb\hat{n}^Q\x^{a+2m} F^{\paren{2m}}\targ \nonumber\\  
&&\!\!\!\!-\frac{1}{c^2}\xb\hat{n}^Q\x^{a+2+2m}F^{\paren{2m+2}}\targ.
\end{eqnarray}
Since
$F\targ$
is past-zero, both real and imaginary parts of
$\xb \hat{n}^Q\x^{a+2+2m}F^{\paren{2m}}\targ$
satisfy the no-incoming radiation condition. Therefore, provided that the retarded integrals converge, we can write
\begin{eqnarray}\label{2.2.2.33a}
\lefteqn{\re\left[\xb\right] \hat{n}^Q\x^{a+2+2m}F^{\paren{2m}}\targ} \nonumber \\ 
&\!\!\!\!\!\!\!\!=\!\!\!\!\!\!\!\!& \retint{ \re\left[\left(B\!+\!a\!-\!q\!+\!2\!+\!2m\right)\left(B\!+\!a\!+\!q\!+\!3\!+\!2m\right) \xpb\right]\hat{n}'^Q\xp^{a+2m} F^{\paren{2m}}\tparg} \nonumber\\  
&&\!\!\!\! -\frac{1}{c^2}\retint{\re\left[\xpb\right]\hat{n}'^Q\xp^{a+2+2m}F^{\paren{2m+2}}\tparg},
\end{eqnarray}
\begin{eqnarray}\label{2.2.2.33b}
\lefteqn{\im\left[\xb\right] \hat{n}^Q\x^{a+2+2m}F^{\paren{2m}}\targ} \nonumber \\ 
&\!\!\!\!\!\!\!\!=\!\!\!\!\!\!\!\!& \retint{ \im\left[\left(B\!+\!a\!-\!q\!+\!2\!+\!2m\right)\left(B\!+\!a\!+\!q\!+\!3\!+\!2m\right) \xpb\right]\hat{n}'^Q\xp^{a+2m} F^{\paren{2m}}\tparg} \nonumber\\  
&&\!\!\!\! -\frac{1}{c^2}\retint{\im\left[\xpb\right]\hat{n}'^Q\xp^{a+2+2m}F^{\paren{2m+2}}\tparg}.
\end{eqnarray}
Based on Theorem 2.1, the first integral on the right-hand side of each of
Eqs. (\ref{2.2.2.33a}) and (\ref{2.2.2.33b})
converges when
$\re B+\left(a+2m\right)>-q-3$
and the second one when
$\re B+\left(a+2+2m\right)>-q-3$.
Thus, in the region
$\re B>-\left(a+2m\right)-q-3$,
we have
\begin{eqnarray}\label{2.2.2.34}
\lefteqn{\xb\hat{n}^Q\x^{a+2+2m}F^{\paren{2m}}\targ} \nonumber \\ 
&\!\!\!\!\!\!\!\!=\!\!\!\!\!\!\!\!&\retint{\left(B\!+\!a\!-\!q\!+\!2\!+\!2m\right)\left(B\!+\!a\!+\!q\!+\!3\!+\!2m\right)\xpb\hat{n}'^Q\xp^{a+2m} F^{\paren{2m}}\tparg} \nonumber\\  
&&\!\!\!\!-\frac{1}{c^2}\retint{\xpb\hat{n}'^Q\xp^{a+2+2m}F^{\paren{2m+2}}\tparg}.
\end{eqnarray}
However, one can observe that the right-hand side of
Eq. (\ref{2.2.2.34})
is nothing but the retarded integral
$\retint{\Box'\left[\xpb \hat{n}'^Q\xp^{a+2+2m}F^{\paren{2m}}\tparg\right]}$.
Hence, we reach
\begin{IEEEeqnarray}{rcl}\label{2.2.2.35}
\retint{\Box'\left[\xpb \hat{n}'^Q\xp^{a+2+2m}F^{\paren{2m}}\tparg\right]} & = & \xb\hat{n}^Q\x^{a+2+2m}F^{\paren{2m}}\targ.\quad
\end{IEEEeqnarray}
Furthermore, it is obvious that the retarded integral of the second term on the right-hand side of
Eq. (\ref{2.2.2.32})
is defined in the half-plane
$\re B>-\left(a+2+2s\right)-q-3$.
Therefore, we can say that in the region
\begin{IEEEeqnarray}{rcl}\label{2.2.2.36}
D & = &\left\{B\left|\right. \re B>\!-a\!-\!q\!-\!3\right\}-\left\{B_{+}\!=\!-a\!+\!q\!-\!2\!-\!2k,B_{-}\!=\!-a\!-\!q\!-\!3\!-\!2k\left|\right. 0\le k \le s\right\}\nonumber
\end{IEEEeqnarray}
we have
\begin{eqnarray}\label{2.2.2.37}
\lefteqn{\retint{\xpb\hat{n}'^Q\xp^a F\tparg}}\nonumber \\ 
&\!\!\!\!\!\!\!\!=\!\!\!\!\!\!\!\!&\sum_{m=0}^{s}\frac{1}{c^{2m}}\xb\frac{\hat{n}^Q\x^{a+2+2m}F^{\paren{2m}}\targ}{\mathop{\prod}\limits_{k=0}^{m}\left(B\!+\!a\!-\!q\!+\!2\!+\!2k\right)\left(B\!+\!a\!+\!q\!+\!3\!+\!2k\right)}\nonumber\\  
&&\!\!\!\!+\frac{1}{c^{2s+2}}\retint{\xpb\frac{\hat{n}'^Q\xp^{a+2+2s}F^{\paren{2s+2}}\tparg}{\mathop{\prod}\limits_{k=0}^{s}\left(B\!+\!a\!-\!q\!+\!2\!+\!2k\right)\left(B\!+\!a\!+\!q\!+\!3\!+\!2k\right)}}.
\end{eqnarray}
$B_{+}$
can be rewritten as
\begin{equation}\label{2.2.2.38}
B_{+}=\left(-a-q-3\right)+2q+1-2k \qquad \text{where} \;0 \le k \le s.
\end{equation}
From
Eq. (\ref{2.2.2.38})
it can be observed that
$\mathrm{min}\left\{\paren{q+1},\paren{s+1}\right\}$
of
$B_{+}$'s
are located inside the half-plane
$\re B>-a-q-3$
and rest of them, if there exists any, on the left-hand side of that. On the other hand, there is no
$B_{-}$
in the aforementioned half-plane because, considering
$0 \le k \le s$,
we have
\begin{equation}\label{2.2.2.39}
-a-q-3-2s \le B_{-} \le -a-q-3.
\end{equation}
Based on what we argued above, the region
$D$
can be expressed as
\begin{IEEEeqnarray}{rcl}\label{2.2.2.40}
D & = &\left\{B\left|\right. \re B>\!-a\!-\!q\!-\!3\right\}-\left\{B_0\!=\!-a\!+\!q\!-\!2\!-\!2k\left|\right. 0\le k \le\mathrm{min}\left\{q,s\right\} \right\}.\nonumber
\end{IEEEeqnarray}
Since both sides of
Eq. (\ref{2.2.2.37})
are also analytic in the region
$D$,
we can differentiate them with respect to
$B$
as many times as is necessary. By doing so
$p$
times, we get
\begin{eqnarray}\label{2.2.2.41}
\lefteqn{\retint{\xpb\hat{n}'^Q\xp^a \lnxp^p F\tparg}} \nonumber \\ 
&\!\!\!\!\!\!\!\!=\!\!\!\!\!\!\!\!&\frac{\partial^p}{\partial B^p}\bigg[\sum_{m=0}^{s}\frac{1}{c^{2m}}\xb\frac{\hat{n}^Q\x^{a+2+2m}F^{\paren{2m}}\targ}{\mathop{\prod}\limits_{k=0}^{m}\left(B\!+\!a\!-\!q\!+\!2\!+\!2k\right)\left(B\!+\!a\!+\!q\!+\!3\!+\!2k\right)}\nonumber\\  
&&\!\!\!\!+\frac{1}{c^{2s+2}}\Box^{-1}_R\bigg[\xpb\frac{\hat{n}'^Q\xp^{a+2+2s}F^{\paren{2s+2}}\tparg}{\mathop{\prod}\limits_{k=0}^{s}\left(B\!+\!a\!-\!q\!+\!2\!+\!2k\right)\left(B\!+\!a\!+\!q\!+\!3\!+\!2k\right)}\bigg]\Bigg].
\end{eqnarray}
By virtue of identity theorem, the equality between analytic continuations of each side of
Eq. (\ref{2.2.2.41})
must also hold wherever they both exist. Thus, By choosing
$s$
so that
$2s>-a-q-5$
(in this way, the domain of analyticity of
$\Box^{-1}_R\bigg[\xpb\hat{n}'^Q\xp^{a+2+2s}F^{\paren{2s+2}}\tparg\bigg]$,
which is the region
$\re B>-\left(a+2+2s\right)-q-3$,
extends to a region that contains a neighborhood of
$B=0$),
down to a neighborhood of
$B=0$
we can write
\begin{eqnarray}\label{2.2.2.44}
\lefteqn{\mathrm{A}\retint{\xpb\hat{n}'^Q\xp^a \lnxp^p F\tparg}} \nonumber \\ 
&\!\!\!\!\!\!\!\!=\!\!\!\!\!\!\!\!&\frac{\partial^p}{\partial B^p}\bigg[\sum_{m=0}^{s}\frac{1}{c^{2m}}\xb\frac{\hat{n}^Q\x^{a+2+2m}F^{\paren{2m}}\targ}{\mathop{\prod}\limits_{k=0}^{m}\left(B\!+\!a\!-\!q\!+\!2\!+\!2k\right)\left(B\!+\!a\!+\!q\!+\!3\!+\!2k\right)}\nonumber\\  
&&\!\!\!\!+\frac{1}{c^{2s+2}}\Box^{-1}_R\bigg[\xpb\frac{\hat{n}'^Q\xp^{a+2+2s}F^{\paren{2s+2}}\tparg}{\mathop{\prod}\limits_{k=0}^{s}\left(B\!+\!a\!-\!q\!+\!2\!+\!2k\right)\left(B\!+\!a\!+\!q\!+\!3\!+\!2k\right)}\bigg]\Bigg].
\end{eqnarray}
Eq. (\ref{2.2.2.44})
implies that the singularities of
$\mathrm{A}\retint{\xpb\hat{n}'^Q\xp^a \lnxp^p F\tparg}$
are at the points with
$B \in \mathbb{Z}$.
Thus,
$\mathrm{A}\retint{\xpb\hat{n}'^Q\xp^a \lnxp^p F\tparg}$
is analytic in some neighborhood of
$B=0$,
or, in other words,
$\retint{\xpb\hat{n}'^Q\xp^a \lnxp^p F\tparg}$
can be analytically continued down to a neighborhood of
$B=0$
and we can thereby say that
$\fpa\retint{\xpb\hat{n}'^Q\xp^a \lnxp^p F\tparg}$
is a particular solution to the equation
$\Box f\finsarg=\hat{n}^Q \x^a \paren{\ln{\frac{\x}{r_0}}}^p F\targ$.

To determine the structure of
$\fpa\retint{\xpb\hat{n}'^Q\xp^a \lnxp^p F\tparg}$
we must first assess whether or not
$B=0$
is a singularity of
$\mathrm{A}\retint{\xpb\hat{n}'^Q\xp^a \lnxp^p F\tparg}$.
From
Eq. (\ref{2.2.2.44})
it is evident that
$B=0$
is a singular point when at least one of the following equalities holds:
\begin{equation}\label{2.2.2.45}
-a+q-2-2k=0\longrightarrow q-a=2k+2\qquad\text{where}\;0\le k \le s,
\end{equation}
\begin{equation}\label{2.2.2.46}
-a-q-3-2k=0\longrightarrow q+a=-2k-3\qquad\text{where\;}0\le k \le s.
\end{equation}
It  can be shown that, for given values of
$a$
and
$q$,
any two of these
$2s+2$
equalities cannot hold simultaneously. Another important point is that, in the case where either
Eq. (\ref{2.2.2.45})
or
Eq. (\ref{2.2.2.46})
holds for some value of
$k$,
although
$B=0$
is definitely a singularity of the second term in square brackets in
Eq. (\ref{2.2.2.44})
it is not a singularity of first
$m_0$
terms of
$\sum_{m=0}^{s}\frac{1}{c^{2m}}\xb\frac{\hat{n}^Q\x^{a+2+2m}F^{\paren{2m}}\targ}{\mathop{\prod}\limits_{k=0}^{m}\left(B+a-q+2+2k\right)\left(B+a+q+3+2k\right)}$,
where
$m_0=\frac{1}{2}\paren{q-a-2}$
if
Eq. (\ref{2.2.2.45})
holds and
$m_0=-\frac{1}{2}\paren{q+a+3}$
if
Eq. (\ref{2.2.2.46}).

Considering
Eq. (\ref{2.2.2.44}),
the general form of
$\mathrm{A}\retint{\xpb\hat{n}'^Q\xp^a \lnxp^p F\tparg}$
in the case where it is analytic at
$B=0$
can be written as
\begin{eqnarray}\label{2.2.2.47}
\lefteqn{\mathrm{A}\retint{\xpb\hat{n}'^Q\xp^a \lnxp^p F\tparg}} \nonumber \\ 
&\!\!\!\!\!\!\!\!=\!\!\!\!\!\!\!\!&\frac{\partial^p}{\partial B^p}\Bigg[\sum_{m=0}^{s}\xb D_m(B)\hat{n}^Q\x^{a+2+2m} F^{\paren{2m}}\targ \nonumber\\  
&&\!\!\!\!+E(B)\retint{\xpb\hat{n}'^Q\xp^{a+2+2s} F^{\paren{2s+2}}\tparg}\Bigg],
\end{eqnarray}
where
$D_m(B)$
and
$E(B)$
are complex functions analytic at
$B=0$.
Using Leibniz formula, this takes the form
\begin{eqnarray}\label{2.2.2.48}
\lefteqn{\mathrm{A}\retint{\xpb\hat{n}'^Q\xp^a \lnxp^p F\tparg}} \nonumber \\ 
&\!\!\!\!\!\!\!\!=\!\!\!\!\!\!\!\!&\sum_{m=0}^{s}\sum_{k=0}^{p}\binom{p}{k}\xb D_m^{\paren{p-k}}(B)\hat{n}^Q\x^{a+2+2m}\lnx^k F^{\paren{2m}}\targ \nonumber\\  
&&\!\!\!\!+\sum_{k=0}^{p}\binom{p}{k}E^{\paren{p-k}}(B)\retint{\xpb\hat{n}'^Q\xp^{a+2+2s}\lnxp^k F^{\paren{2s+2}}\tparg}.\quad
\end{eqnarray}
The coefficient of
$B^0$
in Taylor expansion of the right-hand side of
Eq. (\ref{2.2.2.48})
can be obtained by taking into account the coefficient of
$B^0$
in Taylor expansion of each of the complex functions appearing on this side of the equation. Then we get
\begin{eqnarray}\label{2.2.2.49}
\lefteqn{\fpa\retint{\xpb\hat{n}'^Q\xp^a \lnxp^p F\tparg}} \nonumber \\ 
&\!\!\!\!\!\!\!\!=\!\!\!\!\!\!\!\!&\sum_{m=0}^{s}\sum_{k=0}^{p}\hat{n}^Q\x^{a+2+2m}\lnx^k F_{m,k}\targ \nonumber\\  
&&\!\!\!\!+\sum_{k=0}^{p}\retint{\hat{n}'^Q\xp^{a+2+2s}\lnxp^k F_k\tparg},
\end{eqnarray}
where
$F_{m,k}\targ$
and
$F_k\tparg$
are in terms of
$ F^{\paren{2m}}\targ$
and
$ F^{\paren{2s+2}}\tparg$
respectively.

On the other hand, if
$B=0$
is a singularity of
$\mathrm{A}\retint{\xpb\hat{n}'^Q\xp^a \lnxp^p F\tparg}$,\footnote{It is straightforward to show that, in this case, this singularity is removable if
Eq. (\ref{2.2.2.45})
holds and a pole if
Eq. (\ref{2.2.2.46}).}
considering the remarks about this singularity following
Eqs. (\ref{2.2.2.45})
and
(\ref{2.2.2.46}),
the general form of
$\mathrm{A}\retint{\xpb\hat{n}'^Q\xp^a \lnxp^p F\tparg}$
can be expressed as
\begin{eqnarray}\label{2.2.2.50}
\lefteqn{\mathrm{A}\retint{\xpb\hat{n}'^Q\xp^a \lnxp^p F\tparg}} \nonumber \\ 
&\!\!\!\!\!\!\!\!=\!\!\!\!\!\!\!\!&\frac{\partial^p}{\partial B^p}\Bigg[\sum_{m=0}^{m_0-1}\xb G_m(B)\hat{n}^Q\x^{a+2+2m} F^{\paren{2m}}\targ\nonumber\\  
&&\!\!\!\!+\sum_{m=m_0}^{s}\xb\paren{\frac{H_m(B)}{B}}\hat{n}^Q\x^{a+2+2m} F^{\paren{2m}}\targ\nonumber\\  
&&\!\!\!\!+\paren{\frac{I(B)}{B}}\retint{\xpb\hat{n}'^Q\xp^{a+2+2s}F^{\paren{2s+2}}\tparg}\Bigg],
\end{eqnarray}
where
$G_m(B)$,
$H_m(B)$
and
$I(B)$
are complex functions analytic at
$B=0$.
By means of Leibniz formula
Eq. (\ref{2.2.2.50})
becomes
\begin{eqnarray}\label{2.2.2.51}
\lefteqn{\mathrm{A}\retint{\xpb\hat{n}'^Q\xp^a \lnxp^p F\tparg}} \nonumber \\ 
&\!\!\!\!\!\!\!\!=\!\!\!\!\!\!\!\!&\sum_{m=0}^{m_0-1}\sum_{k=0}^{p}\binom{p}{k}\xb G_m^{\paren{p-k}}(B)\hat{n}^Q\x^{a+2+2m}\lnx^k F^{\paren{2m}}\targ\nonumber\\  
&&\!\!\!\!+\sum_{m=m_0}^{s}\sum_{k=0}^{p}\sum_{\ell=0}^{p-k}\binom{p}{k}\binom{p-k}{\ell}\paren{-1}^\ell \paren{\ell!}\xb\paren{\frac{H_m^{\paren{p-k-\ell}}(B)}{B^{\ell+1}}}\nonumber\\  
&&\!\!\!\!\times\hat{n}^Q\x^{a+2+2m}\lnx^k F^{\paren{2m}}\targ\nonumber\\  
&&\!\!\!\!+\sum_{k=0}^{p}\sum_{\ell=0}^{p-k}\binom{p}{k}\binom{p-k}{\ell}\paren{-1}^\ell \paren{\ell!}\paren{\frac{I^{\paren{p-k-\ell}}(B)}{B^{\ell+1}}}\nonumber\\  
&&\!\!\!\!\times\retint{\xpb\hat{n}'^Q\xp^{a+2+2s}\lnxp^k F^{\paren{2s+2}}\tparg}.
\end{eqnarray}
the coefficient of
$B^0$
in Laurent expansion of the right-hand side
Eq. (\ref{2.2.2.51})
can be obtained as follows:\\
1. In the first term, the coefficient of
$B^0$
in Taylor expansion of each of two complex functions must be taken into account.\\
2. In the second term, the coefficient of
$B^n$
in Taylor expansion of
$H_m^{\paren{p-k-\ell}}(B)$
and the coefficient of
$B^{n'}$
in Taylor expansion of
$\xb$,
where
$n+n'=\ell+1$,
must be taken into account. If
$k=p$,
then
$\ell=0$
and the power of
$\ln{\frac{\x}{r_0}}$
can be either
$p$
or
$p+1$.
If
$k=p-1$,
then
$\ell=0,1$
and the power of
$\ln{\frac{\x}{r_0}}$
can be
$p-1$,
$p$
or
$p+1$.
If
$k=p-2$,
then
$\ell=0,1,2$
and the power of
$\ln{\frac{\x}{r_0}}$
can be
$p-2$,
$p-1$,
$p$
or
$p+1$.
Finally, if
$k=0$,
then
$\ell=0,1,...,p$
and the power of
$\ln{\frac{\x}{r_0}}$
can take any integer value between
$0$
and
$p+1$.\\
3. In the third term, the coefficient of
$B^n$
in Taylor expansion of
$I^{\paren{p-k-\ell}}(B)$
and the coefficient of
$B^{n'}$
in Taylor expansion of
$\xpb$,
where
$n+n'=\ell+1$,
must be taken into account. By arguments similar to those employed in 2 above, it is straightforward to show that if
$k=p-p_0$,
then the power of
$\ln{\frac{\xp}{r_0}}$
can take any integer value between
$p-p_0$
and
$p+1$.\\
Considering 1, 2 and 3 above, we can write
\begin{eqnarray}\label{2.2.2.52}
\lefteqn{\fpa\retint{\xpb\hat{n}'^Q\xp^a \lnxp^p F\tparg}} \nonumber \\ 
&\!\!\!\!\!\!\!\!=\!\!\!\!\!\!\!\!&\sum_{m=0}^{s}\sum_{k=0}^{p+1}\hat{n}^Q\x^{a+2+2m}\lnx^k F_{m,k}\targ \nonumber\\  
&&\!\!\!\!+\sum_{k=0}^{p+1}\retint{\hat{n}'^Q\xp^{a+2+2s}\lnxp^k F_k\tparg},
\end{eqnarray}
where
$F_{m,k}\targ$
and
$F_k\tparg$
are in terms of
$ F^{\paren{2m}}\targ$
and
$ F^{\paren{2s+2}}\tparg$
respectively. Needless to say,
$F_{m,k}\targ$
and
$F_k\tparg$
in this equation are not identical to the corresponding ones in
Eq. (\ref{2.2.2.49}).

The structure of the first term on the right-hand side of each of
Eqs. (\ref{2.2.2.49}) and (\ref{2.2.2.52})
cannot be simplified more. However, by using Lemma 3.3 in
\cite{BD1986},
which is obtained by means of Taylor formula with integral remainder for functions of three variables (see Appendix A), the structure of the other term becomes simpler. That lemma is stated in the followng theorem.

\persiantheorem{2.4}
If
$f\finsarg$
is
$O\!\paren{\x^N}$
as
$\x \to 0$,
then its retarded integral can be written as
\begin{equation}\label{2.2.2.53a}
\retint{f\retarg}\finsarg=\sum_{\ell=0}^{N-1}n^L\x^\ell F_L\targ+g\finsarg,
\end{equation}
where
$g\finsarg=O\!\paren{\x^N}$
when
$\x \to 0$.
\hfill
$\blacksquare$

Since
$\hat{n}^Q\x^{a+2+2s}\lnx^k F_k\targ$
is
$O\!\paren{\x^{2s+a+2}}$
as
$\x \to 0$,
using Theorem 2.4 and rewriting the first term on the right-hand side of
Eq. (\ref{2.2.2.53a})
in terms of STF tensors, the structure of
$\fpa\retint{\xpb\hat{n}'^Q\xp^a \lnxp^p F\tparg}$
takes the form
\begin{eqnarray}\label{2.2.2.66}
\lefteqn{\fpa\retint{\xpb\hat{n}'^Q\xp^a \lnxp^p F\tparg}} \nonumber \\ 
&\!\!\!\!\!\!\!\!=\!\!\!\!\!\!\!\!&\sum_{\ell=0}^{\mathrm{max}\left\{q,2s+a+1\right\}}\sum_{j=0}^{2s+a+2}\sum_{k=0}^{k_{\mathrm{max}}}\hat{n}^L\x^j\lnx^k \hat{F}_{L,j,k}\targ+R\finsarg,
\end{eqnarray}
where
$2s+a+1\ge0$,
$\hat{F}_{L,j,k}\targ$
and
$R\finsarg$
are past-zero and
$R\finsarg$
is
$O\!\paren{\x^{2s+a+2}}$
when
$\x \to 0$.
If
$\mathrm{A}\retint{\xpb\hat{n}'^Q\xp^a \lnxp^p F\tparg}$
is analytic at
$B=0$,
$k_{\mathrm{max}}$
is equal to
$p$
and if
$B=0$
is a singularity of
$\mathrm{A}\retint{\xpb\hat{n}'^Q\xp^a \lnxp^p F\tparg}$,
it equals
$p+1$.

Finally, we examine the equation
$\Box f\finsarg=\x^{N+1}g\finsarg$.
$\x^{N+1}g\finsarg$
is a bounded function owing to
$g\finsarg$
being smooth and past-zero. Since
$\x^{N+1}g\finsarg$
is also a past-zero function, using Theorem 2.1 and considering remark 2 following Theorem 2.3, one finds that the original domain of analyticity of
$\retint{\xpb \xp^{N+1}g\retarg}$
is the half-plane
$\re B > -N-4$.
Because
$B=0$
lies inside this region,\footnote{If we hadn't taken
$N_0$
to be large enough before, instead of
$N+1$
the power of
$\xp$
would have been less than or equal to
$-3$.
In that case,
$B=0$
wouldn't have been inside the original domain of analyticity, and since the functional dependence of
$g\retarg$
is not simple, the examination of the possibility of the analytic continuation of
$\retint{\xpb \xp^{N+1}g\retarg}$
would have been difficult.}
a particular solution to the equation
$\Box f\finsarg=\x^{N+1}g\finsarg$
is
$\fpa\retint{\xpb \xp^{N+1}g\retarg}$.
We have
\begin{eqnarray}\label{2.2.2.69}
\lefteqn{\fpa\retint{\xpb \xp^{N+1}g\retarg}} \nonumber \\ 
&\!\!\!\!\!\!\!\!=\!\!\!\!\!\!\!\!&\fp\retint{\xpb \xp^{N+1}g\retarg} \nonumber \\ 
&\!\!\!\!\!\!\!\!=\!\!\!\!\!\!\!\!&\retint{\xp^{N+1}g\retarg}.
\end{eqnarray}
Accordingly, since
$\x^{N+1}g\finsarg$
is
$O\!\paren{\x^{N+1}}$
as
$\x \to 0$,
using Theorem 2.4 and rewriting the first term on the right-hand side of
Eq. (\ref{2.2.2.53a})
in terms of STF tensors, the structure of
$\fpa\retint{\xpb \xp^{N+1}g\retarg}$
reads
\begin{IEEEeqnarray}{rcl}\label{2.2.2.73}
\fpa\retint{\xpb \xp^{N+1}g\retarg} & = & \sum_{\ell=0}^{N}\sum_{j=0}^{N}\hat{n}^L\x^j \hat{G}_{L,j}\targ + R''\finsarg,\quad
\end{IEEEeqnarray}
where
$\hat{G}_{L,j,k}\targ$
and
$R''\finsarg$
are past-zero and
$R''\finsarg=O\!\paren{\x^{N+1}}$
when
$\x \to 0$.

In light of foregoing investigation, a particular solution to the equation
$\Box  h^{\mu\nu}_{\paren{2}\mathrm{PZ}}\finsarg=\Lambda^{\mu\nu}_{\paren{2}\mathrm{PZ}}\finsarg$
is given by
\begin{IEEEeqnarray}{rcl}\label{2.2.2.74}
u^{\mu\nu}_{\paren{2}\mathrm{PZ}}\finsarg & = &  \sum_{q=0}^{q_{\mathrm{max}}}\sum_{a=a_{\mathrm{min}}}^{a_{\mathrm{max}}}\fpa\retint{\xpb \hat{n}'^Q\xp^a \hat{F}^{\mu\nu}_{\paren{2}Q,a}\tparg} \nonumber \\  
&& \negmedspace {}+\fpa\retint{\xpb R^{\mu\nu}_{\paren{2}}\retarg},
\end{IEEEeqnarray}
which, due to existence of a common original region of analyticity  of all its constituents appearing on the right-hand side (some right half-plane in the complex plane), can be rewritten as
\begin{equation}\label{2.2.2.75}
 u^{\mu\nu}_{\paren{2}\mathrm{PZ}}\finsarg=\fpa\retint{\xpb \Lambda^{\mu\nu}_{\paren{2}\mathrm{PZ}}\retarg}.
\end{equation}
It can be shown that some of
$a$'s
and
$q$'s
appearing in the structure of
$\Lambda^{\mu\nu}_{\paren{2}\mathrm{PZ}}\finsarg$
satisfy either
Eq. (\ref{2.2.2.45})
or
Eq. (\ref{2.2.2.46}).
Therefore, the structure of
$u^{\mu\nu}_{\paren{2}\mathrm{PZ}}\finsarg$
is given by
\begin{equation}\label{2.2.2.76}
u^{\mu\nu}_{\paren{2}\mathrm{PZ}}\finsarg=\sum_{q=0}^{q'_{\mathrm{max}}}\sum_{a=a'_{\mathrm{min}}}^{a'_{\mathrm{max}}}\sum_{p=0}^{1}\hat{n}^Q\x^a\lnx^p \hat{E}^{\mu\nu}_{\paren{2}Q,a,p}+R'^{\;\mu\nu}_{\paren{2}}\finsarg.
\end{equation}

Having constructed particular solutions to the equations
$\Box  h^{\mu\nu}_{\paren{2}\mathrm{AS}}\finsarg=\Lambda^{\mu\nu}_{\paren{2}\mathrm{AS}}\xarg$
and
$\Box  h^{\mu\nu}_{\paren{2}\mathrm{PZ}}\finsarg=\Lambda^{\mu\nu}_{\paren{2}\mathrm{PZ}}\finsarg$,
we are now in a position to write down a paticular solution to the equation
$\Box  h^{\mu\nu}_{\paren{2}}\finsarg=\Lambda^{\mu\nu}_{\paren{2}}\finsarg$.
That solution is
\begin{equation}\label{2.2.2.77}
u^{\mu\nu}_{\paren{2}}\finsarg=u^{\mu\nu}_{\paren{2}\mathrm{AS}}\xarg+u^{\mu\nu}_{\paren{2}\mathrm{PZ}}\finsarg \stareq \fpa\retint{\xpb \Lambda^{\mu\nu}_{\paren{2}}\retarg}
\end{equation}
where
$u^{\mu\nu}_{\paren{2}\mathrm{AS}}\xarg$
and
$u^{\mu\nu}_{\paren{2}\mathrm{PZ}}\finsarg$
are given by
Eqs. (\ref{2.2.2.25b}) and (\ref{2.2.2.75})
respectively. Further, the structure of this solution is the sum of the structures given in
Eqs. (\ref{2.2.2.26}) and (\ref{2.2.2.76}).

\subsection{Harmonic Gauge Condition}

In previous subsection we observed that a particular solution to the equation
$\Box  h^{\mu\nu}_{\paren{2}}\finsarg=\Lambda^{\mu\nu}_{\paren{2}}\finsarg$
is given by
Eq. (\ref{2.2.2.77}).
The equation governing
$ \partial_\mu u_{\paren{2}}^{\mu\nu}\finsarg$
is
\begin{equation}\label{2.2.3.1}
\Box\left(\partial_\mu u_{\paren{2}}^{\mu\nu}\finsarg\right) =0.
\end{equation}
Taking into account that
$ u_{\paren{2}}^{\mu\nu}\finsarg$
is past-zero (we will discuss this property later), the general solution to
Eq. (\ref{2.2.3.1})
is given by
\begin{equation}\label{2.2.3.2}
 \partial_\mu u_{\paren{2}}^{\mu\nu}\finsarg=\sum_{\ell=0}^{\infty}\hat{\partial}_L\left(\frac{\hat{U}^{\nu}_L( t-\frac{ |\mathbf{x}|}{c})}{ |\mathbf{x}|}\right).
\end{equation}
Thus, there is a possibility that
$\partial_\mu u_{\paren{2}}^{\mu\nu}\finsarg$
is nonzero. To decide whether
$\partial_\mu u_{\paren{2}}^{\mu\nu}\finsarg$
vanishes or not, we examine
$\partial_\mu u^{\mu\nu}_{\paren{2}\mathrm{AS}}\xarg$
and
$\partial_\mu u^{\mu\nu}_{\paren{2}\mathrm{PZ}}\finsarg$
separately.

Using
Eq. (\ref{2.2.2.25b}),
we get
\begin{IEEEeqnarray}{rcl}\label{2.2.3.4}
 \partial_\mu u^{\mu\nu}_{\paren{2}\mathrm{AS}}\xarg &=&\partial_\mu\fpa\left[\gnearint{\Lambda^{\mu\nu}_{\paren{2}\mathrm{AS}}(\mathbf{x'})}\right] \nonumber \\  
&& \negmedspace {}+ \partial_\mu\fpa\left[\gfarint{\Lambda^{\mu\nu}_{\paren{2}\mathrm{AS}}(\mathbf{x'})}\right].
\end{IEEEeqnarray}
One can rewrite both terms on the right-hand side of the above equation. In the original domain of analyticity of
$-\frac{1}{4\pi}\int_{D_i} \paren{\frac{\xp}{r_0}}^B \frac{\Lambda^{\mu\nu}_{\paren{2}\mathrm{AS}}(\mathbf{x'})}{|\mathbf{x}-\mathbf{x'}|}\ud^3\mathbf{x'}$,
where
\begin{equation}
D_i=
\begin{cases}
\{\mathbf{x'}\in \mathbb{R }^3 \left|\right. \xp<\mathcal{R}\}, &\qquad i=1,\\
\{\mathbf{x'}\in \mathbb{R }^3 \left|\right. \mathcal{R}<\xp\}, &\qquad i=2,
\end{cases}\nonumber
\end{equation}
namely in either some right or left half-plane depending on whether
$i$
is, we can write
\begin{eqnarray}\label{2.2.3.7}
\lefteqn{\partial_\mu\mathrm{A}\left[-\frac{1}{4\pi}\int_{D_i} \paren{\frac{\xp}{r_0}}^B \frac{\Lambda^{\mu\nu}_{\paren{2}\mathrm{AS}}(\mathbf{x'})}{|\mathbf{x}-\mathbf{x'}|}\ud^3\mathbf{x'}\right]} \nonumber \\ 
&\!\!\!\!\!\!\!\!=\!\!\!\!\!\!\!\!& \partial_\mu\left[-\frac{1}{4\pi}\int_{D_i} \paren{\frac{\xp}{r_0}}^B \frac{\Lambda^{\mu\nu}_{\paren{2}\mathrm{AS}}(\mathbf{x'})}{|\mathbf{x}-\mathbf{x'}|}\ud^3\mathbf{x'}\right] \nonumber\\
&\!\!\!\!\!\!\!\!=\!\!\!\!\!\!\!\!& -\frac{1}{4\pi}\int_{D_i} \paren{\frac{\xp}{r_0}}^B\Lambda^{k\nu}_{\paren{2}\mathrm{AS}}(\mathbf{x'})\partial_k\paren{\frac{1}{|\mathbf{x}-\mathbf{x'}|}}\ud^3\mathbf{x'} \nonumber\\
&\!\!\!\!\!\!\!\!=\!\!\!\!\!\!\!\!& \frac{1}{4\pi}\int_{D_i} \paren{\frac{\xp}{r_0}}^B\Lambda^{k\nu}_{\paren{2}\mathrm{AS}}(\mathbf{x'})\partial'_k\paren{\frac{1}{|\mathbf{x}-\mathbf{x'}|}}\ud^3\mathbf{x'} \nonumber\\
&\!\!\!\!\!\!\!\!=\!\!\!\!\!\!\!\!& \frac{1}{4\pi}\int_{D_i}\partial'_k\left[ \paren{\frac{\xp}{r_0}}^B \frac{\Lambda^{k\nu}_{\paren{2}\mathrm{AS}}(\mathbf{x'})}{|\mathbf{x}-\mathbf{x'}|}\right]\ud^3\mathbf{x'}\nonumber\\
&&\!\!\!\!-\frac{1}{4\pi}\int_{D_i} \paren{\frac{\xp}{r_0}}^B \frac{\partial'_k\Lambda^{k\nu}_{\paren{2}\mathrm{AS}}(\mathbf{x'})}{|\mathbf{x}-\mathbf{x'}|}\ud^3\mathbf{x'}\nonumber\\
&&\!\!\!\!-\frac{1}{4\pi}\int_{D_i}\partial'_k\left[ \paren{\frac{\xp}{r_0}}^B\right] \frac{\Lambda^{k\nu}_{\paren{2}\mathrm{AS}}(\mathbf{x'})}{|\mathbf{x}-\mathbf{x'}|}\ud^3\mathbf{x'}.
\end{eqnarray}
The first term on the right-hand side of the last equality in the above equation can be rewritten by using Gauss' theorem. The second term is zero due to conservation equation,
$\Lambda^{\mu\nu}_{\paren{2}\mathrm{AS}}(\mathbf{x})$
being always stationary and
$\Lambda^{\mu\nu}_{\paren{2}\mathrm{PZ}}\finsarg$
being past-zero. Therefore,
Eq. (\ref{2.2.3.7})
takes the form
\begin{eqnarray}\label{2.2.3.8}
\lefteqn{\partial_\mu\mathrm{A}\left[-\frac{1}{4\pi}\int_{D_i} \paren{\frac{\xp}{r_0}}^B \frac{\Lambda^{\mu\nu}_{\paren{2}\mathrm{AS}}(\mathbf{x'})}{|\mathbf{x}-\mathbf{x'}|}\ud^3\mathbf{x'}\right]} \nonumber \\ 
&\!\!\!\!\!\!\!\!=\!\!\!\!\!\!\!\!&\frac{1}{4\pi}\int_{\partial D_i} \paren{\frac{\xp}{r_0}}^B \frac{\Lambda^{k\nu}_{\paren{2}\mathrm{AS}}(\mathbf{x'})}{|\mathbf{x}-\mathbf{x'}|}\ud\sigma'_{\paren{i}k}\nonumber\\
&&\!\!\!\!-\frac{1}{4\pi}\int_{D_i}B\paren{\frac{\xp}{r_0}}^B \xp^{-1}n'^k\frac{\Lambda^{k\nu}_{\paren{2}\mathrm{AS}}(\mathbf{x'})}{|\mathbf{x}-\mathbf{x'}|}\ud^3\mathbf{x'}.
\end{eqnarray}
For
$i=1$
we have
\begin{eqnarray}\label{2.2.3.9}
\lefteqn{\partial_\mu\mathrm{A}\left[-\frac{1}{4\pi}\int_{|\mathbf{x'}|<\mathcal{R}} \paren{\frac{\xp}{r_0}}^B \frac{\Lambda^{\mu\nu}_{\paren{2}\mathrm{AS}}(\mathbf{x'})}{|\mathbf{x}-\mathbf{x'}|}\ud^3\mathbf{x'}\right]} \nonumber \\ 
&\!\!\!\!\!\!\!\!=\!\!\!\!\!\!\!\!&\frac{1}{4\pi}\int_{\xp=\mathcal{R}} \paren{\frac{\xp}{r_0}}^B \frac{\Lambda^{k\nu}_{\paren{2}\mathrm{AS}}(\mathbf{x'})}{|\mathbf{x}-\mathbf{x'}|}\ud\sigma'_{\paren{1}k}\nonumber\\
&&\!\!\!\!+B\left[-\frac{1}{4\pi}\int_{|\mathbf{x'}|<\mathcal{R}}\paren{\frac{\xp}{r_0}}^B \frac{\xp^{-1}n'^k\Lambda^{k\nu}_{\paren{2}\mathrm{AS}}(\mathbf{x'})}{|\mathbf{x}-\mathbf{x'}|}\ud^3\mathbf{x'}\right].
\end{eqnarray}
As a result of identity theorem,
$\partial_\mu\mathrm{A}\left[-\frac{1}{4\pi}\int_{|\mathbf{x'}|<\mathcal{R}} \paren{\frac{\xp}{r_0}}^B \frac{\Lambda^{\mu\nu}_{\paren{2}\mathrm{AS}}(\mathbf{x'})}{|\mathbf{x}-\mathbf{x'}|}\ud^3\mathbf{x'}\right]$
equates to the analytic continuation of the right-hand side of
Eq. (\ref{2.2.3.9})
wherever they are both defined. Thus, considering that the surface integral and the function
$f(B)=B$
are entire, we find
\begin{eqnarray}\label{2.2.3.10}
\lefteqn{\partial_\mu\mathrm{A}\left[-\frac{1}{4\pi}\int_{|\mathbf{x'}|<\mathcal{R}} \paren{\frac{\xp}{r_0}}^B \frac{\Lambda^{\mu\nu}_{\paren{2}\mathrm{AS}}(\mathbf{x'})}{|\mathbf{x}-\mathbf{x'}|}\ud^3\mathbf{x'}\right]} \nonumber \\ 
&\!\!\!\!\!\!\!\!=\!\!\!\!\!\!\!\!&\frac{1}{4\pi}\int_{\xp=\mathcal{R}} \paren{\frac{\xp}{r_0}}^B \frac{\Lambda^{k\nu}_{\paren{2}\mathrm{AS}}(\mathbf{x'})}{|\mathbf{x}-\mathbf{x'}|}\ud\sigma'_{\paren{1}k}\nonumber\\
&&\!\!\!\!+B\cdot\mathrm{A}\left[-\frac{1}{4\pi}\int_{|\mathbf{x'}|<\mathcal{R}}\paren{\frac{\xp}{r_0}}^B \frac{\xp^{-1}n'^k\Lambda^{k\nu}_{\paren{2}\mathrm{AS}}(\mathbf{x'})}{|\mathbf{x}-\mathbf{x'}|}\ud^3\mathbf{x'}\right].
\end{eqnarray}
For
$i=2$
we get
\begin{eqnarray}\label{2.2.3.11}
\lefteqn{\partial_\mu\mathrm{A}\left[-\frac{1}{4\pi}\int_{\mathcal{R}<|\mathbf{x'}|} \paren{\frac{\xp}{r_0}}^B \frac{\Lambda^{\mu\nu}_{\paren{2}\mathrm{AS}}(\mathbf{x'})}{|\mathbf{x}-\mathbf{x'}|}\ud^3\mathbf{x'}\right]} \nonumber \\ 
&\!\!\!\!\!\!\!\!=\!\!\!\!\!\!\!\!&\frac{1}{4\pi}\int_{\xp=\mathcal{R}} \paren{\frac{\xp}{r_0}}^B \frac{\Lambda^{k\nu}_{\paren{2}\mathrm{AS}}(\mathbf{x'})}{|\mathbf{x}-\mathbf{x'}|}\ud\sigma'_{\paren{2}k}\nonumber\\
&&\!\!\!\!+\frac{1}{4\pi}\int_{\xp\to\infty} \paren{\frac{\xp}{r_0}}^B \frac{\Lambda^{k\nu}_{\paren{2}\mathrm{AS}}(\mathbf{x'})}{|\mathbf{x}-\mathbf{x'}|}\ud\sigma'_{\paren{2}k}\nonumber\\
&&\!\!\!\!+B\left[-\frac{1}{4\pi}\int_{\mathcal{R}<|\mathbf{x'}|}\paren{\frac{\xp}{r_0}}^B \frac{\xp^{-1}n'^k\Lambda^{k\nu}_{\paren{2}\mathrm{AS}}(\mathbf{x'})}{|\mathbf{x}-\mathbf{x'}|}\ud^3\mathbf{x'}\right].
\end{eqnarray}
Taking
$\re B$
to be a large enough negative number, the second surface integral in
Eq. (\ref{2.2.3.11})
vanishes. By virtue of identity theorem,
$\partial_\mu\mathrm{A}\left[-\frac{1}{4\pi}\int_{\mathcal{R}<|\mathbf{x'}|} \paren{\frac{\xp}{r_0}}^B \frac{\Lambda^{\mu\nu}_{\paren{2}\mathrm{AS}}(\mathbf{x'})}{|\mathbf{x}-\mathbf{x'}|}\ud^3\mathbf{x'}\right]$
and the sum of the analytic continuations of the remaining terms on the right-hand side of
Eq. (\ref{2.2.3.11})
can be equated in their common region of definition. Therefore, taking into account that the surface integral and the function
$f(B)=B$
are entire, we reach
\begin{eqnarray}\label{2.2.3.12}
\lefteqn{\partial_\mu\mathrm{A}\left[-\frac{1}{4\pi}\int_{\mathcal{R}<|\mathbf{x'}|} \paren{\frac{\xp}{r_0}}^B \frac{\Lambda^{\mu\nu}_{\paren{2}\mathrm{AS}}(\mathbf{x'})}{|\mathbf{x}-\mathbf{x'}|}\ud^3\mathbf{x'}\right]} \nonumber \\ 
&\!\!\!\!\!\!\!\!=\!\!\!\!\!\!\!\!&\frac{1}{4\pi}\int_{\xp=\mathcal{R}} \paren{\frac{\xp}{r_0}}^B \frac{\Lambda^{k\nu}_{\paren{2}\mathrm{AS}}(\mathbf{x'})}{|\mathbf{x}-\mathbf{x'}|}\ud\sigma'_{\paren{2}k}\nonumber\\
&&\!\!\!\!+B\cdot\mathrm{A}\left[-\frac{1}{4\pi}\int_{\mathcal{R}<|\mathbf{x'}|}\paren{\frac{\xp}{r_0}}^B \frac{\xp^{-1}n'^k\Lambda^{k\nu}_{\paren{2}\mathrm{AS}}(\mathbf{x'})}{|\mathbf{x}-\mathbf{x'}|}\ud^3\mathbf{x'}\right].
\end{eqnarray}
Now, considering
Eqs. (\ref{2.2.3.10}) and (\ref{2.2.3.12}),
we obtain
\begin{eqnarray}\label{2.2.3.13}
\lefteqn{\partial_\mu\mathrm{A}\left[-\frac{1}{4\pi}\int_{D_i} \paren{\frac{\xp}{r_0}}^B \frac{\Lambda^{\mu\nu}_{\paren{2}\mathrm{AS}}(\mathbf{x'})}{|\mathbf{x}-\mathbf{x'}|}\ud^3\mathbf{x'}\right]} \nonumber \\ 
&\!\!\!\!\!\!\!\!=\!\!\!\!\!\!\!\!&\frac{1}{4\pi}\int_{\xp=\mathcal{R}} \paren{\frac{\xp}{r_0}}^B \frac{\Lambda^{k\nu}_{\paren{2}\mathrm{AS}}(\mathbf{x'})}{|\mathbf{x}-\mathbf{x'}|}\ud\sigma'_{\paren{i}k}\nonumber\\
&&\!\!\!\!+B\cdot\mathrm{A}\left[-\frac{1}{4\pi}\int_{D_i}\paren{\frac{\xp}{r_0}}^B \frac{\xp^{-1}n'^k\Lambda^{k\nu}_{\paren{2}\mathrm{AS}}(\mathbf{x'})}{|\mathbf{x}-\mathbf{x'}|}\ud^3\mathbf{x'}\right].
\end{eqnarray}
Irrespective of whether
$i$
is, we can straightforwardly show that all the terms appearing in the above equation are analytic in some punctured neighborhood of
$B=0$.
Thus, each of them possesses a Laurent expansion about
$B=0$
(the term containing surface integral is analytic at
$B=0$,
thereby possessing a Taylor expansion about this point). Since the coefficients of
$B^n$
on both sides of
Eq. (\ref{2.2.3.13})
must be equal for each
$n$,
considering the coefficients of
$B^0$,
we find
\begin{eqnarray}\label{2.2.3.15}
\lefteqn{\partial_\mu\fpa\left[-\frac{1}{4\pi}\int_{D_i} \paren{\frac{\xp}{r_0}}^B \frac{\Lambda^{\mu\nu}_{\paren{2}\mathrm{AS}}(\mathbf{x'})}{|\mathbf{x}-\mathbf{x'}|}\ud^3\mathbf{x'}\right]} \nonumber \\ 
&\!\!\!\!\!\!\!\!=\!\!\!\!\!\!\!\!& \frac{1}{4\pi}\int_{\xp=\mathcal{R}} \frac{\Lambda^{k\nu}_{\paren{2}\mathrm{AS}}(\mathbf{x'})}{|\mathbf{x}-\mathbf{x'}|}\ud{\sigma'}_{\paren{i}k}\nonumber\\
&&\!\!\!\!+ \resa \left[-\frac{1}{4\pi}\int_{D_i}\paren{\frac{\xp}{r_0}}^B \frac{\xp^{-1}n'^k\Lambda^{k\nu}_{\paren{2}\mathrm{AS}}(\mathbf{x'})}{|\mathbf{x}-\mathbf{x'}|}\ud^3\mathbf{x'}\right];
\end{eqnarray}
hence, the sum of
$\partial_\mu\fpa\left[-\frac{1}{4\pi}\int_{D_i} \paren{\frac{\xp}{r_0}}^B \frac{\Lambda^{\mu\nu}_{\paren{2}\mathrm{AS}}(\mathbf{x'})}{|\mathbf{x}-\mathbf{x'}|}\ud^3\mathbf{x'}\right]$'s
with
$i=1$
and
$i=2$,
or, considering
Eq. (\ref{2.2.3.4}),
$\partial_\mu u_{\paren{2}\mathrm{AS}}^{\mu\nu}\xarg$
equals
\begin{IEEEeqnarray}{rcl}\label{2.2.3.16}
\partial_\mu u_{\paren{2}\mathrm{AS}}^{\mu\nu}\xarg & = & \resa \left[-\frac{1}{4\pi}\int_{|\mathbf{x'}|<\mathcal{R}}\paren{\frac{\xp}{r_0}}^B \frac{\xp^{-1}n'^i\Lambda^{i\nu}_{\paren{2}\mathrm{AS}}(\mathbf{x'})}{|\mathbf{x}-\mathbf{x'}|}\ud^3\mathbf{x'}\right] \nonumber \\  
&& \negmedspace {} + \resa \left[-\frac{1}{4\pi}\int_{\mathcal{R}<|\mathbf{x'}|}\paren{\frac{\xp}{r_0}}^B \frac{\xp^{-1}n'^i\Lambda^{i\nu}_{\paren{2}\mathrm{AS}}(\mathbf{x'})}{|\mathbf{x}-\mathbf{x'}|}\ud^3\mathbf{x'}\right],\quad
\end{IEEEeqnarray}
where the disappearance of the surface integrals is due to
$\vec{\ud\sigma'_{\paren{1}}}$
being the opposite of
$\vec{\ud\sigma'_{\paren{2}}}$
at each point of the surface
$\xp=\mathcal{R}$.
The maximal power of
$\x$
in the structure of
$\Lambda^{\mu\nu}_{\paren{2}\mathrm{AS}}\xarg$
is
$-4$.
Therefore, considering remark 2 following Theorem 2.3 and using Theorem 2.2, it is obvious that
$-\frac{1}{4\pi}\int_{\mathcal{R}<|\mathbf{x'}|}\paren{\frac{\xp}{r_0}}^B \frac{\xp^{-1}n'^i\Lambda^{i\nu}_{\paren{2}\mathrm{AS}}(\mathbf{x'})}{|\mathbf{x}-\mathbf{x'}|}\ud^3\mathbf{x'}$
is analytic at
$B=0$
and the second term on the right-hand side of
Eq. (\ref{2.2.3.16})
thereby vanishes. Moreover, it can be shown the structure of
$\Lambda^{\mu\nu}_{\paren{2}\mathrm{AS}}(\mathbf{x})$
is such that
$-\frac{1}{4\pi}\int_{|\mathbf{x'}|<\mathcal{R}}\paren{\frac{\xp}{r_0}}^B \frac{\xp^{-1}n'^i\Lambda^{i\nu}_{\paren{2}\mathrm{AS}}(\mathbf{x'})}{|\mathbf{x}-\mathbf{x'}|}\ud^3\mathbf{x'}$
is analytic at
$B=0$,
i.e.,
the first term vanishes too. Thus,
$\partial_\mu u_{\paren{2}}^{\mu\nu}\finsarg$
is equal to zero.

Now we examine
$\partial_\mu u_{\paren{2}\mathrm{PZ}}^{\mu\nu}\finsarg$.
First, note that in the original domain of analyticity of
$\retint{\xpb \Lambda^{\mu\nu}_{\paren{2}\mathrm{PZ}}\retarg}$
(which is some right half-plane), we can write
\begin{eqnarray}\label{2.2.3.19}
\lefteqn{\partial_\mu\mathrm{A}\retint{\xpb \Lambda^{\mu\nu}_{\paren{2}\mathrm{PZ}}\retarg}} \nonumber \\ 
&\!\!\!\!\!\!\!\!=\!\!\!\!\!\!\!\!& \partial_\mu\retint{\xpb \Lambda^{\mu\nu}_{\paren{2}\mathrm{PZ}}\retarg} \nonumber\\
&\!\!\!\!\!\!\!\!=\!\!\!\!\!\!\!\!& -\frac{1}{4\pi}\int_{\mathbb{R }^3} \paren{\frac{\xp}{r_0}}^B \partial_\mu\left[\frac{ \Lambda^{\mu\nu}_{\paren{2}\mathrm{PZ}}\retarg}{|\mathbf{x}-\mathbf{x'}|}\right]\ud^3\mathbf{x'} \nonumber\\
&\!\!\!\!\!\!\!\!=\!\!\!\!\!\!\!\!& \gwholeint{\partial'_0 \Lambda^{0\nu}_{\paren{2}\mathrm{PZ}}\retarg} \nonumber\\
&&\!\!\!\!- \frac{1}{4\pi}\int_{\mathbb{R }^3} \paren{\frac{\xp}{r_0}}^B \partial'_i\paren{\frac{|\mathbf{x}-\mathbf{x'}|}{c}}\frac{\partial_{t'} \Lambda^{i\nu}_{\paren{2}\mathrm{PZ}}\retarg}{|\mathbf{x}-\mathbf{x'}|}\ud^3\mathbf{x'}\nonumber\\
&&\!\!\!\!+\frac{1}{4\pi}\int_{\mathbb{R }^3}\partial'_i\left[ \paren{\frac{\xp}{r_0}}^B \frac{\Lambda^{i\nu}_{\paren{2}\mathrm{PZ}}\retarg}{|\mathbf{x}-\mathbf{x'}|}\right]\ud^3\mathbf{x'}\nonumber\\
&&\!\!\!\!-\frac{1}{4\pi}\int_{\mathbb{R }^3} \paren{\frac{\xp}{r_0}}^B \frac{\partial'_i\Lambda^{i\nu}_{\paren{2}\mathrm{PZ}}\retarg}{|\mathbf{x}-\mathbf{x'}|}\ud^3\mathbf{x'}\nonumber\\
&&\!\!\!\!-\frac{1}{4\pi}\int_{\mathbb{R }^3}\partial'_i\left[ \paren{\frac{\xp}{r_0}}^B\right] \frac{\Lambda^{i\nu}_{\paren{2}\mathrm{PZ}}\retarg}{|\mathbf{x}-\mathbf{x'}|}\ud^3\mathbf{x'}.
\end{eqnarray}
The third term on the right-hand side of the last equality in the above equation can be rewritten as a surface integral by means of Gauss' theorem. This surface integral vanishes since integration is over a surface at infinity, a region in which the integrand equals zero due to being past-zero. Further, the fourth term can be rewritten as
\begin{eqnarray}\label{2.2.3.21}
\lefteqn{-\frac{1}{4\pi}\int_{\mathbb{R }^3} \paren{\frac{\xp}{r_0}}^B \frac{\partial'_i\Lambda^{i\nu}_{\paren{2}\mathrm{PZ}}\retarg}{|\mathbf{x}-\mathbf{x'}|}\ud^3\mathbf{x'}} \nonumber \\ 
&\!\!\!\!\!\!\!\!=\!\!\!\!\!\!\!\!& \frac{1}{4\pi}\int_{\mathbb{R }^3} \paren{\frac{\xp}{r_0}}^B \partial'_i\paren{\frac{|\mathbf{x}-\mathbf{x'}|}{c}}\frac{\partial_{t'} \Lambda^{i\nu}_{\paren{2}\mathrm{PZ}}\retarg}{|\mathbf{x}-\mathbf{x'}|}\ud^3\mathbf{x'} \nonumber\\
&&\!\!\!\!-\frac{1}{4\pi}\int_{\mathbb{R }^3} \paren{\frac{\xp}{r_0}}^B \frac{\paren{\partial'_i\Lambda^{i\nu}_{\paren{2}\mathrm{PZ}}\retarg}_{t'}}{|\mathbf{x}-\mathbf{x'}|}\ud^3\mathbf{x'},
\end{eqnarray}
where
$\paren{\partial'_i\Lambda^{i\nu}_{\paren{2}\mathrm{PZ}}\retarg}_{t'}$
denotes the partial derivative of
$\Lambda^{i\nu}_{\paren{2}\mathrm{PZ}}\retarg$
with respect to
$x'^i$,
ignoring the contribution from the variable
$t'$
(regarding
$t'$
as a constant). In light of these considerations,
Eq. (\ref{2.2.3.19})
takes the form
\begin{eqnarray}\label{2.2.3.22}
\lefteqn{\partial_\mu\mathrm{A}\retint{\xpb \Lambda^{\mu\nu}_{\paren{2}\mathrm{PZ}}\retarg}} \nonumber \\ 
&\!\!\!\!\!\!\!\!=\!\!\!\!\!\!\!\!&\gwholeint{\partial'_0 \Lambda^{0\nu}_{\paren{2}\mathrm{PZ}}\retarg+\paren{\partial'_i\Lambda^{i\nu}_{\paren{2}\mathrm{PZ}}\retarg}_{t'}}\nonumber\\
&&\!\!\!\!-\frac{1}{4\pi}\int_{\mathbb{R }^3}\partial'_i\left[ \paren{\frac{\xp}{r_0}}^B\right] \frac{\Lambda^{i\nu}_{\paren{2}\mathrm{PZ}}\retarg}{|\mathbf{x}-\mathbf{x'}|}\ud^3\mathbf{x'}.
\end{eqnarray}
Since
$\partial_\mu\Lambda^{\mu\nu}_{\paren{2}\mathrm{AS}}\xarg=\partial_k\Lambda^{k\nu}_{\paren{2}\mathrm{AS}}\xarg=0$,
from conservation equation we can deduce that
$\partial_\mu\Lambda^{\mu\nu}_{\paren{2}\mathrm{PZ}}\finsarg=0$
too. Therefore, the first term on the right-hand side of
Eq. (\ref{2.2.3.22})
vanishes and we thereby get
\begin{IEEEeqnarray}{rcl}\label{2.2.3.23}
\partial_\mu\mathrm{A}\retint{\xpb \Lambda^{\mu\nu}_{\paren{2}\mathrm{PZ}}\retarg} & = & -\frac{1}{4\pi}\int_{\mathbb{R }^3}\partial'_i\left[ \paren{\frac{\xp}{r_0}}^B\right] \frac{\Lambda^{i\nu}_{\paren{2}\mathrm{PZ}}\retarg}{|\mathbf{x}-\mathbf{x'}|}\ud^3\mathbf{x'} \nonumber \\  
& = & B\cdot\retint{\xpb\xp^{-1}n'^i\Lambda^{i\nu}_{\paren{2}\mathrm{PZ}}\retarg}.\quad
\end{IEEEeqnarray}
As a consequence of identity theorem, we can equate
$\partial_\mu\mathrm{A}\retint{\xpb \Lambda^{\mu\nu}_{\paren{2}\mathrm{PZ}}\retarg}$
and the analytic continuation of the right-hand side of
Eq. (\ref{2.2.3.23})
wherever they are both defined. Thus, taking into account that
$f(B)=B$
is an entire function, we reach
\begin{IEEEeqnarray}{rcl}\label{2.2.3.24}
\partial_\mu\mathrm{A}\retint{\xpb \Lambda^{\mu\nu}_{\paren{2}\mathrm{PZ}}\retarg} & = &
 B\cdot\mathrm{A}\retint{\xpb\xp^{-1}n'^i\Lambda^{i\nu}_{\paren{2}\mathrm{PZ}}\retarg}.\qquad
\end{IEEEeqnarray}
Owing to the particular structure of
$ \Lambda^{\mu\nu}_{\paren{2}\mathrm{PZ}}\finsarg$,
it is straightforward to show that both sides of
Eq. (\ref{2.2.3.24})
are analytic in some punctured neighborhood of
$B=0$
and each of them thereby possesses a Laurent expansion around
$B=0$.
Since the coefficients of
$B^n$
on the two sides of
Eq. (\ref{2.2.3.24})
must be equal for each
$n$,
considering the coefficients of
$B^0$, we find
\begin{IEEEeqnarray}{rcl}\label{2.2.3.26}
\partial_\mu\fpa\retint{\xpb \Lambda^{\mu\nu}_{\paren{2}\mathrm{PZ}}\retarg} & = &
\resa\retint{\xpb\xp^{-1}n'^i\Lambda^{i\nu}_{\paren{2}\mathrm{PZ}}\retarg}.\nonumber \\  
\end{IEEEeqnarray}
Note that
$\partial_\mu\fpa\retint{\xpb \Lambda^{\mu\nu}_{\paren{2}\mathrm{PZ}}\retarg}$
is nothing but
$\partial_\mu u_{\paren{2}\mathrm{PZ}}^{\mu\nu}\finsarg$.
Thus, we have
\begin{equation}\label{2.2.3.27}
\partial_\mu u_{\paren{2}\mathrm{PZ}}^{\mu\nu}\finsarg=\resa\retint{\xpb\xp^{-1}n'^i\Lambda^{i\nu}_{\paren{2}\mathrm{PZ}}\retarg}.
\end{equation}

The structure of
$\x^{-1}n^i\Lambda^{i\nu}_{\paren{2}\mathrm{PZ}}\finsarg$
is of the form
\begin{equation}\label{2.2.3.28}
\x^{-1}n^i\Lambda^{i\nu}_{\paren{2}\mathrm{PZ}}\finsarg =\sum_{q=0}^{q_{\mathrm{max}}}\sum_{a=a_{\mathrm{min}}}^{a_{\mathrm{max}}}n^i\hat{n}^Q\x^{a-1}\hat{F}^{i\nu}_{\paren{2}Q,a}\targ +\x^{-1}n^i\ R^{i\nu}_{\paren{2}}\finsarg.
\end{equation}
It can be argued that the structure given in
Eq. (\ref{2.2.3.28})
can be rewritten as
\begin{IEEEeqnarray}{rcl}\label{2.2.3.29}
\x^{-1}n^i\Lambda^{i\nu}_{\paren{2}\mathrm{PZ}}\finsarg &=&\sum_{\ell=0}^{\ell_{\mathrm{max}}}\sum_{k=k_{\mathrm{min}}}^{k_{\mathrm{max}}}\sum_{p=0}^{p_{\mathrm{max}}}\hat{n}^L\x^k\lnx^p\hat{G}^\nu_{\paren{2}L,k,p}\targ + R^\nu_{\paren{2}}\finsarg,\quad
\end{IEEEeqnarray}
where
$\hat{G}^\nu_{\paren{2}L,k,p}\targ $
and
$ R^\nu_{\paren{2}}\finsarg$
are past-zero, and
$ R^\nu_{\paren{2}}\finsarg$
is
$\sum_{N=N_{\mathrm{min}}}^{N_{\mathrm{max}}}O\!\paren{\x^N}$
when
$\x\to0$
(in this structure
$p_{\mathrm{max}}$
must be zero but other values of it appear at higher orders).
$\gretint{R^\nu_{\paren{2}}\finsarg}$
doesn't contribute to the residue owing to being analytic at
$B=0$.
Therefore, we have
\begin{eqnarray}\label{2.2.3.30}
\lefteqn{\partial_\mu u_{\paren{2}\mathrm{PZ}}^{\mu\nu}\finsarg} \nonumber \\ 
&\!\!\!\!\!\!\!\!=\!\!\!\!\!\!\!\!& \resa\gretint{\sum_{\ell=0}^{\ell_{\mathrm{max}}}\sum_{k=k_{\mathrm{min}}}^{k_{\mathrm{max}}}\sum_{p=0}^{p_{\mathrm{max}}}\hat{n}'^L\xp^k\!\lnxp^{\!p}\!\!\hat{G}^\nu_{\paren{2}L,k,p}\tparg}  \nonumber\\
&\!\!\!\!\!\!\!\!=\!\!\!\!\!\!\!\!& \sum_{\ell=0}^{\ell_{\mathrm{max}}}\sum_{k=k_{\mathrm{min}}}^{k_{\mathrm{max}}}\sum_{p=0}^{p_{\mathrm{max}}}\resa\left[\!-\frac{1}{4\pi}\int_{\xp<\x} \paren{\frac{\xp}{r_0}}^B \frac{\hat{n}'^L\xp^k\!\lnxp^{\!p}\!\!\hat{G}^\nu_{\paren{2}L,k,p}\tparg}{|\mathbf{x}-\mathbf{x'}|}\ud^3\mathbf{x'}\!\right]\nonumber\\
&&\!\!\!\!+ \sum_{\ell=0}^{\ell_{\mathrm{max}}}\sum_{k=k_{\mathrm{min}}}^{k_{\mathrm{max}}}\sum_{p=0}^{p_{\mathrm{max}}}\resa\left[\!-\frac{1}{4\pi}\int_{\x<\xp} \paren{\frac{\xp}{r_0}}^B \frac{\hat{n}'^L\xp^k\!\lnxp^{\!p}\!\!\hat{G}^\nu_{\paren{2}L,k,p}\tparg}{|\mathbf{x}-\mathbf{x'}|}\ud^3\mathbf{x'}\!\right]. \nonumber\\
\end{eqnarray}
The second integral on the right-hand side of the last equality in
Eq. (\ref{2.2.3.30})
makes no contribution to the residue since it is analytic at
$B=0$.
Hence, we can write
\begin{eqnarray}\label{2.2.3.31}
\lefteqn{\partial_\mu u_{\paren{2}\mathrm{PZ}}^{\mu\nu}\finsarg} \nonumber \\ 
&\!\!\!\!\!\!\!\!=\!\!\!\!\!\!\!\!&  \sum_{\ell=0}^{\ell_{\mathrm{max}}}\sum_{k=k_{\mathrm{min}}}^{k_{\mathrm{max}}}\sum_{p=0}^{p_{\mathrm{max}}}\resa\left[\!-\frac{1}{4\pi}\int_{\xp<\x} \paren{\frac{\xp}{r_0}}^B \frac{\hat{n}'^L\xp^k\!\lnxp^{\!p}\!\!\hat{G}^\nu_{\paren{2}L,k,p}\tparg}{|\mathbf{x}-\mathbf{x'}|}\ud^3\mathbf{x'}\!\right]\nonumber\\
&\!\!\!\!\!\!\!\!=\!\!\!\!\!\!\!\!& -\frac{1}{4\pi}\sum_{\ell=0}^{\ell_{\mathrm{max}}}\sum_{k=k_{\mathrm{min}}}^{k_{\mathrm{max}}}\sum_{p=0}^{p_{\mathrm{max}}}\resa\left[\!\frac{\partial^p}{\partial B^p}\int_{\xp<\x} \paren{\frac{\xp}{r_0}}^B \frac{\hat{n}'^L\xp^k\hat{G}^\nu_{\paren{2}L,k,p}\tparg}{|\mathbf{x}-\mathbf{x'}|}\ud^3\mathbf{x'}\!\right].\nonumber\\
\end{eqnarray}
Using Taylor expansion for functions of three variables (see Appendix A), at
$\xp<\x$
we have
\begin{IEEEeqnarray}{rcl}\label{2.2.3.32}
\frac{\hat{G}^\nu_{\paren{2}{I_\ell},k,p}\tparg}{|\mathbf{x}-\mathbf{x'}|} &=&\frac{\hat{G}^\nu_{\paren{2}{I_\ell},k,p}\rettarg}{|\mathbf{x}-\mathbf{x'}|}\nonumber\\ &=& \sum_{j=0}^{\infty}\frac{\paren{-1}^j}{j!}x'^{I'_j}\partial_{I'_j}\frac{\hat{G}^\nu_{\paren{2}{I_\ell},k,p}\uarg}{\x};
\end{IEEEeqnarray}
combining
Eqs. (\ref{2.2.3.31}) and (\ref{2.2.3.32}),
some straightforward manipulation leads us to
\begin{IEEEeqnarray}{rcl}\label{2.2.3.33}
\partial_\mu u_{\paren{2}\mathrm{PZ}}^{\mu\nu}\finsarg & = &   -\frac{1}{4\pi}\sum_{j=0}^{\infty}\sum_{\ell=0}^{\ell_{\mathrm{max}}}\sum_{k=k_{\mathrm{min}}}^{k_{\mathrm{max}}}\sum_{p=0}^{p_{\mathrm{max}}}\frac{\paren{-1}^j}{j!}\bigg[\partial_{I'_j}\frac{\hat{G}^\nu_{\paren{2}{I_\ell},k,p}\uarg}{\x}\bigg]\x^{k+j+3}\nonumber \\  
&& \negmedspace {}\times\int n'^{I_\ell}n'^{I'_j}\ud\Omega'\res \frac{\partial^p}{\partial B^p}\xb \paren{B\!+\!k\!+\!j\!+\!3}^{-1}.
\end{IEEEeqnarray}
Using
Eq. (\ref{B.5}),
we get
\begin{equation}\label{2.2.3.34}
\int n'^{I_\ell}n'^{I'_j}\ud\Omega'=
\begin{cases}
\frac{4\pi}{\paren{j+\ell+1}!!}\delta^{\{ i_1 i_2}\cdots\delta^{ i'_{j-1} i'_{j}\}} &\qquad  \text{for} \; j+\ell \;\text{even},\\
0 &\qquad \text{for} \; j+\ell\;\text{odd};
\end{cases}
\end{equation}
therefore, terms with
$j+\ell$
odd don't contribute to the
$j$
sum in
Eq. (\ref{2.2.3.33}).
In addition, there is no contribution to this sum from terms with
$j<\ell$.
The reason behind this is that, if
$j<\ell$,
then each term of
$\delta^{\{ i_1 i_2}\cdots\delta^{ i'_{j-1} i'_{j}\}}$
have some Kronecker deltas with both indices belonging to the set of indices of
$I_\ell$
and contraction of such deltas with
$\hat{G}^\nu_{\paren{2}{I_\ell},k,p}$,
which is a STF tensor with respect to
$I_\ell$,
is thereby zero. In sum, only terms with
$j=\ell+2m$
contribute to the sum over
$j$.
Furthermore, since all the indices of
$I'_j$
are dummy and
$\partial_{I'_j}$
is a symmetric tensor, in computation of
$\partial_\mu u_{\paren{2}\mathrm{PZ}}^{\mu\nu}\finsarg$
we can replace all the terms of
$\delta^{\{ i_1 i_2}\cdots\delta^{ i'_{j-1} i'_{j}\}}$
that contribute to the
$j$
sum by only one term of them times their number. Choosing
$\frac{j!}{2^m m!}\paren{\delta^{ i_1 i'_1}\cdots\delta^{ i_\ell i'_\ell}\delta^{ i'_{\ell+1} i'_{\ell+2}}\cdots\delta^{ i'_{j-1} i'_j}}$
for such a replacement, we find
\begin{IEEEeqnarray}{rcl}\label{2.2.3.35}
\partial_\mu u_{\paren{2}\mathrm{PZ}}^{\mu\nu}\finsarg & = & \sum_{m=0}^{\infty}\sum_{\ell=0}^{\ell_{\mathrm{max}}}\sum_{k=k_{\mathrm{min}}}^{k_{\mathrm{max}}}\sum_{p=0}^{p_{\mathrm{max}}}\frac{\paren{-1}^{\ell+1}}{2^m m! \paren{2\ell+2m+1}!!}\bigg[\partial_L\Delta^m\frac{\hat{G}^\nu_{\paren{2}L,k,p}\uarg}{\x}\bigg]\nonumber \\  
&& \negmedspace {}\times\x^{k+\ell+2m+3}\res \frac{\partial^p}{\partial B^p}\xb \paren{B\!+\!k\!+\!\ell\!+\!2m\!+\!3}^{-1}.
\end{IEEEeqnarray}
It is apparent that only terms with
$m$,
$\ell$
and
$k$
fulfilling
$k+\ell+2m+3=0$
contribute to
$\sum_{m=0}^{\infty}\sum_{\ell=0}^{\ell_{\mathrm{max}}}\sum_{k=k_{\mathrm{min}}}^{k_{\mathrm{max}}}$
appearing in
Eq. (\ref{2.2.3.35}). Hence, we can say:\\
1. Since
$m$
is nonnegative, only terms with
$k \le -\ell-3$
contribute.\\
2. The only
$m$
that contributes to
$\sum_{m=0}^{\infty}$
is
$m_0=-\frac{1}{2}\paren{k\!+\!\ell\!+\!3}$.\\
In light of these considerations we can write
\begin{IEEEeqnarray}{rcl}\label{2.2.3.36}
\partial_\mu u_{\paren{2}\mathrm{PZ}}^{\mu\nu}\finsarg & = & \sum_{\ell=0}^{\ell_{\mathrm{max}}}\sum_{k=k_{\mathrm{min}}}^{-\ell-3}\sum_{p=0}^{p_{\mathrm{max}}}\frac{\paren{-1}^{\ell+1}}{2^{m_0} m_0! \paren{2\ell+2m_0+1}!!}\bigg[\partial_L\Delta^{m_0}\frac{\hat{G}^\nu_{\paren{2}L,k,p}\uarg}{\x}\bigg]\nonumber \\  
&& \negmedspace {}\times\res \frac{\partial^p}{\partial B^p}\frac{1}{B}\xb \nonumber\\
& = & \sum_{\ell=0}^{\ell_{\mathrm{max}}}\sum_{k=k_{\mathrm{min}}}^{-\ell-3}\sum_{p=0}^{p_{\mathrm{max}}}\frac{\paren{-1}^{\ell+1}}{2^{m_0}c^{2m_0} m_0! \paren{2\ell+2m_0+1}!!}\bigg[\partial_L\frac{{}^{\paren{2m_0}}\hat{G}^\nu_{\paren{2}L,k,p}\uarg}{\x}\bigg]\nonumber \\  
&& \negmedspace {}\times\res \frac{\partial^p}{\partial B^p}\frac{1}{B}\xb ,
\end{IEEEeqnarray}
where we have used
Eq. (\ref{2.1.3.7})
$m_0$
times in the last step. By means of Leibniz formula and Taylor expansion of
$\xb$,
we have
\begin{equation}\label{2.2.3.37}
\frac{\partial^p}{\partial B^p}\frac{1}{B}\xb = \sum_{n=0}^{p}\sum_{n'=0}^{\infty}\binom{p}{n}\frac{\paren{-1}^n n!}{n'!}\lnx^{p+n'-n}B^{n'-n-1}.
\end{equation}
Only terms with
$n'$
equal to
$n$
contribute to the residue from
$\sum_{n'=0}^{\infty}$
(terms with different values of
$n'$
don't bring about
$\frac{1}{B}$).
Therefore, using binomial expansion (see Appendix A), we obtain
\begin{equation}\label{2.2.3.38}
\res\frac{\partial^p}{\partial B^p}\frac{1}{B}\xb=\lnx^p\sum_{n=0}^{p}\binom{p}{n}\paren{-1}^n=
\begin{cases}
1, &\qquad  p=0, \\
0, &\qquad p\ge1,
\end{cases}
\end{equation}
thereby getting
\begin{IEEEeqnarray}{rcl}\label{2.2.3.39}
\partial_\mu u_{\paren{2}\mathrm{PZ}}^{\mu\nu}\finsarg&=&\sum_{\ell=0}^{\ell_{\mathrm{max}}}\sum_{k=k_{\mathrm{min}}}^{-\ell-3}\frac{\paren{-1}^{\ell+1}}{2^{m_0}c^{2m_0} m_0! \paren{2\ell+2m_0+1}!!}\partial_L\paren{\frac{{}^{\paren{2m_0}}\hat{G}^\nu_{\paren{2}L,k,0}\uarg}{\x}}.\nonumber\\
\end{IEEEeqnarray}
It can be shown the structure of
$\Lambda_{\paren{2}\mathrm{PZ}}^{\mu\nu}\finsarg$
is such that there exist nonzero
$\hat{G}^\nu_{\paren{2}L,k,0}\targ$'s
with
$k\le-\ell-3$.
Thus,
$\partial_\mu u_{\paren{2}\mathrm{PZ}}^{\mu\nu}\finsarg$
is not zero. We denote
$\partial_\mu u_{\paren{2}\mathrm{PZ}}^{\mu\nu}\finsarg$
by
$\omega_{\paren{2}}^\nu\finsarg$
(needless to say,
$\omega_{\paren{2}}^\nu\finsarg$
is past-zero). If it is possible to define some
$v_{\paren{2}}^{\mu\nu}\finsarg$
so that
\begin{equation}\label{2.2.3.40}
\Box v_{\paren{2}}^{\mu\nu}\finsarg=0,
\end{equation}
\begin{equation}\label{2.2.3.41}
\partial_\mu v_{\paren{2}}^{\mu\nu}\finsarg=-\omega_{\paren{2}}^\nu\finsarg,
\end{equation}
then
$u_{\paren{2}\mathrm{PZ}}^{\mu\nu}\finsarg+v_{\paren{2}}^{\mu\nu}\finsarg$
is a solution to the equation
$\Box h_{\paren{2}\mathrm{PZ}}^{\mu\nu}\finsarg=\Lambda_{\paren{2}\mathrm{PZ}}^{\mu\nu}\finsarg$
which also fulfills the harmonic gauge condition. To examine this, first note that, considering
Eq. (\ref{2.2.3.39}),
$\omega_{\paren{2}}^\nu\finsarg$
is of the general form
$\sum_{\ell=0}^{\infty}\hat{\partial}_L\paren{\frac{\hat{U}_L^\nu(u)}{\x}}$,
where to write this we have also used
Eq. (\ref{B.2})
(of course, the moments
$\hat{U}_L^\nu$
with
$\ell>\ell_{\mathrm{max}}$
are zero). Since
$\hat{U}_L^\nu(u)$'s
and hence
$\omega_{\paren{2}}^\nu\finsarg$
are past-zero, so too is
$v_{\paren{2}}^{\mu\nu}\finsarg$
(due to
Eq. (\ref{2.2.3.41})).
This, together with
Eq. (\ref{2.2.3.40}),
imply
$v_{\paren{2}}^{\mu\nu}\finsarg=\sum_{\ell=0}^{\infty}\hat{\partial}_L\paren{\frac{\hat{U}_L^{\mu\nu}(u)}{\x}}$.
In order to determine
$\hat{U}_L^{\mu\nu}(u)$'s
from
Eq. (\ref{2.2.3.41}),
thereby obtaining
$v_{\paren{2}}^{\mu\nu}\finsarg$,
we rewrite
$v_{\paren{2}}^{0i}\finsarg$,
$v_{\paren{2}}^{ij}\finsarg$
and
$\omega_{\paren{2}}^i\finsarg$
in terms of STF tensors
($v_{\paren{2}}^{00}\finsarg$
and
$\omega_{\paren{2}}^{0}\finsarg$
are already in terms of STF tensors). The results for
$v_{\paren{2}}^{0i}\finsarg$
and
$v_{\paren{2}}^{ij}\finsarg$
are the same as the ones given in
Eqs. (\ref{2.1.3.5}) and (\ref{2.1.3.6})
for
$u_{\paren{2}}^{0i}\finsarg$
and
$u_{\paren{2}}^{ij}\finsarg$,
i.e.,
\begin{IEEEeqnarray}{rcl}\label{2.1.3.5v}
 v^{0i}\finsarg & = & \sum_{\ell=0}^{\infty}\partial_{iL} \left( |\mathbf{x}|^{-1}B_L(u)\right) + \sum_{\ell=1}^{\infty}\partial_{L-1} \left( |\mathbf{x}|^{-1}C_{iL-1}(u)\right) \nonumber \\  
&& \negmedspace {}  + \sum_{\ell=1}^{\infty}\varepsilon_{iab}\partial_{aL-1} \left( |\mathbf{x}|^{-1}D_{bL-1}(u)\right),
\end{IEEEeqnarray}
\begin{IEEEeqnarray}{rcl}\label{2.1.3.6v}
v^{ij}\finsarg & = & \sum_{\ell=0}^{\infty}\partial_{ijL} \left( |\mathbf{x}|^{-1}E_L(u)\right) + \sum_{\ell=0}^{\infty}\delta_{ij}\partial_L \left( |\mathbf{x}|^{-1}F_L(u)\right) \nonumber \\  
&& \negmedspace {}+ \sum_{\ell=1}^{\infty}\partial_{L-1(i} \left( |\mathbf{x}|^{-1}G_{j)L-1}(u)\right) + \sum_{\ell=1}^{\infty}\varepsilon_{ab(i}\partial_{j)aL-1} \left( |\mathbf{x}|^{-1}H_{bL-1}(u)\right) \nonumber \\  
&& \negmedspace {} + \sum_{\ell=2}^{\infty}\partial_{L-2} \left( |\mathbf{x}|^{-1}I_{ijL-2}(u)\right) +  \sum_{\ell=2}^{\infty}\partial_{aL-2} \left( |\mathbf{x}|^{-1}\varepsilon_{ab(i}J_{j)bL-2}(u)\right).
\end{IEEEeqnarray}
The desired form of
$\omega_{\paren{2}}^i\finsarg$
can be obtained in a manner similar to the one resulting in
Eqs. (\ref{2.1.3.5}) and (\ref{2.1.3.6})
(namely by decomposing
$\hat{U}^{i}_L$
into the terms containing moments symmetric and trace-free with respect to all their indices as well as using
Eq. (\ref{2.1.3.7})).
The result is
\begin{IEEEeqnarray}{rcl}\label{2.2.3.43}
\omega_{\paren{2}}^i\finsarg & = & \sum_{\ell=0}^{\infty}\partial_{iL} \left( |\mathbf{x}|^{-1}P_L(u)\right) + \sum_{\ell=1}^{\infty}\partial_{L-1} \left( |\mathbf{x}|^{-1}Q_{iL-1}(u)\right) \nonumber \\  
&& \negmedspace {}  + \sum_{\ell=1}^{\infty}\varepsilon_{iab}\partial_{aL-1} \left( |\mathbf{x}|^{-1}R_{bL-1}(u)\right).
\end{IEEEeqnarray}
We also rewrite
$v_{\paren{2}}^{00}\finsarg$
and
$\omega_{\paren{2}}^0\finsarg$
as
\begin{equation}\label{2.1.3.4}
v^{00}\finsarg=\sum_{\ell=0}^{\infty}\partial_L \left( |\mathbf{x}|^{-1}A_L(u)\right),
\end{equation}
\begin{equation}\label{2.2.3.42}
\omega_{\paren{2}}^0\finsarg=\sum_{\ell=0}^{\infty}\partial_L \left( |\mathbf{x}|^{-1}N_L(u)\right),
\end{equation}
where
$A_L=\hat{U}^{00}_L$
and
$N_L=\hat{U}^{0}_L$.
It is obvious that the STF tensors
$A_L,...,J_L,N_L,P_L,Q_L$
and
$R_L$
are past-zero. More importantly, they are algebraically independent (due to the fact that the set of all STF Cartesian tensors of rank
$\ell$
generates a
($2\ell+1$)-dimensional irreducible representation of SO(3)). Taking the latter into account,
Eq. (\ref{2.2.3.41})
gives
\begin{equation}\label{2.2.3.44}
\frac{A^{\left(1\right)}}{c} + \frac{B^{\left(2\right)}}{c^2}=-N,
\end{equation}
\begin{equation}\label{2.2.3.45}
\frac{A^{\left(1\right)}_L}{c}+\frac{B^{\left(2\right)}_L}{c^2}+C_L=-N_L \qquad \text{for}\;\ell \ge 1,
\end{equation}
\begin{equation}\label{2.2.3.46}
\frac{B^{\left(1\right)}}{c}+\frac{E^{\left(2\right)}}{c^2}+F=-P,
\end{equation}
\begin{equation}\label{2.2.3.47}
\frac{B^{\left(1\right)}_L}{c}+\frac{E^{\left(2\right)}_L}{c^2}+F_L+\frac{G_L}{2}=-P_L \qquad \text{for}\;\ell \ge 1,
\end{equation}
\begin{equation}\label{2.2.3.48}
\frac{C^{\left(1\right)}_i}{c}+\frac{G^{\left(2\right)}_i}{2c^2}=-Q_i,
\end{equation}
\begin{equation}\label{2.2.3.49}
\frac{C^{\left(1\right)}_L}{c}+\frac{G^{\left(2\right)}_L}{2c^2}+I_L=-Q_L \qquad \text{for}\;\ell \ge 2,
\end{equation}
\begin{equation}\label{2.2.3.50}
\frac{D^{\left(1\right)}_i}{c}+\frac{H^{\left(2\right)}_i}{2c^2}=-R_i,
\end{equation}
\begin{equation}\label{2.2.3.51}
\frac{D^{\left(1\right)}_L}{c}+\frac{H^{\left(2\right)}_L}{2c^2}+\frac{J_L}{2}=-R_L \qquad \text{for}\;\ell \ge 2.
\end{equation}
As it can be seen, for each
$\ell$
we have a system of four equations in ten unknowns
$A_L,...,J_L$.
Since the number of unknowns is greater than the number of equations, the unknowns and hence
$v^{\mu\nu}_{\paren{2}}\finsarg$
cannot be uniquely determined in terms of
$N_L$,
$P_L$,
$Q_L$
and
$R_L$.
We consider a case in which
$v^{00}_{\paren{2}}\finsarg$
contains only monopolar or dipolar terms, i.e.,
\begin{equation}\label{2.2.3.52}
A_L=0 \qquad \text{for}\;\ell \ge 2.
\end{equation}
Furthermore, we demand that the spatial trace of
$v^{\mu\nu}_{\paren{2}}\finsarg$,
$v^{ii}_{\paren{2}}\finsarg$,
be monopolar. After some simple manipulation, it can be shown that this restriction is equivalent to
\begin{equation}\label{2.2.3.53}
\frac{E^{\left(2\right)}_L}{c^2}+3F_L+G_L=0 \qquad  \text{for}\;\ell \ge 1.
\end{equation}
Adding
Eqs. (\ref{2.2.3.52}) and (\ref{2.2.3.53})
to the former set of four equations, the number of equations reaches six which is still less than the number of unknowns. In order that the number of equations and unknowns become equal, we add the simplest possible equations, equations of the form ``unknown equals zero''. We choose them to be
\begin{equation}\label{2.2.3.54}
B_L=0 \qquad  \text{for}\;\ell \ge 0,
\end{equation}
\begin{equation}\label{2.2.3.55}
D_L=0 \qquad  \text{for}\;\ell \ge 2,
\end{equation}
\begin{equation}\label{2.2.3.56}
E_L=0 \qquad  \text{for}\;\ell \ge 0,
\end{equation}
\begin{equation}\label{2.2.3.57}
H_L=0 \qquad  \text{for}\;\ell \ge 1,
\end{equation}
thereby obtaining the remaining unknowns as follows:
\begin{equation}\label{2.2.3.58}
A=-cN^{\paren{-1}},
\end{equation}
\begin{equation}\label{2.2.3.59}
A_i=-cN_i^{\paren{-1}}+c^2 Q_i^{\paren{-2}}-3P_i,
\end{equation}
\begin{equation}\label{2.2.3.60}
A_L=0 \qquad  \text{for}\;\ell \ge 2,
\end{equation}
\begin{equation}\label{2.2.3.61}
C_i=-cQ_i^{\paren{-1}}+\frac{3}{c}P_i^{\paren{1}},
\end{equation}
\begin{equation}\label{2.2.3.62}
C_L=-N_L \qquad  \text{for}\;\ell \ge 2,
\end{equation}
\begin{equation}\label{2.2.3.63}
D_i=-cR_i^{\paren{-1}},
\end{equation}
\begin{equation}\label{2.2.3.64}
F=-P,
\end{equation}
\begin{equation}\label{2.2.3.65}
F_L=2P_L \qquad  \text{for}\;\ell \ge 1,
\end{equation}
\begin{equation}\label{2.2.3.66}
G_L=-6P_L \qquad  \text{for}\;\ell \ge 1,
\end{equation}
\begin{equation}\label{2.2.3.67}
I_L=\frac{1}{c}N_L^{\paren{1}}+\frac{3}{c^2}P_L^{\paren{2}}-Q_L \qquad  \text{for}\;\ell \ge 2,
\end{equation}
\begin{equation}\label{2.2.3.68}
J_L=-2R_L \qquad  \text{for}\;\ell \ge 2,
\end{equation}
where
$k^{\paren{-1}}(u)$
and
$k^{\paren{-2}}(u)$
denote
$\int_{-\infty}^{u}k(v) \ud v$
and
$\int_{-\infty}^{u}k^{\paren{-1}}(v) \ud v$
respectively. Since
$N_L$,
$P_L$,
$Q_L$
and
$R_L$
are past-zero, taking the lower limit of these integrals as
$-\infty$
doesn't make them divergent. Considering
Eqs. (\ref{2.2.3.54})-(\ref{2.2.3.68}),
the spatial trace of
$v^{\mu\nu}_{\paren{2}}\finsarg$
becomes
\begin{equation}\label{2.2.3.69}
v^{ii}_{\paren{2}}\finsarg=-3\x^{-1}P(u),
\end{equation}
and
$v^{\mu\nu}_{\paren{2}}\finsarg$
reads
\begin{IEEEeqnarray}{rcl}\label{2.2.3.70}
 v^{00}_{\paren{2}}\finsarg&=&-c\x^{-1}N^{\paren{-1}}(u)+\partial_i\left[\x^{-1}\paren{-cN_i^{\paren{-1}}(u)+c^2 Q_i^{\paren{-2}}(u)-3P_i(u)}\right],\nonumber\\
\end{IEEEeqnarray}
\begin{IEEEeqnarray}{rcl}\label{2.2.3.71}
v^{0i}_{\paren{2}}\finsarg & = & \x^{-1}\paren{-cQ_i^{\paren{-1}}(u)+\frac{3}{c}P_i^{\paren{1}}(u)}-c\varepsilon_{iab}\partial_a\left[\x^{-1}R_b^{\paren{-1}}(u)\right]\nonumber \\  
&& \negmedspace {}-\sum_{\ell=2}^{\infty}\partial_{L-1}\left[\x^{-1}N_{iL-1}(u)\right],
\end{IEEEeqnarray}
\begin{IEEEeqnarray}{rcl}\label{2.2.3.72}
 v^{ij}_{\paren{2}}\finsarg & = &-\delta_{ij}\x^{-1}P(u) +  \sum_{\ell=2}^{\infty}\bigg\{2\delta_{ij}\partial_{L-1}\left[\x^{-1}P_{L-1}(u)\right]-6\partial_{L-2(i} \left[ |\mathbf{x}|^{-1}P_{j)L-2}(u)\right]\nonumber \\  
&& \negmedspace {}+\partial_{L-2} \left[\x^{-1}\left(\frac{1}{c}N_{ijL-2}^{\paren{1}}(u)+\frac{3}{c^2}P_{ijL-2}^{\paren{2}}(u)-Q_{ijL-2}(u)\right)\right]\nonumber \\  
&& \negmedspace {}-2\partial_{aL-2}\left[\x^{-1}\varepsilon_{ab(i}R_{j)bL-2}(u)\right]\bigg\}.
\end{IEEEeqnarray}
Having succeeded in constructing a
$v^{\mu\nu}_{\paren{2}}\finsarg$
fulfilling
Eqs.(\ref{2.2.3.40}) and (\ref{2.2.3.41}),
we can say that
$u_{\paren{2}\mathrm{PZ}}^{\mu\nu}\finsarg+v_{\paren{2}}^{\mu\nu}\finsarg$
is a solution to the equation
$\Box h_{\paren{2}\mathrm{PZ}}^{\mu\nu}\finsarg=\Lambda_{\paren{2}\mathrm{PZ}}^{\mu\nu}\finsarg$
which also satisfies the harmonic gauge condition, and more importantly, taking
$\partial_\mu u_{\paren{2}\mathrm{AS}}^{\mu\nu}\finsarg$
into account,
$u_{\paren{2}}^{\mu\nu}\finsarg+v_{\paren{2}}^{\mu\nu}\finsarg$
is a solution to the equation
$\Box h_{\paren{2}}^{\mu\nu}\finsarg=\Lambda_{\paren{2}}^{\mu\nu}\finsarg$
fulfilling the harmonic gauge condition.

\subsection{General Solution of the Second-Order Problem and Its Structure}

As mentioned in previous subsection,
$u_{\paren{2}}^{\mu\nu}\finsarg+v_{\paren{2}}^{\mu\nu}\finsarg=u_{\paren{2}\mathrm{AS}}^{\mu\nu}\xarg+u_{\paren{2}\mathrm{PZ}}^{\mu\nu}\finsarg+v_{\paren{2}}^{\mu\nu}\finsarg$
is solution to the equation
$\Box h_{\paren{2}}^{\mu\nu}\finsarg=\Lambda_{\paren{2}}^{\mu\nu}\finsarg$
which also fulfills the harmonic gauge condition. From the structure of
$u_{\paren{2}\mathrm{AS}}^{\mu\nu}\xarg$
given by
Eq. (\ref{2.2.2.26}),
it is obvious that
$u_{\paren{2}\mathrm{AS}}^{\mu\nu}\xarg$
is an always stationary function. Therefore, in particular,
$u_{\paren{2}\mathrm{AS}}^{\mu\nu}\xarg$
is stationary in the past. This structure also indicates that
$u_{\paren{2}\mathrm{AS}}^{\mu\nu}\xarg$
is smooth everywhere in
$\mathbb{R }^3$
except at
$\x=0$,
and
its limit as
$\x\to\infty$,
always and in particular at
$t\le-T$,
is zero. The structure of
$u_{\paren{2}\mathrm{PZ}}^{\mu\nu}\finsarg$
given in
Eq. (\ref{2.2.2.76})
shows that
$u_{\paren{2}\mathrm{PZ}}^{\mu\nu}\finsarg$
is past-stationary (particularly, it is past-zero). Considering this structure, it is evident that
$u_{\paren{2}\mathrm{PZ}}^{\mu\nu}\finsarg$ is smooth everywhere in
$\mathbb{R }^3$
except at
$\x=0$.
$v_{\paren{2}}^{\mu\nu}\finsarg$
is past-zero and hence past-stationary due to
$N_L$,
$P_L$,
$Q_L$
and
$R_L$
being past-zero, and smoothness of these moments implies that
$v_{\paren{2}}^{\mu\nu}\finsarg$
is smooth everywhere in
$\mathbb{R }^3$
(except at
$\x=0$).
Furthermore, since
$u_{\paren{2}\mathrm{PZ}}^{\mu\nu}\finsarg$
and
$v_{\paren{2}}^{\mu\nu}\finsarg$
are both past-zero, their limits as
$\x\to\infty$
vanish. In light of these considerations,
$u_{\paren{2}}^{\mu\nu}\finsarg+v_{\paren{2}}^{\mu\nu}\finsarg$
fulfills not only post-Minkowskian equations, but also conditions discussed in Section 1.3, thereby being a solution to the second-order problem. However, this is not the most general solution. The most general solution is the sum of
$u_{\paren{2}}^{\mu\nu}\finsarg+v_{\paren{2}}^{\mu\nu}\finsarg$
and the general solution of homogeneous d'Alembertian equation which fulfills the conditions of harmonicity, past-stationarity and being asymptotically Minkowskian in the past, i.e.,
\begin{IEEEeqnarray}{rcl}\label{2.2.4.1}
h^{\mu\nu}_{\paren{2}}\left[I_{\paren{1}},I_{\paren{2}},W_{\paren{1}},W_{\paren{2}}\right]&=& u_{\paren{2}}^{\mu\nu}\left[I_{\paren{1}},W_{\paren{1}}\right]+v_{\paren{2}}^{\mu\nu}\left[I_{\paren{1}},W_{\paren{1}}\right]+h^{\mu\nu}_{\paren{1}}\left[I_{\paren{2}},W_{\paren{2}}\right],\qquad
\end{IEEEeqnarray}
where
$I_{\paren{1}}$
and
$W_{\paren{1}}$
are the same as
$I$
and
$W$
in
Eqs. (\ref{2.1.4.13})-(\ref{2.1.4.17}),
and
$I_{\paren{2}}$
and
$W_{\paren{2}}$
are smooth and past-stationary. The structures of
$u_{\paren{2}\mathrm{AS}}^{\mu\nu}\xarg$
and
$u_{\paren{2}\mathrm{PZ}}^{\mu\nu}\finsarg$
are given by
Eqs. (\ref{2.2.2.26}) and (\ref{2.2.2.76})
respectively. Since
$v_{\paren{2}}^{\mu\nu}\finsarg$
and
$h^{\mu\nu}_{\paren{1}}\left[I_{\paren{2}},W_{\paren{2}}\right]$
are both solutions to the source-free d'Alembertian equation, their structures are similar to the one given in
Eq. (\ref{2.1.4.26})
(with
$\ell_{\mathrm{max}}$
and
$N_0$
different from the corresponding ones in that equation). Therefore, we can write the structure of
$h_{\paren{2}}^{\mu\nu}\finsarg$
as
\begin{IEEEeqnarray}{rcl}\label{2.2.4.2}
h^{\mu\nu}_{\paren{2}}\finsarg&=&\sum_{\ell=0}^{\ell'_{\mathrm{max}}}\sum_{k=k''_{\mathrm{min}}}^{-1}\hat{n}^L \x^k\hat{C}''^{\;\mu\nu}_{\paren{2}L,k} +\sum_{q=0}^{q''_{\mathrm{max}}}\sum_{a=a''_{\mathrm{min}}}^{a''_{\mathrm{max}}}\sum_{p=0}^{1}\hat{n}^Q\x^a\lnx^p\hat{G}'^{\;\mu\nu}_{\paren{2}Q,a,p}\targ\nonumber \\  
&& \negmedspace {}+ R''^{\;\mu\nu}_{\paren{2}}\finsarg,
\end{IEEEeqnarray}
where
$\hat{C}''^{\;\mu\nu}_{\paren{2}L,k}$
is a constant,
$\hat{G}'^{\;\mu\nu}_{\paren{2}Q,a,p}\targ$
and
$ R''^{\;\mu\nu}_{\paren{2}}\finsarg$
are past-zero, and
$ R''^{\;\mu\nu}_{\paren{2}}\finsarg$
is 
$\sum_{N=N_{\mathrm{min}}}^{N_{\mathrm{max}}}O\!\paren{\x^N}$
as
$\x \to 0$.

\section{$n$th-Order Problem}

Considering the general form of the terms in
$\Lambda^{\mu\nu}$
(discussed in Section 1.2), providing that
$N_0$'s
and
$s$
are chosen to be large enough numbers, the structure of
$\Lambda^{\mu\nu}_{\paren{3}}\finsarg$
can be found as
\begin{IEEEeqnarray}{rcl}\label{2.3.1}
\Lambda^{\mu\nu}_{\paren{3}}\finsarg&=&\sum_{\ell=0}^{\ell_{\mathrm{max}}\paren{3}}\sum_{k=k_{\mathrm{min}}\paren{3}}^{-4}\hat{n}^L \x^k\hat{C}^{\mu\nu}_{\paren{3}L,k} \nonumber \\  
&& \negmedspace {}+\sum_{q=0}^{q_{\mathrm{max}}\paren{3}}\sum_{a=a_{\mathrm{min}}\paren{3}}^{a_{\mathrm{max}}\paren{3}}\sum_{p=0}^{p_{\mathrm{max}}\paren{3}}\hat{n}^Q\x^a\lnx^p\hat{F}^{\mu\nu}_{\paren{3}Q,a,p}\targ+ R^{\mu\nu}_{\paren{3}}\finsarg,\quad
\end{IEEEeqnarray}
where
$a_{\mathrm{max}}\paren{3}>0$,
$p_{\mathrm{max}}\paren{3}=1$,
$\hat{C}^{\mu\nu}_{\paren{3}L,k}$
is a constant,
$\hat{F}^{\mu\nu}_{\paren{3}Q,a,p}\targ$
and
$ R^{\mu\nu}_{\paren{3}}\finsarg$
are past-zero, and
$ R^{\mu\nu}_{\paren{3}}\finsarg=\sum_{N=N_{\mathrm{min}}\paren{3}}^{N_{\mathrm{max}}\paren{3}}O\!\paren{\x^N}$
when
$\x \to 0$.
Further, it should be noted that the terms containing the maximal power of
$\ln{\frac{\x}{r_0}}$
come from
$N^{\mu\nu}(h_{\paren{1}},h_{\paren{2}})$
and
$N^{\mu\nu}(h_{\paren{2}},h_{\paren{1}})$.
Now, by arguments similar to those employed in the second-order problem, it is straightforward to show that the general solution to the third-order problem is
\begin{eqnarray}\label{2.3.2}
\lefteqn{h^{\mu\nu}_{\paren{3}}\left[I_{\paren{1}},...,I_{\paren{3}},W_{\paren{1}},...,W_{\paren{3}}\right]} \nonumber \\ 
&\!\!\!\!\!\!\!\!=\!\!\!\!\!\!\!\!& u_{\paren{3}}^{\mu\nu}\left[I_{\paren{1}},I_{\paren{2}},W_{\paren{1}},W_{\paren{2}}\right]+v_{\paren{3}}^{\mu\nu}\left[I_{\paren{1}},I_{\paren{2}},W_{\paren{1}},W_{\paren{2}}\right]+h^{\mu\nu}_{\paren{1}}\left[I_{\paren{3}},W_{\paren{3}}\right] \nonumber\\
&\!\!\!\!\!\!\!\!=\!\!\!\!\!\!\!\!& u_{\paren{3}\mathrm{AS}}^{\mu\nu}\left[I_{\paren{1}},I_{\paren{2}},W_{\paren{1}},W_{\paren{2}}\right]+ u_{\paren{3}\mathrm{PZ}}^{\mu\nu}\left[I_{\paren{1}},I_{\paren{2}},W_{\paren{1}},W_{\paren{2}}\right] \nonumber\\
&&\!\!\!\!+v_{\paren{3}}^{\mu\nu}\left[I_{\paren{1}},I_{\paren{2}},W_{\paren{1}},W_{\paren{2}}\right]+h^{\mu\nu}_{\paren{1}}\left[I_{\paren{3}},W_{\paren{3}}\right],
\end{eqnarray}
where
\begin{IEEEeqnarray}{rcl}\label{2.3.3}
 u^{\mu\nu}_{\paren{3}\mathrm{AS}}(\mathbf{x}) & = &\fpa\left[\gnearint{\Lambda^{\mu\nu}_{\paren{3}\mathrm{AS}}(\mathbf{x'})}\right] \nonumber \\  
&& \negmedspace {}+\fpa\left[\gfarint{\Lambda^{\mu\nu}_{\paren{3}\mathrm{AS}}(\mathbf{x'})}\right]\nonumber \\  
&& \negmedspace {} \stareq \fpa\gpoisint{\Lambda^{\mu\nu}_{\paren{3}\mathrm{AS}}(\mathbf{x'})},
\end{IEEEeqnarray}
\begin{IEEEeqnarray}{rcl}\label{2.3.4}
\partial_\mu u_{\paren{3}\mathrm{AS}}^{\mu\nu}(\mathbf{x}) & = &\resa \left[-\frac{1}{4\pi}\int_{|\mathbf{x'}|<\mathcal{R}}\paren{\frac{\xp}{r_0}}^B \frac{\xp^{-1}n'^i\Lambda^{i\nu}_{\paren{3}\mathrm{AS}}(\mathbf{x'})}{|\mathbf{x}-\mathbf{x'}|}\ud^3\mathbf{x'}\right] \nonumber \\  
&& \negmedspace {} + \resa \left[-\frac{1}{4\pi}\int_{\mathcal{R}<|\mathbf{x'}|}\paren{\frac{\xp}{r_0}}^B \frac{\xp^{-1}n'^i\Lambda^{i\nu}_{\paren{3}\mathrm{AS}}(\mathbf{x'})}{|\mathbf{x}-\mathbf{x'}|}\ud^3\mathbf{x'}\right] \nonumber \\ 
& = &\; 0+0=0,
\end{IEEEeqnarray}
\begin{equation}\label{2.3.5}
 u_{\paren{3}\mathrm{PZ}}^{\mu\nu}\finsarg=\fpa\gretint{\Lambda^{\mu\nu}_{\paren{3}\mathrm{PZ}}\retarg},
\end{equation}
\begin{IEEEeqnarray}{rcl}\label{2.3.6}
\partial_\mu u_{\paren{3}\mathrm{PZ}}^{\mu\nu}\finsarg & = & \omega^\nu_{\paren{3}}\finsarg=\resa \gretint{\xp^{-1}n'^i\Lambda^{i\nu}_{\paren{3}\mathrm{PZ}}\retarg} \nonumber \\  
& = &\resa \left[-\frac{1}{4\pi}\int_{|\mathbf{x'}|<\mathcal{R}}\paren{\frac{\xp}{r_0}}^B \frac{\xp^{-1}n'^i\Lambda^{i\nu}_{\paren{3}\mathrm{PZ}}\retarg}{|\mathbf{x}-\mathbf{x'}|}\ud^3\mathbf{x'}\right], \qquad
\end{IEEEeqnarray}
\begin{IEEEeqnarray}{rcl}\label{2.3.7}
 u^{\mu\nu}_{\paren{3}}\finsarg & = &\fpa\left[\gnearint{\Lambda^{\mu\nu}_{\paren{3}}\retarg}\right] \nonumber \\  
&& \negmedspace {}+\fpa\left[\gfarint{\Lambda^{\mu\nu}_{\paren{3}}\retarg}\right]\nonumber \\  
&& \negmedspace {} \stareq \fpa\gretint{\Lambda^{\mu\nu}_{\paren{3}}\retarg},
\end{IEEEeqnarray}
\begin{equation}\label{2.3.8}
 \partial_\mu u_{\paren{3}}^{\mu\nu}\finsarg=\partial_\mu u_{\paren{3}\mathrm{AS}}^{\mu\nu}(\mathbf{x})+\partial_\mu u_{\paren{3}\mathrm{PZ}}^{\mu\nu}\finsarg=\omega^\nu_{\paren{3}}\finsarg,
\end{equation}
\begin{equation}\label{2.3.9}
 \Box v_{\paren{3}}^{\mu\nu}\finsarg=0,
\end{equation}
\begin{equation}\label{2.3.10}
 \partial_\mu v_{\paren{3}}^{\mu\nu}\finsarg=-\omega^\nu_{\paren{3}}\finsarg.
\end{equation}
The structure of
$h^{\mu\nu}_{\paren{3}}\finsarg$
can be obtained as
\begin{IEEEeqnarray}{rcl}\label{2.3.11}
h^{\mu\nu}_{\paren{3}}\finsarg&=&\sum_{\ell=0}^{\ell'_{\mathrm{max}}\paren{3}}\sum_{k=k'_{\mathrm{min}}\paren{3}}^{-1}\hat{n}^L \x^k\hat{C}'^{\;\mu\nu}_{\paren{3}L,k} \nonumber \\  
&& \negmedspace {}+\sum_{q=0}^{q'_{\mathrm{max}}\paren{3}}\sum_{a=a'_{\mathrm{min}}\paren{3}}^{a'_{\mathrm{max}}\paren{3}}\sum_{p=0}^{p'_{\mathrm{max}}\paren{3}}\hat{n}^Q\x^a\lnx^p\hat{G}'^{\;\mu\nu}_{\paren{3}Q,a,p}\targ+ R'^{\;\mu\nu}_{\paren{3}}\finsarg,\quad
\end{IEEEeqnarray}
where
$p'_{\mathrm{max}}\paren{3}=2$,
$\hat{C}'^{\;\mu\nu}_{\paren{3}L,k}$
is a constant,
$\hat{G}'^{\;\mu\nu}_{\paren{3}Q,a,p}\targ$
and
$ R'^{\;\mu\nu}_{\paren{3}}\finsarg$
are past-zero, and
$ R'^{\;\mu\nu}_{\paren{3}}\finsarg$
is
$\sum_{N=N'_{\mathrm{min}}\paren{3}}^{N'_{\mathrm{max}}\paren{3}}O\!\paren{\x^N}$
as
$\x \to 0$.

Analogously, it can be argued that, providing all
$N_0$'s
and
$s$'s
at lower orders are taken as large enough numbers. for any arbitrary
$n$
the structure of
$\Lambda^{\mu\nu}_{\paren{n}}\finsarg$
reads
\begin{IEEEeqnarray}{rcl}\label{2.3.12}
\Lambda^{\mu\nu}_{\paren{n}}\finsarg&=&\sum_{\ell=0}^{\ell_{\mathrm{max}}\paren{n}}\sum_{k=k_{\mathrm{min}}\paren{n}}^{-4}\hat{n}^L \x^k\hat{C}^{\mu\nu}_{\paren{n}L,k} \nonumber \\  
&& \negmedspace {}+\sum_{q=0}^{q_{\mathrm{max}}\paren{n}}\sum_{a=a_{\mathrm{min}}\paren{n}}^{a_{\mathrm{max}}\paren{n}}\sum_{p=0}^{p_{\mathrm{max}}\paren{n}}\hat{n}^Q\x^a\lnx^p\!\!\hat{F}^{\mu\nu}_{\paren{n}Q,a,p}\targ+ R^{\mu\nu}_{\paren{n}}\finsarg,\quad
\end{IEEEeqnarray}
where
$a_{\mathrm{max}}\paren{n}>0$,
$p_{\mathrm{max}}\paren{n}=n-2$,
$\hat{C}^{\mu\nu}_{\paren{n}L,k}$
is a constant,
$\hat{F}^{\mu\nu}_{\paren{n}Q,a,p}\targ$
and
$ R^{\mu\nu}_{\paren{n}}\finsarg$
are past-zero, and
$ R^{\mu\nu}_{\paren{n}}\finsarg=\sum_{N=N_{\mathrm{min}}\paren{n}}^{N_{\mathrm{max}}\paren{n}}O\!\paren{\x^N}$
when
$\x \to 0$.
The reasoning that shows the maximal power of
$\ln{\frac{\x}{r_0}}$
equals
$n-2$
is as follows. Considering the structures of
$h^{\mu\nu}_{\paren{1}}\finsarg$,
$h^{\mu\nu}_{\paren{2}}\finsarg$
and
$h^{\mu\nu}_{\paren{3}}\finsarg$,
it seems that the maximal power of
$\ln{\frac{\x}{r_0}}$
that appears in the structure of
$h^{\mu\nu}_{\paren{n}}\finsarg$
is equal to
$n-1$.
Since the terms of
$\Lambda^{\mu\nu}_{\paren{n}}$
are of the general forms
$h_{\paren{n_2}}\partial^2 h_{\paren{n_1}}$
and
$h_{\paren{m_{k+2}}}\;...\;h_{\paren{m_3}}\partial h_{\paren{m_2}}\partial h_{\paren{m_1}}$
with
$k=0,1,...,n-2$,
it is obvious that the maximal power of
$\ln{\frac{\x}{r_0}}$
appearing in each type of these
$n$
terms is either
$\sum_{i=1}^{2}n_i-2$
or
$\sum_{i=1}^{k+2}m_i-k-2$.
However,
$\sum_{i=1}^{k+2}m_i=\sum_{i=1}^{2}n_i=n$,
and therefore, the maximal power of
$\ln{\frac{\x}{r_0}}$
appearing in the structure of
$\Lambda^{\mu\nu}_{\paren{n}}\finsarg$
is
$n-2$
and
$\ln{\frac{\x}{r_0}}$
in
$h^{\mu\nu}_{\paren{n}}\finsarg$
thereby has the maximal power of
$n-1$,
as assumed (it is worth noting that the terms containing the maximal power of
$\ln{\frac{\x}{r_0}}$
in
$\Lambda^{\mu\nu}_{\paren{n}}\finsarg$
come from the terms
$N^{\mu\nu}(h_{\paren{m}},h_{\paren{m'}})$
with
$m+m'=n$).
The structure of
$\Lambda^{\mu\nu}_{\paren{n}}\finsarg$
given by
Eq. (\ref{2.3.12})
guarantees that, for any arbitrary
$n$,
the general solution of the
$n$th-order problem is
\begin{eqnarray}\label{2.3.13}
\lefteqn{h^{\mu\nu}_{\paren{n}}\left[I_{\paren{1}},...,I_{\paren{n}},W_{\paren{1}},...,W_{\paren{n}}\right]} \nonumber \\ 
&\!\!\!\!\!\!\!\!=\!\!\!\!\!\!\!\!& u_{\paren{n}}^{\mu\nu}\left[I_{\paren{1}},...,I_{\paren{n-1}},W_{\paren{1}},...,W_{\paren{n-1}}\right]+v_{\paren{n}}^{\mu\nu}\left[I_{\paren{1}},...,I_{\paren{n-1}},W_{\paren{1}},...,W_{\paren{n-1}}\right]\nonumber\\
&&\!\!\!\!+h^{\mu\nu}_{\paren{1}}\left[I_{\paren{n}},W_{\paren{n}}\right] \nonumber\\
&\!\!\!\!\!\!\!\!=\!\!\!\!\!\!\!\!& u_{\paren{n}\mathrm{AS}}^{\mu\nu}\left[I_{\paren{1}},...,I_{\paren{n-1}},W_{\paren{1}},...,W_{\paren{n-1}}\right]\nonumber\\
&&\!\!\!\!+ u_{\paren{n}\mathrm{PZ}}^{\mu\nu}\left[I_{\paren{1}},...,I_{\paren{n-1}},W_{\paren{1}},...,W_{\paren{n-1}}\right] \nonumber\\
&&\!\!\!\!+v_{\paren{n}}^{\mu\nu}\left[I_{\paren{1}},...,I_{\paren{n-1}},W_{\paren{1}},...,W_{\paren{n-1}}\right]+h^{\mu\nu}_{\paren{1}}\left[I_{\paren{n}},W_{\paren{n}}\right],
\end{eqnarray}
where
\begin{IEEEeqnarray}{rcl}\label{2.3.14}
 u^{\mu\nu}_{\paren{n}\mathrm{AS}}(\mathbf{x}) & = &\fpa\left[\gnearint{\Lambda^{\mu\nu}_{\paren{n}\mathrm{AS}}(\mathbf{x'})}\right] \nonumber \\  
&& \negmedspace {}+\fpa\left[\gfarint{\Lambda^{\mu\nu}_{\paren{n}\mathrm{AS}}(\mathbf{x'})}\right]\nonumber \\  
&& \negmedspace {} \stareq \fpa\gpoisint{\Lambda^{\mu\nu}_{\paren{n}\mathrm{AS}}(\mathbf{x'})},
\end{IEEEeqnarray}
\begin{IEEEeqnarray}{rcl}\label{2.3.15}
\partial_\mu u_{\paren{n}\mathrm{AS}}^{\mu\nu}(\mathbf{x}) & = &\resa \left[-\frac{1}{4\pi}\int_{|\mathbf{x'}|<\mathcal{R}}\paren{\frac{\xp}{r_0}}^B \frac{\xp^{-1}n'^i\Lambda^{i\nu}_{\paren{n}\mathrm{AS}}(\mathbf{x'})}{|\mathbf{x}-\mathbf{x'}|}\ud^3\mathbf{x'}\right] \nonumber \\  
&& \negmedspace {} + \resa \left[-\frac{1}{4\pi}\int_{\mathcal{R}<|\mathbf{x'}|}\paren{\frac{\xp}{r_0}}^B \frac{\xp^{-1}n'^i\Lambda^{i\nu}_{\paren{n}\mathrm{AS}}(\mathbf{x'})}{|\mathbf{x}-\mathbf{x'}|}\ud^3\mathbf{x'}\right] \nonumber \\ 
& = &\; 0+0=0,
\end{IEEEeqnarray}
\begin{equation}\label{2.3.16}
 u_{\paren{n}\mathrm{PZ}}^{\mu\nu}\finsarg=\fpa\gretint{\Lambda^{\mu\nu}_{\paren{n}\mathrm{PZ}}\retarg},
\end{equation}
\begin{IEEEeqnarray}{rcl}\label{2.3.17}
\partial_\mu u_{\paren{n}\mathrm{PZ}}^{\mu\nu}\finsarg & = & \omega^\nu_{\paren{n}}\finsarg=\resa \gretint{\xp^{-1}n'^i\Lambda^{i\nu}_{\paren{n}\mathrm{PZ}}\retarg} \nonumber \\  
& = &\resa \left[-\frac{1}{4\pi}\int_{|\mathbf{x'}|<\mathcal{R}}\paren{\frac{\xp}{r_0}}^B \frac{\xp^{-1}n'^i\Lambda^{i\nu}_{\paren{n}\mathrm{PZ}}\retarg}{|\mathbf{x}-\mathbf{x'}|}\ud^3\mathbf{x'}\right],\qquad
\end{IEEEeqnarray}
\begin{IEEEeqnarray}{rcl}\label{2.3.18}
 u^{\mu\nu}_{\paren{n}}\finsarg & = &\fpa\left[\gnearint{\Lambda^{\mu\nu}_{\paren{n}}\retarg}\right] \nonumber \\  
&& \negmedspace {}+\fpa\left[\gfarint{\Lambda^{\mu\nu}_{\paren{n}}\retarg}\right]\nonumber \\  
&& \negmedspace {} \stareq \fpa\gretint{\Lambda^{\mu\nu}_{\paren{n}}\retarg},
\end{IEEEeqnarray}
\begin{equation}\label{2.3.19}
 \partial_\mu u_{\paren{n}}^{\mu\nu}\finsarg=\partial_\mu u_{\paren{n}\mathrm{AS}}^{\mu\nu}(\mathbf{x})+\partial_\mu u_{\paren{n}\mathrm{PZ}}^{\mu\nu}\finsarg=\omega^\nu_{\paren{n}}\finsarg,
\end{equation}
\begin{equation}\label{2.3.20}
 \Box v_{\paren{n}}^{\mu\nu}\finsarg=0,
\end{equation}
\begin{equation}\label{2.3.21}
 \partial_\mu v_{\paren{n}}^{\mu\nu}\finsarg=-\omega^\nu_{\paren{n}}\finsarg.
\end{equation}
In addition, the structure of
$h^{\mu\nu}_{\paren{n}}\finsarg$
can be written as
\begin{IEEEeqnarray}{rcl}\label{2.3.22}
h^{\mu\nu}_{\paren{n}}\finsarg&=&\sum_{\ell=0}^{\ell'_{\mathrm{max}}\paren{n}}\sum_{k=k'_{\mathrm{min}}\paren{n}}^{-1}\hat{n}^L \x^k\hat{C}'^{\;\mu\nu}_{\paren{n}L,k} \targ\nonumber \\  
&& \negmedspace {}+\sum_{q=0}^{q'_{\mathrm{max}}\paren{n}}\sum_{a=a'_{\mathrm{min}}\paren{n}}^{a'_{\mathrm{max}}\paren{n}}\sum_{p=0}^{p'_{\mathrm{max}}\paren{n}}\hat{n}^Q\x^a\lnx^p\!\!\hat{G}'^{\;\mu\nu}_{\paren{n}Q,a,p}+ R'^{\;\mu\nu}_{\paren{n}}\finsarg,\quad
\end{IEEEeqnarray}
where
$p'_{\mathrm{max}}\paren{n}=n-1$,
$\hat{C}'^{\;\mu\nu}_{\paren{n}L,k}$
is a constant,
$\hat{G}'^{\;\mu\nu}_{\paren{n}Q,a,p}\targ$
and
$ R'^{\;\mu\nu}_{\paren{n}}\finsarg$
are past-zero, and
$ R'^{\;\mu\nu}_{\paren{n}}\finsarg$
is
$\sum_{N=N'_{\mathrm{min}}\paren{n}}^{N'_{\mathrm{max}}\paren{n}}O\!\paren{\x^N}$
as
$\x \to 0$.

\section{Post-Minkowskian Expansion}

If at each order
$N_0\to\infty$
and
$s\to\infty$,
then
$h^{\mu\nu}_{\paren{n}}\finsarg$
can be determined for any
$n\ge1$.
Therefore, the (untruncated) post-Minkowskian expansion of
$h^{\mu\nu}\finsarg$
and
$\Lambda^{\mu\nu}\finsarg$,
where from now on we denote them by
$\mathcal{M}\paren{h^{\mu\nu}}\finsarg$
and
$\Lambda^{\mu\nu}(\mathcal{M}\paren{h})\finsarg$,
can be obtained. The structures of
$\mathcal{M}\paren{h^{\mu\nu}}\finsarg$
and
$\Lambda^{\mu\nu}(\mathcal{M}\paren{h})\finsarg$
are given by
\begin{IEEEeqnarray}{rcl}\label{2.4.1}
\mathcal{M}\paren{h^{\mu\nu}}\finsarg&=&\sum_{n=1}^{\infty}G^n h^{\mu\nu}_{\paren{n}}\finsarg=\sum_{\ell=0}^{\infty}\sum_{a=-\infty}^{-1}\hat{n}^L \x^k\hat{C}'^{\;\mu\nu}_{L,k} \nonumber \\  
&& \negmedspace {}+\sum_{q=0}^{\infty}\sum_{a=-\infty}^{\infty}\sum_{p=0}^{\infty}\hat{n}^Q\x^a\lnx^p\hat{G}'^{\;\mu\nu}_{Q,a,p}\targ+ R'^{\;\mu\nu}\finsarg,
\end{IEEEeqnarray}
\begin{IEEEeqnarray}{rcl}\label{2.4.2}
\Lambda^{\mu\nu}(\mathcal{M}\paren{h})\finsarg&=&\sum_{n=1}^{\infty}G^n\Lambda^{\mu\nu}_{\paren{n}}\finsarg=\sum_{\ell=0}^{\infty}\sum_{a=-\infty}^{-4}\hat{n}^L \x^k\hat{C}^{\mu\nu}_{L,k}\nonumber \\  
&& \negmedspace {} +\sum_{q=0}^{\infty}\sum_{a=-\infty}^{\infty}\sum_{p=0}^{\infty}\hat{n}^Q\x^a\lnx^p\hat{F}^{\mu\nu}_{Q,a,p}\targ+ R^{\mu\nu}\finsarg,
\end{IEEEeqnarray}
and the set of equations governing them are as follows:
\begin{equation}\label{2.4.3}
\Box \mathcal{M}\paren{h^{\mu\nu}}\finsarg=\Lambda^{\mu\nu}(\mathcal{M}\paren{h})\finsarg,
\end{equation}
\begin{equation}\label{2.4.4}
 \partial_\mu \mathcal{M}\paren{h^{\mu\nu}}\finsarg=0,
\end{equation}
\begin{equation}\label{2.4.5}
 \partial_\mu \Lambda^{\mu\nu}(\mathcal{M}\paren{h})\finsarg=0,
\end{equation}
\begin{eqnarray}\label{2.4.6}
\lefteqn{\mathcal{M}\paren{h^{\mu\nu}}\left[I_{\paren{1}},I_{\paren{2}},...,W_{\paren{1}},W_{\paren{2}},...\right]} \nonumber \\ 
&\!\!\!\!\!\!\!\!=\!\!\!\!\!\!\!\!& u^{\mu\nu}\left[I_{\paren{1}},I_{\paren{2}},...,W_{\paren{1}},W_{\paren{2}},...\right]+v^{\mu\nu}\left[I_{\paren{1}},I_{\paren{2}},...,W_{\paren{1}},W_{\paren{2}},...\right]\nonumber\\
&&\!\!\!\!+Gh^{\mu\nu}_{\paren{1}}\left[I,W\right] \qquad \text{where}\;I=\sum_{n=1}^{\infty}G^{n-1}I_{\paren{n}}\;\text{and}\;W=\sum_{n=1}^{\infty}G^{n-1}W_{\paren{n}},
\end{eqnarray}
\begin{IEEEeqnarray}{rcl}\label{2.4.7}
 u^{\mu\nu}\finsarg & = &\fpa\left[\gnearint{\Lambda^{\mu\nu}(\mathcal{M}\paren{h})\retarg}\right] \nonumber \\  
&& \negmedspace {}+\fpa\left[\gfarint{\Lambda^{\mu\nu}(\mathcal{M}\paren{h})\retarg}\right]\nonumber \\  
&& \negmedspace {} \stareq \fpa\gretint{\Lambda^{\mu\nu}(\mathcal{M}\paren{h})\retarg},
\end{IEEEeqnarray}
\begin{IEEEeqnarray}{rcl}\label{2.4.8}
\partial_\mu u^{\mu\nu}\finsarg & = &\omega^\nu\finsarg \nonumber\\
&=&\resa \left[-\frac{1}{4\pi}\int_{|\mathbf{x'}|<\mathcal{R}}\paren{\frac{\xp}{r_0}}^B \frac{\xp^{-1}n'^i\Lambda^{i\nu}(\mathcal{M}\paren{h})\retarg}{|\mathbf{x}-\mathbf{x'}|}\ud^3\mathbf{x'}\right]\nonumber \\  
&& \negmedspace {}\stareq \resa \gretint{\xp^{-1}n'^i\Lambda^{i\nu}(\mathcal{M}\paren{h})\retarg},
\end{IEEEeqnarray}
\begin{equation}\label{2.4.9}
 \Box v^{\mu\nu}\finsarg=0,
\end{equation}
\begin{equation}\label{2.4.10}
 \partial_\mu v^{\mu\nu}\finsarg=-\omega^\nu\finsarg.
\end{equation}


\chapter{Post-Newtonian Approximation}

This chapter is devoted to constructing the general solution to the post-Newtonian equations,
Eqs.(\ref{1.2.32}) and (\ref{1.2.31}),
subject to the conditions
(\ref{1.3.12}) and (\ref{1.3.13}).
The general solution at each order is the sum of a particular solution to
Eq. (\ref{1.2.31})
and the general solution to the corresponding source-free equation, a Laplace equation. At second and third orders, that particular solution can be constructed by using Poisson integral as usual. However, we need a new method to find a particular solution to
Eq. (\ref{1.2.31}) at higher orders. As we will see, this new method is based on Theorem 2.3 and the similarity of the structures of
$\bar{h}^{\mu\nu}_{\paren{n}}\finsarg$'s.

\section{Second-Order Problem}

\subsection{General Solution of the Relaxed Einstein Field Equation at Second Order} 

The relaxed Einstein field equation at second order is
\begin{equation}\label{3.1.1.1}
 \Delta\bar{h}^{\mu\nu}_{\paren{2}}\finsarg=16\pi G \bar{\tau}^{\mu\nu}_{\paren{-2}}\finsarg,
\end{equation}
where, considering
Eq. (\ref{1.1.23}),
$\bar{\tau}^{\mu\nu}_{\paren{-2}}$
appearing in that is given by
\begin{equation}\label{3.1.1.2}
\bar{\tau}^{\mu\nu}_{\paren{-2}} =  \left(-\bar{\mathfrak{g}}_{\paren{n}} \right) \bar{T}^{\mu\nu}_{\paren{m}}+\frac{1}{16\pi G}\bar{\Lambda}^{\mu\nu}_{\paren{2}} \qquad\text{with}\; n+m=-2,
\end{equation}
which, taking
Eqs. (\ref{1.2.16}), (\ref{1.2.17}) and (\ref{1.2.21})
into account, reads
\begin{equation}\label{3.1.1.3}
\bar{\tau}^{\mu\nu}_{\paren{-2}} =  \bar{T}^{\mu\nu}_{\paren{-2}}.
\end{equation}
The general solution to
Eq. (\ref{3.1.1.1})
consists of a particular solution to this equation and the general solution to the Laplace equation.
$\bar{T}^{\mu\nu}_{\paren{-2}}\finsarg$
is a smooth, compactly supported and hence bounded function. Therefore, based on Theorem  2.1, the Poisson integral of
$16\pi G \bar{\tau}^{\mu\nu}_{\paren{-2}}\finsarg$,
$\poisint{16\pi G \bar{\tau}^{\mu\nu}_{\paren{-2}}\sinsarg}$,
is convergent, thereby being a particular solution to
Eq. (\ref{3.1.1.1}).
Since the source emitting gravitational wave is entirely inside the near zone (whose bounary is at
$\mathcal{R}$),
this particular solution can be rewritten as
\begin{equation}\label{3.1.1.4}
\bar{u}^{\mu\nu}_{\paren{2}<}\finsarg = \nearint{16\pi G \bar{\tau}^{\mu\nu}_{\paren{-2}}\sinsarg}.
\end{equation}
To obtain the general solution to the Laplace equation, we use the method of separation of variables. Substituting a separated solution of the form
$\bar{x}^{\mu\nu}_{\paren{2}}\finsarg=f^{\mu\nu}_{\paren{2}}(t, |\mathbf{x}|) h(\theta, \varphi )$
into the Laplace equation, we find
\begin{equation}\label{3.1.1.7}
\frac{\partial^2}{\partial \theta^2}h + \frac{1}{\tan \theta}\frac{\partial}{\partial \theta}h+\frac{1}{\sin^2 \theta}\frac{\partial^2}{\partial \varphi^2}h=-\ell\left(\ell +1\right)h,
\end{equation}
\begin{equation}\label{3.1.1.8}
\frac{\partial^2}{\partial  |\mathbf{x}|^2}f^{\mu\nu}_{\paren{2}} + \frac{2}{ |\mathbf{x}|}\frac{\partial}{\partial  |\mathbf{x}|}f^{\mu\nu}_{\paren{2}}=\frac{f^{\mu\nu}_{\paren{2}}}{ |\mathbf{x}|^2}\ell \left(\ell +1\right).
\end{equation}
The solution of
Eq. (\ref{3.1.1.7})
is
$\ylm$.
Further, the solutions to
Eq. (\ref{3.1.1.8})
are
$A^{\mu\nu}_{\paren{2}\ell}\targ\x^{-\ell-1}$
and
$B^{\mu\nu}_{\paren{2}\ell}\targ\x^\ell$
where
$A^{\mu\nu}_{\paren{2}\ell}\targ$
and
$B^{\mu\nu}_{\paren{2}\ell}\targ$
are arbitrary smooth functions of time. Therefore, the general solution to the Laplace equation is given by
\begin{IEEEeqnarray}{rcl}\label{3.1.1.9}
\bar{x}^{\mu\nu}_{\paren{2}}\finsarg & = & \sum_{\ell=0}^{\infty}\sum_{m=-\ell}^{\ell}A^{\mu\nu}_{\paren{2}\ell m}\targ\x^{-\ell-1}\ylm+\sum_{\ell=0}^{\infty}\sum_{m=-\ell}^{\ell}B^{\mu\nu}_{\paren{2}\ell m}\targ\x^\ell\ylm, \nonumber \\ 
\end{IEEEeqnarray}
which by means of
Eq. (\ref{B.6})
can be rewritten as
\begin{equation}\label{3.1.1.10}
\bar{x}^{\mu\nu}_{\paren{2}}\finsarg=\sum_{\ell=0}^{\infty}\hat{n}^L \x^{-\ell-1}\hat{A}^{\mu\nu}_{\paren{2}L}\targ+\sum_{\ell=0}^{\infty}\hat{n}^L\x^\ell \hat{B}^{\mu\nu}_{\paren{2}L}\targ,
\end{equation}
where
$\hat{A}^{\mu\nu}_{\paren{2}L}=\sum_{m=-\ell}^{\ell}A^{\mu\nu}_{\paren{2}\ell m}\hat{\mathcal{Y}}^{\ell m}_L$
and
$\hat{B}^{\mu\nu}_{\paren{2}L}=\sum_{m=-\ell}^{\ell}B^{\mu\nu}_{\paren{2}\ell m}\hat{\mathcal{Y}}^{\ell m}_L$.
From now on, we denote the first term on the right-hand side of
Eq. (\ref{3.1.1.10})
by
$\bar{y}^{\mu\nu}_{\paren{2}}\finsarg$
and the second term by
$\bar{z}^{\mu\nu}_{\paren{2}}\finsarg$.

Having obtained the required constituents, the general solution to the relaxed Einstein equation at second order is
\begin{IEEEeqnarray}{rcl}\label{3.1.1.11}
\bar{u}^{\mu\nu}_{\paren{2}<}\finsarg+\bar{y}^{\mu\nu}_{\paren{2}}\finsarg +\bar{z}^{\mu\nu}_{\paren{2}}\finsarg & = & \nearint{16\pi G \bar{\tau}^{\mu\nu}_{\paren{-2}}\sinsarg} \nonumber \\  
&& \negmedspace {}+\sum_{\ell=0}^{\infty}\hat{n}^L \x^{-\ell-1}\hat{A}^{\mu\nu}_{\paren{2}L}\targ+\sum_{\ell=0}^{\infty}\hat{n}^L\x^\ell \hat{B}^{\mu\nu}_{\paren{2}L}\targ.\qquad
\end{IEEEeqnarray}

\subsection{Past-Stationarity and Smoothness Conditions}

The general solution to the relaxed Einstein field equation at second order consists of three parts
$\bar{u}^{\mu\nu}_{\paren{2}<}\finsarg$,
$\bar{y}^{\mu\nu}_{\paren{2}}\finsarg$
and
$\bar{z}^{\mu\nu}_{\paren{2}}\finsarg$.
Three following arguments concerning them can be made:\\
1.
$\bar{u}^{\mu\nu}_{\paren{2}<}\finsarg$
is past-stationary and smooth in the near zone due to past-stationarity and smoothness of
$\bar{T}^{\mu\nu}_{\paren{-2}}\finsarg$.\\
2. For
$\bar{y}^{\mu\nu}_{\paren{2}}\finsarg$
to be past-stationary, all
$\hat{A}^{\mu\nu}_{\paren{2}L}\targ$'s
must be stationary in the past. However, the condition of smoothness is fulfilled only if all
$\hat{A}^{\mu\nu}_{\paren{2}L}\targ$'s
are taken to be zero (owing to
$\x^{-\ell-1}$
being singular at
$\x=0$).\\
3.
$\bar{z}^{\mu\nu}_{\paren{2}}\finsarg$
satisfies the past-stationarity condition if we take all
$\hat{B}^{\mu\nu}_{\paren{2}L}\targ$'s
as past-stationary, but for its smoothness in the near zone we don't need any further restriction.\\
In light of these arguments, one can deduce that the general solution to the relaxed Einstein field equation at second order which is also past-stationary and smooth in the near zone reads
\begin{IEEEeqnarray}{rcl}\label{3.1.2.1}
\bar{u}^{\mu\nu}_{\paren{2}<}\finsarg+\bar{z}^{\mu\nu}_{\paren{2}}\finsarg & = & \nearint{16\pi G \bar{\tau}^{\mu\nu}_{\paren{-2}}\sinsarg}+\sum_{\ell=0}^{\infty}\hat{n}^L\x^\ell \hat{B}^{\mu\nu}_{\paren{2}L}\targ. \qquad
\end{IEEEeqnarray}

\subsection{General Solution of the Second-Order Problem and Its Structure Outside the Near Zone}

It can be shown that by imposing some more restrictions on
$\hat{B}^{\mu\nu}_{\paren{2}L}\targ$'s,
$\bar{u}^{\mu\nu}_{\paren{2}<}\finsarg+\bar{z}^{\mu\nu}_{\paren{2}}\finsarg $
will fulfill the harmonic gauge condition. Therefore, the general solution of the second-order problem can be written as
\begin{IEEEeqnarray}{rcl}\label{3.1.3.1}
\bar{h}^{\mu\nu}_{\paren{2}}\finsarg &=& \bar{u}^{\mu\nu}_{\paren{2}<}\finsarg+\bar{z}^{\mu\nu}_{\paren{2}}\finsarg\nonumber \\
& = &\nearint{16\pi G \bar{\tau}^{\mu\nu}_{\paren{-2}}\sinsarg}+\sum_{\ell=0}^{\infty}\hat{n}^L\x^\ell \hat{B}^{\mu\nu}_{\paren{2}L}\targ,
\end{IEEEeqnarray}
where, as stated above, the moments
$\hat{B}^{\mu\nu}_{\paren{2}L}\targ$
are subject to some restrictions other than past-stationarity condition. We won't deal with the determination of these restrictions since it is more convenient to impose the harmonic gauge condition after performing matching procedure. What is important for us is the structure of
$\bar{h}^{\mu\nu}_{\paren{2}}\finsarg$
outside the near zone. In order to determine it, we must determine the structures of
$\bar{u}^{\mu\nu}_{\paren{2}<}\finsarg$
and
$\bar{z}^{\mu\nu}_{\paren{2}}\finsarg$
outside the near zone. Noting that
$\bar{T}^{\mu\nu}_{\paren{-2}}\finsarg$
can be written as
$\sum_{q=0}^{\infty}\hat{n}^Q\bar{T}^{\mu\nu}_{\paren{-2}Q}(t,\x)$
(due to smoothness of
$\bar{T}^{\mu\nu}_{\paren{-2}}\finsarg$)
and using
Eq. (\ref{2.2.1.7}),
we get
\begin{IEEEeqnarray}{rcl}\label{3.1.1.5}
\bar{u}^{\mu\nu}_{\paren{2}<}\finsarg & = & \sum_{q=0}^{\infty}\sum_{\ell=0}^{\infty}\frac{-16\pi G}{2\ell+1}\frac{1}{\x^{\ell+1}}\int_{0}^{\mathcal{R}}\xp^{\ell+2}\bar{T}^{\mu\nu}_{\paren{-2}Q}(t,\xp)\ud\xp \nonumber \\  
&& \negmedspace {}\times\sum_{m=-\ell}^{\ell}\ylm \int \hat{n}'^Q\left[\ylmp\right]^*\ud\Omega' ,
\end{IEEEeqnarray}
which by means of
Eq. (\ref{B.3})
takes the form
\begin{IEEEeqnarray}{rcl}\label{3.1.1.6}
\bar{u}^{\mu\nu}_{\paren{2}<}\finsarg & = & \sum_{q=0}^{\infty}\frac{-16\pi G}{2q+1}\frac{\hat{n}^Q}{\x^{q+1}}\int_{0}^{\mathcal{R}}\xp^{q+2}\bar{T}^{\mu\nu}_{\paren{-2}Q}(t,\xp)\ud\xp=  \sum_{q=0}^{\infty}\hat{n}^Q\x^{-q-1}\hat{F}^{\mu\nu}_{\paren{2}Q}\targ.  \nonumber \\ 
\end{IEEEeqnarray}
To determine the structure of
$\bar{z}^{\mu\nu}_{\paren{2}}\finsarg$
outside the near zone, there is no need for any further attempt since
$\bar{z}^{\mu\nu}_{\paren{2}}\finsarg$
is already written in the appropriate form. We only assume that the moments
$\hat{B}^{\mu\nu}_{\paren{2}L}\targ$
with
$\ell>\ell_{\mathrm{max}}$
are zero as in the case of the general solution to the source-free d'Alembertian equation discussed in Chapter 2. All in all, the structure of
$\bar{h}^{\mu\nu}_{\paren{2}}\finsarg$
outside the near zone reads
\begin{equation}\label{3.1.3.2}
\bar{h}^{\mu\nu}_{\paren{2}}\finsarg = \sum_{q=0}^{\infty}\sum_{a=-\infty}^{\ell_{\mathrm{max}}}\hat{n}^Q\x^a\hat{G}^{\mu\nu}_{\paren{2}Q,a}\targ,
\end{equation}
where
$\hat{G}^{\;\mu\nu}_{\paren{2}Q,a}\targ $
is a past-stationary function.

\section{Third-Order Problem}

The relaxed Einstein field equation at third order is
\begin{equation}\label{3.2.1}
 \Delta\bar{h}^{\mu\nu}_{\paren{3}}\finsarg=16\pi G \bar{\tau}^{\mu\nu}_{\paren{-1}}\finsarg,
\end{equation}
where, considering
Eq. (\ref{1.1.23}),
$\bar{\tau}^{\mu\nu}_{\paren{-1}}$
is given by
\begin{equation}\label{3.2.2}
\bar{\tau}^{\mu\nu}_{\paren{-1}} =  \left(-\bar{\mathfrak{g}}_{\paren{n}} \right) \bar{T}^{\mu\nu}_{\paren{m}}+\frac{1}{16\pi G}\bar{\Lambda}^{\mu\nu}_{\paren{3}} \qquad \text{with}\; n+m=-1,
\end{equation}
which, taking into account
Eqs. (\ref{1.2.16}), 
(\ref{1.2.17})
and
(\ref{1.2.21}),
reads
\begin{equation}\label{3.2.3}
\bar{\tau}^{\mu\nu}_{\paren{-1}} =  \bar{T}^{\mu\nu}_{\paren{-1}}.
\end{equation}
As it can be observed, the relaxed Einstein field equations at second and third orders are exactly of the same type. Thus, by using the same reasoning involved in the second-order problem, the general solution to the third-order problem can be obtained as
\begin{IEEEeqnarray}{rcl}\label{3.2.4}
\bar{h}^{\mu\nu}_{\paren{3}}\finsarg &=& \bar{u}^{\mu\nu}_{\paren{3}<}\finsarg+\bar{z}^{\mu\nu}_{\paren{3}}\finsarg\nonumber \\
& = & \nearint{16\pi G \bar{\tau}^{\mu\nu}_{\paren{-1}}\sinsarg}+\sum_{\ell=0}^{\infty}\hat{n}^L\x^\ell \hat{B}^{\mu\nu}_{\paren{3}L}\targ,
\end{IEEEeqnarray}
whose structure outside the near zone, with the assumption that
$\hat{B}^{\mu\nu}_{\paren{3}L}\targ$
is zero for
$\ell>\ell'_{\mathrm{max}}$,
is of the form
\begin{equation}\label{3.2.5}
\bar{h}^{\mu\nu}_{\paren{3}}\finsarg = \sum_{q=0}^{\infty}\sum_{a=-\infty}^{\ell'_{\mathrm{max}}}\hat{n}^Q\x^a\hat{G}^{\;\mu\nu}_{\paren{3}Q,a}\targ ,
\end{equation}
where
$\hat{G}^{\mu\nu}_{\paren{3}Q,a}\targ $
is past-stationary.

\section{Fourth-Order Problem}

\subsection{Particular Solution of the Relaxed Einsein Field Equation at Fourth Order and Its Structure Outside the Near Zone} 

The relaxed Einstein field equation at fourth order is
\begin{equation}\label{3.3.1.1}
 \Delta\bar{h}^{\mu\nu}_{\paren{4}}\finsarg=16\pi G \bar{\tau}^{\mu\nu}_{\paren{0}}\finsarg+\partial^2_t \bar{h}^{\mu\nu}_{\paren{2}}\finsarg,
\end{equation}
where, considering
Eq. (\ref{1.1.23}),
$\bar{\tau}^{\mu\nu}_{\paren{0}}$
is given by
\begin{equation}\label{3.3.1.2}
\bar{\tau}^{\mu\nu}_{\paren{0}} =  \left(-\bar{\mathfrak{g}}_{\paren{n}} \right) \bar{T}^{\mu\nu}_{\paren{m}}+\frac{1}{16\pi G}\bar{\Lambda}^{\mu\nu}_{\paren{4}} \qquad\text{with}\; n+m=0,
\end{equation}
which, taking
Eqs. (\ref{1.2.16}), (\ref{1.2.17}) and (\ref{1.2.21})
into account, reads
\begin{IEEEeqnarray}{rcl}\label{3.3.1.3}
\bar{\tau}^{00}_{\paren{0}} & = & \left[ \bar{T}^{00}_{\paren{0}}+\bar{h}_{\paren{2}}\bar{T}^{00}_{\paren{-2}}\right]+\frac{1}{16\pi G}\bigg[-\bar{h}^{ij}_{\paren{2}}\partial_{ij}\bar{h}^{00}_{\paren{2}}+\partial_i\bar{h}^{0\rho}_{\paren{2}}\partial^i\bar{h}^0_{\paren{2}\rho}+\partial_i\bar{h}^{0j}_{\paren{2}}\partial_j\bar{h}^{0i}_{\paren{2}} \nonumber \\  
&& \negmedspace {}-\big(-\frac{1}{4}\partial_i\bar{h}_{\paren{2}\rho\sigma}\partial^i\bar{h}^{\rho\sigma}_{\paren{2}}+\frac{1}{8}\partial_i\bar{h}_{\paren{2}}\partial^i\bar{h}_{\paren{2}}+\frac{1}{2}\partial_i\bar{h}_{\paren{2}j\tau}\partial^j\bar{h}^{i\tau}_{\paren{2}}\big)\bigg],
\end{IEEEeqnarray}
\begin{IEEEeqnarray}{rcl}\label{3.3.1.4}
\bar{\tau}^{0k}_{\paren{0}} & = & \left[ \bar{T}^{0k}_{\paren{0}}+\bar{h}_{\paren{2}}\bar{T}^{0k}_{\paren{-2}}\right]+\frac{1}{16\pi G}\bigg[-\bar{h}^{ij}_{\paren{2}}\partial_{ij}\bar{h}^{0k}_{\paren{2}}+\partial_i\bar{h}^{0\rho}_{\paren{2}}\partial^i\bar{h}^k_{\paren{2}\rho}+\partial_i\bar{h}^{0j}_{\paren{2}}\partial_j\bar{h}^{ki}_{\paren{2}} \nonumber \\  
&& \negmedspace {}+\partial^k\bar{h}_{\paren{2}i\sigma}\partial^i\bar{h}^{0\sigma}_{\paren{2}}\bigg],
\end{IEEEeqnarray}
\begin{IEEEeqnarray}{rcl}\label{3.3.1.5}
\bar{\tau}^{\ell k}_{\paren{0}} & = & \left[ \bar{T}^{\ell k}_{\paren{0}}+\bar{h}_{\paren{2}}\bar{T}^{\ell k}_{\paren{-2}}\right]+\frac{1}{16\pi G}\bigg[-\bar{h}^{ij}_{\paren{2}}\partial_{ij}\bar{h}^{\ell k}_{\paren{2}}+\frac{1}{2}\partial^\ell\bar{h}_{\paren{2}\rho\sigma}\partial^k\bar{h}^{\rho\sigma}_{\paren{2}}-\frac{1}{4}\partial^\ell\bar{h}_{\paren{2}}\partial^k\bar{h}_{\paren{2}}\nonumber \\  
&& \negmedspace {}+\partial_i\bar{h}^{\ell\rho}_{\paren{2}}\partial^i\bar{h}^k_{\paren{2}\rho}+\partial_i\bar{h}^{\ell j}_{\paren{2}}\partial_j\bar{h}^{ki}_{\paren{2}}-\partial^\ell\bar{h}_{\paren{2}i\sigma}\partial^i\bar{h}^{k\sigma}_{\paren{2}}-\partial^k\bar{h}_{\paren{2}i\sigma}\partial^i\bar{h}^{\ell\sigma}_{\paren{2}} \nonumber \\  
&& \negmedspace {}+\delta^{ij}\big(-\frac{1}{4}\partial_i\bar{h}_{\paren{2}\rho\sigma}\partial^i\bar{h}^{\rho\sigma}_{\paren{2}}+\frac{1}{8}\partial_i\bar{h}_{\paren{2}}\partial^i\bar{h}_{\paren{2}}+\frac{1}{2}\partial_i\bar{h}_{\paren{2}j\tau}\partial^j\bar{h}^{i\tau}_{\paren{2}}\big)\bigg].
\end{IEEEeqnarray}
$\left[ \bar{T}^{\mu\nu}_{\paren{0}}+\bar{h}_{\paren{2}}\bar{T}^{\mu\nu}_{\paren{-2}}\right]$
part of
$\bar{\tau}^{\mu\nu}_{\paren{0}}$
is with compact support and smooth at
$\x<\mathcal{R}$.
Considering
Eq. (\ref{3.1.3.2}), the structure of
$\bar{\Lambda}^{\mu\nu}_{\paren{4}}\finsarg$,
which appears in the other part of
$\bar{\tau}^{\mu\nu}_{\paren{0}}$,
outside the near zone is of the form
$\sum_{p=0}^{\infty}\sum_{a=-\infty}^{a_{\mathrm{max}}}n^P\x^a E^{\mu\nu}_{P,a}\targ$
where
$a_{\mathrm{max}}$
depends on
$\ell_{\mathrm{max}}$.
Since
$n^P$
is a well-behaved function of
$\theta$
and
$ \varphi$,
one can rewrite this structure as
\begin{equation}\label{3.3.1.6}
\bar{\Lambda}^{\mu\nu}_{\paren{4}}\finsarg=\sum_{q=0}^{\infty}\sum_{a=-\infty}^{a_{\mathrm{max}}}\hat{n}^Q\x^a\hat{E}^{\mu\nu}_{\paren{4}Q,a}\targ.
\end{equation}
Taking
Eqs. (\ref{3.1.3.2}) and (\ref{3.3.1.6})
into account and using Theorem 2.2, the far-zone integral
$I_>\finsarg$ in
\begin{IEEEeqnarray}{rcl}\label{3.3.1.7}
I\finsarg& = & I_<\finsarg+I_>\finsarg \nonumber \\  
& = &\nearint{16\pi G \left[ \bar{T}^{\mu\nu}_{\paren{0}}\sinsarg+\bar{h}_{\paren{2}}\sinsarg\bar{T}^{\mu\nu}_{\paren{-2}}\sinsarg+\frac{1}{16\pi G}\bar{\Lambda}^{\mu\nu}_{\paren{4}}\sinsarg\right]}\nonumber \\  
&& \negmedspace {}\nearint{\partial^2_t \bar{h}^{\mu\nu}_{\paren{2}}\sinsarg}\nonumber \\  
&& \negmedspace {}\farint{16\pi G \left[ \bar{T}^{\mu\nu}_{\paren{0}}\sinsarg+\bar{h}_{\paren{2}}\sinsarg\bar{T}^{\mu\nu}_{\paren{-2}}\sinsarg+\frac{1}{16\pi G}\bar{\Lambda}^{\mu\nu}_{\paren{4}}\sinsarg\right]}\nonumber \\  
&& \negmedspace {}\farint{\partial^2_t \bar{h}^{\mu\nu}_{\paren{2}}\sinsarg}\nonumber \\  
&=&\nearint{16\pi G \left[ \bar{T}^{\mu\nu}_{\paren{0}}\sinsarg+\bar{h}_{\paren{2}}\sinsarg\bar{T}^{\mu\nu}_{\paren{-2}}\sinsarg+\frac{1}{16\pi G}\bar{\Lambda}^{\mu\nu}_{\paren{4}}\sinsarg\right]}\nonumber \\  
&& \negmedspace {}\nearint{\partial^2_t \bar{h}^{\mu\nu}_{\paren{2}}\sinsarg}\nonumber \\  
&& \negmedspace {}\farint{\bar{\Lambda}^{\mu\nu}_{\paren{4}}\sinsarg+\partial^2_t \bar{h}^{\mu\nu}_{\paren{2}}\sinsarg}\nonumber
\end{IEEEeqnarray}
diverges. Thus, despite the convergence of the near-zone integral
$I_<\finsarg$,
$I\finsarg=\poisint{16\pi G \bar{\tau}^{\mu\nu}_{\paren{0}}\sinsarg+\partial^2_t \bar{h}^{\mu\nu}_{\paren{2}}\sinsarg}$
is divergent, thereby not being a particular solution to
Eq. (\ref{3.3.1.1}).
Similarly,
$\fpa\gpoisint{\paren{16\pi G \bar{\tau}^{\mu\nu}_{\paren{0}}\sinsarg+\partial^2_t \bar{h}^{\mu\nu}_{\paren{2}}\sinsarg}}$
is not a particular solution to that equation since, due to the fact that the original domains of analyticity of the near-zone and far-zone integrals appearing in it don't overlap, the preliminary hypothesis of Theorem 2.3 is not satisfied. However, fortunately
$\Delta$
is a linear operator. Therefore, if
$f_{\mathrm{part,1}}\finsarg$
and
$f_{\mathrm{part,2}}\finsarg$
are particular solutions to the equations
$\Delta f\finsarg=g_1\finsarg$
and
$\Delta f\finsarg=g_2\finsarg$
respectively, then
$f_{\mathrm{part,1}}\finsarg+f_{\mathrm{part,2}}\finsarg$
is a particular solution to the equation
$\Delta f\finsarg=g_1\finsarg+g_2\finsarg$.
Hence, since the constituent parts of
$16\pi G \bar{\tau}^{\mu\nu}_{\paren{0}}\finsarg+\partial^2_t \bar{h}^{\mu\nu}_{\paren{2}}\finsarg$
are of the general forms
$g\finsarg$
at
$\x<\mathcal{R}$
and
$\hat{n}^Q\x^a F\targ$
at
$\x>\mathcal{R}$
($g\finsarg$
and
$ F\targ$
both are smooth and past-stationary), in order to obtain a particular solution of
Eq. (\ref{3.3.1.1}),
it is enough to find a particular solution to each of the two equations
$\Delta f\finsarg=g\finsarg\theta(\mathcal{R}-\x)$
and
$\Delta f\finsarg=\hat{n}^Q\x^a\lnx^p F\targ\theta(\x-\mathcal{R})$
(where
$r_0$
is the same constant as the one appearing in Theorem 2.3 and
$\theta$
the unit step function) for arbitrary values of
$a$,
$q$
and
$p$
where
$a \in \mathbb{Z}$
and
$p,q \in \mathbb{N}$
(in the structure of
$16\pi G \bar{\tau}^{\mu\nu}_{\paren{0}}\finsarg+\partial^2_t \bar{h}^{\mu\nu}_{\paren{2}}\finsarg$
outside the near zone, the values of
$a$,
$q$
and
$p$
are not arbitrary [in particular,
$p$
is equal to zero], but other values of them appear at higher orders).

Since
$g\finsarg$
is a smooth function, considering remark 2 following Theorem 2.3 and using Theorem 2.1, it is obvious that the original domain of analyticity of the Poisson integral
$\gpoisint{g\sinsarg\theta(\mathcal{R}-\xp)}$
is the half-plane
$\re B>-3$.
Due to
$B=0$
lying in this region,
$\fpa\gpoisint{g\sinsarg\theta(\mathcal{R}-\xp)}$
is a particular solution to the equation
$\Delta f\finsarg=g\finsarg\theta(\mathcal{R}-\x)$.
Now note that
\begin{IEEEeqnarray}{rcl}\label{3.3.1.8}
 \fpa\gpoisint{g\sinsarg\theta(\mathcal{R}-\xp)}& = & \fp\gpoisint{g\sinsarg\theta(\mathcal{R}-\xp)}\nonumber \\  
& = & \poisint{g\sinsarg\theta(\mathcal{R}-\xp)}\nonumber \\  
& = &\nearint{g\sinsarg}.
\end{IEEEeqnarray}
As it can be seen,
$\bar{u}^{\mu\nu}_{\paren{2}<}\finsarg$
and
$\fpa\gpoisint{g\sinsarg\theta(\mathcal{R}-\xp)}$
are of the same type. Therefore, the structure of
$\fpa\gpoisint{g\sinsarg\theta(\mathcal{R}-\xp)}$
outside the near-zone, by reasoning along the same lines as in the case of
$\bar{u}^{\mu\nu}_{\paren{2}<}\finsarg$,
can be obtained as
\begin{equation}\label{3.3.1.9}
\fpa\gpoisint{g\sinsarg\theta(\mathcal{R}-\xp)}=\sum_{q=0}^{\infty}\hat{n}^Q\x^{-q-1}\hat{G}^{\mu\nu}_Q\targ.
\end{equation}

Based on the foregoing discussion, a particualr solution to the equation
$ \Delta\bar{h}^{\mu\nu}_{\paren{4}<}\finsarg=\paren{16\pi G \bar{\tau}^{\mu\nu}_{\paren{0}}\finsarg+\partial^2_t \bar{h}^{\mu\nu}_{\paren{2}}\finsarg}\theta(\mathcal{R}-\x)$
is given by
\begin{IEEEeqnarray}{rcl}\label{3.3.1.10}
\bar{u}^{\mu\nu}_{\paren{4}<}\finsarg& = & \fpa\gpoisint{\paren{16\pi G \bar{\tau}^{\mu\nu}_{\paren{0}}\sinsarg+\partial^2_t \bar{h}^{\mu\nu}_{\paren{2}}\sinsarg}\theta(\mathcal{R}-\xp)}\nonumber \\  
& = & \fpa\left[\gnearint{16\pi G \bar{\tau}^{\mu\nu}_{\paren{0}}\sinsarg+\partial^2_t \bar{h}^{\mu\nu}_{\paren{2}}\sinsarg}\right],\qquad
\end{IEEEeqnarray}
whose structure outside the near zone reads
\begin{equation}\label{3.3.1.11}
\bar{u}^{\mu\nu}_{\paren{4}<}\finsarg=\sum_{q=0}^{\infty}\hat{n}^Q\x^{-q-1}\hat{F}^{\mu\nu}_{\paren{4}Q}\targ.
\end{equation}

Next, we examine the equation
$\Delta f\finsarg=\hat{n}^Q\x^a\lnx^p F\targ\theta(\x-\mathcal{R})$.
Considering remark 2 following Theorem 2.3 and using Theorem 2.2, one finds that the original domain of analyticity of
$\gpoisint{\hat{n}'^Q\xp^a\lnxp^p F\targ\theta(\xp-\mathcal{R})}$
is the half-plane
$\re B<-a+q-2$.
Now note that with the use of
Eqs. (\ref{2.2.1.25}) and (\ref{2.2.1.28}),
in the region
\begin{equation}\label{3.3.1.14}
D= \left\{B\left|\right. \re B<\!-a\!+\!q\!-\!2\right\}-\left\{\paren{-a\!-\!q\!-\!3,0}\right\}\nonumber
\end{equation}
we can write
\begin{eqnarray}\label{3.3.1.15}
\lefteqn{\gpoisint{\hat{n}'^Q\xp^a F\targ\theta(\xp-\mathcal{R})}} \nonumber \\ 
&\!\!\!\!\!\!\!\!=\!\!\!\!\!\!\!\!&
\begin{cases}
\frac{1}{2q+1}\paren{\frac{\mathcal{R}}{r_0}}^B\hat{n}^Q\x^q F\targ\frac{\mathcal{R}^{a-q+2}}{B+a-q+2}, &\!\quad \x\!<\!\mathcal{R},\\
\\
\xb\frac{\hat{n}^Q \x^{a+2}F\targ}{\paren{B+a-q+2}\paren{B+a+q+3}}\!+\!\frac{1}{2q+1}\paren{\frac{\mathcal{R}}{r_0}}^B\hat{n}^Q\x^{-q-1} F\targ\frac{\mathcal{R}^{a+q+3}}{B+a+q+3},&\!\quad \mathcal{R}\!<\!\x\!<\!\infty.
\end{cases} \nonumber\\
\end{eqnarray}
Since the right-hand side of the above equation is also analytic in the region
$D$,
we can differentiate both sides of it with respect to
$B$
as many times as is necessary. By doing so
$p$
times, we reach
\begin{eqnarray}\label{3.3.1.16}
\lefteqn{\gpoisint{\hat{n}'^Q\xp^a\lnxp^p F\targ\theta(\xp-\mathcal{R})}}
\nonumber \\ 
&\!\!\!\!\!\!\!\!=\!\!\!\!\!\!\!\!&
\begin{cases}
\frac{\partial^p}{\partial B^p}\left[\frac{\mathcal{R}^{a-q+2}}{2q+1}\paren{\frac{\mathcal{R}}{r_0}}^B\frac{\hat{n}^Q\x^q F\targ}{B+a-q+2}\right], &\!\quad \x\!<\!\mathcal{R},\\
\\
\frac{\partial^p}{\partial B^p}\left[\xb\frac{\hat{n}^Q \x^{a+2}F\targ}{\paren{B+a-q+2}\paren{B+a+q+3}}+\frac{\mathcal{R}^{a+q+3}}{2q+1}\paren{\frac{\mathcal{R}}{r_0}}^B\frac{\hat{n}^Q\x^{-q-1}F\targ}{B+a+q+3}\right],&\!\quad \mathcal{R}\!<\!\x\!<\!\infty.
\end{cases} \nonumber\\
\end{eqnarray}
As a consequence of identity theorem, we can equate the analytic continuations of each side of
Eq. (\ref{3.3.1.16})
wherever they are both defined. Therefore, we have
\begin{eqnarray}\label{3.3.1.18}
\lefteqn{\mathrm{A}\gpoisint{\hat{n}'^Q\xp^a\lnxp^p F\targ\theta(\xp-\mathcal{R})}}
\nonumber \\ 
&\!\!\!\!\!\!\!\!=\!\!\!\!\!\!\!\!&
\begin{cases}
\frac{\partial^p}{\partial B^p}\left[\frac{\mathcal{R}^{a-q+2}}{2q+1}\paren{\frac{\mathcal{R}}{r_0}}^B\frac{\hat{n}^Q\x^q F\targ}{B+a-q+2}\right], &\!\quad \x\!<\!\mathcal{R},\\
\\
\frac{\partial^p}{\partial B^p}\left[\xb\frac{\hat{n}^Q \x^{a+2}F\targ}{\paren{B+a-q+2}\paren{B+a+q+3}}+\frac{\mathcal{R}^{a+q+3}}{2q+1}\paren{\frac{\mathcal{R}}{r_0}}^B\frac{\hat{n}^Q\x^{-q-1}F\targ}{B+a+q+3}\right],&\!\quad \mathcal{R}\!<\!\x\!<\!\infty.
\end{cases} \nonumber\\
\end{eqnarray}
From
Eq. (\ref{3.3.1.18}),
it can be observed that
$\mathrm{A}\gpoisint{\hat{n}'^Q\xp^a\lnxp^p F\targ\theta(\xp-\mathcal{R})}$
is singular at the points with
$B\in \mathbb{Z}$.
$\mathrm{A}\gpoisint{\hat{n}'^Q\xp^a\lnxp^p F\targ\theta(\xp-\mathcal{R})}$
is therefore analytic in some punctured neighborhood of
$B=0$,
or, in other words, one can analytically continue
$\gpoisint{\hat{n}'^Q\xp^a\lnxp^p F\targ\theta(\xp-\mathcal{R})}$
down to a punctured neghborhood of
$B=0$.
Hence,
$\fpa\gpoisint{\hat{n}'^Q\xp^a\lnxp^p F\targ\theta(\xp-\mathcal{R})}$
is a particular solution for the equation
 $\Delta f\finsarg=\hat{n}^Q\x^a\lnx^p F\targ\theta(\x-\mathcal{R})$.

To determine the structure of
$\fpa\gpoisint{\hat{n}'^Q\xp^a\lnxp^p F\targ\theta(\xp-\mathcal{R})}$,
we must first assess whether
$\mathrm{A}\gpoisint{\hat{n}'^Q\xp^a\lnxp^p F\targ\theta(\xp-\mathcal{R})}$
is singular at
$B=0$
or not.
Eq. (\ref{3.3.1.18})
implies that, for
$B=0$
to be a singularity, at least one of the following equalities must hold:
\begin{equation}\label{3.3.1.19}
a-q+2=0 \longrightarrow a=q-2,
\end{equation}
\begin{equation}\label{3.3.1.20}
a+q+3=0 \longrightarrow a=-q-3.
\end{equation}
The crucial point is that these two equalities cannot be fulfilled simultaneously because if they can, for given values of
$a$
and
$q$
we deduce the false proposition that
$q$
(or
$a$)
is not an integer.

Considering
Eq. (\ref{3.3.1.18}),
if
$\mathrm{A}\gpoisint{\hat{n}'^Q\xp^a\lnxp^p F\targ\theta(\xp-\mathcal{R})}$
is analytic at
$B=0$,
its general form can be expressed as follows:\\
1. For
$\x<\mathcal{R}$
:
\begin{IEEEeqnarray}{rcl}\label{3.3.1.21a}
\mathrm{A}\gpoisint{\hat{n}'^Q\xp^a\lnxp^p F\targ\theta(\xp-\mathcal{R})} & = &\frac{\partial^p}{\partial B^p}\left[C(B)\hat{n}^Q\x^q F\targ\right].\qquad 
\end{IEEEeqnarray}
2. For
$\mathcal{R}<\x<\infty$
:
\begin{eqnarray}\label{3.3.1.21b}
\lefteqn{\mathrm{A}\gpoisint{\hat{n}'^Q\xp^a\lnxp^p F\targ\theta(\xp-\mathcal{R})}} \nonumber \\ 
&\!\!\!\!\!\!\!\!=\!\!\!\!\!\!\!\!&\frac{\partial^p}{\partial B^p}\left[\xb D(B)\hat{n}^Q\x^{a+2}
F\targ+ E(B)\hat{n}^Q\x^{-q-1} F\targ\right].
\end{eqnarray}
In these equations,
$C(B)$,
$D(B)$
and
$E(B)$
are complex functions analytic at
$B=0$.
Using Leibniz formula, we have:\\
1. For
$\x<\mathcal{R}$
:
\begin{equation}\label{3.3.1.22a}
\mathrm{A}\gpoisint{\hat{n}'^Q\xp^a\lnxp^p F\targ\theta(\xp-\mathcal{R})} =
C^{\paren{p}}(B)\hat{n}^Q\x^q F\targ.
\end{equation}
2. For
$\mathcal{R}<\x<\infty$
:
\begin{eqnarray}\label{3.3.1.22b}
\lefteqn{\mathrm{A}\gpoisint{\hat{n}'^Q\xp^a\lnxp^p F\targ\theta(\xp-\mathcal{R})}} \nonumber \\ 
&\!\!\!\!\!\!\!\!=\!\!\!\!\!\!\!\!&\sum_{k=0}^p\binom{p}{k}\xb D^{\paren{p-k}}(B)\hat{n}^Q\x^{a+2}\lnx^k F\targ+ E^{\paren{p}}(B)\hat{n}^Q\x^{-q-1} F\targ.\nonumber\\
\end{eqnarray}
The coefficients of
$B^0$
in Taylor expansions of right-hand sides of
Eqs. (\ref{3.3.1.22a}) and (\ref{3.3.1.22b}) can be obtained by taking into account the coefficient of
$B^0$
in Taylor expansion of each of the complex functions on the right-hand sides. Then we get
\begin{eqnarray}\label{3.3.1.23}
\lefteqn{\fpa\gpoisint{\hat{n}'^Q\xp^a\lnxp^p F\targ\theta(\xp-\mathcal{R})}}
\nonumber \\ 
&\!\!\!\!\!\!\!\!=\!\!\!\!\!\!\!\!&
\begin{cases}
\hat{n}^Q\x^q F'\targ,&\qquad \x<\mathcal{R},\\
\\
\sum_{k=0}^p\hat{n}^Q\x^{a+2}\lnx^k F_k\targ+\hat{n}^Q\x^{-q-1} F''\targ,&\qquad \mathcal{R}<\x<\infty,
\end{cases}
\end{eqnarray}
where
$F'\targ$,
$F''\targ$
and
$F_k\targ$
are in terms of
$F\targ$.

On the other hand, in the case of
$\mathrm{A}\gpoisint{\hat{n}'^Q\xp^a\lnxp^p F\targ\theta(\xp-\mathcal{R})}$
being singular at
$B=0$,\footnote{In this case, it is straightforward to show that this singularity is a pole if
Eq. (\ref{3.3.1.19})
holds and removable if
Eq. (\ref{3.3.1.20}).}
taking into account the remark follwing
Eqs. (\ref{3.3.1.19}) and (\ref{3.3.1.20}),
the general form of
$\mathrm{A}\gpoisint{\hat{n}'^Q\xp^a\lnxp^p F\targ\theta(\xp-\mathcal{R})}$
can be written as follows:\\
1. For
$a=q-2$
and
$\x<\mathcal{R}$
:
\begin{IEEEeqnarray}{rcl}\label{3.3.1.24a}
\mathrm{A}\gpoisint{\hat{n}'^Q\xp^a\lnxp^p F\targ\theta(\xp-\mathcal{R})} & = &\frac{\partial^p}{\partial B^p}\left[\paren{\frac{G_1(B)}{B}}\hat{n}^Q\x^q F\targ \right].\nonumber \\  
\end{IEEEeqnarray}
2. For
$a=q-2$
and
$\mathcal{R}<\x<\infty$
:
\begin{eqnarray}\label{3.3.1.24b}
\lefteqn{\mathrm{A}\gpoisint{\hat{n}'^Q\xp^a\lnxp^p F\targ\theta(\xp-\mathcal{R})}} \nonumber \\ 
&\!\!\!\!\!\!\!\!=\!\!\!\!\!\!\!\!&\frac{\partial^p}{\partial B^p}\left[\xb\paren{\frac{H_1(B)}{B}}\hat{n}^Q\x^{a+2} F\targ+ I_1(B)\hat{n}^Q\x^{-q-1} F\targ\right].
\end{eqnarray}
3. For
$a=-q-3$
and
$\x<\mathcal{R}$
:
\begin{IEEEeqnarray}{rcl}\label{3.3.1.24c}
\mathrm{A}\gpoisint{\hat{n}'^Q\xp^a\lnxp^p F\targ\theta(\xp-\mathcal{R})} & = &\frac{\partial^p}{\partial B^p}\left[G_2(B)\hat{n}^Q\x^q F\targ\right]. \qquad
\end{IEEEeqnarray}
4. For
$a=-q-3$
and
$\mathcal{R}<\x<\infty$
:
\begin{eqnarray}\label{3.3.1.24d}
\lefteqn{\mathrm{A}\gpoisint{\hat{n}'^Q\xp^a\lnxp^p F\targ\theta(\xp-\mathcal{R})}} \nonumber \\ 
&\!\!\!\!\!\!\!\!=\!\!\!\!\!\!\!\!&\frac{\partial^p}{\partial B^p}\left[\xb\paren{\frac{H_2(B)}{B}}\hat{n}^Q\x^{a+2}F\targ+\paren{\frac{I_2(B)}{B}}\hat{n}^Q\x^{-q-1} F\targ\right].
\end{eqnarray}
In the above equations,
$G_1(B)$,
$G_2(B)$,
$H_1(B)$,
$H_2(B)$,
$I_1(B)$
and
$I_2(B)$
are complex functions analytic at
$B=0$.
With the use of Leibniz formula, we get:\\
1. For
$a=q-2$
and
$\x<\mathcal{R}$
:
\begin{eqnarray}\label{3.3.1.25a}
\lefteqn{\mathrm{A}\gpoisint{\hat{n}'^Q\xp^a\lnxp^p F\targ\theta(\xp-\mathcal{R})}} \nonumber \\ 
&\!\!\!\!\!\!\!\!=\!\!\!\!\!\!\!\!& \sum_{k=0}^p\binom{p}{k}{\paren{-1}}^k\paren{k!}\paren{\frac{G_1^{\paren{p-k}}(B)}{B^{k+1}}}\hat{n}^Q\x^q F\targ.
\end{eqnarray}
2. For
$a=q-2$
and
$\mathcal{R}<\x<\infty$
:
\begin{eqnarray}\label{3.3.1.25b}
\lefteqn{\mathrm{A}\gpoisint{\hat{n}'^Q\xp^a\lnxp^p F\targ\theta(\xp-\mathcal{R})}} \nonumber \\ 
&\!\!\!\!\!\!\!\!=\!\!\!\!\!\!\!\!&\sum_{k=0}^{p}\sum_{\ell=0}^{p-k}\binom{p}{k}\binom{p-k}{\ell}\paren{-1}^\ell \paren{\ell!}\xb\paren{\frac{H_1^{\paren{p-k-\ell}}(B)}{B^{\ell+1}}}\hat{n}^Q\x^{a+2}\lnx^k F\targ\nonumber\\
&&\!\!\!\!+ I_1^{\paren{p}}(B)\hat{n}^Q\x^{-q-1} F\targ.
\end{eqnarray}
3. For
$a=-q-3$
and
$\x<\mathcal{R}$
:
\begin{equation}\label{3.3.1.25c}
\mathrm{A}\gpoisint{\hat{n}'^Q\xp^a\lnxp^p F\targ\theta(\xp-\mathcal{R})}=G_2^{\paren{p}}(B)\hat{n}^Q\x^q F\targ.
\end{equation}
4. For
$a=-q-3$
and
$\mathcal{R}<\x<\infty$
:
\begin{eqnarray}\label{3.3.1.25d}
\lefteqn{\mathrm{A}\gpoisint{\hat{n}'^Q\xp^a\lnxp^p F\targ\theta(\xp-\mathcal{R})}} \nonumber \\ 
&\!\!\!\!\!\!\!\!=\!\!\!\!\!\!\!\!&\sum_{k=0}^{p}\sum_{\ell=0}^{p-k}\binom{p}{k}\binom{p-k}{\ell}\paren{-1}^\ell \paren{\ell!}\xb\paren{\frac{H_2^{\paren{p-k-\ell}}(B)}{B^{\ell+1}}}\hat{n}^Q\x^{a+2}\lnx^k F\targ\nonumber\\
&&\!\!\!\!+ \sum_{k=0}^p\binom{p}{k}{\paren{-1}}^k\paren{k!}\paren{\frac{I_2^{\paren{p-k}}(B)}{B^{k+1}}}\hat{n}^Q\x^{-q-1} F\targ.
\end{eqnarray}
The coefficients of
$B^0$
in Laurent expansions of the right-hand sides of the above equations are as follows:\\
1. If
$\x<\mathcal{R}$,
whether
$a=q-2$
or
$a=-q-3$,
the coefficients of
$B^0$
are proportional to
$\hat{n}^Q\x^q F\targ$.\\
2. If
$\mathcal{R}<\x<\infty$,
the coefficients of
$B^0$
in Laurent expansions of the first terms can be determined by reasoning similar to that expressed in 2 following
Eq. (\ref{2.2.2.51}).
Further, the coefficients of
$B^0$
in Laurent expansions of the second terms, whether
$a=q-2$
or
$a=-q-3$,
are proportional to
$\hat{n}^Q\x^{-q-1} F\targ$.\\
Considering 1 and 2 above, we reach
\begin{eqnarray}\label{3.3.1.26}
\lefteqn{\fpa\gpoisint{\hat{n}'^Q\xp^a\lnxp^p F\targ\theta(\xp-\mathcal{R})}}
\nonumber \\ 
&\!\!\!\!\!\!\!\!=\!\!\!\!\!\!\!\!&
\begin{cases}
\hat{n}^Q\x^q F'\targ,&\qquad \x<\mathcal{R},\\
\\
\sum_{k=0}^{p+1}\hat{n}^Q\x^{a+2}\lnx^k F_k\targ+\hat{n}^Q\x^{-q-1} F''\targ,&\qquad \mathcal{R}<\x<\infty,
\end{cases}
\end{eqnarray}
where
$F'\targ$,
$F''\targ$
and
$F_k\targ$
are in terms of
$F\targ$
(it is clear that
$F'\targ$,
$F''\targ$
and
$F_k\targ$
differ from the corresponding ones in
Eq. (\ref{3.3.1.23})).

Based on the foregoing investigation, a particular solution to the equation under study,
$\Delta\bar{h}^{\mu\nu}_{\paren{4}>}\finsarg=\paren{16\pi G \bar{\tau}^{\mu\nu}_{\paren{0}}\finsarg+\partial^2_t \bar{h}^{\mu\nu}_{\paren{2}}\finsarg}\theta(\x-\mathcal{R})$,
is given by
\begin{IEEEeqnarray}{rcl}\label{3.3.1.27}
\bar{u}^{\mu\nu}_{\paren{4}>}\finsarg& = & \fpa\gpoisint{\paren{16\pi G \bar{\tau}^{\mu\nu}_{\paren{0}}\sinsarg+\partial^2_t \bar{h}^{\mu\nu}_{\paren{2}}\sinsarg}\theta(\xp-\mathcal{R})}\nonumber \\  
& = & \fpa\left[\gfarint{16\pi G \bar{\tau}^{\mu\nu}_{\paren{0}}\sinsarg+\partial^2_t \bar{h}^{\mu\nu}_{\paren{2}}\sinsarg}\right].\quad
\end{IEEEeqnarray}
It can be shown that some of
$a$'s
and
$q$'s
appearing in the structure of
$16\pi G \bar{\tau}^{\mu\nu}_{\paren{0}}\finsarg+\partial^2_t \bar{h}^{\mu\nu}_{\paren{2}}\finsarg$
outside the near zone satisfy either
Eq. (\ref{3.3.1.19})
or
Eq. (\ref{3.3.1.20}).
Therefore, the structure of
$\bar{u}^{\mu\nu}_{\paren{4}>}\finsarg$
reads
\begin{IEEEeqnarray}{rcl}\label{3.3.1.28}
\bar{u}^{\mu\nu}_{\paren{4}>}\finsarg&=&
\begin{cases}
\sum_{q=0}^\infty\hat{n}^Q\x^q \hat{F}'^{\;\mu\nu}_{\paren{4}Q}\targ,&\qquad \x<\mathcal{R},\\
\\
\sum_{q=0}^{\infty}\sum_{a=-\infty}^{a'_{\mathrm{max}}}\sum_{p=0}^{1}\hat{n}^Q\x^a\lnx^p \hat{F}^{\mu\nu}_{\paren{4}Q,a,p}\targ,&\qquad \mathcal{R}<\x<\infty.
\end{cases}\nonumber\\
\end{IEEEeqnarray}

Particular solutions of
$ \Delta\bar{h}^{\mu\nu}_{\paren{4}<}\finsarg=\paren{16\pi G \bar{\tau}^{\mu\nu}_{\paren{0}}\finsarg+\partial^2_t \bar{h}^{\mu\nu}_{\paren{2}}\finsarg}\theta(\mathcal{R}-\x)$
and
$\Delta\bar{h}^{\mu\nu}_{\paren{4}>}\finsarg=\paren{16\pi G \bar{\tau}^{\mu\nu}_{\paren{0}}\finsarg+\partial^2_t \bar{h}^{\mu\nu}_{\paren{2}}\finsarg}\theta(\x-\mathcal{R})$
having been obtained, we are now in a position to write down a particular solution for the equation
$ \Delta\bar{h}^{\mu\nu}_{\paren{4}}\finsarg=16\pi G \bar{\tau}^{\mu\nu}_{\paren{0}}\finsarg+\partial^2_t \bar{h}^{\mu\nu}_{\paren{2}}\finsarg$.
That solution is
\begin{IEEEeqnarray}{rcl}\label{3.3.1.29}
\bar{u}^{\mu\nu}_{\paren{4}}\finsarg& = & \bar{u}^{\mu\nu}_{\paren{4}<}\finsarg+\bar{u}^{\mu\nu}_{\paren{4}>}\finsarg\nonumber \\  
& \stareq & \fpa\gpoisint{\paren{16\pi G \bar{\tau}^{\mu\nu}_{\paren{0}}\sinsarg+\partial^2_t \bar{h}^{\mu\nu}_{\paren{2}}\sinsarg}},
\end{IEEEeqnarray}
where
$\bar{u}^{\mu\nu}_{\paren{4}<}\finsarg$
and
$\bar{u}^{\mu\nu}_{\paren{4}>}\finsarg$
are given by
Eqs. (\ref{3.3.1.10})
and
(\ref{3.3.1.27})
respectively, and we have used
the sign
$\stareq$
due to 
$\gpoisint{\paren{16\pi G \bar{\tau}^{\mu\nu}_{\paren{0}}\sinsarg+\partial^2_t \bar{h}^{\mu\nu}_{\paren{2}}\sinsarg}\theta(\mathcal{R}-\xp)}$
and
$\gpoisint{\paren{16\pi G \bar{\tau}^{\mu\nu}_{\paren{0}}\sinsarg+\partial^2_t \bar{h}^{\mu\nu}_{\paren{2}}\sinsarg}\theta(\xp-\mathcal{R})}$
not possessing common original domain of analyticity, and hence,
$\fpa\gpoisint{\paren{16\pi G \bar{\tau}^{\mu\nu}_{\paren{0}}\sinsarg+\partial^2_t \bar{h}^{\mu\nu}_{\paren{2}}\sinsarg}}$
being meaningless. Moreover, the structure of
$\bar{u}^{\mu\nu}_{\paren{4}}\finsarg$
outside the near zone reads
\begin{equation}\label{3.3.1.30}
\bar{u}^{\mu\nu}_{\paren{4}}\finsarg=\sum_{q=0}^{\infty}\sum_{a=-\infty}^{a'_{\mathrm{max}}}\sum_{p=0}^{1}\hat{n}^Q\x^a\lnx^p \hat{F}'^{\;\mu\nu}_{\paren{4}Q,a,p}\targ,
\end{equation}
where
$\hat{F}'^{\;\mu\nu}_{\paren{4}Q,a,p}\targ$
is past-stationary.

\subsection{General Solution of the Fourth-Order Problem and Its Structure Outside the Near Zone}

In previous subsection we found that
$\bar{u}^{\mu\nu}_{\paren{4}}\finsarg=\bar{u}^{\mu\nu}_{\paren{4}<}\finsarg+\bar{u}^{\mu\nu}_{\paren{4}>}\finsarg$
is a particular solution to the relaxed Einstein field equation at fourth order.
$\bar{u}^{\mu\nu}_{\paren{4}<}\finsarg$
is past-stationary and smooth in the near zone owing to past-stationarity and smootness of
$ \bar{\tau}^{\mu\nu}_{\paren{0}}\finsarg$
and
$ \bar{h}^{\mu\nu}_{\paren{2}}\finsarg$.
Additionally, taking into account  the structure of
$\bar{u}^{\mu\nu}_{\paren{4}>}\finsarg$
given in
Eq. (\ref{3.3.1.28}),
past-stationarity of
$\bar{u}^{\mu\nu}_{\paren{4}>}\finsarg$
and its smoothness in the near zone are evident. Therefore, the general solution to
Eq. (\ref{3.3.1.1})
subject to the conditions of being past-stationary and smooth in the near zone is the sum of
$\bar{u}^{\mu\nu}_{\paren{4}}\finsarg=\bar{u}^{\mu\nu}_{\paren{4}<}\finsarg+\bar{u}^{\mu\nu}_{\paren{4}>}\finsarg$
and
$\bar{z}^{\mu\nu}_{\paren{4}}\finsarg= \sum_{\ell=0}^{\infty}\hat{n}^L\x^\ell \hat{B}^{\mu\nu}_{\paren{4}L}\targ $
provided that the moments
$\hat{B}^{\mu\nu}_{\paren{4}L}\targ$
are stationary in the past. After imposing some more restrictions on
$\hat{B}^{\mu\nu}_{\paren{4}L}\targ$'s,
$\bar{u}^{\mu\nu}_{\paren{4}}\finsarg+\bar{z}^{\mu\nu}_{\paren{4}}\finsarg $
will also satisfy the harmonic gauge condition. Thus, the general solution to the fourth-order problem can be expressed as
\begin{IEEEeqnarray}{rcl}\label{3.3.3.1}
\bar{h}^{\mu\nu}_{\paren{4}}\finsarg & = &  \bar{u}^{\mu\nu}_{\paren{4}}\finsarg+\bar{z}^{\mu\nu}_{\paren{4}}\finsarg=\bar{u}^{\mu\nu}_{\paren{4}<}\finsarg+\bar{u}^{\mu\nu}_{\paren{4}>}\finsarg+\bar{z}^{\mu\nu}_{\paren{4}}\finsarg  \nonumber \\  
& = & \fpa\gpoisint{\paren{16\pi G \bar{\tau}^{\mu\nu}_{\paren{0}}\sinsarg+\partial^2_t \bar{h}^{\mu\nu}_{\paren{2}}\sinsarg}\theta(\mathcal{R}-\xp)} \nonumber \\  
&& \negmedspace {} +\fpa\gpoisint{\paren{16\pi G \bar{\tau}^{\mu\nu}_{\paren{0}}\sinsarg+\partial^2_t \bar{h}^{\mu\nu}_{\paren{2}}\sinsarg}\theta(\xp-\mathcal{R})} \nonumber \\  
&& \negmedspace {} +\sum_{\ell=0}^{\infty}\hat{n}^L\x^\ell \hat{B}^{\mu\nu}_{\paren{4}L}\targ,
\end{IEEEeqnarray}
where, as mentioned above, the moments
$\hat{B}^{\mu\nu}_{\paren{4}L}\targ$
are subject to some restrictions other than the condition of past-stationarity. We won't try to determine these restrictions since, as stated in Subsection 3.1.3, imposing the harmonic gauge condition after performing matching procedure is more convenient. What is important for us is the structure of
$\bar{h}^{\mu\nu}_{\paren{4}}\finsarg$
outside the near zone which, by considering
Eq. (\ref{3.3.1.30})
and taking
$\hat{B}^{\mu\nu}_{\paren{4}L}\targ$'s
with
$\ell>\ell''_{\mathrm{max}}$
to be zero, reads
\begin{equation}\label{3.3.3.2}
\bar{h}^{\mu\nu}_{\paren{4}}\finsarg= \sum_{q=0}^{\infty}\sum_{a=-\infty}^{a''_{\mathrm{max}}}\sum_{p=0}^{1}\hat{n}^Q\x^a\lnx^p\hat{G}^{\mu\nu}_{\paren{4}Q,a,p}\targ,
\end{equation}
where
$\hat{G}^{\mu\nu}_{\paren{4}Q,a,p}\targ $
is stationary in the past.

\section{$n$th-Order Problem}

Reasoning along the same lines as in previous section, we can say that, for any arbitrary
$n$,
$\bar{\Lambda}^{\mu\nu}_{\paren{n}}\finsarg$
is smooth in the near zone, and outside the near zone its structure reads
\begin{equation}\label{3.4.1}
\bar{\Lambda}^{\mu\nu}_{\paren{n}}\finsarg= \sum_{q=0}^{\infty}\sum_{a=-\infty}^{a_{\mathrm{max}}\paren{n}}\sum_{p=0}^{p_{\mathrm{max}}\paren{n}}\hat{n}^Q\x^a\lnx^p\hat{E}^{\mu\nu}_{\paren{n}Q,a,p}\targ,
\end{equation}
where
$a_{\mathrm{max}}\paren{n}>0$
and
$\hat{E}^{\mu\nu}_{\paren{n}Q,a,p}\targ$
is past-stationary. These structural properties of
$\bar{\Lambda}^{\mu\nu}_{\paren{n}}\finsarg$
in the near zone and outside of it guarantee that the general solution to
$n$th-order
problem is given by
\begin{equation}\label{3.4.2}
\bar{h}^{\mu\nu}_{\paren{n}}\finsarg=\bar{u}^{\mu\nu}_{\paren{n}}\finsarg+\bar{z}^{\mu\nu}_{\paren{n}}\finsarg=\bar{u}^{\mu\nu}_{\paren{n}<}\finsarg+\bar{u}^{\mu\nu}_{\paren{n}>}\finsarg+\bar{z}^{\mu\nu}_{\paren{n}}\finsarg,
\end{equation}
where
\begin{IEEEeqnarray}{rcl}\label{3.4.3}
\bar{u}^{\mu\nu}_{\paren{n}<}\finsarg&=&\fpa\gpoisint{\paren{16\pi G \bar{\tau}^{\mu\nu}_{\paren{n-4}}\sinsarg+\partial^2_t \bar{h}^{\mu\nu}_{\paren{n-2}}\sinsarg}\theta(\mathcal{R}-\xp)},\nonumber \\  
\end{IEEEeqnarray}
\begin{IEEEeqnarray}{rcl}\label{3.4.4}
\bar{u}^{\mu\nu}_{\paren{n}>}\finsarg&=&\fpa\gpoisint{\paren{16\pi G \bar{\tau}^{\mu\nu}_{\paren{n-4}}\sinsarg+\partial^2_t \bar{h}^{\mu\nu}_{\paren{n-2}}\sinsarg}\theta(\xp-\mathcal{R})},\nonumber \\  
\end{IEEEeqnarray}
\begin{equation}\label{3.4.5}
\bar{z}^{\mu\nu}_{\paren{n}}\finsarg=\sum_{\ell=0}^{\infty}\hat{n}^L\x^\ell \hat{B}^{\mu\nu}_{\paren{n}L}\targ,
\end{equation}
\begin{IEEEeqnarray}{rcl}\label{3.4.6}
\bar{u}^{\mu\nu}_{\paren{n}}\finsarg& \stareq &\fpa\gpoisint{\paren{16\pi G \bar{\tau}^{\mu\nu}_{\paren{n-4}}\sinsarg+\partial^2_t \bar{h}^{\mu\nu}_{\paren{n-2}}\sinsarg}}.
\end{IEEEeqnarray}
$\bar{h}^{\mu\nu}_{\paren{n}}\finsarg$
is smooth in the near zone. Further, its structure outside the near zone, provided that
$\hat{B}^{\mu\nu}_{\paren{n}L}\targ$'s
with
$\ell>\ell_{\mathrm{max}}\paren{n}$
are taken as zero, can be written as
\begin{equation}\label{3.4.7}
\bar{h}^{\mu\nu}_{\paren{n}}\finsarg= \sum_{q=0}^{\infty}\sum_{a=-\infty}^{a'_{\mathrm{max}}\paren{n}}\sum_{p=0}^{p'_{\mathrm{max}}\paren{n}}\hat{n}^Q\x^a\lnx^p\hat{G}^{\mu\nu}_{\paren{n}Q,a,p}\targ,
\end{equation}
where
$\hat{G}^{\mu\nu}_{\paren{n}Q,a,p}\targ$
is past-stationary.

\section{Post-Newtonian Expansion}

It is obvious that
$\bar{h}^{\mu\nu}\finsarg$
and
$\bar{\Lambda}^{\mu\nu}\finsarg$,
the (untruncated) post-Newtonian expansions of
$h^{\mu\nu}\finsarg$
and
$\Lambda^{\mu\nu}\finsarg$,
are smooth in the near zone. In addition, considering the structures given in
Eqs. (\ref{3.4.1})
and
(\ref{3.4.7})
which are  valid for any arbitrary
$n$,
the structures of
$\bar{h}^{\mu\nu}\finsarg$
and
$\bar{\Lambda}^{\mu\nu}\finsarg$
outside the near zone read
\begin{equation}\label{3.5.1}
\bar{h}^{\mu\nu}\finsarg=\sum_{n=2}^{\infty} \frac{1}{c^n} \bar{h}^{\mu\nu}_{\left(n\right)}\finsarg= \sum_{q=0}^{\infty}\sum_{a=-\infty}^{\infty}\sum_{p=0}^{\infty}\hat{n}^Q\x^a\lnx^p\hat{G}^{\mu\nu}_{Q,a,p}\targ,
\end{equation}
\begin{equation}\label{3.5.2}
\bar{\Lambda}^{\mu\nu}\finsarg=\sum_{n=4}^{\infty} \frac{1}{c^n} \bar{\Lambda}^{\mu\nu}_{\left(n\right)}\finsarg= \sum_{q=0}^{\infty}\sum_{a=-\infty}^{\infty}\sum_{p=0}^{\infty}\hat{n}^Q\x^a\lnx^p\hat{E}^{\mu\nu}_{Q,a,p}\targ.
\end{equation}
Finally, the set of equations governing them are as follows:
\begin{equation}\label{3.5.3}
\Box \bar{h}^{\mu\nu}\finsarg=\frac{16\pi G}{c^4}\bar{\tau}^{\mu\nu}\finsarg,
\end{equation}
\begin{equation}\label{3.5.4}
 \partial_\mu \bar{h}^{\mu\nu}\finsarg=0,
\end{equation}
\begin{equation}\label{3.5.5}
 \partial_\mu \bar{\tau}^{\mu\nu}\finsarg=0.
\end{equation}


\chapter{Matching Procedure}

In this final chapter, we discuss matching procedure, the last step in Blanchet-Damour approach. This procedure is based on the so-called ``asymptotic matching'' appearing on the subject of asymptotic expansions. There is also an ``intermediate matching''. These are two different methods used to match ``inner'' and ``outer'' asymptotic expansions. The explanation of these concepts is beyond the scope of this thesis. For more details one can refer to
\cite{N2000,H2013}.

\section{Matching Equation}

Considering the structures of
$\mathcal{M}\paren{h^{\mu\nu}}\finsarg$
and
$\bar{h}^{\mu\nu}\finsarg$
given in
Eqs. (\ref{2.4.1}) and (\ref{3.5.1})
respectively, it is clear that we can write
\begin{IEEEeqnarray}{rcl}\label{4.1.1}
\overline{\mathcal{M}\paren{h^{\mu\nu}}}\finsarg&=&\lim_{\substack{|\mathbf{x}|\to 0 \\ t=\mathrm{const}}}\mathcal{M}\paren{h^{\mu\nu}}\finsarg=  \sum_{q=0}^{\infty}\sum_{a=-\infty}^{\infty}\sum_{p=0}^{\infty}\hat{n}^Q\x^a\lnx^p\hat{G}'^{\;\mu\nu}_{Q,a,p}\targ,\quad
\end{IEEEeqnarray}
\begin{equation}\label{4.1.2}
\mathcal{M}\paren{\bar{h}^{\mu\nu}}\finsarg=\lim_{\substack{|\mathbf{x}|\to \infty \\ t=\mathrm{const}}}\bar{h}^{\mu\nu}\finsarg= \sum_{q=0}^{\infty}\sum_{a=-\infty}^{\infty}\sum_{p=0}^{\infty}\hat{n}^Q\x^a\lnx^p\hat{G}^{\mu\nu}_{Q,a,p}\targ.
\end{equation}
If
Eqs. (\ref{4.1.1}) and (\ref{4.1.2})
had been written in terms of ``rescaled'' variables (see
\cite{N2000,H2013}),
it would have been obvious that
$\overline{\mathcal{M}\paren{h^{\mu\nu}}}\finsarg$
and
$\mathcal{M}\paren{\bar{h}^{\mu\nu}}\finsarg$
are in fact the post-Minkowskian and post-Newtonian expansions of
$h^{\mu\nu}\finsarg$ within the common region of validity of these expansions. Therefore, according to the assumption of existence of matching region discussed at the end of Chapter 1, we have
\begin{equation}\label{4.1.3}
\overline{\mathcal{M}\paren{h^{\mu\nu}}}\finsarg=\mathcal{M}\paren{\bar{h}^{\mu\nu}}\finsarg,
\end{equation}
which is in full agreement with the similarity of the structures of
$\overline{\mathcal{M}\paren{h^{\mu\nu}}}\finsarg$
and
$\mathcal{M}\paren{\bar{h}^{\mu\nu}}\finsarg$.
Eq. (\ref{4.1.3}) is called the ``matching equation''. In the next section we use this equation to determine
$\mathcal{M}\paren{h^{\mu\nu}}\finsarg$
and
$\bar{h}^{\mu\nu}\finsarg$.

\section{Determination of $\mathcal{M}\paren{h^{\mu\nu}}\finsarg$ and $\bar{h}^{\mu\nu}\finsarg$}

Consider the following equation:
\begin{IEEEeqnarray}{rcl}\label{4.2.1}
\Box \left[\xb\mathcal{M}\paren{h^{\mu\nu}}\finsarg\right]&=&\xb\Box\mathcal{M}\paren{h^{\mu\nu}}\finsarg\nonumber \\  
&& \negmedspace {}+\xb\bigg[2B\x^{-1}\partial_{\x}\mathcal{M}\paren{h^{\mu\nu}}\finsarg\nonumber \\  
&& \negmedspace {}+B\paren{B+1}\x^{-2}\mathcal{M}\paren{h^{\mu\nu}}\finsarg\bigg].
\end{IEEEeqnarray}
$\mathcal{M}\paren{h^{\mu\nu}}\finsarg$
consists of an always stationary and a past-zero part. Since the past-zero part vanishes at
$t\le-T$,
in this interval, and due to
$\mathcal{M}\paren{h^{\mu\nu}_\mathrm{AS}}\xarg$
being always stationary, at any time, we have
\begin{IEEEeqnarray}{rcl}\label{4.2.2}
\Delta \left[\xb\mathcal{M}\paren{h^{\mu\nu}_\mathrm{AS}}\xarg\right]&=&\xb\Delta\mathcal{M}\paren{h^{\mu\nu}_\mathrm{AS}}\xarg\nonumber \\  
&& \negmedspace {}+\xb\bigg[2B\x^{-1}\partial_{\x}\mathcal{M}\paren{h^{\mu\nu}_\mathrm{AS}}\xarg\nonumber \\  
&& \negmedspace {}+B\paren{B+1}\x^{-2}\mathcal{M}\paren{h^{\mu\nu}_\mathrm{AS}}\xarg\bigg].
\end{IEEEeqnarray}
Taking into account the linearity of the operators
$\Box$
and
$\partial_{\x}$,
after subtracting the above equation from
Eq. (\ref{4.2.1}),
one finds
\begin{IEEEeqnarray}{rcl}\label{4.2.3}
\Box\left[\xb\mathcal{M}\paren{h^{\mu\nu}_\mathrm{PZ}}\finsarg\right]&=&\xb\Box\mathcal{M}\paren{h^{\mu\nu}_\mathrm{PZ}}\finsarg\nonumber \\  
&& \negmedspace {}+\xb\bigg[2B\x^{-1}\partial_{\x}\mathcal{M}\paren{h^{\mu\nu}_\mathrm{PZ}}\finsarg\nonumber \\  
&& \negmedspace {}+B\paren{B+1}\x^{-2}\mathcal{M}\paren{h^{\mu\nu}_\mathrm{PZ}}\finsarg\bigg],
\end{IEEEeqnarray}
which is valid at any time as
Eq. (\ref{4.2.2}).

Consider
Eq. (\ref{4.2.2}).
For any
$k$
and
$\ell$
appearing in the structure of
$\mathcal{M}\paren{h^{\mu\nu}_\mathrm{AS}}\xarg$
(which can be determined from
Eq. (\ref{2.4.1})),
we can write
\begin{IEEEeqnarray}{rcl}\label{4.2.4}
\Delta \left[\xb\hat{n}^L\x^k\hat{C}'^{\;\mu\nu}_{L,k}\right]&=&\xb\Delta\paren{\hat{n}^L\x^k\hat{C}'^{\;\mu\nu}_{L,k}}\nonumber \\  
&& \negmedspace {}+\xb\bigg[2B\x^{-1}\partial_{\x}\paren{\hat{n}^L\x^k\hat{C}'^{\;\mu\nu}_{L,k}}\nonumber \\  
&& \negmedspace {}+B\paren{B+1}\x^{-2}\paren{\hat{n}^L\x^k\hat{C}'^{\;\mu\nu}_{L,k}}\bigg].
\end{IEEEeqnarray}
If
$\re B+k<0$,
both real and imaginary parts of
$\xb\hat{n}^L\x^k\hat{C}'^{\;\mu\nu}_{L,k}$
tend to zero as
$\x \to \infty$.
Therefore, providing that the Poisson integrals converge, we can write
\begin{IEEEeqnarray}{rcl}\label{4.2.5}
\re\left[\xb\right] \hat{n}^L\x^k\hat{C}'^{\;\mu\nu}_{L,k}&=&\poisint{\re\left[\xpb\right]\Delta'\paren{\hat{n}'^L\xp^k\hat{C}'^{\;\mu\nu}_{L,k}}}\nonumber \\  
&& \negmedspace {}+\poisint{2\;\re\left[B\xpb\right]\xp^{-1}\partial_{\xp}\paren{\hat{n}'^L\xp^k\hat{C}'^{\;\mu\nu}_{L,k}}}\nonumber \\  
&& \negmedspace {}+\poisint{\re\left[B\paren{B+1}\xpb\right]\xp^{-2}\paren{\hat{n}'^L\xp^k\hat{C}'^{\;\mu\nu}_{L,k}}},\nonumber \\  
\end{IEEEeqnarray}
\begin{IEEEeqnarray}{rcl}\label{4.2.6}
\im\left[\xb\right] \hat{n}^L\x^k\hat{C}'^{\;\mu\nu}_{L,k}&=&\poisint{\im\left[\xpb\right]\Delta'\paren{\hat{n}'^L\xp^k\hat{C}'^{\;\mu\nu}_{L,k}}}\nonumber \\  
&& \negmedspace {}+\poisint{2\;\im\left[B\xpb\right]\xp^{-1}\partial_{\xp}\paren{\hat{n}'^L\xp^k\hat{C}'^{\;\mu\nu}_{L,k}}}\nonumber \\  
&& \negmedspace {}+\poisint{\im\left[B\paren{B+1}\xpb\right]\xp^{-2}\paren{\hat{n}'^L\xp^k\hat{C}'^{\;\mu\nu}_{L,k}}}.\nonumber \\  
\end{IEEEeqnarray}
Taking Theorems 2.1 and 2.2 into account, all the Poisson integrals appearing in
Eqs. (\ref{4.2.5}) and (\ref{4.2.6})
are convergent provided that
$-\ell-3<\re B +k -2<\ell-2$.
Thus, we can say that the equalities in both
Eqs. (\ref{4.2.5}) and (\ref{4.2.6})
hold if
$-\ell-1<\re B +k <0$,
thereby in the vertical strip
$-k-\ell-3<\re B <-k$
having
\begin{IEEEeqnarray}{rcl}\label{4.2.7}
\xb\hat{n}^L\x^k\hat{C}'^{\;\mu\nu}_{L,k}&=&\poisint{\xpb\Delta'\paren{\hat{n}'^L\xp^k\hat{C}'^{\;\mu\nu}_{L,k}}}\nonumber \\  
&& \negmedspace {}+\poisint{\xpb\left[2B\xp^{-1}\partial_{\xp}\paren{\hat{n}'^L\xp^k\hat{C}'^{\;\mu\nu}_{L,k}}\right]}\nonumber \\  
&& \negmedspace {}+\poisint{\xpb\left[B\paren{B+1}\xp^{-2}\paren{\hat{n}'^L\xp^k\hat{C}'^{\;\mu\nu}_{L,k}}\right]}.\quad
\end{IEEEeqnarray}
We can easily show that all the terms appearing in the above equation are analytic in the aforementioned region. Therefore, by virtue of identity theorem, the equality between analytic continuations of each side of
Eq. (\ref{4.2.7}) must hold wherever both are defined. This, together with the fact that
$\xb\hat{n}^L\x^k\hat{C}'^{\;\mu\nu}_{L,k}$
is entire, result in
\begin{IEEEeqnarray}{rcl}\label{4.2.8}
\xb\hat{n}^L\x^k\hat{C}'^{\;\mu\nu}_{L,k}&=&\mathrm{A}\poisint{\xpb\Delta'\paren{\hat{n}'^L\xp^k\hat{C}'^{\;\mu\nu}_{L,k}}}\nonumber \\  
&& \negmedspace {}+\mathrm{A}\poisint{\xpb\left[2B\xp^{-1}\partial_{\xp}\paren{\hat{n}'^L\xp^k\hat{C}'^{\;\mu\nu}_{L,k}}\right]}\nonumber \\  
&& \negmedspace {}+\mathrm{A}\poisint{\xpb\left[B\paren{B+1}\xp^{-2}\paren{\hat{n}'^L\xp^k\hat{C}'^{\;\mu\nu}_{L,k}}\right]}.\quad
\end{IEEEeqnarray}
Summing
Eq. (\ref{4.2.8})
over all values of
$k$
and
$\ell$
appearing in the structure of
$\mathcal{M}\paren{h^{\mu\nu}_\mathrm{AS}}\xarg$,
we get
\begin{eqnarray}\label{4.2.9}
\lefteqn{\xb\mathcal{M}\paren{h^{\mu\nu}_\mathrm{AS}}\xarg=\sum_{\ell=0}^{\infty}\sum_{k=-\infty}^{-1}\xb\hat{n}^L\x^k\hat{C}'^{\;\mu\nu}_{L,k}} \nonumber \\ 
&\!\!\!\!\!\!\!\!=\!\!\!\!\!\!\!\!&\sum_{\ell=0}^{\infty}\sum_{k=-\infty}^{-1}\A\left[\gnearint{\Delta'\paren{\hat{n}'^L\xp^k\hat{C}'^{\;\mu\nu}_{L,k}}}\right]\nonumber\\
&&\!\!\!\!+\sum_{\ell=0}^{\infty}\sum_{k=-\infty}^{-1}\A\left[\gfarint{\Delta'\paren{\hat{n}'^L\xp^k\hat{C}'^{\;\mu\nu}_{L,k}}}\right] \nonumber\\
&&\!\!\!\!+\sum_{\ell=0}^{\infty}\sum_{k=-\infty}^{-1}\A\left[\gnearint{2B\xp^{-1}\partial_{\xp}\paren{\hat{n}'^L\xp^k\hat{C}'^{\;\mu\nu}_{L,k}}}\right] \nonumber\\
&&\!\!\!\!+\sum_{\ell=0}^{\infty}\sum_{k=-\infty}^{-1}\A\left[\gfarint{2B\xp^{-1}\partial_{\xp}\paren{\hat{n}'^L\xp^k\hat{C}'^{\;\mu\nu}_{L,k}}}\right] \nonumber\\
&&\!\!\!\!+ \sum_{\ell=0}^{\infty}\sum_{k=-\infty}^{-1}\A\left[\gnearint{B\paren{B+1}\xp^{-2}\paren{\hat{n}'^L\xp^k\hat{C}'^{\;\mu\nu}_{L,k}}}\right] \nonumber\\
&&\!\!\!\!+ \sum_{\ell=0}^{\infty}\sum_{k=-\infty}^{-1}\A\left[\gfarint{B\paren{B+1}\xp^{-2}\paren{\hat{n}'^L\xp^k\hat{C}'^{\;\mu\nu}_{L,k}}}\right].\qquad
\end{eqnarray}
Since there exists a common region in which all the integrals in each term of
Eq. (\ref{4.2.9})
are analytic (depending on whether the region of integration is, this common region is either some right or left half-plane),
Eq. (\ref{4.2.9})
can be rewritten as
\begin{eqnarray}\label{4.2.10}
\lefteqn{\xb\mathcal{M}\paren{h^{\mu\nu}_\mathrm{AS}}\xarg} \nonumber \\ 
&\!\!\!\!\!\!\!\!=\!\!\!\!\!\!\!\!&\A\left[\gnearint{\Lambda^{\mu\nu}_{\mathrm{AS}}(\mathcal{M}\paren{h})\xparg}\right]\nonumber\\
&&\!\!\!\!+\A\left[\gfarint{\Lambda^{\mu\nu}_{\mathrm{AS}}(\mathcal{M}\paren{h})\xparg}\right] \nonumber\\
&&\!\!\!\!+\A\left[\gnearint{2B\xp^{-1}\partial_{\xp}\mathcal{M}\paren{h^{\mu\nu}_\mathrm{AS}}\xparg}\right] \nonumber\\
&&\!\!\!\!+\A\left[\gfarint{2B\xp^{-1}\partial_{\xp}\mathcal{M}\paren{h^{\mu\nu}_\mathrm{AS}}\xparg}\right] \nonumber\\
&&\!\!\!\!+ \A\left[\gnearint{B\paren{B+1}\xp^{-2}\mathcal{M}\paren{h^{\mu\nu}_\mathrm{AS}}\xparg}\right] \nonumber\\
&&\!\!\!\!+ \A\left[\gfarint{B\paren{B+1}\xp^{-2}\mathcal{M}\paren{h^{\mu\nu}_\mathrm{AS}}\xparg}\right].
\end{eqnarray}
The structure of
$\mathcal{M}\paren{h^{\mu\nu}_\mathrm{AS}}\xarg$
is such that it can be shown each of the analytic continuations on the right-hand side of
Eq. (\ref{4.2.10})
is defined in some punctured neighborhood of
$B=0$,
thereby possessing a Laurent expansion about
$B=0$.
Further, it is obvious that the left-hand side of this equation has a Taylor expansion about
$B=0$.
Therefore, since the coefficients of
$B^n$
on both sides of the equation must be equal for each
$n$,
taking the coefficients of
$B^0$
into account, we can write
\begin{eqnarray}\label{4.2.13}
\lefteqn{\mathcal{M}\paren{h^{\mu\nu}_\mathrm{AS}}\xarg} \nonumber \\ 
&\!\!\!\!\!\!\!\!=\!\!\!\!\!\!\!\!&\fpa\left[\gnearint{\Lambda^{\mu\nu}_{\mathrm{AS}}(\mathcal{M}\paren{h})\xparg}\right]\nonumber\\
&&\!\!\!\!+\fpa\left[\gfarint{\Lambda^{\mu\nu}_{\mathrm{AS}}(\mathcal{M}\paren{h})\xparg}\right] \nonumber\\
&&\!\!\!\!+\fpa\left[\gnearint{2B\xp^{-1}\partial_{\xp}\mathcal{M}\paren{h^{\mu\nu}_\mathrm{AS}}\xparg}\right] \nonumber\\
&&\!\!\!\!+\fpa\left[\gfarint{2B\xp^{-1}\partial_{\xp}\mathcal{M}\paren{h^{\mu\nu}_\mathrm{AS}}\xparg}\right] \nonumber\\
&&\!\!\!\!+ \fpa\left[\gnearint{B\paren{B+1}\xp^{-2}\mathcal{M}\paren{h^{\mu\nu}_\mathrm{AS}}\xparg}\right] \nonumber\\
&&\!\!\!\!+ \fpa\left[\gfarint{B\paren{B+1}\xp^{-2}\mathcal{M}\paren{h^{\mu\nu}_\mathrm{AS}}\xparg}\right].
\end{eqnarray}
Noting that the maximal power of
$\xp$
in the structures of both
$\xp^{-1}\partial_{\xp}\mathcal{M}\paren{h^{\mu\nu}_\mathrm{AS}}\xparg$
and
$\xp^{-2}\mathcal{M}\paren{h^{\mu\nu}_\mathrm{AS}}\xparg$
is
$-3$
and using Theorem 2.2, it is straightforward to show that
$\gfarint{\xp^{-1}\partial_{\xp}\mathcal{M}\paren{h^{\mu\nu}_\mathrm{AS}}\xparg}$
and
$\gfarint{\xp^{-2}\mathcal{M}\paren{h^{\mu\nu}_\mathrm{AS}}\xparg}$
are both analytic at
$B=0$,
and hence, the fourth and sixth terms on the right-hand side of
Eq. (\ref{4.2.13})
are zero due to the coefficients
$B$
and
$B\paren{B+1}$.
Moreover,
$\Delta\mathcal{M}\paren{h^{\mu\nu}_\mathrm{AS}}\xarg=\Lambda^{\mu\nu}_{\mathrm{AS}}(\mathcal{M}\paren{h})\xarg$.
Thus,
Eq. (\ref{4.2.13})
takes the form
\begin{eqnarray}\label{4.2.14}
\lefteqn{\mathcal{M}\paren{h^{\mu\nu}_\mathrm{AS}}\xarg} \nonumber \\ 
&\!\!\!\!\!\!\!\!=\!\!\!\!\!\!\!\!&\fpa\left[\gnearint{\Lambda^{\mu\nu}_{\mathrm{AS}}(\mathcal{M}\paren{h})\xparg}\right]\nonumber\\
&&\!\!\!\!+\fpa\left[\gfarint{\Lambda^{\mu\nu}_{\mathrm{AS}}(\mathcal{M}\paren{h})\xparg}\right] \nonumber\\
&&\!\!\!\!+\fpa\left[\gnearint{2B\xp^{-1}\partial_{\xp}\mathcal{M}\paren{h^{\mu\nu}_\mathrm{AS}}\xparg}\right] \nonumber\\
&&\!\!\!\!+ \fpa\left[\gnearint{B\paren{B+1}\xp^{-2}\mathcal{M}\paren{h^{\mu\nu}_\mathrm{AS}}\xparg}\right].
\end{eqnarray}

Now consider
Eq. (\ref{4.2.3}).
Since
$\mathcal{M}\paren{h^{\mu\nu}_\mathrm{PZ}}\finsarg$
is past-zero, both real and imaginary parts of
$\xb\mathcal{M}\paren{h^{\mu\nu}_\mathrm{PZ}}\finsarg$
satisfy the no-incoming radiation condition. Therefore, providing that the retarded integrals are convergent, we have
\begin{IEEEeqnarray}{rcl}\label{4.2.15}
\re\left[\xb\right] \mathcal{M}\paren{h^{\mu\nu}_\mathrm{PZ}}&=&\retint{\re\left[\xpb\right]\Box'\mathcal{M}\paren{h^{\mu\nu}_\mathrm{PZ}}\retarg}\nonumber \\  
&& \negmedspace {}+\retint{2\;\re\left[B\xpb\right]\xp^{-1}\paren{\partial_{\xp}\mathcal{M}\paren{h^{\mu\nu}_\mathrm{PZ}}\retarg}_{t'}}\nonumber \\  
&& \negmedspace {}+\retint{\re\left[B\paren{B+1}\xpb\right]\xp^{-2}\mathcal{M}\paren{h^{\mu\nu}_\mathrm{PZ}}\retarg},\nonumber \\  
\end{IEEEeqnarray}
\begin{IEEEeqnarray}{rcl}\label{4.2.16}
\im\left[\xb\right] \mathcal{M}\paren{h^{\mu\nu}_\mathrm{PZ}}&=&\retint{\im\left[\xpb\right]\Box'\mathcal{M}\paren{h^{\mu\nu}_\mathrm{PZ}}\retarg}\nonumber \\  
&& \negmedspace {}+\retint{2\;\im\left[B\xpb\right]\xp^{-1}\paren{\partial_{\xp}\mathcal{M}\paren{h^{\mu\nu}_\mathrm{PZ}}\retarg}_{t'}}\nonumber \\  
&& \negmedspace {}+\retint{\im\left[B\paren{B+1}\xpb\right]\xp^{-2}\mathcal{M}\paren{h^{\mu\nu}_\mathrm{PZ}}\retarg}.\nonumber \\  
\end{IEEEeqnarray}
Using Theorem 2.1, we can show that all the retarded integrals in the above equatioans converge for
$\re B$
a large enough positive number. Thus, there exists some right half-palne in which both
Eqs. (\ref{4.2.15}) and (\ref{4.2.16})
are valid and we can thereby write
\begin{IEEEeqnarray}{rcl}\label{4.2.17}
\xb\mathcal{M}\paren{h^{\mu\nu}_\mathrm{PZ}}\finsarg&=&\retint{\xpb\Box'\mathcal{M}\paren{h^{\mu\nu}_\mathrm{PZ}}\retarg}\nonumber \\  
&& \negmedspace {}+\retint{\xpb\left[2B\xp^{-1}\paren{\partial_{\xp}\mathcal{M}\paren{h^{\mu\nu}_\mathrm{PZ}}\retarg}_{t'}\right]}\nonumber \\  
&& \negmedspace {}+\retint{\xpb\left[B\paren{B+1}\xp^{-2}\mathcal{M}\paren{h^{\mu\nu}_\mathrm{PZ}}\retarg\right]}.\qquad
\end{IEEEeqnarray}
It can be straightforwardly shown that all the terms on both sides of the above equation are also analytic in that right half-plane. Thus, as a result of identity theorem, the equality between analytic continuations of each side of
Eq. (\ref{4.2.17})
must hold in the region where both exist. This, together with the fact that
$\xb\mathcal{M}\paren{h^{\mu\nu}_\mathrm{PZ}}\finsarg$
is an entire function, yield
\begin{eqnarray}\label{4.2.18}
\lefteqn{\xb\mathcal{M}\paren{h^{\mu\nu}_\mathrm{PZ}}\finsarg} \nonumber \\ 
&\!\!\!\!\!\!\!\!=\!\!\!\!\!\!\!\!&\A\retint{\xpb\Box'\mathcal{M}\paren{h^{\mu\nu}_\mathrm{PZ}}\retarg}\nonumber\\
&&\!\!\!\!+\A\retint{\xpb\left[2B\xp^{-1}\paren{\partial_{\xp}\mathcal{M}\paren{h^{\mu\nu}_\mathrm{PZ}}\retarg}_{t'}\right]} \nonumber\\
&&\!\!\!\!+\A\retint{\xpb\left[B\paren{B+1}\xp^{-2}\mathcal{M}\paren{h^{\mu\nu}_\mathrm{PZ}}\retarg\right]}\nonumber \\  
&\!\!\!\!\!\!\!\!=\!\!\!\!\!\!\!\!&\A\retint{\xpb\Box'\mathcal{M}\paren{h^{\mu\nu}_\mathrm{PZ}}\retarg}\nonumber\\
&&\!\!\!\!+\A\left[\gnearint{2B\xp^{-1}\paren{\partial_{\xp}\mathcal{M}\paren{h^{\mu\nu}_\mathrm{PZ}}\retarg}_{t'}}\right] \nonumber\\
&&\!\!\!\!+\A\left[\gfarint{2B\xp^{-1}\paren{\partial_{\xp}\mathcal{M}\paren{h^{\mu\nu}_\mathrm{PZ}}\retarg}_{t'}}\right] \nonumber\\
&&\!\!\!\!+ \A\left[\gnearint{B\paren{B+1}\xp^{-2}\mathcal{M}\paren{h^{\mu\nu}_\mathrm{PZ}}\retarg}\right] \nonumber\\
&&\!\!\!\!+ \A\left[\gfarint{B\paren{B+1}\xp^{-2}\mathcal{M}\paren{h^{\mu\nu}_\mathrm{PZ}}\retarg}\right].
\end{eqnarray}
The structure of
$\mathcal{M}\paren{h^{\mu\nu}_\mathrm{PZ}}\finsarg$
is so that we can show each of the analytic continuations on the right-hand side of the above equation is defined in some punctured neighborhood of
$B=0$
and thereby has a Laurent expansion around
$B=0$.
In addition, it is clear that the left-hand side of this equation possesses a Taylor expansion around
$B=0$.
Since, for each
$n$,
the coefficients of
$B^n$
on the two sides of
Eq. (\ref{4.2.18})
must be equal, considering the coefficients of
$B^0$,
we obtain
\begin{eqnarray}\label{4.2.19}
\lefteqn{\mathcal{M}\paren{h^{\mu\nu}_\mathrm{PZ}}\finsarg} \nonumber \\ 
&\!\!\!\!\!\!\!\!=\!\!\!\!\!\!\!\!&\fpa\retint{\xpb\Box'\mathcal{M}\paren{h^{\mu\nu}_\mathrm{PZ}}\retarg}\nonumber\\
&&\!\!\!\!+\fpa\left[\gnearint{2B\xp^{-1}\paren{\partial_{\xp}\mathcal{M}\paren{h^{\mu\nu}_\mathrm{PZ}}\retarg}_{t'}}\right] \nonumber\\
&&\!\!\!\!+\fpa\left[\gfarint{2B\xp^{-1}\paren{\partial_{\xp}\mathcal{M}\paren{h^{\mu\nu}_\mathrm{PZ}}\retarg}_{t'}}\right] \nonumber\\
&&\!\!\!\!+ \fpa\left[\gnearint{B\paren{B+1}\xp^{-2}\mathcal{M}\paren{h^{\mu\nu}_\mathrm{PZ}}\retarg}\right] \nonumber\\
&&\!\!\!\!+ \fpa\left[\gfarint{B\paren{B+1}\xp^{-2}\mathcal{M}\paren{h^{\mu\nu}_\mathrm{PZ}}\retarg}\right].
\end{eqnarray}
Because
$\mathcal{M}\paren{h^{\mu\nu}_\mathrm{PZ}}\finsarg$
is a past-zero function, it can be straightforwardly shown that both
$\gfarint{\!\xp^{\!-1}\!\paren{\partial_{\xp}\mathcal{M}\paren{h^{\mu\nu}_\mathrm{PZ}}\!\retarg\!}_{t'}\!\!}$
and
$\gfarint{\!\xp^{\!-2}\!\mathcal{M}\paren{h^{\mu\nu}_\mathrm{PZ}}\!\retarg\!\!}$
are analytic at
$B=0$
and the third and fifth terms on the right-hand side of
Eq. (\ref{4.2.19})
thereby equal zero due to the coefficients
$B$
and
$B\paren{B+1}$.
Further,
$\Box\mathcal{M}\paren{h^{\mu\nu}_\mathrm{PZ}}\finsarg=\Lambda^{\mu\nu}_{\mathrm{PZ}}(\mathcal{M}\paren{h})\finsarg$.
Then we reach
\begin{eqnarray}\label{4.2.20}
\lefteqn{\mathcal{M}\paren{h^{\mu\nu}_\mathrm{PZ}}\finsarg} \nonumber \\ 
&\!\!\!\!\!\!\!\!=\!\!\!\!\!\!\!\!&\fpa\retint{\xpb\Lambda^{\mu\nu}_{\mathrm{PZ}}(\mathcal{M}\paren{h})\retarg}\nonumber\\
&&\!\!\!\!+\fpa\left[\gnearint{2B\xp^{-1}\paren{\partial_{\xp}\mathcal{M}\paren{h^{\mu\nu}_\mathrm{PZ}}\retarg}_{t'}}\right] \nonumber\\
&&\!\!\!\!+ \fpa\left[\gnearint{B\paren{B+1}\xp^{-2}\mathcal{M}\paren{h^{\mu\nu}_\mathrm{PZ}}\retarg}\right].
\end{eqnarray}

Summing
Eqs. (\ref{4.2.14}) and (\ref{4.2.20}),
we obtain
$\mathcal{M}\paren{h^{\mu\nu}}\finsarg$
as
\begin{IEEEeqnarray}{rcl}\label{4.2.21}
\mathcal{M}\paren{h^{\mu\nu}}\finsarg & = &  u^{\mu\nu}\finsarg \nonumber \\  
&& \negmedspace {} +\fpa\left[\gnearint{2B\xp^{-1}\paren{\partial_{\xp}\mathcal{M}\paren{h^{\mu\nu}}\retarg}_{t'}}\right]\nonumber \\  
&& \negmedspace {} + \fpa\left[\gnearint{B\paren{B+1}\xp^{-2}\mathcal{M}\paren{h^{\mu\nu}}\retarg}\right],\nonumber \\  
\end{IEEEeqnarray}
where
$u^{\mu\nu}\finsarg$
is the same as the one given by
Eq. (\ref{2.4.7}).
It is worth noting that without the knowledge of the structure of
$\mathcal{M}\paren{h^{\mu\nu}}\finsarg$
obtained in Chapter 2, it would have been impossible to reach
Eq. (\ref{4.2.21}).

Since
$\mathcal{M}\paren{h^{\mu\nu}}\finsarg=\overline{\mathcal{M}\paren{h^{\mu\nu}}}\finsarg+R'^{\;\mu\nu}\finsarg$,
Eq. (\ref{4.2.21})
can be rewritten as
\begin{IEEEeqnarray}{rcl}\label{4.2.22}
\mathcal{M}\paren{h^{\mu\nu}}\finsarg & = &  u^{\mu\nu}\finsarg \nonumber \\  
&& \negmedspace {} +\fpa\left[\gnearint{2B\xp^{-1}\paren{\partial_{\xp}\overline{\mathcal{M}\paren{h^{\mu\nu}}}\retarg}_{t'}}\right]\nonumber \\  
&& \negmedspace {} + \fpa\left[\gnearint{B\paren{B+1}\xp^{-2}\overline{\mathcal{M}\paren{h^{\mu\nu}}}\retarg}\right],\nonumber \\  
\end{IEEEeqnarray}
where
$\gfarint{\!\xp^{\!-1}\!\paren{\partial_{\xp}R'^{\;\mu\nu}\retarg}_{t'}\!}$
and
$\gfarint{\!\xp^{\!-2}\!R'^{\;\mu\nu}\retarg\!}$
haven't appeared due to  being analytic at
$B=0$
and having the coefficients
$B$
and
$B\paren{B+1}$.
If
$\x>\mathcal{R}$,
by using
Eq. (\ref{4.1.1})
and Taylor expansion of functions of three variables, we get
\begin{IEEEeqnarray}{rcl}\label{4.2.23}
\mathcal{M}\paren{h^{\mu\nu}} & = & u^{\mu\nu}-\frac{1}{4\pi}\sum_{j=0}^{\infty}\sum_{q=0}^{\infty}\sum_{a=-\infty}^{\infty}\sum_{p=0}^{\infty}\frac{\paren{-1}^j}{j!}\bigg[\partial_{I'_j}\frac{\hat{G}'^{\;\mu\nu}_{{I_q},a,p}\uarg}{\x}\bigg]\nonumber \\  
&& \negmedspace {}\times\int \hat{n}'^{I_q}n'^{I'_j}\ud\Omega' \fp\paren{2aB\!+\!B\paren{B\!+\!1}} \frac{\partial^p}{\partial B^p}\left[\frac{1}{r_{0}^{B}}\cdot\A \int_{0}^{\mathcal{R}}\xp^{B+a+j}\ud\xp\right]\nonumber\\
&& \negmedspace {}-\frac{1}{4\pi}\sum_{j=0}^{\infty}\sum_{q=0}^{\infty}\sum_{a=-\infty}^{\infty}\sum_{p=1}^{\infty}\frac{\paren{-1}^j}{j!}\bigg[\partial_{I'_j}\frac{\hat{G}'^{\;\mu\nu}_{{I_q},a,p}\uarg}{\x}\bigg]\nonumber \\  
&& \negmedspace {}\times\int \hat{n}'^{I_q}n'^{I'_j}\ud\Omega' \fp\paren{2pB} \frac{\partial^{p-1}}{\partial B^{p-1}}\left[\frac{1}{r_{0}^{B}}\cdot\A \int_{0}^{\mathcal{R}}\xp^{B+a+j}\ud\xp\right].
\end{IEEEeqnarray}
In addition, whereas the equality of
$\int_{0}^{\mathcal{R}}\xp^{B+a+j}\ud\xp$
and
$-\int_{\mathcal{R}}^{\infty}\xp^{B+a+j}\ud\xp$
is meaningless owing to their original domains of analyticity not overlapping, we have
\begin{equation}\label{4.2.24}
\A\int_{0}^{\mathcal{R}}\xp^{B+a+j}\ud\xp=\frac{{\mathcal{R}}^{B+a+j+1}}{B\!+\!a\!+\!j\!+\!1}=-\A\int_{\mathcal{R}}^{\infty}\xp^{B+a+j}\ud\xp.
\end{equation}
Replacing
$\A\int_{0}^{\mathcal{R}}\xp^{B+a+j}\ud\xp$
by
$-\A\int_{\mathcal{R}}^{\infty}\xp^{B+a+j}\ud\xp$
in
Eq. (\ref{4.2.23})
and then reversing the order of computations resulting in that equation, we get
\begin{IEEEeqnarray}{rcl}\label{4.2.26}
\mathcal{M}\paren{h^{\mu\nu}}\finsarg & = & u^{\mu\nu}\finsarg+\frac{1}{4\pi}\sum_{j=0}^{\infty}\frac{\paren{-1}^j}{j!}\partial_{J}\bigg(\x^{-1}\fpa\int_{\mathcal{R}<\xp}\xpb x'^{J}\nonumber \\  
&& \negmedspace {}\times\bigg[2B\xp^{-1}\partial_{\xp}\overline{\mathcal{M}\paren{h^{\mu\nu}}}(t-\frac{\x}{c},\mathbf{x'})\nonumber \\  
&& \negmedspace {}+B\paren{B+1}\xp^{-2}\overline{\mathcal{M}\paren{h^{\mu\nu}}}(t-\frac{\x}{c},\mathbf{x'})\bigg]\ud^3\mathbf{x'}\bigg).
\end{IEEEeqnarray}
In the case
$0<\x<\mathcal{R}$,
the same equation as above can be obtained. Therefore,
Eq. (\ref{4.2.26})
is valid everywhere in
$\mathbb{R }^3$
(except at
$\x=0$).

Now it's time to use the matching equation. Replacing
$\overline{\mathcal{M}\paren{h^{\mu\nu}}}$
by
$\mathcal{M}\paren{\bar{h}^{\mu\nu}}$
in
Eq. (\ref{4.2.26}),
we find
\begin{IEEEeqnarray}{rcl}\label{4.2.30}
\mathcal{M}\paren{h^{\mu\nu}}\finsarg & = & u^{\mu\nu}\finsarg+\frac{1}{4\pi}\sum_{j=0}^{\infty}\frac{\paren{-1}^j}{j!}\partial_{J}\bigg(\x^{-1}\fpa\int_{\mathcal{R}<\xp}\xpb x'^{J}\nonumber \\  
&& \negmedspace {}\times\bigg[2B\xp^{-1}\partial_{\xp}\mathcal{M}\paren{\bar{h}^{\mu\nu}}(t-\frac{\x}{c},\mathbf{x'})\nonumber \\  
&& \negmedspace {}+B\paren{B+1}\xp^{-2}\mathcal{M}\paren{\bar{h}^{\mu\nu}}(t-\frac{\x}{c},\mathbf{x'})\bigg]\ud^3\mathbf{x'}\bigg).
\end{IEEEeqnarray}
Comparing
Eq. (\ref{4.1.2})
with
Eq. (\ref{3.5.1}),
it can be observed that
$\mathcal{M}\paren{\bar{h}^{\mu\nu}}\finsarg={\bar{h}}^{\mu\nu}\finsarg$
holds not only at
$\x \to \infty$,
but also everywhere outside the near zone. Thus,
Eq. (\ref{4.2.30})
takes the form
\begin{IEEEeqnarray}{rcl}\label{4.2.31}
\mathcal{M}\paren{h^{\mu\nu}}\finsarg & = & u^{\mu\nu}\finsarg+\frac{1}{4\pi}\sum_{j=0}^{\infty}\frac{\paren{-1}^j}{j!}\partial_{J}\bigg(\x^{-1}\fpa\int_{\mathcal{R}<\xp}\xpb x'^{J}\nonumber \\  
&& \negmedspace {}\times\bigg[2B\xp^{-1}\partial_{\xp}{\bar{h}}^{\mu\nu}(t-\frac{\x}{c},\mathbf{x'})\nonumber \\  
&& \negmedspace {}+B\paren{B+1}\xp^{-2}{\bar{h}}^{\mu\nu}(t-\frac{\x}{c},\mathbf{x'})\bigg]\ud^3\mathbf{x'}\bigg).
\end{IEEEeqnarray}
Moreover, we have
\begin{eqnarray}\label{4.2.32}
\lefteqn{\fpa\int_{\xp<\mathcal{R}}\xpb x'^{J}\bigg[2B\xp^{-1}\partial_{\xp}{\bar{h}}^{\mu\nu}(t-\frac{\x}{c},\mathbf{x'})} \nonumber \\ 
&&\!\!\!\!+B\paren{B+1}\xp^{-2}{\bar{h}}^{\mu\nu}(t-\frac{\x}{c},\mathbf{x'})\bigg]\ud^3\mathbf{x'}=0,
\end{eqnarray}
which is due to
$\int_{\xp<\mathcal{R}}\!\left(\frac{\xp}{r_0}\right)^{\!\!B}\!\! x'^{J}\!\xp^{\!-1}\partial_{\xp}{\bar{h}}^{\mu\nu}\!(t-\frac{\x}{c},\!\mathbf{x'})\!\ud^3\mathbf{x'}$
and
$\int_{\xp<\mathcal{R}}\!\left(\frac{\xp}{r_0}\right)^{\!\!B}\!\! x'^{J}\!\xp^{\!-2}{\bar{h}}^{\mu\nu}\!(t-\frac{\x}{c},\!\mathbf{x'})\!\ud^3\mathbf{x'}$
being analytic at
$B=0$
(as a consequence of smoothness of
${\bar{h}}^{\mu\nu}\finsarg$
in the near zone) and having the coefficients
$B$
and
$B\paren{B+1}$.
Using this equation, we can rewrite
Eq. (\ref{4.2.31})
as
\begin{IEEEeqnarray}{rcl}\label{4.2.33}
\mathcal{M}\paren{h^{\mu\nu}}\finsarg & = & u^{\mu\nu}\finsarg\nonumber \\  
&& \negmedspace {}+\frac{1}{4\pi}\sum_{j=0}^{\infty}\frac{\paren{-1}^j}{j!}\partial_{J}\bigg(\x^{-1}\fpa\int_{\xp<\mathcal{R}}\xpb x'^{J}\nonumber \\  
&& \negmedspace {}\times\bigg[2B\xp^{-1}\partial_{\xp}{\bar{h}}^{\mu\nu}(t-\frac{\x}{c},\mathbf{x'})\nonumber \\  
&& \negmedspace {}+B\paren{B+1}\xp^{-2}{\bar{h}}^{\mu\nu}(t-\frac{\x}{c},\mathbf{x'})\bigg]\ud^3\mathbf{x'}\bigg)\nonumber\\
&& \negmedspace {}+\frac{1}{4\pi}\sum_{j=0}^{\infty}\frac{\paren{-1}^j}{j!}\partial_{J}\bigg(\x^{-1}\fpa\int_{\mathcal{R}<\xp}\xpb x'^{J}\nonumber \\  
&& \negmedspace {}\times\bigg[2B\xp^{-1}\partial_{\xp}{\bar{h}}^{\mu\nu}(t-\frac{\x}{c},\mathbf{x'})\nonumber \\  
&& \negmedspace {}+B\paren{B+1}\xp^{-2}{\bar{h}}^{\mu\nu}(t-\frac{\x}{c},\mathbf{x'})\bigg]\ud^3\mathbf{x'}\bigg).
\end{IEEEeqnarray}

Now consider the following equation:
\begin{IEEEeqnarray}{rcl}
\Box \left[\xb{\bar{h}}^{\mu\nu}\finsarg\right]&=&\xb\Box{\bar{h}}^{\mu\nu}\finsarg+\xb\bigg[2B\x^{-1}\partial_{\x}{\bar{h}}^{\mu\nu}\finsarg\nonumber \\  
&& \negmedspace {}+B\paren{B+1}\x^{-2}{\bar{h}}^{\mu\nu}\finsarg\bigg]\longrightarrow\nonumber
\end{IEEEeqnarray}
\vspace*{-25mm}
\begin{eqnarray}\label{4.2.34}
\lefteqn{\xb\bigg[2B\x^{-1}\partial_{\x}{\bar{h}}^{\mu\nu}\finsarg+B\paren{B+1}\x^{-2}{\bar{h}}^{\mu\nu}\finsarg\bigg]} \nonumber \\ 
&\!\!\!\!\!\!\!\!=\!\!\!\!\!\!\!\!&\Box \left[\xb{\bar{h}}^{\mu\nu}\finsarg\right]-\frac{16\pi G}{c^4}\xb{\bar{\tau}}^{\mu\nu}\finsarg,
\end{eqnarray}
where to obtain the final result we have also used
Eq. (\ref{3.5.3}).
Making the change
$t \to t-t_0$,
we get
\begin{eqnarray}\label{4.2.35}
\lefteqn{\xb\bigg[2B\x^{-1}\partial_{\x}{\bar{h}}^{\mu\nu}(t-t_0,\mathbf{x})+B\paren{B+1}\x^{-2}{\bar{h}}^{\mu\nu}(t-t_0,\mathbf{x})\bigg]} \nonumber \\ 
&\!\!\!\!\!\!\!\!=\!\!\!\!\!\!\!\!&\Box_{t-t_0} \left[\xb{\bar{h}}^{\mu\nu}(t-t_0,\mathbf{x})\right]-\frac{16\pi G}{c^4}\xb{\bar{\tau}}^{\mu\nu}(t-t_0,\mathbf{x}).
\end{eqnarray}
Using above equation,
Eq. (\ref{4.2.33})
can be written as
\begin{eqnarray}\label{4.2.36}
\lefteqn{\mathcal{M}\paren{h^{\mu\nu}}\finsarg} \nonumber \\ 
&\!\!\!\!\!\!\!\!=\!\!\!\!\!\!\!\!& u^{\mu\nu}\finsarg \nonumber\\
&&\!\!\!\!-\frac{4G}{c^4}\sum_{j=0}^{\infty}\frac{\paren{-1}^j}{j!}\partial_{J}\bigg(\x^{-1}\fpa\int_{\xp<\mathcal{R}}\xpb x'^{J}{\bar{\tau}}^{\mu\nu}(u,\mathbf{x}')\ud^3\mathbf{x'}\bigg) \nonumber\\
&&\!\!\!\!-\frac{4G}{c^4}\sum_{j=0}^{\infty}\frac{\paren{-1}^j}{j!}\partial_{J}\bigg(\x^{-1}\fpa\int_{\mathcal{R}<\xp}\xpb x'^{J}{\bar{\tau}}^{\mu\nu}(u,\mathbf{x}')\ud^3\mathbf{x'}\bigg) \nonumber\\
&&\!\!\!\!+\frac{1}{4\pi}\sum_{j=0}^{\infty}\frac{\paren{-1}^j}{j!}\partial_{J}\bigg(\x^{-1}\fpa\int_{\xp<\mathcal{R}}x'^{J}\Box'_{u} \left[\xpb{\bar{h}}^{\mu\nu}(u,\mathbf{x}')\right]\ud^3\mathbf{x'}\bigg) \nonumber\\
&&\!\!\!\!+\frac{1}{4\pi}\sum_{j=0}^{\infty}\frac{\paren{-1}^j}{j!}\partial_{J}\bigg(\x^{-1}\fpa\int_{\mathcal{R}<\xp}x'^{J}\Box'_{u} \left[\xpb{\bar{h}}^{\mu\nu}(u,\mathbf{x}')\right]\ud^3\mathbf{x'}\bigg),\quad
\end{eqnarray}
where
$u=t-\frac{\x}{c}$
and
$\Box'_{u}=\Delta'-\frac{1}{c^2}\partial^{2}_{u}$.
Now note that in the original domain of analyticity of
$\int_{D_i}x'^{J}\Delta' \left[\xpb{\bar{h}}^{\mu\nu}(u,\mathbf{x}')\right]\ud^3\mathbf{x'}$,
where
\begin{equation}
D_i=
\begin{cases}
\{\mathbf{x'}\in \mathbb{R }^3 \left|\right. \xp<\mathcal{R}\}, &\qquad i=1,\\
\{\mathbf{x'}\in \mathbb{R }^3 \left|\right. \mathcal{R}<\xp\}, &\qquad i=2,
\end{cases}\nonumber
\end{equation}
namely in either some right or left half-plane depending on whether
$i$
is, by using Green's theorem we can write
\begin{eqnarray}\label{4.2.38}
\lefteqn{\A\int_{D_i}x'^{J}\Delta' \left[\xpb{\bar{h}}^{\mu\nu}(u,\mathbf{x}')\right]\ud^3\mathbf{x'}} \nonumber \\ 
&\!\!\!\!\!\!\!\!=\!\!\!\!\!\!\!\!& \int_{D_i}x'^{J}\Delta' \left[\xpb{\bar{h}}^{\mu\nu}(u,\mathbf{x}')\right]\ud^3\mathbf{x'} \nonumber\\
&\!\!\!\!\!\!\!\!=\!\!\!\!\!\!\!\!& \int_{D_i}\left[\Delta' x'^{J}\right]\xpb{\bar{h}}^{\mu\nu}(u,\mathbf{x}')\ud^3\mathbf{x'} \nonumber\\
&&\!\!\!\!+\int_{\partial D_i}x'^{J}\partial'_{k} \left[\xpb{\bar{h}}^{\mu\nu}(u,\mathbf{x}')\right]\ud\sigma'_{\paren{i}k}\nonumber\\
&&\!\!\!\!-\int_{\partial D_i}\left[\partial'_{k}x'^{J}\right]\xpb{\bar{h}}^{\mu\nu}(u,\mathbf{x}')\ud\sigma'_{\paren{i}k}.
\end{eqnarray}
For
$i=1$
we have
\begin{eqnarray}\label{4.2.39}
\lefteqn{\A\int_{\xp<\mathcal{R}}x'^{J}\Delta' \left[\xpb{\bar{h}}^{\mu\nu}(u,\mathbf{x}')\right]\ud^3\mathbf{x'}} \nonumber \\ 
&\!\!\!\!\!\!\!\!=\!\!\!\!\!\!\!\!& \int_{\xp<\mathcal{R}}\left[\Delta' x'^{J}\right]\xpb{\bar{h}}^{\mu\nu}(u,\mathbf{x}')\ud^3\mathbf{x'} \nonumber\\
&&\!\!\!\!+\int_{\xp=\mathcal{R}}x'^{J}\partial'_{k} \left[\xpb{\bar{h}}^{\mu\nu}(u,\mathbf{x}')\right]\ud\sigma'_{\paren{1}k}\nonumber\\
&&\!\!\!\!-\int_{\xp=\mathcal{R}}\left[\partial'_{k}x'^{J}\right]\xpb{\bar{h}}^{\mu\nu}(u,\mathbf{x}')\ud\sigma'_{\paren{1}k}.
\end{eqnarray}
As a result of identity theorem, we can equate
$\A\int_{\xp<\mathcal{R}}x'^{J}\Delta' \left[\xpb{\bar{h}}^{\mu\nu}(u,\mathbf{x}')\right]\ud^3\mathbf{x'}$
and the analytic continuation of the right-hand side of
Eq. (\ref{4.2.39})
wherever they are both defined. This, together with the fact that the surface integrals are entire, yield
\begin{eqnarray}\label{4.2.40}
\lefteqn{\A\int_{\xp<\mathcal{R}}x'^{J}\Delta' \left[\xpb{\bar{h}}^{\mu\nu}(u,\mathbf{x}')\right]\ud^3\mathbf{x'}} \nonumber \\ 
&\!\!\!\!\!\!\!\!=\!\!\!\!\!\!\!\!&\A\int_{\xp<\mathcal{R}}\left[\Delta' x'^{J}\right]\xpb{\bar{h}}^{\mu\nu}(u,\mathbf{x}')\ud^3\mathbf{x'} \nonumber\\
&&\!\!\!\!+\int_{\xp=\mathcal{R}}x'^{J}\partial'_{k} \left[\xpb{\bar{h}}^{\mu\nu}(u,\mathbf{x}')\right]\ud\sigma'_{\paren{1}k}\nonumber\\
&&\!\!\!\!-\int_{\xp=\mathcal{R}}\left[\partial'_{k}x'^{J}\right]\xpb{\bar{h}}^{\mu\nu}(u,\mathbf{x}')\ud\sigma'_{\paren{1}k}.
\end{eqnarray}
For
$i=2$
we get
\begin{eqnarray}\label{4.2.41}
\lefteqn{\A\int_{\mathcal{R}<\xp}x'^{J}\Delta' \left[\xpb{\bar{h}}^{\mu\nu}(u,\mathbf{x}')\right]\ud^3\mathbf{x'}} \nonumber \\ 
&\!\!\!\!\!\!\!\!=\!\!\!\!\!\!\!\!& \int_{\mathcal{R}<\xp}\left[\Delta' x'^{J}\right]\xpb{\bar{h}}^{\mu\nu}(u,\mathbf{x}')\ud^3\mathbf{x'} \nonumber\\
&&\!\!\!\!+\int_{\xp=\mathcal{R}}x'^{J}\partial'_{k} \left[\xpb{\bar{h}}^{\mu\nu}(u,\mathbf{x}')\right]\ud\sigma'_{\paren{2}k}\nonumber\\
&&\!\!\!\!+\int_{\xp\to\infty}x'^{J}\partial'_{k} \left[\xpb{\bar{h}}^{\mu\nu}(u,\mathbf{x}')\right]\ud\sigma'_{\paren{2}k}\nonumber\\
&&\!\!\!\!-\int_{\xp=\mathcal{R}}\left[\partial'_{k}x'^{J}\right]\xpb{\bar{h}}^{\mu\nu}(u,\mathbf{x}')\ud\sigma'_{\paren{2}k}\nonumber\\
&&\!\!\!\!-\int_{\xp\to\infty}\left[\partial'_{k}x'^{J}\right]\xpb{\bar{h}}^{\mu\nu}(u,\mathbf{x}')\ud\sigma'_{\paren{2}k}.
\end{eqnarray}
Taking
$\re B$
to be a sufficiently large negative number, the surface integrals over a surface at
$\xp \to \infty$
vanish. As a consequence of identity theorem, the sum of the analytic continuations of the remaining terms on the right-hand side of
Eq. (\ref{4.2.41})
and
$\A\int_{\mathcal{R}<\xp}x'^{J}\Delta' \left[\xpb{\bar{h}}^{\mu\nu}(u,\mathbf{x}')\right]\ud^3\mathbf{x'}$
must be equal wherever they are all defined. Thus, considering that the surface integrals are entire, we find
\begin{eqnarray}\label{4.2.42}
\lefteqn{\A\int_{\mathcal{R}<\xp}x'^{J}\Delta' \left[\xpb{\bar{h}}^{\mu\nu}(u,\mathbf{x}')\right]\ud^3\mathbf{x'}} \nonumber \\ 
&\!\!\!\!\!\!\!\!=\!\!\!\!\!\!\!\!&\A\int_{\mathcal{R}<\xp}\left[\Delta' x'^{J}\right]\xpb{\bar{h}}^{\mu\nu}(u,\mathbf{x}')\ud^3\mathbf{x'} \nonumber\\
&&\!\!\!\!+\int_{\xp=\mathcal{R}}x'^{J}\partial'_{k} \left[\xpb{\bar{h}}^{\mu\nu}(u,\mathbf{x}')\right]\ud\sigma'_{\paren{2}k}\nonumber\\
&&\!\!\!\!-\int_{\xp=\mathcal{R}}\left[\partial'_{k}x'^{J}\right]\xpb{\bar{h}}^{\mu\nu}(u,\mathbf{x}')\ud\sigma'_{\paren{2}k}.
\end{eqnarray}
Now, taking
Eqs. (\ref{4.2.40}) and (\ref{4.2.42}) into account,
we reach
\begin{eqnarray}\label{4.2.43}
\lefteqn{\A\int_{D_i}x'^{J}\Delta' \left[\xpb{\bar{h}}^{\mu\nu}(u,\mathbf{x}')\right]\ud^3\mathbf{x'}} \nonumber \\ 
&\!\!\!\!\!\!\!\!=\!\!\!\!\!\!\!\!&\A\int_{D_i}\left[\Delta' x'^{J}\right]\xpb{\bar{h}}^{\mu\nu}(u,\mathbf{x}')\ud^3\mathbf{x'} \nonumber\\
&&\!\!\!\!+\int_{\xp=\mathcal{R}}x'^{J}\partial'_{k} \left[\xpb{\bar{h}}^{\mu\nu}(u,\mathbf{x}')\right]\ud\sigma'_{\paren{i}k}\nonumber\\
&&\!\!\!\!-\int_{\xp=\mathcal{R}}\left[\partial'_{k}x'^{J}\right]\xpb{\bar{h}}^{\mu\nu}(u,\mathbf{x}')\ud\sigma'_{\paren{i}k}.
\end{eqnarray}
Furthermore, with the use of
Eq. (\ref{A.2})
we have
\begin{eqnarray}\label{4.2.44}
\lefteqn{\sum_{j=0}^{\infty}\frac{\paren{-1}^j}{j!}\partial_{J}\bigg(\x^{-1}\A\int_{D_i}\left[\Delta' x'^{J}\right]\xpb{\bar{h}}^{\mu\nu}(u,\mathbf{x}')\ud^3\mathbf{x'}\bigg)} \nonumber\\ 
&\!\!\!\!\!\!\!\!=\!\!\!\!\!\!\!\!&2\sum_{j=2}^{\infty}\frac{\paren{-1}^j}{j!}\partial_{J}\bigg(\x^{-1}\delta^{\{ i_1 i_2}\A\int_{D_i} x'^{ i_3... i_{j}\}}\xpb{\bar{h}}^{\mu\nu}(u,\mathbf{x}')\ud^3\mathbf{x'}\bigg) \nonumber\\
&\!\!\!\!\!\!\!\!=\!\!\!\!\!\!\!\!&\sum_{j=0}^{\infty}\frac{\paren{-1}^j}{j!}\partial_{J}\left[\Delta\bigg(\x^{-1}\A\int_{D_i}  x'^{J}\xpb{\bar{h}}^{\mu\nu}(u,\mathbf{x}')\ud^3\mathbf{x'}\bigg)\right],
\end{eqnarray}
where the key reasoning behind transition from first equality to the second one is as follows. Since all the indices of
$J$
are dummy and
$\partial_J$
is a symmetric tensor, we can replace
$\delta^{\{ i_1 i_2} x'^{ i_3... i_{j}\}}$
by only one of its terms times the number of its terms. Due to
$x'^{J-2}$
and Kronecker delta being totally symmetric tensors, the number of terms of
$\delta^{\{ i_1 i_2} x'^{ i_3... i_{j}\}}$
is equal to the number of ways in which we can select 2 indices from
$j$,
i.e.,
$\binom{j}{2}$.
Hence, we have replaced
$\delta^{\{ i_1 i_2} x'^{ i_3... i_{j}\}}$
by
$\binom{j}{2}\delta^{ i_1 i_2} x'^{ i_3... i_{j}}$
in
Eq. (\ref{4.2.44})
to obtain the final result. Moreover,
$\A\int_{D_i}  x'^{J}\xpb{\bar{h}}^{\mu\nu}(u,\mathbf{x}')\ud^3\mathbf{x'}$
is a function of
$u=t-\frac{\x}{c}$.
Thus, taking
Eq. (\ref{2.1.3.7})
into account,
Eq. (\ref{4.2.44})
can be written as
\begin{eqnarray}\label{4.2.45a}
\lefteqn{\sum_{j=0}^{\infty}\frac{\paren{-1}^j}{j!}\partial_{J}\bigg(\x^{-1}\A\int_{D_i}\left[\Delta' x'^{J}\right]\xpb{\bar{h}}^{\mu\nu}(u,\mathbf{x}')\ud^3\mathbf{x'}\bigg)} \nonumber\\ 
&\!\!\!\!\!\!\!\!=\!\!\!\!\!\!\!\!&\sum_{j=0}^{\infty}\frac{\paren{-1}^j}{j!}\partial_{J}\bigg(\x^{-1}\A\int_{D_i}  x'^{J}\xpb\left[\frac{1}{c^2}\partial^2_{u}{\bar{h}}^{\mu\nu}(u,\mathbf{x}')\right]\ud^3\mathbf{x'}\bigg).
\end{eqnarray}
Combining
Eqs. (\ref{4.2.43}) and (\ref{4.2.45a}),
we find
\begin{eqnarray}\label{4.2.45b}
\lefteqn{\frac{1}{4\pi}\sum_{j=0}^{\infty}\frac{\paren{-1}^j}{j!}\partial_{J}\bigg(\x^{-1}\A\int_{D_i}x'^{J}\Box'_{u} \left[\xpb{\bar{h}}^{\mu\nu}(u,\mathbf{x}')\right]\ud^3\mathbf{x'}\bigg)} \nonumber\\ 
&\!\!\!\!\!\!\!\!=\!\!\!\!\!\!\!\!&\frac{1}{4\pi}\sum_{j=0}^{\infty}\frac{\paren{-1}^j}{j!}\partial_{J}\bigg(\x^{-1}\int_{\xp=\mathcal{R}}x'^{J}\partial'_{k} \left[\xpb{\bar{h}}^{\mu\nu}(u,\mathbf{x}')\right]\ud\sigma'_{\paren{i}k}\bigg) \nonumber\\
&&\!\!\!\! - \frac{1}{4\pi}\sum_{j=0}^{\infty}\frac{\paren{-1}^j}{j!}\partial_{J}\bigg(\x^{-1}\int_{\xp=\mathcal{R}}\left[\partial'_{k}x'^{J}\right]\xpb{\bar{h}}^{\mu\nu}(u,\mathbf{x}')\ud\sigma'_{\paren{i}k}\bigg).
\end{eqnarray}
Regardless of whether
$i$
is, it can be readily shown that all the terms appearing in the above equation  are analytic in some punctured neighborhood of
$B=0$.
Therefore, each of them has a Laurent expansion around
$B=0$
(each term containing a surface integral is analytic at
$B=0$
and thereby possesses a Taylor expansion around this point). Because the coefficients of
$B^n$
on both sides of
Eq. (\ref{4.2.45b})
must be equal for each
$n$,
considering the coefficients of
$B^0$,
we get
\begin{eqnarray}\label{4.2.45c}
\lefteqn{\frac{1}{4\pi}\sum_{j=0}^{\infty}\frac{\paren{-1}^j}{j!}\partial_{J}\bigg(\x^{-1}\fpa\int_{D_i}x'^{J}\Box'_{u} \left[\xpb{\bar{h}}^{\mu\nu}(u,\mathbf{x}')\right]\ud^3\mathbf{x'}\bigg)} \nonumber\\ 
&\!\!\!\!\!\!\!\!=\!\!\!\!\!\!\!\!&\frac{1}{4\pi}\sum_{j=0}^{\infty}\frac{\paren{-1}^j}{j!}\partial_{J}\bigg(\x^{-1}\int_{\xp=\mathcal{R}}x'^{J}\partial'_{k}{\bar{h}}^{\mu\nu}(u,\mathbf{x}')\ud\sigma'_{\paren{i}k}\bigg) \nonumber\\
&&\!\!\!\! - \frac{1}{4\pi}\sum_{j=0}^{\infty}\frac{\paren{-1}^j}{j!}\partial_{J}\bigg(\x^{-1}\int_{\xp=\mathcal{R}}\left[\partial'_{k}x'^{J}\right]{\bar{h}}^{\mu\nu}(u,\mathbf{x}')\ud\sigma'_{\paren{i}k}\bigg).
\end{eqnarray}
The sum of
$\frac{1}{4\pi}\sum_{j=0}^{\infty}\frac{\paren{-1}^j}{j!}\partial_{J}\bigg(\x^{-1}\fpa\int_{D_i}x'^{J}\Box'_{u} \left[\xpb{\bar{h}}^{\mu\nu}(u,\mathbf{x}')\right]\ud^3\mathbf{x'}\bigg)$'s
with
$i=1$
and
$i=2$
is zero due to
$\vec{\ud\sigma'_{\paren{1}}}$
being the opposite of
$\vec{\ud\sigma'_{\paren{2}}}$
at each point on the surface
$\xp=\mathcal{R}$.
Thus, we have
\begin{equation}\label{4.2.46a}
\mathcal{M}\paren{h^{\mu\nu}}\finsarg=u^{\mu\nu}\finsarg -\frac{4G}{c^4}\sum_{j=0}^{\infty}\frac{\paren{-1}^j}{j!}\partial_{J}\paren{{\x}^{-1}H^{\mu\nu}_{J}(u)},
\end{equation}
where
\begin{IEEEeqnarray}{rcl}\label{4.2.46b}
H^{\mu\nu}_{J}(u) & = &\fpa\int_{\xp<\mathcal{R}}\xpb x'^{J}{\bar{\tau}}^{\mu\nu}(u,\mathbf{x}')\ud^3\mathbf{x'} \nonumber \\  
&& \negmedspace {}+\fpa\int_{\mathcal{R}<\xp}\xpb x'^{J}{\bar{\tau}}^{\mu\nu}(u,\mathbf{x}')\ud^3\mathbf{x'}\nonumber \\  
&& \negmedspace {} \stareq \fpa\int_{\mathbb{R }^3}\xpb x'^{J}{\bar{\tau}}^{\mu\nu}(u,\mathbf{x}')\ud^3\mathbf{x'}.
\end{IEEEeqnarray}

$\mathcal{M}\paren{h^{\mu\nu}}\finsarg$
given by
Eqs. (\ref{4.2.46a}) and (\ref{4.2.46b})
is acceptable if it fulfills the harmonic gauge condition. Since
$\Lambda^{\mu\nu}(\mathcal{M}\paren{h})\finsarg=\Lambda^{\mu\nu}(\overline{\mathcal{M}\paren{h}})\finsarg+R^{\;\mu\nu}\finsarg$,
by means of
Eq. (\ref{2.4.8})
we have
\begin{IEEEeqnarray}{rcl}\label{4.2.47}
\partial_\mu u^{\mu\nu}\finsarg &=&\resa \left[-\frac{1}{4\pi}\int_{|\mathbf{x'}|<\mathcal{R}}\paren{\frac{\xp}{r_0}}^B \frac{\xp^{-1}n'^i\Lambda^{i\nu}(\overline{\mathcal{M}\paren{h}})\retarg}{|\mathbf{x}-\mathbf{x'}|}\ud^3\mathbf{x'}\right],\qquad 
\end{IEEEeqnarray}
where the disappearance of
$R^{\;\mu\nu}\retarg$
is due to analyticity of the integral containing
$R^{\;\mu\nu}\retarg$
at
$B=0$.
Because the structure of
$\Lambda^{\mu\nu}(\overline{\mathcal{M}\paren{h}})\finsarg$
is similar to the structure of
$\overline{\mathcal{M}\paren{h^{\mu\nu}}}\finsarg$,
in a manner analogous to the one used to obtain
Eq. (\ref{4.2.26}),
we can rewrite
Eq. (\ref{4.2.47})
as
\begin{IEEEeqnarray}{rcl}\label{4.2.48}
\partial_\mu u^{\mu\nu}\finsarg & = &\frac{1}{4\pi}\sum_{j=0}^{\infty}\frac{\paren{-1}^j}{j!}\partial_{J}\bigg(\x^{-1}\resa\int_{\mathcal{R}<\xp}\xpb x'^{J} \nonumber \\  
&& \negmedspace {}\times\xp^{-1}n'^i\Lambda^{i\nu}(\overline{\mathcal{M}\paren{h}})(u,\mathbf{x}')\ud^3\mathbf{x'}\bigg).
\end{IEEEeqnarray}
Replacing
$\overline{\mathcal{M}\paren{h^{\mu\nu}}}$
by
$\mathcal{M}\paren{\bar{h}^{\mu\nu}}$
and considering the equality of
$\mathcal{M}\paren{\bar{h}^{\mu\nu}}\finsarg$
and
$\bar{h}^{\mu\nu}\finsarg$
outside the near zone, we get
\begin{IEEEeqnarray}{rcl}\label{4.2.49}
\partial_\mu u^{\mu\nu}\finsarg & = &\frac{1}{4\pi}\sum_{j=0}^{\infty}\frac{\paren{-1}^j}{j!}\partial_{J}\bigg(\x^{-1}\resa\int_{\mathcal{R}<\xp}\xpb x'^{J} \nonumber \\  
&& \negmedspace {}\times\xp^{-1}n'^i\Lambda^{i\nu}(\bar{h})(u,\mathbf{x}')\ud^3\mathbf{x'}\bigg)\nonumber\\
 & = &\frac{1}{4\pi}\sum_{j=0}^{\infty}\frac{\paren{-1}^j}{j!}\partial_{J}\bigg(\x^{-1}\resa\int_{\mathcal{R}<\xp}\xpb x'^{J} \nonumber \\  
&& \negmedspace {}\times\xp^{-1}n'^i\bar{\Lambda}^{i\nu}(u,\mathbf{x}')\ud^3\mathbf{x'}\bigg).
\end{IEEEeqnarray}
$\paren{-\bar{\mathfrak{g}}} \bar{T}^{\mu\nu}$
is zero outside the near zone, and hence, in the above equation we can replace
$\bar{\Lambda}^{i\nu}$
by
$\frac{16\pi G}{c^4}\bar{\tau}^{i\nu}$.
Further, since
$\frac{16\pi G}{c^4}\resa\int_{\xp<\mathcal{R}}\xpb x'^{J}\xp^{-1}n'^{i}\bar{\tau}^{i\nu}(u,\mathbf{x}')\ud^3\mathbf{x'}$
vanishes (owing to
$\int_{\xp<\mathcal{R}}\xpb x'^{J}\xp^{-1}n'^{i}\bar{\tau}^{i\nu}(u,\mathbf{x}')\ud^3\mathbf{x'}$
being analytic at
$B=0$
which is itself a consequence of smoothness of
$\bar{\tau}^{\mu\nu}\finsarg$
in the near zone), we can add it to the right-hand side of
Eq. (\ref{4.2.49}).
In this way we find
\begin{equation}\label{4.2.50a}
\partial_\mu u^{\mu\nu}\finsarg=\frac{4G}{c^4}\sum_{j=0}^{\infty}\frac{\paren{-1}^j}{j!}\partial_{J}\paren{{\x}^{-1}G^{\nu}_{J}(u)},
\end{equation}
where
\begin{IEEEeqnarray}{rcl}\label{4.2.50b}
G^{\nu}_{J}(u) & = &\resa\int_{\xp<\mathcal{R}}\xpb x'^{J}\xp^{-1}n'^{i}\bar{\tau}^{i\nu}(u,\mathbf{x}')\ud^3\mathbf{x'} \nonumber \\  
&& \negmedspace {}+\resa\int_{\mathcal{R}<\xp}\xpb x'^{J}\xp^{-1}n'^{i}\bar{\tau}^{i\nu}(u,\mathbf{x}')\ud^3\mathbf{x'}\nonumber \\  
&& \negmedspace {} \stareq \resa\int_{\mathbb{R }^3}\xpb x'^{J}\xp^{-1}n'^{i}\bar{\tau}^{i\nu}(u,\mathbf{x}')\ud^3\mathbf{x'}.
\end{IEEEeqnarray}
Now note that we have
\begin{eqnarray}\label{4.2.51}
\lefteqn{\partial_\mu\left[\frac{4G}{c^4}\sum_{j=0}^{\infty}\frac{\paren{-1}^j}{j!}\partial_{J}\paren{{\x}^{-1}H^{\mu\nu}_{J}(u)}\right]} \nonumber \\ 
&\!\!\!\!\!\!\!\!=\!\!\!\!\!\!\!\!& \frac{4G}{c^4}\sum_{j=0}^{\infty}\frac{\paren{-1}^j}{j!}\partial_{J}\paren{{\x}^{-1}\partial_{0}\fpa\int_{\xp<\mathcal{R}}\xpb x'^{J}{\bar{\tau}}^{0\nu}(u,\mathbf{x}')\ud^3\mathbf{x'}} \nonumber\\
&&\!\!\!\!+\frac{4G}{c^4}\sum_{j=0}^{\infty}\frac{\paren{-1}^j}{j!}\partial_{J}\paren{{\x}^{-1}\partial_{0}\fpa\int_{\mathcal{R}<\xp}\xpb x'^{J}{\bar{\tau}}^{0\nu}(u,\mathbf{x}')\ud^3\mathbf{x'}} \nonumber\\
&&\!\!\!\!+\frac{4G}{c^4}\sum_{j=0}^{\infty}\frac{\paren{-1}^j}{j!}\partial_{J}\left[\partial_{k}\paren{{\x}^{-1}H^{k\nu}_{J}(u)}\right].
\end{eqnarray}
In the original domain of analyticity of
$\int_{D_i}\xpb x'^{J}{\bar{\tau}}^{0\nu}(u,\mathbf{x}')\ud^3\mathbf{x'}$,
where
\begin{equation}
D_i=
\begin{cases}
\{\mathbf{x'}\in \mathbb{R }^3 \left|\right. \xp<\mathcal{R}\}, &\qquad i=1,\\
\{\mathbf{x'}\in \mathbb{R }^3 \left|\right. \mathcal{R}<\xp\}, &\qquad i=2,
\end{cases}\nonumber
\end{equation}
namely in either some right or left half-plane depending on whether
$i$
is, by using
Eq. (\ref{3.5.5})
and Gauss' theorem we can write
\begin{eqnarray}\label{4.2.53}
\lefteqn{\partial_{0}\A\int_{D_i}\xpb x'^{J}{\bar{\tau}}^{0\nu}(u,\mathbf{x}')\ud^3\mathbf{x'}} \nonumber \\ 
&\!\!\!\!\!\!\!\!=\!\!\!\!\!\!\!\!& \partial_{0}\int_{D_i}\xpb x'^{J}{\bar{\tau}}^{0\nu}(u,\mathbf{x}')\ud^3\mathbf{x'} \nonumber\\
&\!\!\!\!\!\!\!\!=\!\!\!\!\!\!\!\!& -\int_{\partial D_i}\xpb x'^{J}{\bar{\tau}}^{k\nu}(u,\mathbf{x}')\ud\sigma'_{\paren{i}k} \nonumber\\
&&\!\!\!\!+\int_{D_i}\left[\partial'_{k}x'^{J}\right]\xpb{\bar{\tau}}^{k\nu}(u,\mathbf{x}')\ud^3\mathbf{x'}\nonumber\\
&&\!\!\!\!+B\int_{D_i}\xpb x'^{J}\xp^{-1}n'^{k}\bar{\tau}^{k\nu}(u,\mathbf{x}')\ud^3\mathbf{x'}.
\end{eqnarray}
For
$i=1$
we have
\begin{eqnarray}\label{4.2.54}
\lefteqn{\partial_{0}\A\int_{\xp<\mathcal{R}}\xpb x'^{J}{\bar{\tau}}^{0\nu}(u,\mathbf{x}')\ud^3\mathbf{x'}} \nonumber \\ 
&\!\!\!\!\!\!\!\!=\!\!\!\!\!\!\!\!& -\int_{\xp=\mathcal{R}} \xpb x'^{J}{\bar{\tau}}^{k\nu}(u,\mathbf{x}')\ud\sigma'_{\paren{1}k} \nonumber\\
&&\!\!\!\!+\int_{\xp<\mathcal{R}}\left[\partial'_{k}x'^{J}\right]\xpb{\bar{\tau}}^{k\nu}(u,\mathbf{x}')\ud^3\mathbf{x'}\nonumber\\
&&\!\!\!\!+B\int_{\xp<\mathcal{R}} \xpb x'^{J}\xp^{-1}n'^{k}\bar{\tau}^{k\nu}(u,\mathbf{x}')\ud^3\mathbf{x'}.
\end{eqnarray}
By virtue of identity theorem, we can equate
$\partial_{0}\A\int_{\xp<\mathcal{R}}\xpb x'^{J}{\bar{\tau}}^{0\nu}(u,\mathbf{x}')\ud^3\mathbf{x'}$
and the analytic continuation of the right-hand side of the above equation in the region where they are both defined. Therefore, considering that the surface integral appearing in
Eq. (\ref{4.2.54})
and the function
$f(B)=B$
are entire, we find
\begin{eqnarray}\label{4.2.55}
\lefteqn{\partial_{0}\A\int_{\xp<\mathcal{R}} \xpb x'^{J}{\bar{\tau}}^{0\nu}(u,\mathbf{x}')\ud^3\mathbf{x'}} \nonumber \\ 
&\!\!\!\!\!\!\!\!=\!\!\!\!\!\!\!\!& -\int_{\xp=\mathcal{R}} \xpb x'^{J}{\bar{\tau}}^{k\nu}(u,\mathbf{x}')\ud\sigma'_{\paren{1}k} \nonumber\\
&&\!\!\!\!+\A\int_{\xp<\mathcal{R}}\left[\partial'_{k}x'^{J}\right]\xpb{\bar{\tau}}^{k\nu}(u,\mathbf{x}')\ud^3\mathbf{x'}\nonumber\\
&&\!\!\!\!+B\cdot\A\int_{\xp<\mathcal{R}} \xpb x'^{J}\xp^{-1}n'^{k}\bar{\tau}^{k\nu}(u,\mathbf{x}')\ud^3\mathbf{x'}.
\end{eqnarray}
For
$i=2$
we reach
\begin{eqnarray}\label{4.2.56}
\lefteqn{\partial_{0}\A\int_{\mathcal{R}<\xp}\xpb x'^{J}{\bar{\tau}}^{0\nu}(u,\mathbf{x}')\ud^3\mathbf{x'}} \nonumber \\ 
&\!\!\!\!\!\!\!\!=\!\!\!\!\!\!\!\!& -\int_{\xp=\mathcal{R}}\xpb x'^{J}{\bar{\tau}}^{k\nu}(u,\mathbf{x}')\ud\sigma'_{\paren{2}k} \nonumber\\
&&\!\!\!\! -\int_{\xp\to \infty}\xpb x'^{J}{\bar{\tau}}^{k\nu}(u,\mathbf{x}')\ud\sigma'_{\paren{2}k}\nonumber\\
&&\!\!\!\!+\int_{\mathcal{R}<\xp}\left[\partial'_{k}x'^{J}\right]\xpb{\bar{\tau}}^{k\nu}(u,\mathbf{x}')\ud^3\mathbf{x'}\nonumber\\
&&\!\!\!\!+B\int_{\mathcal{R}<\xp}\xpb x'^{J}\xp^{-1}n'^{k}\bar{\tau}^{k\nu}(u,\mathbf{x}')\ud^3\mathbf{x'}.
\end{eqnarray}
Taking
$\re B$
to be a large enough negative number, the second surface integral on the right-hand side of the above equation vanishes. As a consequence of identity theorem, the sum of the analytic continuations of the remaining terms on the right-hand side of
Eq. (\ref{4.2.56})
and
$\partial_{0}\A\int_{\mathcal{R}<\xp}\xpb x'^{J}{\bar{\tau}}^{0\nu}(u,\mathbf{x}')\ud^3\mathbf{x'}$
must be equal wherever they are all defined. Hence, taking into account that the surface integral and
$f(B)=B$
are entire functions, we get
\begin{eqnarray}\label{4.2.57}
\lefteqn{\partial_{0}\A\int_{\mathcal{R}<\xp}\xpb x'^{J}{\bar{\tau}}^{0\nu}(u,\mathbf{x}')\ud^3\mathbf{x'}} \nonumber \\ 
&\!\!\!\!\!\!\!\!=\!\!\!\!\!\!\!\!& -\int_{\xp=\mathcal{R}} \xpb x'^{J}{\bar{\tau}}^{k\nu}(u,\mathbf{x}')\ud\sigma'_{\paren{2}k} \nonumber\\
&&\!\!\!\!+\A\int_{\mathcal{R}<\xp}\left[\partial'_{k}x'^{J}\right]\xpb{\bar{\tau}}^{k\nu}(u,\mathbf{x}')\ud^3\mathbf{x'}\nonumber\\
&&\!\!\!\!+B\cdot\A\int_{\mathcal{R}<\xp} \xpb x'^{J}\xp^{-1}n'^{k}\bar{\tau}^{k\nu}(u,\mathbf{x}')\ud^3\mathbf{x'}.
\end{eqnarray}
Considering
Eqs. (\ref{4.2.55}) and (\ref{4.2.57}), we can write
\begin{eqnarray}\label{4.2.58}
\lefteqn{\partial_{0}\A\int_{D_i}\xpb x'^{J}{\bar{\tau}}^{0\nu}(u,\mathbf{x}')\ud^3\mathbf{x'}} \nonumber \\ 
&\!\!\!\!\!\!\!\!=\!\!\!\!\!\!\!\!& -\int_{\xp=\mathcal{R}} \xpb x'^{J}{\bar{\tau}}^{k\nu}(u,\mathbf{x}')\ud\sigma'_{\paren{i}k} \nonumber\\
&&\!\!\!\!+\A\int_{D_i}\left[\partial'_{k}x'^{J}\right]\xpb{\bar{\tau}}^{k\nu}(u,\mathbf{x}')\ud^3\mathbf{x'}\nonumber\\
&&\!\!\!\!+B\cdot\A\int_{D_i} \xpb x'^{J}\xp^{-1}n'^{k}\bar{\tau}^{k\nu}(u,\mathbf{x}')\ud^3\mathbf{x'}.
\end{eqnarray}
Additionally, by using
Eq. (\ref{A.3})
we have
\begin{eqnarray}\label{4.2.59}
\lefteqn{\sum_{j=0}^{\infty}\frac{\paren{-1}^j}{j!}\partial_{J}\bigg(\x^{-1}\A\int_{D_i}\left[\partial'_{k}x'^{J}\right]\xpb{\bar{\tau}}^{k\nu}(u,\mathbf{x}')\ud^3\mathbf{x'}\bigg)} \nonumber\\ 
&\!\!\!\!\!\!\!\!=\!\!\!\!\!\!\!\!&\sum_{j=1}^{\infty}\frac{\paren{-1}^j}{j!}\partial_{J}\bigg(\x^{-1}\delta^{k\{ i_1}\A\int_{D_i} x'^{ i_2... i_{j}\}}\xpb\bar{\tau}^{k\nu}(u,\mathbf{x}')\ud^3\mathbf{x'}\bigg) \nonumber\\
&\!\!\!\!\!\!\!\!=\!\!\!\!\!\!\!\!&-\sum_{j=0}^{\infty}\frac{\paren{-1}^j}{j!}\partial_{J}\left[\partial_{k}\bigg(\x^{-1}\A\int_{D_i} x'^{J} \xpb \bar{\tau}^{k\nu}(u,\mathbf{x}')\ud^3\mathbf{x'}\bigg)\right],
\end{eqnarray}
where the key reasoning behind transition from first equality to the second one is as follows. Since all the indices of
$J$
are dummy and
$\partial_J$
is a symmetric tensor, we can replace
$\delta^{k\{ i_1} x'^{ i_2... i_{j}\}}$
by only one of its terms times the total number of its terms. Owing to
$x'^{J-1}$
and Kronecker delta being totally symmetric tensors, the number of terms of
$\delta^{k\{ i_1} x'^{ i_2... i_{j}\}}$
is equal to the number of ways in which we can choose 1 index from
$j$,
i.e.,
$\binom{j}{1}=j$.
Thus, we have replaced
$\delta^{k\{ i_1} x'^{ i_2... i_{j}\}}$
by
$j\delta^{k i_1} x'^{ i_2... i_{j}}$
in
Eq. (\ref{4.2.58})
to derive the final result. Combining
Eqs. (\ref{4.2.58}) and (\ref{4.2.59}), we reach
\begin{eqnarray}\label{4.2.60a}
\lefteqn{\frac{4G}{c^4}\sum_{j=0}^{\infty}\frac{\paren{-1}^j}{j!}\partial_{J}\left[{\x}^{-1}\partial_{0}\A\int_{D_i}\xpb x'^{J}{\bar{\tau}}^{0\nu}(u,\mathbf{x}')\ud^3\mathbf{x'}\right]} \nonumber\\ 
&\!\!\!\!\!\!\!\!=\!\!\!\!\!\!\!\!& - \frac{4G}{c^4}\sum_{j=0}^{\infty}\frac{\paren{-1}^j}{j!}\partial_{J}\left[{\x}^{-1}\partial_{0}\int_{\xp=\mathcal{R}} \xpb x'^{J}{\bar{\tau}}^{k\nu}(u,\mathbf{x}')\ud\sigma'_{\paren{i}k}\right] \nonumber\\
&&\!\!\!\! - \frac{4G}{c^4}\sum_{j=0}^{\infty}\frac{\paren{-1}^j}{j!}\partial_{J}\left[\partial_{k}\bigg(\x^{-1}\A\int_{D_i} \xpb x'^{J}\bar{\tau}^{k\nu}(u,\mathbf{x}')\ud^3\mathbf{x'}\bigg)\right] \nonumber\\
&&\!\!\!\! + \frac{4G}{c^4}\sum_{j=0}^{\infty}\frac{\paren{-1}^j}{j!}\partial_{J}\left[{\x}^{-1}B\cdot\A\int_{D_i} \xpb x'^{J}\xp^{-1}n'^{k}\bar{\tau}^{k\nu}(u,\mathbf{x}')\ud^3\mathbf{x'}\right].
\end{eqnarray}
Irrespective of whether
$i$
is, it is straightforward to show that each term in the above equation is analytic in some punctured neighborhood of
$B=0$,
thereby possessing a Laurent expansion about
$B=0$
(the term containing a surface integral is analytic at
$B=0$
 and hence has a Taylor expansion about this point). Since, for each
$n$,
the coefficients of
$B^n$
on the two sides of
Eq. (\ref{4.2.60a})
must be equal, taking the coefficients of
$B^0$
into account, we get
\begin{eqnarray}\label{4.2.60b}
\lefteqn{\frac{4G}{c^4}\sum_{j=0}^{\infty}\frac{\paren{-1}^j}{j!}\partial_{J}\left[{\x}^{-1}\partial_{0}\fpa\int_{D_i}\xpb x'^{J}{\bar{\tau}}^{0\nu}(u,\mathbf{x}')\ud^3\mathbf{x'}\right]} \nonumber\\ 
&\!\!\!\!\!\!\!\!=\!\!\!\!\!\!\!\!& - \frac{4G}{c^4}\sum_{j=0}^{\infty}\frac{\paren{-1}^j}{j!}\partial_{J}\left[{\x}^{-1}\partial_{0}\int_{\xp=\mathcal{R}} x'^{J}{\bar{\tau}}^{k\nu}(u,\mathbf{x}')\ud\sigma'_{\paren{i}k}\right] \nonumber\\
&&\!\!\!\! - \frac{4G}{c^4}\sum_{j=0}^{\infty}\frac{\paren{-1}^j}{j!}\partial_{J}\left[\partial_{k}\bigg(\x^{-1}\fpa\int_{D_i} \xpb x'^{J}\bar{\tau}^{k\nu}(u,\mathbf{x}')\ud^3\mathbf{x'}\bigg)\right] \nonumber\\
&&\!\!\!\! + \frac{4G}{c^4}\sum_{j=0}^{\infty}\frac{\paren{-1}^j}{j!}\partial_{J}\left[{\x}^{-1}\resa\int_{D_i} \xpb x'^{J}\xp^{-1}n'^{k}\bar{\tau}^{k\nu}(u,\mathbf{x}')\ud^3\mathbf{x'}\right].\qquad
\end{eqnarray}
The sum of
$\frac{4G}{c^4}\sum_{j=0}^{\infty}\frac{\paren{-1}^j}{j!}\partial_{J}\left[{\x}^{-1}\partial_{0}\fpa\int_{D_i}\xpb x'^{J}{\bar{\tau}}^{0\nu}(u,\mathbf{x}')\ud^3\mathbf{x'}\right]$'s
with
$i=1$
and
$i=2$
is as follows:
\begin{eqnarray}\label{4.2.60c}
\lefteqn{\frac{4G}{c^4}\sum_{j=0}^{\infty}\frac{\paren{-1}^j}{j!}\partial_{J}\left[{\x}^{-1}\partial_{0}\fpa\int_{D_i}\xpb x'^{J}{\bar{\tau}}^{0\nu}(u,\mathbf{x}')\ud^3\mathbf{x'}\right]} \nonumber\\ 
&&\!\!\!\! + \frac{4G}{c^4}\sum_{j=0}^{\infty}\frac{\paren{-1}^j}{j!}\partial_{J}\left[{\x}^{-1}\partial_{0}\fpa\int_{D_i}\xpb x'^{J}{\bar{\tau}}^{0\nu}(u,\mathbf{x}')\ud^3\mathbf{x'}\right] \nonumber\\
&\!\!\!\!\!\!\!\!=\!\!\!\!\!\!\!\!& -\frac{4G}{c^4}\sum_{j=0}^{\infty}\frac{\paren{-1}^j}{j!}\partial_{J}\left[\partial_{k}\paren{{\x}^{-1}H^{k\nu}_{J}(u)}\right] + \frac{4G}{c^4}\sum_{j=0}^{\infty}\frac{\paren{-1}^j}{j!}\partial_{J}\paren{{\x}^{-1}G^{\nu}_{J}(u)},\quad 
\end{eqnarray}
where the surface integrals haven't appeared owing to the fact that
$\vec{\ud\sigma'_{\paren{1}}}$
is the opposite of
$\vec{\ud\sigma'_{\paren{2}}}$
at each point of the surface
$\xp=\mathcal{R}$.
Using
Eq. (\ref{4.2.60c}),
Eq. (\ref{4.2.51})
can be rewritten as
\begin{equation}\label{4.2.61}
\partial_\mu\left[\frac{4G}{c^4}\sum_{j=0}^{\infty}\frac{\paren{-1}^j}{j!}\partial_{J}\paren{{\x}^{-1}H^{\mu\nu}_{J}(u)}\right]=\frac{4G}{c^4}\sum_{j=0}^{\infty}\frac{\paren{-1}^j}{j!}\partial_{J}\paren{{\x}^{-1}G^{\nu}_{J}(u)},
\end{equation}
which means that
$\mathcal{M}\paren{h^{\mu\nu}}\finsarg$
given by
Eqs. (\ref{4.2.46a}) and (\ref{4.2.46b})
satisfies the harmonic gauge condition too. We can also show that this
$\mathcal{M}\paren{h^{\mu\nu}}\finsarg$
is independent of
$\mathcal{R}$
and
$r_0$
\cite{B1998}.

Now it's time to determine
$v^{\mu\nu}\finsarg$
and
$Gh^{\mu\nu}_{\paren{1}}\left[I,W\right]$.
$\partial_\mu v^{\mu\nu}\finsarg$
is minus the right-hand side of
Eq. (\ref{4.2.50a})
which can be rewritten as
\cite{BD1989}
\begin{equation}\label{4.2.62a}
\partial_\mu u^{\mu\nu}\finsarg=\frac{4G}{c^4}\sum_{\ell=0}^{\infty}\frac{\paren{-1}^\ell}{\ell!}\hat{\partial}_{L}\paren{{\x}^{-1}\hat{\mathcal{G}}^{\nu}_{L}(u)},
\end{equation}
where
\begin{IEEEeqnarray}{rcl}\label{4.2.62b}
\hat{\mathcal{G}}^{\nu}_{L}(u) & = &\resa\int_{\xp<\mathcal{R}}\xpb \hat{x}'^{L}\xp^{-1}n'^{i}\int_{-1}^{1}\delta_\ell(z)\bar{\tau}^{i\nu}(u+z\frac{\xp}{c},\mathbf{x}')\ud z\ud^3\mathbf{x'} \nonumber \\  
&& \negmedspace {}+\resa\int_{\mathcal{R}<\xp}\xpb \hat{x}'^{L}\xp^{-1}n'^{i}\int_{-1}^{1}\delta_\ell(z)\bar{\tau}^{i\nu}(u+z\frac{\xp}{c},\mathbf{x}')\ud z\ud^3\mathbf{x'}\nonumber\\  
&& \negmedspace {}\stareq\resa\int_{\mathbb{R }^3}\xpb \hat{x}'^{L}\xp^{-1}n'^{i}\int_{-1}^{1}\delta_\ell(z)\bar{\tau}^{i\nu}(u+z\frac{\xp}{c},\mathbf{x}')\ud z\ud^3\mathbf{x'},\qquad
\end{IEEEeqnarray}
with
$\delta_\ell(z)$
given by
\begin{equation}\label{4.2.62c}
\delta_\ell(z)=\frac{\paren{2\ell+1}!!}{2^{\ell+1}\ell!}{\paren{1-z^2}}^\ell.
\end{equation}
Having rewritten
$\partial_\mu u^{\mu\nu}\finsarg$
in terms of tensors STF with respect to the multi-index
$L$,
by proceeding as we did for
$v^{\mu\nu}_{\paren{2}}\finsarg$,
from
Eqs. (\ref{2.4.9}) and (\ref{2.4.10}),
we can obtain
\begin{IEEEeqnarray}{rcl}\label{4.2.63}
 v^{00}\finsarg &=&\frac{4G}{c^4}\bigg\{\!-\!c\x^{-1}\int\!\hat{\mathcal{G}}^{0}(u)\nonumber \\  
&& \negmedspace {}+\partial_a\left[\x^{-1}\paren{c\int\!\hat{\mathcal{G}}^{0}_{a}(u)+c^2\int\!\!\int\!\hat{\mathcal{G}}^{a}(u)-\hat{\mathcal{G}}^{b}_{ab}(u)}\right]\bigg\},
\end{IEEEeqnarray}
\begin{IEEEeqnarray}{rcl}\label{4.2.64}
v^{0i}\finsarg & = & \frac{4G}{c^4}\bigg\{\x^{-1}\paren{-c\int\!\hat{\mathcal{G}}^{i}(u)+\frac{1}{c}\dot{\hat{\mathcal{G}}}^{a}_{ai}(u)}+\frac{c}{2}\partial_a\left[\x^{-1}\paren{\int\!\hat{\mathcal{G}}^{i}_{a}(u)-\int\!\hat{\mathcal{G}}^{a}_{i}(u)}\right]\nonumber \\  
&& \negmedspace {}-\sum_{\ell=2}^{\infty}\frac{\paren{-1}^\ell}{\ell!}\partial_{L-1}\left[\x^{-1}\hat{\mathcal{G}}^{0}_{iL-1}(u)\right]\bigg\},
\end{IEEEeqnarray}
\begin{IEEEeqnarray}{rcl}\label{4.2.65}
 v^{ij}\finsarg & = &\frac{4G}{c^4}\bigg\{\x^{-1}\hat{\mathcal{G}}^{(i}_{j)}(u) + 2 \sum_{\ell=3}^{\infty}\frac{\paren{-1}^\ell}{\ell!}\partial_{L-3}\left[c^2\x^{-1}\ddot{\hat{\mathcal{G}}}^{a}_{aijL-3}(u)\right] \nonumber \\  
&& \negmedspace {}+\sum_{\ell=2}^{\infty}\frac{\paren{-1}^\ell}{\ell!}\bigg[\partial_{L-2}
\paren{c\x^{-1}\dot{\hat{\mathcal{G}}}^{0}_{ijL-2}(u)}+\partial_{aL-2}\paren{\x^{-1}\hat{\mathcal{G}}^{a}_{ijL-2}(u)} \nonumber \\  
&& \negmedspace {}+2\delta_{ij}\partial_{L-1}\paren{\x^{-1}\hat{\mathcal{G}}^{a}_{aL-1}(u)}
-4\partial_{L-2(i}\paren{\x^{-1}\hat{\mathcal{G}}^{a}_{j)aL-2}(u)} \nonumber \\  
&& \negmedspace {}-2\partial_{L-1}\paren{\x^{-1}\hat{\mathcal{G}}^{(i}_{j)L-1}(u)}\bigg]\bigg\},
\end{IEEEeqnarray}
where
$\int\!\hat{\mathcal{G}}$,
$\int\!\!\int\!\hat{\mathcal{G}}$,
$\dot{\hat{\mathcal{G}}}$
and
$\ddot{\hat{\mathcal{G}}}$
are the same as
$\hat{\mathcal{G}}^{\paren{-1}}$,
$\hat{\mathcal{G}}^{\paren{-2}}$,
$\hat{\mathcal{G}}^{\paren{1}}$
and
$\hat{\mathcal{G}}^{\paren{2}}$.
Now by rewriting the second term on the right-hand side of
Eq. (\ref{4.2.46a})
as
$-\frac{4G}{c^4}\sum_{\ell=0}^{\infty}\frac{\paren{-1}^\ell}{\ell!}\hat{\partial}_{L}\paren{{\x}^{-1}\hat{\mathcal{H}}^{\mu\nu}_{L}(u)}$,
where
\begin{IEEEeqnarray}{rcl}\label{4.2.66}
\hat{\mathcal{H}}^{\mu\nu}_{L}(u) & = &\fpa\int_{\xp<\mathcal{R}}\xpb \hat{x}'^{L}\int_{-1}^{1}\delta_\ell(z){\bar{\tau}}^{\mu\nu}(u+z\frac{\xp}{c},\mathbf{x}')\ud z\ud^3\mathbf{x'} \nonumber \\  
&& \negmedspace {}+\fpa\int_{\mathcal{R}<\xp}\xpb \hat{x}'^{L}\int_{-1}^{1}\delta_\ell(z){\bar{\tau}}^{\mu\nu}(u+z\frac{\xp}{c},\mathbf{x}')\ud z\ud^3\mathbf{x'}\nonumber \\  
&& \negmedspace {} \stareq\fpa\int_{\mathbb{R }^3}\xpb \hat{x}'^{L}\int_{-1}^{1}\delta_\ell(z){\bar{\tau}}^{\mu\nu}(u+z\frac{\xp}{c},\mathbf{x}')\ud z\ud^3\mathbf{x'},
\end{IEEEeqnarray}
and then subtracting
$v^{\mu\nu}\finsarg$
given in
Eqs. (\ref{4.2.63})-(\ref{4.2.65})
from it (with
$\hat{\mathcal{H}}^{00}_{L}$,
$\hat{\mathcal{H}}^{0i}_{L}$
and
$\hat{\mathcal{H}}^{ij}_{L}$
rewritten in terms of tensors which are STF with respect to all their indices), the six moments describing
$Gh^{\mu\nu}_{\paren{1}}\left[I,W\right]$
can be obtained as follows
\cite{B2014}:
\begin{IEEEeqnarray}{rcl}\label{4.2.67}
I_L(u) & \stareq &  \fpa\int_{\mathbb{R }^3}\xpb\int_{-1}^{1}\bigg\{\delta_\ell(z)\hat{x}'^{L}\Sigma-\frac{4\paren{2\ell+1}}{c^2\paren{\ell+1}\!\paren{2\ell+3}}\delta_{\ell+1}(z)\hat{x}'^{iL}\Sigma_i^{\paren{1}}\nonumber \\  
&& \negmedspace {} +\frac{2\paren{2\ell+1}}{c^4\paren{\ell+1}\!\paren{\ell+2}\!\paren{2\ell+5}}\delta_{\ell+2}(z)\hat{x}'^{ijL}\Sigma_{ij}^{\paren{2}}\bigg\}(u\!+\!z\frac{\xp}{c},\mathbf{x}')\ud z\ud^3\mathbf{x'},
\end{IEEEeqnarray}
\begin{IEEEeqnarray}{rcl}\label{4.2.68}
J_L(u) & \stareq &  \fpa\int_{\mathbb{R }^3}\xpb\int_{-1}^{1}\varepsilon_{ab<i_\ell}\bigg\{\delta_\ell(z)\hat{x}'^{L-1>a}\Sigma_b \nonumber \\  
&& \negmedspace {}-\frac{2\ell+1}{c^2\paren{\ell+2}\!\paren{2\ell+3}}\delta_{\ell+1}(z)\hat{x}'^{L-1>ac}\Sigma_{bc}^{\paren{1}}\bigg\}(u\!+\!z\frac{\xp}{c},\mathbf{x}')\ud z \ud^3\mathbf{x'},
\end{IEEEeqnarray}
\begin{IEEEeqnarray}{rcl}\label{4.2.69l}
W_L(u) & \stareq &  \fpa\int_{\mathbb{R }^3}\xpb\int_{-1}^{1}\bigg\{\frac{2\ell+1}{\paren{\ell+1}\!\paren{2\ell+3}}\delta_{\ell+1}(z)\hat{x}'^{iL}\Sigma_i\nonumber \\  
&& \negmedspace {} -\frac{2\ell+1}{2c^2\paren{\ell+1}\!\paren{\ell+2}\!\paren{2\ell+5}}\delta_{\ell+2}(z)\hat{x}'^{ijL}\Sigma_{ij}^{\paren{1}}\bigg\}(u\!+\!z\frac{\xp}{c},\mathbf{x}')\ud z\ud^3\mathbf{x'},\qquad
\end{IEEEeqnarray}
\begin{IEEEeqnarray}{rcl}\label{4.2.70}
X_L(u) & \stareq & \fpa\int_{\mathbb{R }^3}\xpb\int_{-1}^{1}\bigg\{\frac{2\ell+1}{2\paren{\ell+1}\!\paren{\ell+2}\!\paren{2\ell+5}}\delta_{\ell+2}(z)\nonumber\\
&& \negmedspace {}\times\hat{x}'^{ijL}\Sigma_{ij}\bigg\}(u\!+\!z\frac{\xp}{c},\mathbf{x}')\ud z\ud^3\mathbf{x'},
\end{IEEEeqnarray}
\begin{IEEEeqnarray}{rcl}\label{4.2.71}
Y_L(u) & \stareq &  \fpa\int_{\mathbb{R }^3}\xpb\int_{-1}^{1}\bigg\{-\delta_\ell(z)\hat{x}'^{L}\Sigma_{ii}+\frac{3\paren{2\ell+1}}{\paren{\ell+1}\!\paren{2\ell+3}}\delta_{\ell+1}(z)\hat{x}'^{iL}\Sigma_i^{\paren{1}}\nonumber \\  
&& \negmedspace {} -\frac{2\paren{2\ell+1}}{c^2\paren{\ell+1}\!\paren{\ell+2}\!\paren{2\ell+5}}\delta_{\ell+2}(z)\hat{x}'^{ijL}\Sigma_{ij}^{\paren{2}}\bigg\}(u\!+\!z\frac{\xp}{c},\mathbf{x}')\ud z \ud^3\mathbf{x'},
\end{IEEEeqnarray}
\begin{IEEEeqnarray}{rcl}\label{4.2.72}
Z_L(u) & \stareq &  \fpa\int_{\mathbb{R }^3}\xpb\int_{-1}^{1}\varepsilon_{ab<i_\ell}\bigg\{\!\!-\frac{2\ell+1}{\paren{\ell+2}\!\paren{2\ell+3}}\delta_{\ell+1}(z)\nonumber\\
&& \negmedspace {}\times\hat{x}'^{L-1>bc}\Sigma_{ac}\bigg\}(u\!+\!z\frac{\xp}{c},\mathbf{x}')\ud z \ud^3\mathbf{x'},
\end{IEEEeqnarray}
where
\begin{equation}\label{4.2.73}
\Sigma=\frac{\bar{\tau}^{00}+\bar{\tau}^{ii}}{c^2},
\end{equation}
\begin{equation}\label{4.2.74}
\Sigma_i=\frac{\bar{\tau}^{0i}}{c},
\end{equation}
\begin{equation}\label{4.2.75}
\Sigma_{ij}=\bar{\tau}^{ij}.
\end{equation}

In an analogous way, one can find
$\bar{h}^{\mu\nu}\finsarg$
as
\cite{PB2002,B2014}
\begin{IEEEeqnarray}{rcl}\label{4.2.76}
\bar{h}^{\mu\nu}\finsarg & = &  \frac{16\pi G}{c^4}\fpa \retint{\bar{\tau}^{\mu\nu}\retarg}\nonumber\\
&& \negmedspace {}-\frac{4G}{c^4}\sum_{\ell=0}^{\infty}\frac{\paren{-1}^\ell}{\ell!}\hat{\partial}_{L}\left[\frac{\hat{\mathcal{R}}^{\mu\nu}_L\uarg - \hat{\mathcal{R}}^{\mu\nu}_L\varg}{2\x}\right],
\end{IEEEeqnarray}
where
\begin{IEEEeqnarray}{rcl}\label{4.2.77}
\hat{\mathcal{R}}^{\mu\nu}_L\targ & \stareq &\fpa\int_{\mathbb{R }^3}\xpb\hat{x}'^{L}\ellfpa\int_{1}^{\infty}\gamma_j(z)\mathcal{M}\paren{\tau^{\mu\nu}}(t-z\frac{\xp}{c},\mathbf{x}')\ud z\ud^3\mathbf{x'},\qquad
\end{IEEEeqnarray}
with
\begin{equation}\label{4.2.78}
\mathcal{M}\paren{\tau^{\mu\nu}}=\frac{c^4}{16\pi G}\Lambda^{\mu\nu}(\mathcal{M}\paren{h}),
\end{equation}
\begin{equation}\label{4.2.79}
\gamma_j(z)=\frac{\paren{-1}^{j+1}}{2^{2j}}\frac{\Gamma\paren{2j+2}}{{\left[\Gamma\paren{j+1}\right]}^2}{\paren{z^2-1}}^j,
\end{equation}
$j$
a complex number and
$\mathop{\mathrm{FP}}\limits_{j=\ell}$
denoting the coefficient of
$\paren{j-\ell}^0$
in Laurent expansion.
$\Gamma$
is the gamma function, and hence,
$\gamma_j(z)=-2\delta_j(z)$
if
$j \in \mathbb{N}$.

\section{Determination of Orders of Post-Minkowskian and Post-Newtonian Approximations}

Since in the begining
$\bar{\tau}^{\mu\nu}$
doesn't contain any power of
$G$,
only first power of
$G$
appears in
$\mathcal{M}\paren{h^{\mu\nu}}$
after performing matching procedure once. Substituting the resultant
$\mathcal{M}\paren{h^{\mu\nu}}$
into
Eq. (\ref{4.2.77}),
higher powers of
$G$
appear in
$\bar{h}^{\mu\nu}$
and hence in
$\bar{\tau}^{\mu\nu}$.
Therefore, performing matching procedure one more time yields higher powers of
$G$
in
$\mathcal{M}\paren{h^{\mu\nu}}$.
Proceeding similarly, we can obtain
$\mathcal{M}\paren{h^{\mu\nu}}$
up to the desired power of
$G$,
or, in other words, up to the desired order. Post-Newtonian approximation to
$h^{\mu\nu}$,
$\bar{h}^{\mu\nu}$,
can be obtained up to the desired order in an analogous way. The only difference is that we now consider the powers of
$\frac{1}{c}$
instead of powers of
$G$.
In doing so, we must also consider the powers of
$\frac{1}{c}$
coming from the expansion of the retarded terms in such powers.

\section{Conclusion}

What Blanchet and Damour presented is of high importance for the reason that their approach reduces the problem of solving the post-Minkowskian and post-Newtonian equations to the problem of finding analytic continuations of some specific expressions, analytic continuations which are proven to exist.



\newpage
\addcontentsline{toc}{chapter}{Appendices}
\appendix


\chapter{General Requisite Formulae}

binomial expansion:
\begin{equation}
{\paren{a+b}}^p=\sum_{k=0}^p\binom{p}{k}a^k b^{p-k}.
\end{equation}
Leibniz formula:
\begin{equation}
\frac{\mathrm{d}^p}{\mathrm{d}x^p}f(x)g(x)=\sum_{k=0}^p\binom{p}{k}f^{\paren{k}}(x) g^{\paren{p-k}}(x).
\end{equation}
Taylor formula with integral remainder for functions of one variable:
\begin{IEEEeqnarray}{rcl}
f(u) & = &\sum_{j=0}^N\frac{1}{j!}{\paren{u-u_0}}^j f^{\paren{j}}(u){\bigg|}_{u=u_0}\nonumber \\  
&& \negmedspace {} +{\paren{u-u_0}}^{N+1}\int_0^1\frac{{\paren{1-\alpha}}^N}{N!}f^{\paren{N+1}}(u'){\bigg|}_{u'=u_0+\alpha\paren{u-u_0}}\ud\alpha.
\end{IEEEeqnarray}
Taylor formula with integral remainder for functions of three variables:
\begin{IEEEeqnarray}{rcl}
f\xarg & = &\sum_{j=0}^N\frac{1}{j!}{\paren{{\mathbf{x}}^J-{\mathbf{x}}^J_0}} \partial_J f\xarg{\bigg|}_{{\mathbf{x}}={\mathbf{x}}_0}\nonumber \\  
&& \negmedspace {} +{\paren{{\mathbf{x}}^{I_{N+1}}-{\mathbf{x}}^{I_{N+1}}_0}}\int_0^1\frac{{\paren{1-\alpha}}^N}{N!}\partial'_{I_{N+1}}f\xparg{\bigg|}_{\mathbf{x'}={\mathbf{x}}_0+\alpha\paren{\mathbf{x}-{\mathbf{x}}_0}}\ud\alpha.
\end{IEEEeqnarray}
Taylor expansion for functions of three variables:
\begin{equation}
f(\mathbf{x}+\mathbf{a})=\sum_{j=0}^{\infty}\frac{1}{j!}a^J\partial_Jf\xarg.
\end{equation}
\begin{equation}\label{A.3}
\partial_k x^L=
\begin{cases}
0, &\qquad  \ell=0, \\
\delta^{k\{ i_1}x^{ i_2... i_{\ell}\}}, &\qquad \ell \ge 1.
\end{cases}
\end{equation}
\begin{equation}\label{A.2}
\Delta x^L=
\begin{cases}
0, &\qquad  \ell=0,1, \\
2\delta^{\{ i_1 i_2} x^{ i_3... i_{\ell}\}}, &\qquad \ell \ge 2.
\end{cases}
\end{equation}


\chapter{Introduction to Symmetric-Trace-Free Cartesian Tensors}

\section{Spherical Harmoninics}
Spherical harmonics, which are denoted by
$\ylm$,
are the solutions of
\begin{equation}
\frac{\partial^2}{\partial \theta^2}h + \frac{1}{\tan \theta}\frac{\partial}{\partial \theta}h+\frac{1}{\sin^2 \theta}\frac{\partial^2}{\partial \varphi^2}h=-\ell\left(\ell +1\right)h,
\end{equation}
for
$\ell \ge 0$
and
$-\ell \le m \le \ell$.
These functions fulfill the following orthogonality relation:
\begin{equation}
\int \ylmp\left[Y^{\ell' m'}(\theta', \varphi')\right]^*\ud \Omega'=\delta_{\ell \ell'}\delta_{m m'},
\end{equation}
and since they form a complete set of functions, any well-behaved function
$f(\theta,\varphi)$
possesses an expansion in terms of them, i.e.,
\begin{equation}\label{B.0}
f(\theta,\varphi)=\sum_{\ell=0}^{\infty}\sum_{m=-\ell}^{\ell}a_{\ell m}\ylm,
\end{equation}
where
\begin{equation}
a_{\ell m}=\int f(\theta',\varphi') \left[\ylmp\right]^*\ud \Omega'.
\end{equation}
One of  these well-behaved functions of
$\theta$
and
$\varphi$
is
$n^{I_\ell} \equiv n^L$,
which denotes
$n^{i_1}\cdots n^{i_\ell}$
(in this ``multi-index notation'' we also denote
$n^{i_1}\cdots n^{i_{\ell-m}}$
by
$ n^{L-m}$).
Based on the discussion above, one can expand this function in terms of spherical harmonics. Such an expansion can be obtained by using the following formulae:
\begin{equation}
n^1=n^x=\sin\theta\cos\varphi,
\end{equation}
\begin{equation}
n^2=n^y=\sin\theta\sin\varphi,
\end{equation}
\begin{equation}
n^3=n^z=\cos\theta.
\end{equation}
It is also worth remarking that
$n^i$'s
are orthogonal, in the sense that
\begin{equation}\label{B.5}
\int n'^{i_1}\cdots n'^{i_q}\ud\Omega'=
\begin{cases}
\frac{4\pi}{\paren{2p+1}!!}\delta^{\{ i_1 i_2}\cdots\delta^{ i_{2p-1} i_{2p}\}}, &\qquad  q=2p, \\
0, &\qquad q=2p+1.
\end{cases}
\end{equation}

\section{Symmetric-Trace-Free Cartesian Tensors}
A Cartesian tensor is called symmetric-trace-free if any single contraction of it vanishes. We denote (the components of) a symmetric-trace-free Cartesian tensor of rank
$q$
by
$\hat{T}_Q$
or
$T_{<Q>}$.
If
$T_Q$
is a general tensor, we define its symmetric-trace-free part as
\begin{equation}
\hat{T}_Q=\sum_{m=0}^{\left[\frac{q}{2}\right]}a_{qm}\delta_{(i_1 i_2}\cdots\delta_{i_{2m-1}i_{2m}}T_{(i_{2m+1}...i_q))a_1a_1...a_ma_m},
\end{equation}
where
\begin{equation}
a_{qm}={\paren{-1}}^m \frac{q!}{\paren{q-2m}!}\frac{\paren{2q-2m-1}!!}{\paren{2q-1}!!\paren{2m}!!}.
\end{equation}
Two general Cartesian tensors
$F_L$
and
$G_L$
satisfy the following important formula:
\begin{equation}\label{B.2}
\hat{F}^L\hat{G}_L=\hat{F}^LG_L=F^L\hat{G}_L.
\end{equation}
The set of all symmetric-trace-free Cartesian tensors of rank
$\ell$
form a
($2\ell+1$)-dimensional vector space.
$\hat{\mathcal{Y}}^{\ell m}_L$'s,
which form a basis for this vector space, are given by
\begin{equation}
\hat{\mathcal{Y}}^{\ell m}_L=A^{\ell m}E^{\ell m}_{<L>},
\end{equation}
where
\begin{equation}
E^{\ell m}_L=\paren{\delta^{1}_{i_1}+i\delta^{2}_{i_1}}\cdots\paren{\delta^{1}_{i_m}+i\delta^{2}_{i_m}}\delta^{3}_{i_{m+1}}\cdots\delta^{3}_{i_{\ell}},
\end{equation}
\begin{equation}
A^{\ell m}={\paren{-1}}^m\paren{2\ell-1}!!{\left[\frac{2\ell+1}{4\pi\paren{\ell-m}!\paren{\ell+1}!}\right]}^\frac{1}{2},
\end{equation}
if
$m\ge 0$,
and by
\begin{equation}
\hat{\mathcal{Y}}^{\ell m}_L={\paren{-1}}^m\left[\hat{\mathcal{Y}}^{\ell |m|}_L\right]^*,
\end{equation}
if
$m<0$.
$\hat{\mathcal{Y}}^{\ell m}_L$'s are orthogonal, in the sense that
\begin{equation}
\hat{\mathcal{Y}}^{\ell m}_L\left[\hat{\mathcal{Y}}^{\ell m'}_L\right]^*=\frac{\paren{2\ell+1}!!}{4\pi\ell!}\delta_{mm'}.
\end{equation}
Further, they are related to
$\ylm$'s
by
\begin{equation}\label{B.6}
\ylm=\hat{n}^L\hat{\mathcal{Y}}^{\ell m}_L.
\end{equation}
The above equation relates symmetric-trace-free tensors
$\hat{n}^L$
to spherical harmonics. Using this equation,
Eq. (\ref{B.0})
takes the form
\begin{equation}
f(\theta,\varphi)=\sum_{\ell=0}^{\infty}\left[\sum_{m=-\ell}^{\ell}a_{\ell m}\hat{\mathcal{Y}}^{\ell m}_L \right]\hat{n}^L=\sum_{\ell=0}^{\infty}\hat{A}_L\hat{n}^L,
\end{equation}
which means every well-behaved function
$f(\theta,\varphi)$
is expandable in terms of
$\hat{n}^L$'s.
Foremost amongst other formulae on this subject are
\begin{IEEEeqnarray}{rcl}\label{B.1}
\hat{\partial}_L\paren{\frac{F(t-\varepsilon\frac{\x}{c})}{\x}} & = &{\paren{-1}}^\ell\hat{n}^L\sum_{j=0}^{\ell}\frac{\varepsilon^{\ell-j}\paren{\ell+j}!}{2^j j!\paren{\ell-j}!}\frac{F^{\paren{\ell-j}}(t-\varepsilon\frac{\x}{c})}{c^{\ell-j}\x^{j+1}} \qquad \text{for}\;\varepsilon^2=1,\quad
\end{IEEEeqnarray}
\begin{IEEEeqnarray}{rcl}\label{B.3}
\sum_{m=-\ell}^{\ell}\ylm\int\hat{n}'^Q\left[\ylmp\right]^*\ud\Omega' & = &\sum_{m=-\ell}^{\ell}\left[\ylm\right]^*\int\hat{n}'^Q\ylmp\ud\Omega'\nonumber \\  
&=& \hat{n}^Q\delta_{\ell q},
\end{IEEEeqnarray}
\begin{equation}\label{B.10}
\hat{n}_L=\sum_{k=0}^{\left[\frac{\ell}{2}\right]}{\paren{-1}}^k\frac{\paren{2\ell-2k-1}!!}{\paren{2\ell-1}!!}\delta_{\{i_1 i_2}\cdots\delta_{i_{2k-1}i_{2k}}n_{i_{2k+1}...i_\ell\}},
\end{equation}
\begin{equation}\label{B.11}
n_L=\sum_{k=0}^{\left[\frac{\ell}{2}\right]}\frac{\paren{2\ell-4k+1}!!}{\paren{2\ell-2k+1}!!}\delta_{\{i_1 i_2}\cdots\delta_{i_{2k-1}i_{2k}}\hat{n}_{i_{2k+1}...i_\ell\}},
\end{equation}
\begin{equation}\label{B.12}
\hat{\partial}_L=\sum_{k=0}^{\left[\frac{\ell}{2}\right]}{\paren{-1}}^k\frac{\paren{2\ell-2k-1}!!}{\paren{2\ell-1}!!}\delta_{\{i_1 i_2}\cdots\delta_{i_{2k-1}i_{2k}}\partial_{i_{2k+1}...i_\ell\}}\Delta^k,
\end{equation}
\begin{equation}\label{B.13}
\partial_L=\sum_{k=0}^{\left[\frac{\ell}{2}\right]}\frac{\paren{2\ell-4k+1}!!}{\paren{2\ell-2k+1}!!}\delta_{\{i_1 i_2}\cdots\delta_{i_{2k-1}i_{2k}}\hat{\partial}_{i_{2k+1}...i_\ell\}}\Delta^k.
\end{equation}



\begin{thebibliography}{99}

\bibitem{E1916}
Einstein, A.,  ``Approximative Integration of the Field Equations of Gravitation'',
\textit{Sitzungsber. K. Akad. Wiss.},
\textbf{1916},
688-696, (1916).

\bibitem{E1918}
Einstein, A.,  ``Über Gravitationswellen'',
\textit{Sitzungsber. K. Akad. Wiss.},
\textbf{1918},
154-167, (1918).

\bibitem{P1962a}
Pirani, F. A. E.,
\textit{Gravitation: an introduction to current research},
(Wiley, 1962).

\bibitem{P1962b}
Pirani, F. A. E.,
\textit{Recent development in general relativity},
(Pergamon, 1962).

\bibitem{B1959}
Bonner, W. B.,  ``Spherical gravitational waves'',
\textit{Philos. Trans. R. Soc. London, Ser. A},
\textbf{251},
233-271, (1959).

\bibitem{T1980}
Thorne, K. S.,  ``Multipole expansion of gravitational radiation'',
\textit{Rev. Mod. Phys.},
\textbf{52},
299-339, (1980).

\bibitem{BD1986}
Blanchet, L., Damour, T.,  ``Radiative gravitational fields in general relativity I. General structure of the field outside the source'',
\textit{Philos. Trans. R. Soc. London, Ser. A},
\textbf{320},
379-430, (1986).

\bibitem{EW1975}
Epstein, R., Wagoner, R. V.,  ``Post-Newtonian Generation of Gravitational Waves'',
\textit{Astrophys. J.},
\textbf{197},
717-723, (1975).

\bibitem{K1980a}
Kerlick, G. D.,  ``Finite reduced hydrodynamic equations in the slow-motion approximation to general relativity. Part I. First post-Newtonian equations'',
\textit{Gen. Relativ. Gravit.},
\textbf{12},
467-482, (1980).

\bibitem{K1980b}
Kerlick, G. D.,  ``Finite reduced hydrodynamic equations in the slow-motion approximation to general relativity. Part II. Radiation reaction and higher order divergent terms'',
\textit{Gen. Relativ. Gravit.},
\textbf{12},
521-543, (1980).

\bibitem{F1959}
Fock, V. A.,
\textit{Theory of space, time and gravitation},
(Pergamon, 1959).

\bibitem{M2007}
Maggiore, M.,
\textit{Gravitational Waves, Volume 1: Theory and Experiments},
(Oxford University Press, 2007).

\bibitem{PB2002}
Poujade, O., Blanchet, L., ``Post-Newtonian approximation for isolated systems calculated by matched asymptotic expansions'',
\textit{Phys. Rev. D},
\textbf{65},
124020, (2002).

\bibitem{B1998}
Blanchet, L., ``On the multipole expansion  of gravitational field'',
\textit{Class. Quantum Grav.},
\textbf{15},
89–111, (1998).

\bibitem{B2014}
Blanchet, L., ``Gravitational radiation from post-Newtonain sources and inspiralling compact binaries'',
\textit{Living Rev. Relativity},
\textbf{17},
(2014), 2.

\bibitem{P2010}
Padmanabhan, T.,
\textit{Gravitation, Foundation and Frontiers},
(Cambridge University Press, 2010).

\bibitem{PW2014}
Poisson, E., Will, C.,
\textit{Gravity: Newtonian, Post-Newtonian, Relativistic},
(Cambridge University Press, 2014).

\bibitem{BW1999}
Born, M., Wolf, E.,
\textit{Principles of optics},
(Cambridge University Press, 2014), 7th edition.

\bibitem{R1949}
Riesz, M., ``L'intégrale de Riemann-Liouville et le problème de Cauchy'',
\textit{Acta Math.},
\textbf{81},
1-218, (1949).

\bibitem{DI1991}
Damour, T., Iyer, B. R.,``Multipole analysis for electromagnetism and linearized gravity with irreducible Cartesian tensors'',
\textit{Phys. Rev. D},
\textbf{43},
3259-3272, (1991).

\bibitem{N2000}
Nayfeh, A. H.,
\textit{Perturbation Methods},
(Wiley-VCH, 2000).

\bibitem{H2013}
Holmes, M. H.,
\textit{Introduction to Perturbation Methods (Texts in Applied Mathematics)},
(Springer, 2013).

\bibitem{BD1989}
Blanchet, L., Damour, T.,  ``Post-Newtonian generation of gravitational waves'',
\textit{Ann. Inst. Henri Poincare A},
\textbf{50},
377-408, (1989).

\end{thebibliography}
\end{document}